\documentclass{biometrika-noname}

\usepackage{amsmath,bm,bbm}
\usepackage{newtxtext}
\usepackage[subscriptcorrection]{newtxmath}
\usepackage[plain,noend]{algorithm2e}
\usepackage{wrapfig}
\usepackage{xcolor}
\definecolor{darkred}{RGB}{139,0,0}

\makeatletter
\renewcommand{\algocf@captiontext}[2]{#1\algocf@typo. \AlCapFnt{}#2}

\def\@algocf@capt@plain{top}
\renewcommand{\algocf@makecaption}[2]{%
  \addtolength{\hsize}{\algomargin}%
  \sbox\@tempboxa{\algocf@captiontext{#1}{#2}}%
  \ifdim\wd\@tempboxa >\hsize
    \hskip .5\algomargin%
    \parbox[t]{\hsize}{\algocf@captiontext{#1}{#2}}%
  \else
    \global\@minipagefalse
    \hbox to\hsize{\box\@tempboxa}%
  \fi
  \addtolength{\hsize}{-\algomargin}%
}
\makeatother

\def\T{\top}
\DeclareMathOperator{\sgn}{sgn}
\DeclareMathOperator{\expit}{expit}
\DeclareMathOperator{\logit}{logit}

\newcommand{\E}{\mathbb E}
\newcommand{\V}{\operatorname{Var}}
\renewcommand{\P}{\mathbb P}
\newcommand{\Q}{\mathbb Q}
\newcommand{\R}{\mathbb R}
\newcommand{\N}{\mathbb N}
\newcommand{\indicator}{\mathbbm 1}
\newcommand{\independent}{{\perp \! \! \! \perp}}
\newcommand{\iidsim}{\stackrel{\mathrm{i.i.d.}}{\sim}}
\newcommand{\indsim}{\stackrel{\mathrm{ind}}{\sim}}
\newcommand{\convp}{\overset{p}{\longrightarrow}}
\newcommand{\convd}{\overset{d}{\longrightarrow}}
\newcommand{\convdp}{\overset{d,p}{\longrightarrow}}

\newcommand{\dCRT}{\textnormal{dCRT}}
\newcommand{\spacrt}{\textnormal{spaCRT}}
\newcommand{\fastspa}{\textnormal{fastSPA}}

\newcommand{\RR}{\textnormal{RR}}
\newcommand{\LR}{\mathrm{LR}}
\newcommand{\bgam}{\bm{\gamma}}
\newcommand{\bgamhat}{\widehat{\bm{\gamma}}}

\newcommand{\obs}{\bm O_{1:n}}

\makeatletter
\newcommand{\withtwosubscripts}[1]{%
  #1\@ifnextchar\bgroup{\withtwosubscripts@one}{}%
}
\def\withtwosubscripts@one#1{%
  _{#1\@ifnextchar\bgroup{\withtwosubscripts@two}{}}%
}
\def\withtwosubscripts@two#1{,#1}
\newcommand{\withonesubscript}[1]{%
  #1\@ifnextchar\bgroup{\withonesubscript@one}{}%
}
\def\withonesubscript@one#1{_{#1}}
\makeatother
\newcommand{\ex}{\withtwosubscripts{X}}
\newcommand{\exk}{\withtwosubscripts{\tilde X}}
\newcommand{\ey}{\withonesubscript{Y}}
\newcommand{\rz}{\withtwosubscripts{\bm Z}}

\newcommand{\Ber}{\operatorname{Bernoulli}}
\newcommand{\NB}{\operatorname{NB}}
\newcommand{\Unif}{\operatorname{Unif}}
\newcommand{\Phibar}{\overline\Phi}
\newcommand{\eps}{\varepsilon}
\newcommand{\indep}{\perp\!\!\!\perp}

\newcommand{\OP}[1]{O_{#1}}
\newcommand{\PRR}{\mathcal P^{\mathrm{GLM\text{-}RR}}}

\newcommand{\PLL}{\mathcal P^{\mathrm{LL}}}
\newcommand{\PBNB}{\mathcal P^{\mathrm{CRISPR}}}
\newcommand{\PRRnull}{\mathcal P^{\mathrm{GLM\text{-}RR},0}}

\newcommand{\PLLnull}{\mathcal P^{\mathrm{LL},0}}
\newcommand{\PBNBnull}{\mathcal P^{\mathrm{CRISPR},0}}

\let\oldnl\nl
\newcommand{\nonl}{\renewcommand{\nl}{\let\nl\oldnl}}

\usepackage[hidelinks]{hyperref}
\renewcommand\paragraph[1]{\vspace{6pt}\noindent\textbf{#1}}

\usepackage{booktabs,array,longtable,multirow,chngcntr,xparse,enumitem}
\providecommand{\E}{\mathbb E}
\renewcommand{\P}{\mathbb P}
\providecommand{\Var}{\operatorname{Var}}
\providecommand{\Cov}{\operatorname{Cov}}
\providecommand{\R}{\mathbb R}
\providecommand{\cF}{\mathcal F}
\providecommand{\cD}{\mathcal D}
\providecommand{\indicator}{\mathbbm 1}
\providecommand{\norm}[1]{\left\lVert #1\right\rVert}
\providecommand{\abs}[1]{\left|#1\right|}
\providecommand{\diag}{\operatorname{diag}}
\providecommand{\expit}{\operatorname{expit}}
\providecommand{\logit}{\operatorname{logit}}
\providecommand{\LR}{\ensuremath{\mathrm{LR}}}
\providecommand{\Rob}{\ensuremath{\mathrm R}}
\providecommand{\RR}{\textnormal{RR}}
\providecommand{\dCRT}{\textnormal{dCRT}}
\providecommand{\spacrt}{\textnormal{spaCRT}}
\providecommand{\fastspa}{\textnormal{fastSPA}}
\providecommand{\Ber}{\operatorname{Bernoulli}}
\providecommand{\NB}{\operatorname{NB}}
\providecommand{\Unif}{\operatorname{Unif}}
\providecommand{\Phibar}{\overline\Phi}
\providecommand{\eps}{\varepsilon}
\providecommand{\RE}{\operatorname{RE}}
\providecommand{\op}{\mathrm{op}}
\providecommand{\convp}{\overset{p}{\longrightarrow}}
\providecommand{\toD}{\xrightarrow{d}}
\providecommand{\convd}{\xrightarrow{d}}
\providecommand{\Sx}{S_x}
\providecommand{\Sy}{S_y}
\providecommand{\mTwo}{m_2}
\providecommand{\mThree}{m_3}

\providecommand{\OP}[1]{O_{#1}}
\providecommand{\PRR}{\mathcal P^{\mathrm{GLM\text{-}RR}}}
\providecommand{\PLL}{\mathcal P^{\mathrm{LL}}}
\providecommand{\PBNB}{\mathcal P^{\mathrm{CRISPR}}}
\providecommand{\PRRnull}{\mathcal P^{\mathrm{GLM\text{-}RR},0}}
\providecommand{\PLLnull}{\mathcal P^{\mathrm{LL},0}}
\providecommand{\PBNBnull}{\mathcal P^{\mathrm{CRISPR},0}}
\providecommand{\Pnullgen}{\mathcal P^{0}}

\providecommand{\V}{\operatorname{Var}}
\providecommand{\Q}{\mathbb Q}
\providecommand{\Z}{\mathbb Z}
\providecommand{\N}{\mathbb N}
\providecommand{\independent}{{\perp\!\!\!\perp}}
\providecommand{\iidsim}{\stackrel{\mathrm{i.i.d.}}{\sim}}
\providecommand{\indsim}{\stackrel{\mathrm{ind}}{\sim}}
\providecommand{\convdp}{\overset {d,p} \longrightarrow}
\providecommand{\convpp}{\overset {p,p} \longrightarrow}
\providecommand{\bgam}{\bm\gamma}
\providecommand{\bgamhat}{\widehat{\bm\gamma}}
\def\T{\top}   %
\def\sgn{\operatorname{sgn}}
\ProvideDocumentCommand{\ex}{g g}{X\IfValueT{#1}{_{#1\IfValueT{#2}{,#2}}}}
\ProvideDocumentCommand{\ey}{g}{Y\IfValueT{#1}{_{#1}}}
\ProvideDocumentCommand{\rz}{g g}{\bm Z\IfValueT{#1}{_{#1\IfValueT{#2}{,#2}}}}
\ProvideDocumentCommand{\exk}{g g}{\widetilde X\IfValueT{#1}{_{#1\IfValueT{#2}{,#2}}}}
\providecommand{\indep}{\perp\!\!\!\perp}

\hypersetup{bookmarksdepth=2}
\providecommand\paragraph[1]{\vspace{6pt}\noindent\textbf{#1}}
\makeatletter
\newcounter{subsubsection}[subsection]

\providecommand\subsubsection{%
  \@startsection{subsubsection}{3}{\z@}%
    {-6pt \@plus -2pt \@minus -1pt}%
    {3pt}%
    {\normalfont\itshape\raggedright}}
\providecommand\l@subsubsection{\@dottedtocline{3}{7.0em}{4.1em}}
\makeatother

\providecommand{\obs}{\bm O_{1:n}}
\providecommand{\Bin}{\operatorname{Bin}}
\providecommand{\Pois}{\operatorname{Pois}}
\providecommand{\pn}{p_{\infty,n}^{\mathrm{GLM\text{-}RR}}}
\providecommand{\ps}{p_{\mathrm{SPA},n}^{\mathrm{GLM\text{-}RR}}}
\providecommand{\gam}{\bm\gamma}
\providecommand{\gh}{\widehat{\bm\gamma}_n}
\providecommand{\gb}{\overline{\bm\gamma}}
\providecommand{\ab}{\overline{\bm\alpha}}
\providecommand{\ah}{\widehat{\bm\alpha}_n}

\begin{document}
\addtocontents{toc}{\protect\setcounter{tocdepth}{-1}}

\jname{}

\title{Saddlepoint approximations for plug-in resampling}

\author{Z. Niu}
\affil{Department of Statistics and Data Science, University of Pennsylvania,\\ Philadelphia, PA 19104, U.S.A.
\email{ziangniu@wharton.upenn.edu}}

\author{Z. Huang}
\affil{Department of Statistics and Data Science, University of Pennsylvania,\\ Philadelphia, PA 19104, U.S.A. \email{zhihanh@wharton.upenn.edu}}

\author{J. Ray Choudhury}
\affil{School of Industrial and Systems Engineering, Georgia Institute of Technology,\\ Atlanta, GA 30332, U.S.A. \email{jchoudhury3@gatech.edu}}

\author{\and E. Katsevich}
\affil{Department of Statistics and Data Science, University of Pennsylvania,\\
Philadelphia, PA 19104, U.S.A. \email{ekatsevi@wharton.upenn.edu}}

\maketitle

\begin{abstract}
Resampling-based procedures can improve on normal approximations in sparse, large-scale testing problems, but their computational cost can be prohibitive. We recognize that several existing procedures belong to a faster plug-in resampling subclass, fixing fitted nuisance parameters during resampling. When the resampled statistic is a sum of conditionally independent terms, the saddlepoint approximation (SPA) for the resampling $p$-value offers further acceleration, replacing resampling with an analytical tail approximation. However, standard Edgeworth-based approximation-error bounds impose regularity conditions that are hard to verify for plug-in resampling laws. We use an alternative approach to establish a finite-sample relative-error bound for the Lugannani--Rice approximation under more tractable conditions, which we apply in two contexts. In statistical genetics, we identify response resampling procedures as the targets of existing SPAs and establish guarantees in a representative setting. In conditional independence testing, we introduce spaCRT, an SPA for the distilled conditional randomization test (dCRT), which has been applied successfully in biology. Our rates quantify the effects of sparsity and signal strength, with matching lower bounds in special cases. We additionally establish asymptotic Type-I error control of the corresponding plug-in resampling procedures under growing sparsity. In simulations and single-cell CRISPR data analysis, spaCRT closely approximates dCRT \(p\)-values and preserves its statistical performance while accelerating computation by up to 250-fold.

\end{abstract}


\clearpage

\section{Introduction} \label{sec:introduction}

\subsection{Saddlepoint approximations to plug-in resampling procedures} \label{sec:spa-plugin}

Modern applications often require testing large numbers of hypotheses in settings where standard normality-based tests fail to control Type-I error. In high-throughput biological applications, discrete and highly imbalanced variables can make convergence to normality especially slow in the tails relevant after multiplicity correction. This slow convergence has been documented by \citet{Dey2017} in genome-wide association studies (GWAS), which test genetic variants for association with a phenotype while controlling for covariates, and by \citet{Barry2024} in single-cell CRISPR screens, which test perturbation--gene pairs for perturbation-induced changes in gene expression. In both applications, non-normality is most severe when predictor and outcome are simultaneously sparse: rare variants and diseases in GWAS, and rare perturbations and mostly zero expression counts in CRISPR screens. Resampling-based procedures can be more accurate, but their computational cost creates a serious obstacle in such large-scale settings.

One approach to accelerating resampling-based procedures is to avoid the conventional practice of refitting nuisance parameters $\gamma$ for each resample \citep{Efron1979,davidsonBootstrapTestingNonlinear1999,CetL16}, which is sometimes possible while maintaining Type-I error control. We call procedures that estimate $\gamma$ once from the original data and hold the estimate fixed during resampling \textit{plug-in resampling procedures}. Here we focus on an additive subclass. Given independent observations $(O_i)_{i = 1}^n$, an estimate $\hat \gamma$, and functions $s_i$, consider a test statistic
\begin{equation}
S_n \equiv \sum_{i = 1}^n s_i(O_i, \hat \gamma).
\label{eq:test-stat}
\end{equation}
The procedures we study generate resampled contributions $\tilde s_i$ independently conditional on the observed data and construct a reference distribution for $S_n$ through Monte Carlo draws of $\widetilde S_n = \sum_{i = 1}^n \tilde s_i$. Existing instances do so via wild bootstrapping, sign-flipping, or conditional randomization \citep{klineScoreBasedApproach,Hemerik2019a,Liu2022a,Niu2022a}. %

Although plug-in resampling avoids repeated nuisance fitting, it remains cumbersome at scale. Testing many hypotheses requires running the procedure many times, with enough resamples in each run to attain the high $p$-value resolution required by stringent multiplicity adjustments. We address this remaining limitation by exploiting the fact that the resampled statistic $\widetilde S_n = \sum_{i = 1}^n \tilde s_i$ is the sum of independent terms (conditionally on the data $\obs = (O_i)_{i = 1}^n$), making it amenable to the \textit{saddlepoint approximation} (SPA), a classical technique giving fast and accurate tail probability approximations \citep{Daniels1954,Lugannani1980}. We use the SPA to approximate the limiting plug-in resampling $p$-value as the number of resamples grows to infinity:
\begin{equation}
p_{\infty, n}(\obs) \equiv \widetilde\P[\widetilde S_n \geq S_n \mid \obs] = \widetilde\P\left[\sum_{i = 1}^n \tilde s_i \geq S_n \mid \obs\right],
\label{eq:infinite-M}
\end{equation}
where $\widetilde \P[\ \cdot \mid \obs]$ denotes the probability over resampling. Computing the SPA for $p_{\infty, n}$ requires one univariate root search and can be orders of magnitude faster than generating many resamples.

\subsection{Two probability layers and a theoretical gap}\label{sec:two-probability-layers}

To justify replacing $p_{\infty, n}(\obs)$ by its approximation $p_{\text{SPA},n}(\obs)$, we must show that the latter is an accurate approximation of the former, in the sense that
\begin{equation}
\frac{p_{\infty,n}(\obs)}{p_{\text{SPA},n}(\obs)} - 1 = O_{\P}(a_n), \quad \text{for some sequence } a_n \rightarrow 0.
\label{eq:relative-error-convergence}
\end{equation}
The relative-error formulation of this bound ensures reliable approximation even for small $p$-values. The challenge is that SPA theory does not easily justify the guarantee~\eqref{eq:relative-error-convergence} because of a gap that emerges between the two layers of probability involved: the probabilities $\P$ and $\widetilde \P$ over the data $\obs$ and the resamples $(\tilde s_1, \dots, \tilde s_n) \mid \obs$, respectively.

\paragraph{The probability over resamples.} When holding the data $\obs$ fixed at a realization $\bm o_{1:n}$, classical results~\citep{jensen1995saddlepoint} provide asymptotic bounds on the relative error of the SPA, given that the distribution ${\widetilde \P[\ \cdot \mid \obs = \bm o_{1:n}]}$ of the summands $\tilde s_i$ satisfies a set of regularity conditions. Since this distribution is a function of the realized data, the regularity conditions on $\widetilde \P$ impose a constraint on this realization: $\bm o_{1:n} \in \mathcal R_n$. Standard SPA theory then gives an error rate $\varepsilon_n(\bm o_{1:n})$ such that
\begin{equation}
\text{If } \bm o_{1:n} \in \mathcal R_n \text{ for sufficiently large } n, \text{ then} \quad
\frac{p_{\infty,n}(\bm o_{1:n})}
     {p_{\mathrm{SPA},n}(\bm o_{1:n})}
- 1
= O(\varepsilon_n(\bm o_{1:n})).
\label{eq:conditional-spa-bound}
\end{equation}
By itself, this bound falls short of the desired one~\eqref{eq:relative-error-convergence}: it does not address how often $\bm o_{1:n} \in \mathcal R_n$ occurs or how large $\varepsilon_n(\bm o_{1:n})$ typically is.

\paragraph{The probability over the data, and the core challenge.} To upgrade the conditional bound~\eqref{eq:conditional-spa-bound} to the unconditional one~\eqref{eq:relative-error-convergence}, additional statements concerning the data distribution $\P$ are needed:
\begin{equation}
\mathbb P[\obs \in \mathcal R_n] \rightarrow 1 \quad \text{and} \quad \varepsilon_n(\obs) = O_{\P}(a_n), \quad \text{for some sequence } a_n \rightarrow 0.
\label{eq:outer-conditions}
\end{equation}
The first of these statements is challenging to prove because the regularity conditions $\mathcal R_n$ are quite delicate \citep{Albers1976,BoothHallWood1994Validity}, usually stemming from Edgeworth expansions underlying the SPA bounds~\eqref{eq:conditional-spa-bound}. To our knowledge, therefore, guarantees~\eqref{eq:relative-error-convergence} are available only for one- and two-sample testing problems, with either no nuisance parameters or a single unknown variance \citep{wangSaddlepointApproximationsResampling1990,Jing1994,jensen1995saddlepoint,RobinsonSkovgaard1998Bounds,HuRobinsonWang2012FinitePopulation}. By contrast, modern plug-in resampling procedures involve more complex nuisance parameters and fitting procedures, leading to intractable regularity events $\mathcal R_n$. This creates a barrier to theoretically grounding the promising methodological idea presented in Section~\ref{sec:spa-plugin}.

\subsection{Our contributions: verifiable SPA guarantees and two applications}

We overcome this challenge to establish a general alternative route from additive plug-in resampling laws to SPAs with verifiable relative-error guarantees~\eqref{eq:relative-error-convergence}, and showcase two applications in which this approximation theory is combined with procedure-specific target-validity results.

\paragraph{An SPA relative-error bound for plug-in resampling $p$-values.} We observe that the regularity conditions $\mathcal R_n$ are a function of the route taken to establish the SPA bound~\eqref{eq:conditional-spa-bound}. Edgeworth expansions are the standard route in the broader SPA literature because they yield \textit{higher-order} accuracy, but, in our context, lead to the verification difficulties described above. We circumvent this issue by establishing a \textit{lower-order} finite-sample SPA bound (Theorem~\ref{thm:conditional-lr} in Section~\ref{sec:universal-theory}) that results in a more tractable regularity event $\mathcal R_n$, for which the statement $\mathbb P[\obs \in \mathcal R_n] \rightarrow 1$ can be verified for a variety of plug-in resampling procedures. Theorem~\ref{thm:conditional-lr} extends a bound from the large-deviations literature \citep{sakhanenkoBerryEsseenTypeEstimates1991} to the {Lugannani--Rice (LR)} formula. {While Edgeworth-based bounds can give \textit{second-order} accuracy (roughly, $a_n = n^{-1}$ in the guarantee~\eqref{eq:relative-error-convergence}), ours gives \textit{first-order} accuracy (roughly, $a_n = n^{-1/2}$), in a sense made precise in Section~\ref{sec:universal-theory}. Nevertheless, we retain relative-error control in moderate-deviation regimes, where the cutoff moves increasingly far into the tail of the resampling distribution, allowing accurate approximation even for small $p$-values.} Theorem~\ref{thm:conditional-lr} concerns accuracy relative to the resampling $p$-value $p_{\infty,n}$; when $p_{\infty,n}$ is valid, vanishing relative error also transfers fixed-level asymptotic Type-I error control to $p_{\text{SPA},n}$.

\paragraph{Application 1: Response resampling (statistical genetics).} SPA-based methods in statistical genetics, pioneered by fastSPA \citep{Dey2017}, are now widely used to overcome Type-I error inflation in GWAS with rare variants and diseases \citep{Zhou2018,biFastAccurateMethod2019,biFastAccurateMethod2020,biEfficientMixedModel2021,mbatchouComputationallyEfficientWholegenome2021,Jiang_2021,Ko_2022,deyEfficientAccurateFrailty2022,hofFastAccurateRecurrent2023,zhouEfficientAccurateMixed2024,maSPAmixScalableAccurate2025}. Yet the exact tests being approximated in these procedures have not been precisely identified, and neither the Type-I error control of these procedures nor the accuracy of their SPAs has been theoretically established. We show that the procedures proposed by \citet{Dey2017} and the successors cited above are in fact SPAs for plug-in resampling procedures (Table~\ref{tab:genetics-resampling-mappings}). As a representative of such procedures, we propose and study \textit{generalized linear model response resampling} (GLM-RR), which includes the approximation target of fastSPA as a special case (Section~\ref{sec:response-resampling}). We establish conditions under which GLM-RR itself controls Type-I error, and derive a first-order rate for the relative error~\eqref{eq:relative-error-convergence} of the SPA for GLM-RR, which is sharp in a Bernoulli specialization. These results allow increasing sparsity in the predictor and/or response variables, matching the setting in which SPAs are most needed. There is still a gap between the SPA we analyze (of LR form) and that implemented in fastSPA (of Barndorff--Nielsen form, with additional accelerations); however, this gap is not fundamental and is negligible in our numerical comparison (Supplementary Fig.~\ref{fig:fastspa-numerical-comparison}).

\paragraph{Application 2: Conditional randomization (single-cell CRISPR screens).} Next, we apply the SPA to the distilled conditional randomization test (dCRT) \citep{Liu2022a,Niu2022a}, a resampling-based conditional independence test already deployed in scientific applications. These include single-cell CRISPR screen analysis, for which the widely adopted \texttt{sceptre} R package \citep{Katsevich2020c} is based on a dCRT variant. Replacing resampling in the dCRT by the SPA results in the spaCRT, a versatile resampling-free conditional independence test. We prove relative-error bounds~\eqref{eq:relative-error-convergence} in low- and high-dimensional settings and show that one such bound is sharp in a count-data setting (Section~\ref{sec:headline-bnb}). Furthermore, we extend the validity of the dCRT \citep{Niu2022a} to regimes of increasing sparsity in the predictor and/or response variables. We demonstrate that spaCRT matches the dCRT's excellent statistical performance in simulations and on real single-cell CRISPR data, while achieving up to a 250-fold speedup over direct Monte Carlo \dCRT{} (Sections~\ref{sec:simulation} and~\ref{sec:crispr-application}). An R package implementation of spaCRT is publicly available; integrating it into \texttt{sceptre} could substantially accelerate this software.

\subsection{Relation to existing work} \label{sec:existing-work}

Our work builds on ideas from three literatures, while contributing to each in a different way.

\paragraph{Modern resampling.} Several resampling procedures already keep nuisance fits fixed across resamples, and their computational advantage over refitting-based procedures has been recognized in special cases \citep{klineScoreBasedApproach,Hemerik2019a,Liu2022a}. We place these independently developed methods within a common plug-in resampling framework (Section~\ref{sec:procedures}) and demonstrate a common route to further SPA acceleration with provable accuracy guarantees.

\paragraph{Statistical genetics.} Replacing normal approximations with SPAs is routine in statistical genetics \citep{Dey2017}. We identify the tests targeted by these SPAs and, in the representative GLM-RR setting, establish their validity and relative approximation accuracy~\eqref{eq:relative-error-convergence}. \citet{Johnsen2021} give a complementary heuristic interpretation of fastSPA, but without theoretical guarantees.

\paragraph{Classical SPA theory.} SPAs have previously been used to approximate resampling procedures, with relative-error guarantees~\eqref{eq:relative-error-convergence} established for one- and two-sample problems, including one first-order guarantee \citep{HuRobinsonWang2012FinitePopulation}. We identify first-order SPA bounds as a general route to establishing guarantees for resampling procedures involving much more complex nuisance fits. %

\section{Plug-in resampling procedures}\label{sec:procedures}

\subsection{Definition} \label{sec:pirp-def}

Suppose we have independent, but not necessarily identically distributed observations
\begin{equation*}
O_i \indsim \Q_i, \qquad i=1,\dots,n,
\end{equation*}
and write $\P=\bigotimes_{i=1}^n \Q_i\in\mathcal P$ for the joint law of $\obs=(O_1,\dots,O_n) \in \mathcal O^n$. The goal is to test $H_0:\P\in\mathcal P_0$ for some family $\mathcal P_0\subset\mathcal P$ in the presence of a nuisance parameter $\gamma\in\Gamma$. We keep the family $\mathcal P$, null hypothesis $\mathcal P_0$, sample space $\mathcal O$, and nuisance space $\Gamma$ deliberately general in this subsection; we consider special cases of these in Section~\ref{section:examples}.
We write $\P_n$ when indexing the data law by sample size and use $\E$ for expectation under the stipulated law.

Next, we formalize the additive plug-in resampling procedures from Section~\ref{sec:spa-plugin} in Algorithm~\ref{alg:plug-in-resampling}. For each dataset, let $\widetilde\P(\,\cdot\mid\obs)$ denote the conditional joint resampling law of $(\tilde s_1,\dots,\tilde s_n)$. This law may depend on $\widehat\gamma$, which remains fixed across resampling draws. We require that
\begin{equation}
\tilde s_1,\dots,\tilde s_n \text{ are independent under } \widetilde\P(\,\cdot\mid\obs), \qquad \widetilde\E[\tilde s_i\mid\obs]=0, \ \ i=1,\dots,n.
\label{eq:plugin-resampling-law}
\end{equation}
\vspace{-2\baselineskip}
\begin{wrapfigure}[14]{r}{0.53\linewidth}
	\begin{minipage}{\linewidth}
		\makeatletter
		\patchcmd{\@algocf@start}{\addtolength{\hsize}{-1.5em}}{}{}{}
		\makeatother
		\begin{algorithm}[H]
			\nonl \textbf{Input:} Data $(O_i)_{i = 1}^n$, function $s_i$, estimator $\widehat \gamma$, resampling construction $\widetilde\P(\,\cdot\mid\obs)$, number of draws $M$. \\
			$\widehat \gamma \leftarrow \widehat \gamma(\obs)$\;
			$S_n\leftarrow\sum_{i=1}^n s_i(O_i;\widehat\gamma)$\;
			Construct resampling law $\widetilde\P(\,\cdot\mid\obs)$\;
			\For{$m=1,2,\dots,M$}{
				Draw $(\tilde s_1^{(m)},\dots,\tilde s_n^{(m)}) \sim \widetilde\P(\,\cdot\mid\obs)$\;
				$\widetilde S_n^{(m)}\leftarrow\sum_{i=1}^n\tilde s_i^{(m)}$\;
			}
			$p_M\leftarrow\{1+\sum_{m=1}^M\indicator(\widetilde S_n^{(m)}\geq S_n)\}/(M+1)$\;
			\nonl \textbf{Output:} $p$-value $p_M$.
			\caption{\bf Additive plug-in resampling}
			\label{alg:plug-in-resampling}
		\end{algorithm}
	\end{minipage}
\end{wrapfigure}

It may be surprising that a resampling procedure like Algorithm~\ref{alg:plug-in-resampling} can be valid without refitting $\widehat\gamma$, as is conventionally recommended. Existing validity arguments \citep{klineScoreBasedApproach,Hemerik2019a,Niu2022a} for plug-in resampling procedures proceed by establishing an approximation of the form
\begin{equation}
S_n = \sum_{i = 1}^n s_i(O_i, \hat \gamma) \approx \sum_{i = 1}^n s_i(O_i, \gamma),
\end{equation}
which follows under suitable regularity conditions if $s_i$ is defined as an efficient score statistic that is first-order insensitive to errors in $\hat \gamma$. Then, under the null, $S_n$ is approximately the sum of independent mean-zero terms, and can be shown to converge (after rescaling) to a centered normal distribution. Type-I error control follows by also establishing a conditional normal approximation for $\widetilde S_n = \sum_{i = 1}^n \tilde s_i$ with asymptotically matching variance, so that the original and resampled statistics approach the same distribution under a common normalization.
{These validity arguments rely on normal approximations, whose poor finite-sample accuracy motivates the use of resampling in the first place. Nevertheless, they establish a baseline guarantee of fixed-level asymptotic validity. Empirical results suggest Type-I error control beyond regimes where the normal approximation applies, but justifying this theoretically is beyond our scope; our aim is to approximate the resampling $p$-value accurately while avoiding its computational cost.}

\subsection{Examples} \label{section:examples}

A diverse set of additive plug-in resampling procedures exists. Some modify the score contributions $s_i(O_i,\widehat\gamma)$ directly, using i.i.d. mean-zero weights \citep{klineScoreBasedApproach} or random sign flips \citep{Hemerik2019a,andreellaRobustInferenceGeneralized2024}. Others evaluate the score on independently resampled observations from a fitted null law. In predictor resampling, each predictor is drawn from a fitted conditional distribution; the dCRT is a prominent example \citep{Liu2022a,Niu2022a}, and the same mechanism has appeared in genetic association testing \citep{maSPAmixScalableAccurate2025}. In fitted-response resampling, each response is drawn from a fitted observation-specific distribution; fastSPA implicitly targets a Bernoulli instance \citep{Dey2017}, while related work has used Bernoulli, ordinal-categorical, and Poisson working-response laws \citep{Zhou2018,biEfficientMixedModel2021,deyEfficientAccurateFrailty2022,zhouEfficientAccurateMixed2024}. Empirical-residual resampling instead uses the empirical distribution of fitted observation-level residual contributions \citep{biFastAccurateMethod2020,Ko_2022,hofFastAccurateRecurrent2023}. All of these procedures are instances of Algorithm~\ref{alg:plug-in-resampling}; Table~\ref{tab:genetics-resampling-mappings} details the statistical-genetics examples.

We now give two running examples, specifying in each case the instantiations of the general constructs in Section~\ref{sec:pirp-def}. The first is GLM response resampling (GLM-RR), whose logistic regression specialization underlies fastSPA; the second is the dCRT.

\begin{example}[GLM response resampling: generalizing fastSPA's target]
\label{ex:fastspa}

Let $x_i\in\R$ be a fixed scalar predictor and let $\bm z_i\in\R^d$
be fixed covariates, including an intercept. The responses follow the
canonical GLM
\begin{equation}
  \ey{i}\indsim f_Y(\,\cdot\,;\theta_{y,i}),\ \
  \theta_{y,i}=\rho x_i+\bm z_i^{\T}\bgam,\ \
  f_Y(y;\theta)=\exp\{y\theta-b_Y(\theta)\}h_Y(y),\ \ i=1,\ldots,n.
  \label{eq:fastspa-nef-model}
\end{equation}
Write $v_Y=b_Y''$ for the response variance function.
In Section~\ref{sec:pirp-def}'s notation, $O_i=\ey{i}$, $\Q_i$ is the law in~\eqref{eq:fastspa-nef-model}, and
$\bgam$ is the nuisance parameter. We test
\[
  H_0:\rho=0.
\]
Fit the null response model by maximum likelihood, obtaining $\bgamhat$, and
write
\begin{equation}\label{eq:new-glm-projection}
  \widehat\theta_i=\bm z_i^{\T}\bgamhat,\qquad
  \widehat\mu_i=b_Y'(\widehat\theta_i),\qquad
  \widehat{\bm\alpha}=
  \left(\sum_{i=1}^n v_Y(\widehat\theta_i)\bm z_i\bm z_i^{\T}\right)^{-1}
  \sum_{i=1}^n v_Y(\widehat\theta_i)\bm z_i x_i.
\end{equation}
Here, $\bm z_i^{\T}\widehat{\bm\alpha}$ is the $i$th fitted value from the
variance-weighted linear regression of $x_i$ on $\bm z_i$. The summands are then defined as the per-observation efficient scores
\begin{equation}
  s_i(\ey{i};\bgamhat)
  =(x_i-\bm z_i^{\T}\widehat{\bm\alpha})
  (\ey{i}-\widehat\mu_i).
\label{eq:fastSPA-summands}
\end{equation}
GLM-RR is Algorithm~\ref{alg:plug-in-resampling} with these summands. The
resampling law $\widetilde\P(\,\cdot\mid\obs)$ is induced by
independently redrawing the responses from their fitted null laws:
\begin{equation}
  \tilde s_i
  =(x_i-\bm z_i^{\T}\widehat{\bm\alpha})
  (\tilde Y_i-\widehat\mu_i),
  \qquad
  \tilde Y_i\mid\obs\indsim f_Y(\,\cdot\,;\widehat\theta_i),
  \qquad i=1,\ldots,n.
\label{eq:fastSPA-resampling}
\end{equation}
When $f_Y$ is Bernoulli, equations~\eqref{eq:fastSPA-summands} and~\eqref{eq:fastSPA-resampling} recover the statistic and fitted response-resampling law underlying the fastSPA approximation target.

\end{example}

\begin{example}[\protect\dCRT{}: predictor resampling]
\label{ex:dcrt}

Suppose the observations are
\begin{equation*}
\bm O_i = (\ex{i}, \ey{i}, \rz{i}) \iidsim \Q, \quad i = 1, \dots, n,
\end{equation*}
so $\P=\Q^{\otimes n}$.
The goal is to test
\[
  H_0: \ \ex \ \independent\ \ey \mid \rz.
\]
This is a nonparametric analog of the problem in Example~\ref{ex:fastspa}. The nuisance parameter is
\[
  \gamma=\big\{\Q_{X\mid\bm Z},\,\mu_y\big\},
  \qquad \mu_y(\rz{})=\E(\ey{}\mid\rz{}).
\]
Here $\Q_{X\mid\bm Z}$ denotes the conditional law of $\ex{}\mid\rz{}$ under $\Q$.
Fit this conditional law and the conditional mean of
$\ey{}\mid\rz{}$ by arbitrary regression procedures, obtaining
$\widehat\gamma=\{\widehat{\Q}_{X\mid\bm Z},\widehat\mu_y\}$.
Let $\widehat\mu_{x,i}$ be the mean of the fitted conditional law
$\widehat{\Q}_{X\mid\bm Z}(\,\cdot\mid\rz{i})$, and write
$\widehat\mu_{y,i}=\widehat\mu_y(\rz{i})$.
The dCRT is the plug-in form of the conditional randomization test
\citep{CetL16,Liu2022a,Niu2022a} with product-of-residuals summands
\begin{equation}
  s_i(\ex{i}, \ey{i}, \rz{i}; \widehat\gamma)
  =
  (\ex{i}-\widehat\mu_{x,i})
  (\ey{i}-\widehat\mu_{y,i}).
\label{eq:dcrt-summand}
\end{equation}
The resampling law $\widetilde\P(\,\cdot\mid\obs)$ is induced by
independently redrawing the predictors from their fitted conditional laws:
\begin{equation}\label{eq:dcrt-resampling-summand}
  \tilde s_i
  =
  (\exk{i}-\widehat\mu_{x,i})
  (\ey{i}-\widehat\mu_{y,i}),
  \qquad
  \exk{i} \mid \obs \indsim \widehat{\Q}_{X\mid\bm Z}(\,\cdot\mid\rz{i}).
\end{equation}
The dCRT has been applied in a range of domains, including microbiome mediation
and differential-abundance studies \citep{Ma2024,Liu2025}, long COVID
phenotyping \citep{Maripuri2024}, and the analysis of single-cell CRISPR
screens \citep{Katsevich2020c}.

\end{example}

\section{General first-order bounds on SPA relative error for plug-in resampling}
\label{sec:universal-theory}

\subsection{The Lugannani--Rice approximation to the plug-in resampling $p$-value} \label{sec:lugannani-rice}

We define the LR SPA \citep{Lugannani1980} to the infinite-resampling limit of the plug-in resampling $p$-value $p_{\infty, n}$~\eqref{eq:infinite-M}, applied after conditioning. To define this approximation, we construct the average conditional cumulant generating function (CGF)
\begin{equation}
 K_n(t)\equiv\frac1n\sum_{i=1}^n
 \log{\widetilde\E}\{\exp(t\tilde s_i)\mid\obs\}.
\label{eq:avg-cond-cgf}
\end{equation}
We call $\widehat t_n$ an \emph{admissible saddlepoint} if it is the unique
finite solution, in the interior of the effective domain of $K_n$, to the
saddlepoint equation
\begin{equation}\label{eq:saddlepoint-equation}
 K_n'(\widehat t_n)=S_n/n,
\end{equation}
and satisfies $K_n''(\widehat t_n)>0$. For an admissible saddlepoint,
convexity and $K_n(0)=0$ imply
$\widehat t_nS_n-nK_n(\widehat t_n)\ge0$, so we may define
\[
 r_n\equiv\sgn(\widehat t_n)
 \{2(\widehat t_nS_n-nK_n(\widehat t_n))\}^{1/2},
 \qquad
 u_n\equiv\widehat t_n\{nK_n''(\widehat t_n)\}^{1/2}.
\]
For $S_n\ne0$, the conditional LR approximation is
\begin{equation}\label{eq:new-lr}
 p_{\text{SPA},n}(\obs)
 \equiv1-\Phi(r_n)+\phi(r_n)
 \left(\frac1{u_n}-\frac1{r_n}\right).
\end{equation}
When $S_n=0$ and an admissible saddlepoint exists, the LR formula has a removable singularity because $u_n=r_n=0$. We use its continuous extension \citep{Lugannani1980}:
\begin{equation}\label{eq:new-lr-zero}
 p_{\text{SPA},n}(\obs)
 \equiv\frac12-\frac{K_n'''(0)}
 {6\sqrt{2\pi n}\,K_n''(0)^{3/2}}.
\end{equation}
If no admissible saddlepoint exists, we use the conservative convention
$p_{\text{SPA},n}(\obs)=1$.

Next, we turn to the accuracy of this approximation.

\subsection{A finite-sample relative-error theorem} \label{sec:finite-sample-theorem}

\begin{theorem}[Finite-sample Lugannani--Rice bound]
\label{thm:new-universal}\label{thm:conditional-lr}
Conditionally on the data $\obs$, let $\tilde s_1,\ldots,\tilde s_n$ be independent
real-valued random variables satisfying
${\widetilde\E}(\tilde s_i\mid\obs)=0$, and let $S_n$ be an arbitrary function of the data. Define the per-observation and total variances
\[
v_i \equiv v_i(\obs) \equiv {\widetilde\E}[\tilde s_i^2 \mid \obs] \quad \text{and} \quad V_n \equiv V_n(\obs) \equiv  \sum_{i = 1}^n v_i.
\]
Suppose there are functions of the data $b_i \equiv b_i(\obs) \ge 0$ for each $i = 1, \dots, n$ such that
\begin{equation}\label{eq:HB-main}
|{\widetilde\E}(\tilde s_i^k\mid\obs)|
 \le \frac{k!}{2}b_i^{k-2}v_i \quad \text{for every integer } k \geq 3.
\end{equation}
On $\{V_n>0\}$, define
\begin{equation}\label{eq:three-coordinates}
 B_n(\obs)
 \equiv
 \frac{1}{\sqrt{V_n/n}}\sum_{i = 1}^n\frac{v_i}{V_n}b_i,
 \qquad
 Z_n(\obs)\equiv\frac{|S_n|}{\sqrt{V_n}},
 \qquad
 D_n(\obs)\equiv\frac{(\max_{i = 1}^n b_i)|S_n|}{V_n}.
\end{equation}
Then, there are universal constants $c,C>0$ such that, on the event
\begin{equation}\label{eq:scale-adaptive-lr-conditions}
 V_n(\obs)>0,
 \qquad
 \frac{B_n(\obs)\{1+Z_n(\obs)\}}{\sqrt n}\le c,
 \qquad
 D_n(\obs)\le c,
\end{equation}
an admissible saddlepoint exists,
$p_{\textnormal{SPA},n}(\obs) > 0$, and
\begin{equation}\label{eq:master-bound-main}
 \left|
 \frac{p_{\infty, n}(\obs)}{p_{\textnormal{SPA},n}(\obs)}-1
 \right|
 \le C\frac{B_n(\obs)\{1+Z_n(\obs)\}}{\sqrt n}.
\end{equation}
\end{theorem}

\paragraph{Regularity conditions.}
The regularity condition~\eqref{eq:HB-main} required on each summand $\tilde s_i$ is a data-conditional Bernstein condition, which assumes control of higher moments in terms of the variance $v_i$ and \textit{Bernstein scale} $b_i$. For example, bounded summands $|\tilde s_i|\le b_i$ satisfy this condition with the absolute bound $b_i$ as the Bernstein scale. The Bernstein condition also permits unbounded summands; it holds as soon as the CGF (needed to define the SPA~\eqref{eq:avg-cond-cgf}) is finite in an open interval containing the origin. Theorem~\ref{thm:conditional-lr} also avoids the support restrictions that complicate Edgeworth-based arguments. In our discrete resampling examples, finite support rules out Cram\'er's condition, while heterogeneous, data-dependent weights generally preclude a common lattice. Relevant Edgeworth arguments therefore require more delicate restrictions on the arrangement of support points or coefficients \citep{Albers1976,BoothHallWood1994Validity}. For weighted Bernoulli sums, for example, one such condition requires sufficiently many weights to remain well spread at a shrinking scale. When the weights depend on fitted nuisance parameters, standard estimation-error bounds do not by themselves establish this spacing property.

\paragraph{Interpreting the bound.} The bound~\eqref{eq:master-bound-main} on the relative approximation error involves three factors. The first is the \textit{standardized Bernstein scale} $B_n$. An aggregate of the per-observation Bernstein scales $b_i$, $B_n$ measures variance-normalized higher-moment growth: the faster this growth, the larger the relative error bound. The second factor is $1+Z_n$, where the \textit{tail depth} $Z_n$ measures the distance of the cutoff ($S_n$) from the resampling mean (zero) in standard-deviation units. The relative error bound increases as the cutoff moves deeper into the tail. The third factor is $n^{-1/2}$. For i.i.d. summands $\tilde s_i$ with a fixed distribution and bounded tail depths $Z_n$, the bound approaches zero at rate $n^{-1/2}$: this is the first-order benchmark. In general, whether and how quickly the bound approaches zero also depends on the asymptotic behavior of $B_n$ and $Z_n$.

\paragraph{A Bernoulli illustration.}
Drop the conditioning, take a deterministic cutoff $S_n$, and let
$\tilde s_1,\ldots,\tilde s_n\iidsim\operatorname{Ber}(p_n)-p_n$,
where $0<p_n<1$.
Since $|\tilde s_i|\le1$, we may take $b_i=1$, giving
$V_n=np_n(1-p_n)$ and $B_n=\{p_n(1-p_n)\}^{-1/2}$.
Theorem~\ref{thm:new-universal} becomes
\[
 \left|\frac{p_{\infty,n}}{p_{\mathrm{SPA},n}}-1\right|
 \le C\frac{1+Z_n}{\sqrt{np_n(1-p_n)}},
 \qquad \text{for} \quad
 \frac{1 + Z_n}{\sqrt{np_n(1-p_n)}} \leq c.
\]
This result accommodates $p_n \rightarrow 0$, i.e. increasingly rare Bernoulli summands whose variances tend to zero. In that case, the factor $B_n \asymp p_n^{-1/2}$ grows to infinity, but the bound still converges to zero if $np_n \rightarrow \infty$ and $Z_n = o(\sqrt{np_n})$. The latter requirement shows that cutoffs increasingly many standard deviations into the tail ($Z_n \gg 1$), i.e., moderate deviations, are allowed by our bound. At bounded tail depth, the bound is of order $(np_n)^{-1/2}$, so the relevant quantity in this case is the expected number of successes, rather than the number of trials alone.

\paragraph{Relation to earlier bounds.} We highlight three relevant precedents. The first is \citet{sakhanenkoBerryEsseenTypeEstimates1991}, whose results we build on to establish Theorem~\ref{thm:conditional-lr}. \citet{sakhanenkoBerryEsseenTypeEstimates1991} established finite-sample bounds for large-deviation probabilities under an exponentially weighted third-moment condition. These bounds can be expressed as relative-error control for a saddlepoint approximation distinct from LR. Our proof verifies this moment condition under our assumptions and controls the difference between the resulting approximation and the LR formula. In related work on sharp large deviations, \citet{fanSharpLargeDeviation2015} establish relative-error bounds for tail probabilities under a common Bernstein condition. Our theorem provides relative-error control for the LR formula and accommodates observation-specific Bernstein scales, allowing the error bound to depend on their variance-weighted average rather than their maximum. This refinement avoids a logarithmic loss in our count-data application (Section~\ref{sec:headline-bnb}) and can yield even larger improvements in the error bound when the residual multipliers have heavier tails under the data-generating law. Finally, \citet{HuRobinsonWang2012FinitePopulation} give first-order relative tail bounds for finite-population sums and two-sample permutation tests, without Edgeworth-type restrictions on the population. Their guarantees concern sampling without replacement; Theorem~\ref{thm:new-universal} provides a complementary LR bound for heterogeneous independent summands, in a form suited to plug-in resampling laws.

\paragraph{Applying the theorem.} The main benefit of Theorem~\ref{thm:conditional-lr} is its broad applicability to plug-in resampling procedures. Recall from Section~\ref{sec:two-probability-layers} that translating the data-conditional bound~\eqref{eq:master-bound-main} into in-probability bounds over the data distribution~\eqref{eq:relative-error-convergence} requires establishing that the regularity conditions~\eqref{eq:scale-adaptive-lr-conditions} are satisfied with probability approaching one, and that $B_n(1+Z_n)/\sqrt n=O_{\P}(a_n)$ for a deterministic sequence $a_n\to0$. In fact, we establish both statements uniformly over model classes, yielding a uniform version of the in-probability error bound~\eqref{eq:relative-error-convergence}. Let $\text{RE}_n$ denote the relative approximation error in the left-hand side of equation~\eqref{eq:master-bound-main}. In all asymptotic statements, define $p_{\infty,n}/p_{\mathrm{SPA},n}=1$ on $\{p_{\mathrm{SPA},n}\le0\}$, and define $\text{RE}_n$ using this extended ratio. This convention leaves the reported LR approximation unchanged. Under the relevant assumptions, the exceptional event has probability tending to zero, uniformly wherever a uniform conclusion is asserted. For a sequence of model classes $\mathcal P=\{\mathcal P_n\}$ and a deterministic rate $a_n(\P)>0$, write
\begin{equation}
  \text{RE}_n=\OP{\mathcal P}\{a_n(\P)\}
  \quad\text{if}\quad
  \sup_{\P\in\mathcal P_n}
  \P\{|\text{RE}_n|>C(\eps)\,a_n(\P)\}\le\eps
  \quad\text{for all }n\ge N(\eps),
  \label{eq:op-def}
\end{equation}
for every $\eps>0$, with $C(\eps)$ independent of $\P$. In certain cases, we also prove that such bounds are sharp, in the sense that there exists a sequence of laws $\P_n\in\mathcal P_n$ such that $\text{RE}_n \asymp_{\P_n} a_n(\P_n)$, meaning that $\text{RE}_n/a_n(\P_n)=O_{\P_n}(1)$ and $a_n(\P_n)/\text{RE}_n=O_{\P_n}(1)$.

\section{Application 1: Response resampling (statistical genetics)}
\label{sec:response-resampling}\label{sec:saige}

\subsection{Information and assumptions}
\label{sec:glmrr-setup}

In this section, we theoretically explore GLM-RR, a response resampling procedure that includes the target of fastSPA \citep{Dey2017} as a special case (Example~\ref{ex:fastspa}). We establish its asymptotic validity and the relative accuracy of its SPA (Section~\ref{sec:glmrr-spa-accuracy}), then specialize to the GWAS setting to obtain an interpretable rate and a matching sharpness result (Section~\ref{sec:headline-fastSPA}). Prior to that, we establish preliminaries for this model, including the null efficient Fisher information (an important part of our relative error bound) and the assumptions required for our results.

\paragraph{GLM-RR information.} Consider a sequence of fixed-design GLMs~\eqref{eq:fastspa-nef-model}, with natural parameters $\theta_{y,i}=\rho_n x_i+\bm z_i^\top\bm\gamma_n$, association coefficients $\rho_n$, nuisance
parameters $\bm\gamma_n$, and covariates $\bm z_i=(1,\bm u_i^\top)^\top\in\R^d$, where $d$ is fixed. Recall that GLM-RR uses products of response residuals and information-weighted predictor residuals as its score summands~\eqref{eq:fastSPA-summands}, independently resampling the responses from their fitted null laws while holding the fit fixed. Define
\begin{equation}\label{eq:glmrr-baseline-projection}
 \bm\alpha_n^0
 =\left(\sum_{i=1}^n v_Y(\bm z_i^\top\bm\gamma_n)\bm z_i\bm z_i^\top\right)^{-1}
     \sum_{i=1}^n v_Y(\bm z_i^\top\bm\gamma_n)\bm z_ix_i,
\end{equation}
the null counterpart of the fitted projection~\eqref{eq:new-glm-projection} in Example~\ref{ex:fastspa}. For a data-generating law $\P_n$, define $\P_n^0$ as the law obtained by setting $\rho_n=0$. The total efficient information under $\P_n^0$ is
\begin{equation}\label{eq:glmrr-information}
 I_{n,x}
 =\V_{\P_n^0}\!\left[
   \sum_{i=1}^n(x_i-\bm z_i^\top\bm\alpha_n^0)
       \{Y_i-b_Y'(\bm z_i^\top\bm\gamma_n)\}\right]
 =\sum_{i=1}^n v_Y(\bm z_i^\top\bm\gamma_n)(x_i-\bm z_i^\top\bm\alpha_n^0)^2.
\end{equation}
It combines response variance with residual predictor variation. 

\paragraph{Assumptions.} Next, we state the assumptions required for our results, using the notation $\|\bm\gamma_n\|_{\mathrm{nonint}}$ for the Euclidean norm of the nonintercept coefficients of $\bm \gamma_n$:
\begin{assumption}[Fixed-design GLM-RR regularity]\label{assu:fastspa-regularity}
For fixed $K<\infty$ and $\lambda>0$:
\textup{(i)} $\max_i(|x_i|\vee\|\bm u_i\|)\le K$,
$\|\bm\gamma_n\|_{\mathrm{nonint}}\le K$, and
$\lambda_{\min}\{n^{-1}\sum_{i=1}^n\bm z_i\bm z_i^\top\}\ge\lambda$;
\textup{(ii)} the natural parameter space is $\R$, $v_Y(\theta)>0$, and,
with $\alpha_{n,y}$ the intercept in $\bm\gamma_n$,
$0<v_Y(\alpha_{n,y})\le K$ and
$\sup_{\theta\in\R}|v_Y'(\theta)/v_Y(\theta)|\le K$.
\end{assumption}
Part (i) bounds the predictors, covariates and nonintercept coefficients, and prevents the nuisance design from becoming nearly collinear. The intercept may tend to $-\infty$, allowing increasingly rare responses. Part (ii) imposes regularity on the response family, ensuring stable variances and controlled tails; it accommodates known-variance Gaussian and Bernoulli responses, as well as Poisson responses with an intercept bounded from above.

\subsection{Validity of GLM-RR and the accuracy of its SPA}
\label{sec:glmrr-results}\label{sec:glmrr-validity}\label{sec:glmrr-spa-accuracy}

\paragraph{Validity of GLM-RR.} Recall from Section~\ref{sec:introduction} that, despite the prevalence of SPAs in statistical genetics, the tests being approximated had not been explicitly identified. Before examining the accuracies of such SPAs, it therefore is sensible to first establish the Type-I error control of their approximation targets. Let $p_{\infty,n}^{\mathrm{GLM\text{-}RR}}$ denote the infinite-resampling GLM-RR $p$-value~\eqref{eq:infinite-M}. The next result establishes the asymptotic validity of this $p$-value under the assumptions in Section~\ref{sec:glmrr-setup}.

\begin{theorem}[Validity of GLM response resampling]\label{thm:glmrr-target-validity}
For deterministic $\delta_n\to\infty$, define
\begin{equation}\label{eq:glmrr-class}
 \PRR_n=\{\P_n\in\mathcal P_n:\;
 \text{Assumption~\ref{assu:fastspa-regularity} holds with }(K,\lambda,d); \quad I_{n,x}\ge\delta_n,\quad |\rho_n|\le\delta_n^{-1}\}.
\end{equation}
On the null subclass $\PRRnull_n=\{\P_n\in\PRR_n:\rho_n=0\}$, the GLM-RR $p$-value is uniformly asymptotically valid:
\begin{equation}\label{eq:glmrr-target-uniformity}
 \sup_{\P_n\in\PRRnull_n}\sup_{t\in[0,1]}
 \left|\P_n(p_{\infty,n}^{\mathrm{GLM\text{-}RR}}\le t)-t\right|\longrightarrow0.
\end{equation}
\end{theorem}
This result suggests that the types of tests implicitly being approximated by statistical genetics SPAs are indeed valid. Underpinning this validity is the fact that the GLM-RR test statistic is an efficient score (as pointed out by \citet{Johnsen2021} for fastSPA), similarly to other plug-in resampling procedures \citep{klineScoreBasedApproach,Hemerik2019a,Niu2022a}.

\paragraph{Approximation accuracy of SPA for GLM-RR.} Let $p_{\mathrm{SPA},n}^{\mathrm{GLM\text{-}RR}}$ be the LR approximation based on the average conditional CGF~\eqref{eq:avg-cond-cgf} of the GLM-RR resampling distribution,
\begin{equation}\label{eq:new-fastspa-cgf}
 K_n(t)=\frac1n\sum_{i=1}^n\left[
 b_Y\{\widehat\theta_i+t(x_i-\bm z_i^\top\widehat{\bm\alpha})\}
 -b_Y(\widehat\theta_i)
 -t(x_i-\bm z_i^\top\widehat{\bm\alpha})\widehat\mu_i
 \right].
\end{equation}
Under the assumptions of Section~\ref{sec:glmrr-setup}, the regularity conditions of Theorem~\ref{thm:conditional-lr} hold with probability tending to one uniformly over $\PRR_n$, verifying the first requirement in~\eqref{eq:outer-conditions}. Moreover, the Bernstein scales can be chosen so that
\[
 \frac{B_n}{\sqrt n}=O_{\PRR}(I_{n,x}^{-1/2}),
 \qquad
 Z_n=O_{\PRR}\{1+|\rho_n|\sqrt{I_{n,x}}\}.
\]
Consequently, the error scale $B_n(1+Z_n)/\sqrt n$ is $O_{\PRR}\{I_{n,x}^{-1/2}+|\rho_n|\}$, verifying the second requirement in~\eqref{eq:outer-conditions}. Applying Theorem~\ref{thm:conditional-lr} yields the following result.

\begin{theorem}[Relative accuracy of the GLM-RR SPA]\label{thm:new-glmrr}\label{thm:saige-validity}
The maximum likelihood estimator under the null, information-weighted
projection and admissible saddlepoint exist with probability tending to
one uniformly over $\P_n\in\PRR_n$, and
\begin{equation}\label{eq:glmrr-quantitative-main}
 \frac{p_{\infty,n}^{\mathrm{GLM\text{-}RR}}}{p_{\mathrm{SPA},n}^{\mathrm{GLM\text{-}RR}}}-1
 =O_{\PRR}\{I_{n,x}^{-1/2}+|\rho_n|\}.
\end{equation}
Under the null, \eqref{eq:glmrr-target-uniformity} also holds with $p_{\mathrm{SPA},n}^{\mathrm{GLM\text{-}RR}}$ replacing $p_{\infty,n}^{\mathrm{GLM\text{-}RR}}$.
\end{theorem}
The rate bound combines a central approximation term $I_{n,x}^{-1/2}$ with an additional term $|\rho_n|$ accounting for departure from the null. It covers shrinking alternatives $\rho_n \rightarrow 0$, including both local alternatives $\rho_n = O(I_{n,x}^{-1/2})$ and larger alternatives $|\rho_n| \gg I_{n,x}^{-1/2}$.

\subsection{Genetic association testing: rate and sharpness}
\label{sec:headline-fastSPA}

We specialize Theorem~\ref{thm:new-glmrr} to the GWAS setting of fastSPA \citep{Dey2017}. Here, $Y_i \in \{0, 1\}$ is a disease indicator and $x_i \in \{0, 1, 2\}$ is individual $i$'s genotype at a particular single nucleotide polymorphism, coded as the number of copies of the less frequent variant (the minor allele). In this case, the model for the outcome~\eqref{eq:fastspa-nef-model} specializes to a logistic regression
\begin{equation}\label{eq:genotype-model}
 Y_i \indsim \Ber\!\left[\expit\{\alpha_{n,y}+\rho_nx_i+
                         \bm u_i^\top\bm\beta_{n,y}\}\right], \quad i = 1, \dots, n.
\end{equation}
Here $\rho_n$ is the per-allele log odds ratio.
Let $\pi_{n,x}\equiv(2n)^{-1}\sum_{i=1}^n x_i\in(0,1/2]$ be the empirical \textit{minor-allele frequency}.
We assume an asymptotic version of the standard \textit{Hardy--Weinberg equilibrium} assumption: the observed genotype proportions approach those of $\operatorname{Bin}(2,\pi_{n,x})$, corresponding to independent pairing of maternal and paternal alleles within individuals. In particular, we require the empirical genotype variance to approach that of the reference binomial distribution: 
\begin{equation}\label{eq:genotype-hwe}
  \left|
 \frac{n^{-1}\sum_{i=1}^n(x_i-2\pi_{n,x})^2}
      {2\pi_{n,x}(1-\pi_{n,x})}-1
 \right|\le\eta_n,
 \ \ \text{where}
\ \eta_n \rightarrow 0.
\end{equation}

In addition, we require the following regularity conditions, mostly derived from Assumption~\ref{assu:fastspa-regularity}. Let $R_{n,x}^2$ be the ordinary regression $R^2$ of $x_i$ on $\bm z_i$, including the intercept. For fixed $K\ge2$, $\lambda>0$ and $d$, assume
\begin{equation}\label{eq:genotype-regularity}
 R_{n,x}^2\le1-K^{-1}, \quad \max_{i=1,\ldots,n}\|\bm u_i\|\le K,\quad
 \alpha_{n,y}\le K, \quad \|\bm\beta_{n,y}\|\le K,\quad
 \lambda_{\min}\!\left(\frac1n\sum_{i=1}^n\bm z_i\bm z_i^\top\right)\ge\lambda.
\end{equation}
The bound on $R_{n,x}^2$ ensures that covariate adjustment preserves a nonvanishing fraction of genotype variation. If we define the baseline disease prevalence $\pi_{n,y}=\expit(\alpha_{n,y})$, conditions~\eqref{eq:genotype-hwe} and~\eqref{eq:genotype-regularity} yield $I_{n,x}\asymp n\pi_{n,x}\pi_{n,y}$ uniformly. This brings us to our next result.

\begin{theorem}[GWAS: rate and sharpness]\label{thm:headline-fastSPA}\label{cor:new-berber}\label{thm:saige-rate}
For a deterministic sequence $\delta_n\to\infty$, define
\begin{equation}\label{eq:genotype-class}
 \mathcal P^{\mathrm{GWAS}}_n=\{\P_n\in\mathcal P_n:\;
 \eqref{eq:genotype-model}-\eqref{eq:genotype-regularity}\text{ hold with }(K,\lambda,d); \quad n\pi_{n,x}\pi_{n,y}\ge\delta_n,\quad |\rho_n|\le\delta_n^{-1}\}.
\end{equation}
Its null slice is
$\mathcal P^{\mathrm{GWAS},0}_n
=\{\P_n\in\mathcal P^{\mathrm{GWAS}}_n:\rho_n=0\}$.
The conclusions of Theorems~\ref{thm:glmrr-target-validity}--\ref{thm:new-glmrr}
hold uniformly over these classes, and in particular,
\begin{equation}\label{eq:genotype-rate}
 \frac{p_{\infty,n}^{\mathrm{GLM\text{-}RR}}}{p_{\mathrm{SPA},n}^{\mathrm{GLM\text{-}RR}}}-1
 =O_{\mathcal P^{\mathrm{GWAS}}}\{(n\pi_{n,x}\pi_{n,y})^{-1/2}+|\rho_n|\}.
\end{equation}
Moreover, there exist fixed $K,\lambda,d$, a sequence $\delta_n\to\infty$, and a sequence $\P_n\in\mathcal P^{\mathrm{GWAS},0}_n$ such that
\begin{equation}\label{eq:genotype-sharpness}
 \left|\frac{p_{\infty,n}^{\mathrm{GLM\text{-}RR}}}{p_{\mathrm{SPA},n}^{\mathrm{GLM\text{-}RR}}}-1\right|
 \asymp_{\P_n}(n\pi_{n,x}\pi_{n,y})^{-1/2}.
\end{equation}
\end{theorem}

The information scale $n\pi_{n,x}\pi_{n,y}$ measures joint genotype--disease sparsity. Our result allows either frequency to vanish provided $n\pi_{n,x}\pi_{n,y}\to\infty$. Sharpness~\eqref{eq:genotype-sharpness} is attained in an intercept-only model with $\pi_{n,x}=1/2$ and $\pi_{n,y}\to0$, where lattice point masses cause a first-order discrepancy from ordinary LR. A lattice correction could address this example, but a standard common-lattice correction is not directly applicable to general heterogeneous, covariate-adjusted resampling laws.

\section{Application 2: Predictor resampling (single-cell CRISPR screens)}
\label{sec:predictor-resampling}

\subsection{The \spacrt{} procedure and a finite-sample approximation bound}
\label{alg:spacrt}

In this section, we return to the dCRT \citep{Liu2022a,Niu2022a} from Example~\ref{ex:dcrt}. A resampling-free variant of the dCRT was proposed by \citet{Liu2022a} via transformation of $X$ to a normal distribution. This approach was effective for continuous predictors, but resulted in a significant loss of power for discrete predictors. We instead propose the spaCRT (Algorithm~\ref{alg:new-spacrt}), an SPA-based resampling-free dCRT approximation, leveraging the framework developed in Sections~\ref{sec:procedures} and~\ref{sec:universal-theory}. Sections~\ref{alg:spacrt}--\ref{sec:headline-bnb} give approximation and validity theory; Sections~\ref{sec:simulation}--\ref{sec:crispr-application} give empirical results.

\begin{wrapfigure}[10]{r}{0.53\linewidth}
\vspace{-\intextsep}
\begin{minipage}{\linewidth}
\makeatletter
\patchcmd{\@algocf@start}{\addtolength{\hsize}{-1.5em}}{}{}{}
\makeatother
\begin{algorithm}[H]
\nonl \textbf{Input:} Data $(X_i,Y_i,\bm Z_i)_{i = 1}^n$ and family $f_{X}$.\\
Fit $\widehat{\Q}_{X\mid\bm Z} = f_{X}(\ \cdot\ ; \widehat \theta_x(\bm Z))$ and $\widehat{\E}_{\Q}[Y\mid \bm Z]$\;
$S_n\leftarrow\sum_{i=1}^n(X_i-\widehat\mu_{x,i})(Y_i-\widehat\mu_{y,i})$\;
Construct $K_n$, $K_n'$, and $K_n''$ from~\eqref{eq:spacrt-centered-cgf}\;
Solve $K_n'(\widehat t_n)=S_n/n$ and evaluate
\eqref{eq:new-lr} or~\eqref{eq:new-lr-zero}\;
\nonl \textbf{Output:} $p_{\mathrm{SPA},n}^{\dCRT}$.
\caption{\bf The spaCRT.}
\label{alg:new-spacrt}
\end{algorithm}
\end{minipage}
\end{wrapfigure}

\paragraph{spaCRT definition.} Let $f_X(\ \cdot \ ; \theta_x)$ be an arbitrary distributional family for $X$, with mean and variance functions $\mu_X(\theta_x)$ and $v_X(\theta_x)$, respectively, and centered CGF
\begin{equation*}
\smash{G_X(\theta,t)=
 \log\E_{\theta}[\exp\{t(X-\mu_X(\theta))\}].}
\end{equation*}
Let $\widehat{\Q}_{X\mid\bm Z} \equiv f_{X}(\ \cdot\ ; \widehat \theta_x(\bm Z))$ be a fit for the law $\Q_{X\mid\bm Z}$ for some function $\widehat \theta_x(\bm Z)$; denote \[\smash{\widehat \theta_{x,i} \equiv \widehat \theta_x(\bm Z_i) \text{ and } \widehat \mu_{x,i} \equiv \E_{\widehat{\Q}_{X|\bm Z}}[X_i \mid \bm Z_i] \equiv \mu_X(\widehat \theta_{x,i}).}\]
Letting $\widehat \mu_{y,i}$ be conditional means $\widehat{\E}_{\Q}[Y\mid \bm Z = \bm Z_i]$ fitted via an arbitrary procedure, the dCRT test statistic $S_n$ based on the product-of-residual summands~\eqref{eq:dcrt-summand} can be computed. The average conditional cumulant-generating function~\eqref{eq:avg-cond-cgf} of the resampled summands~\eqref{eq:dcrt-resampling-summand} is then
\begin{equation}
 K_n(t)
 =\frac1n\sum_{i=1}^n
 G_X\{\widehat\theta_{x,i},t(Y_i-\widehat\mu_{y,i})\},
 \label{eq:spacrt-centered-cgf}
\end{equation}
from which the LR $p$-value can be computed as in Section~\ref{sec:lugannani-rice}.

\paragraph{Finite-sample approximation bound.} We can derive a finite-sample bound on the approximation error of the spaCRT relative to the dCRT as a corollary of Theorem~\ref{thm:conditional-lr}.

\begin{corollary}[Finite-sample \spacrt{} approximation error bound]
\label{cor:new-finite-spacrt}\label{thm:spacrt-approximation}
Suppose the predictor family $f_X(\cdot;\theta_x)$ has support of diameter at most
$D_0<\infty$, uniformly over $\theta_x$. Define
\[
 A_{kn}=\frac1n\sum_{i=1}^n
 v_X(\widehat\theta_{x,i})
 |Y_i-\widehat\mu_{y,i}|^k,\qquad k=2,3.
\]
Then the conclusions of
Theorem~\ref{thm:conditional-lr} apply, under its stated event conditions, with
\[
 B_n=\frac{D_0A_{3n}}{A_{2n}^{3/2}},
 \qquad
 Z_n=\frac{|S_n|}{\sqrt{nA_{2n}}},
 \qquad
 D_n=\frac{D_0|S_n|}{nA_{2n}}\max_{1\le i\le n}|Y_i-\widehat\mu_{y,i}|.
\]
\end{corollary}
While stated for compactly supported predictors, this result extends to unbounded predictors via Theorem~\ref{thm:conditional-lr}; we omit this extension for brevity. $nA_{2n}$ is the total resampling variance, while $A_{3n}/A_{2n}^{3/2}$ is analogous to a standardized third-moment factor. Larger values give a weaker accuracy guarantee at the same sample size and tail depth, reflecting how heterogeneous response residuals and sparse fitted predictor distributions can make accurate approximation more demanding. The bound also worsens as the cutoff moves deeper into the tail; the condition on $D_n$ additionally restricts the largest residual. These empirical averages can be controlled using moment bounds and guarantees on nuisance-estimation error, as demonstrated in the CRISPR setting (Section~\ref{sec:headline-bnb}) and with high-dimensional nuisance fits (Supplementary Section~\ref{sec:spacrt-high-dimensional}). 

Corollary~\ref{cor:new-finite-spacrt} concerns only the quality of the approximation of the dCRT and therefore requires no correctness assumption on the nuisance fits. The complementary statement of Type-I error control of the dCRT has already been established \citep{Niu2022a} under doubly robust conditions on the estimators $\widehat \theta_{x,i}$ and $\widehat \mu_{y,i}$. However, this result assumed that the average per-observation resampling variance is bounded away from zero asymptotically, precluding sparse regimes in which this variance vanishes. We extend this result to accommodate this sparse regime (Theorem~\ref{thm:spacrt-validity} in Supplementary Section~\ref{sec:proof-spacrt-validity}).

\subsection{Single-cell CRISPR screens: theory}
\label{sec:headline-bnb}

Next, we specialize our theory to single-cell CRISPR screens. Recall that these experiments measure how genetic perturbations alter gene expression \citep{Dixit2016}, and that the dCRT underpins a leading method for analyzing these data \citep{Katsevich2020c}. In this application, $X \in \{0, 1\}$ represents a binary perturbation indicator (predictor), $Y \in \N$ represents a single-cell gene expression (response), and $\bm Z=(1,\bm U^{\T})^{\T}\in\R^d$ is a set of covariates in fixed dimension $d$. Following standard modeling choices in this literature, we model $X|\bm Z$ via logistic regression and $Y|X,\bm Z$ as a negative binomial (NB) regression:
\begin{align}
 X_i\mid \bm Z_i
 &\sim\Ber
 \{\expit(\alpha_{n,x}+\bm U_i^{\T}\bm{\beta}_{n,x})\},\label{eq:new-bnb-x}\\
 Y_i\mid X_i,\bm Z_i
 &\sim\NB
 \{\exp(\alpha_{n,y}+\rho_nX_i+\bm U_i^{\T}\bm{\beta}_{n,y}),\eta_{n,y}\},
 \label{eq:new-bnb-y}
\end{align}
where $\NB(\mu,\eta)$ has mean $\mu$, variance $\mu(1+\eta\mu)$, and dispersion $\eta\geq0$. Under this model, the conditional independence null hypothesis reduces to $H_0: \rho_n = 0$. We consider dCRT and spaCRT where the predictor model is fit by logistic regression and the response model by NB regression on $\bm Z$ alone (setting $\rho=0$), both via maximum likelihood, with the simplification that dispersion $\eta_{n,y}$ is known. We leave the case of estimated dispersion (what is done in practice) to future work.

To translate Corollary~\ref{cor:new-finite-spacrt} into an in-probability bound over the data distribution, we require the following regularity conditions: 
\begin{equation}\label{eq:crispr-regularity-conditions}
\begin{aligned}
&\|\bm U_i\|_2\le B \quad\text{a.s.},\qquad
\lambda_{\min}\{\E(\bm Z_i\bm Z_i^\top)\}\ge\lambda;\\
&\alpha_{n,x}\vee\alpha_{n,y}\le B,\qquad
\|\bm\beta_{n,x}\|_2\vee\|\bm\beta_{n,y}\|_2\le B,\qquad
|\rho_n|\le B.
\end{aligned}
\end{equation}
The quantities controlling SPA error are the baseline perturbation prevalence and gene expression,
\[
  \pi_{n,x}=\expit(\alpha_{n,x}),
  \qquad
  \mu_{n,y}=e^{\alpha_{n,y}}.
\]
Under~\eqref{eq:crispr-regularity-conditions}, with sufficiently rapid information accumulation and sufficiently small departures from the null, the quantities in Corollary~\ref{cor:new-finite-spacrt} satisfy
\[
 \frac{B_n}{\sqrt n}=O_{\P_n}\!\left(\sqrt{\frac{1+\eta_{n,y}\mu_{n,y}}{n\pi_{n,x}\mu_{n,y}}}\right),
 \qquad
 Z_n=O_{\P_n}\!\left(1+|\rho_n|\sqrt{\frac{n\pi_{n,x}\mu_{n,y}}{1+\eta_{n,y}\mu_{n,y}}}\right).
\]
Analogously to our finding in Section~\ref{sec:headline-fastSPA}, the key quantity $n\pi_{n,x}\mu_{n,y}/(1+\eta_{n,y}\mu_{n,y})$ is of the same order as the efficient information for $\rho_n$ under $\rho_n = 0$ in the NB model~\eqref{eq:new-bnb-y}. The following theorem makes these regime requirements precise.
\begin{theorem}[Bernoulli--negative-binomial \spacrt{}: rate and sharpness]
\label{thm:new-bnb}\label{thm:spacrt-rate}\label{thm:headline-bnb}
For fixed $B<\infty$, $\lambda>0$, $d$, and deterministic $\delta_n\to\infty$, define
\begin{equation}\label{eq:crispr-class}
 \PBNB_n=\left\{\P_n\in\mathcal P_n:\;
 \eqref{eq:new-bnb-x}-\eqref{eq:crispr-regularity-conditions}\text{ hold};\;
 \frac{n\pi_{n,x}\mu_{n,y}}{1+\eta_{n,y}\mu_{n,y}}\ge\delta_n\log^4n,\; |\rho_n|\log n\le\delta_n^{-1}\right\}.
\end{equation}
Its null slice is $\PBNBnull_n=\{\P_n\in\PBNB_n:\rho_n=0\}$.
Uniformly over $\P_n\in\PBNB_n$, both nuisance maximum likelihood fits and the
saddlepoint exist with probability tending to one, and
\begin{equation}\label{eq:headline-bnb-rate}
 \frac{p_{\infty,n}^{\dCRT}}{p_{\mathrm{SPA},n}^{\dCRT}}-1
 =\OP{\PBNB}\!\left(\sqrt{\frac{1+\eta_{n,y}\mu_{n,y}}{n\pi_{n,x}\mu_{n,y}}}+|\rho_n|\right).
\end{equation}
Moreover, for suitable fixed $B$, $\lambda$, and $d$ and a suitable
choice of $\delta_n \rightarrow \infty$, there is a null sequence $\P_n\in\PBNBnull_n$ with
$\eta_{n,y}=0$ on which
\[
  \left|\frac{p_{\infty,n}^{\dCRT}}{p_{\mathrm{SPA},n}^{\dCRT}}-1\right|
  \asymp_{\P_n}(n\pi_{n,x}\mu_{n,y})^{-1/2}.
\]
Under the null, the fitted \dCRT{} target and the \spacrt{} are
uniformly asymptotically valid:
\begin{equation}\label{eq:bnb-null-uniformity}
 \sup_{\P_n\in\PBNBnull_n}\sup_{t\in[0,1]}
 |\P_n(p_{\infty,n}^{\dCRT}\le t)-t|\longrightarrow0,
 \quad
 \sup_{\P_n\in\PBNBnull_n}\sup_{t\in[0,1]}
 |\P_n(p_{\mathrm{SPA},n}^{\dCRT}\le t)-t|\longrightarrow0.
\end{equation}
\end{theorem}

As in the GWAS setting, the result permits both perturbation prevalence and
mean gene expression to vanish, provided the dispersion-adjusted overlap
grows sufficiently quickly. Here, the perturbation-expression overlap
$n\pi_{n,x}\mu_{n,y}$ replaces $n\pi_{n,x}\pi_{n,y}$, while NB
dispersion weakens the relative-error guarantee. The $\eta_{n,y} = 0$ lower-bound
example establishes sharpness of the inverse-square-root dependence on
overlap, but not of the dispersion factor.

\subsection{Single-cell CRISPR screens: simulation} \label{sec:simulation}

We use simulations to examine the approximation accuracy and Type-I error control studied in Theorem~\ref{thm:new-bnb} and to compare spaCRT with alternative methods. We estimate dispersion and use a Gaussian covariate, evaluating the practical procedure beyond the theorem's known-dispersion and bounded-covariate assumptions. While these screens involve testing many perturbation--gene pairs, we simulate individual pairs and apply left- and right-sided tests at nominal level $0.005$ (see Supplementary Section~\ref{sec:crispr-sensitivity} for left-sided SPAs), using a stringent threshold to probe multiplicity-relevant tail behavior.

As our data-generating model, we specialize~\eqref{eq:new-bnb-x} and~\eqref{eq:new-bnb-y} to a one-dimensional covariate $\smash{U_i\iidsim N(0,1)}$ and unit covariate slopes in both models, with $n = 5000$ cells. The intercepts $\alpha_{n,x}$ and $\alpha_{n,y}$ control perturbation and expression sparsity, with values chosen to resemble sparse single-cell CRISPR screen data; $\rho_n$ controls the association, with $\rho_n=0$ under the null. We compare the \spacrt{} and finite-resampling \dCRT{} ($M=10{,}000$) with the normality-based generalized covariance measure (GCM) test \citep{Shah2018} for conditional independence and the negative-binomial score test. The \spacrt{}, \dCRT{}, and GCM use the same fitted logistic and negative-binomial regressions for $X\mid\bm Z$ and $Y\mid\bm Z$, respectively. For each combination of model parameters, we run each method on 10,000 independent replicates of the data.

\begin{figure*}[!ht]
\centering
\includegraphics[width=\textwidth]{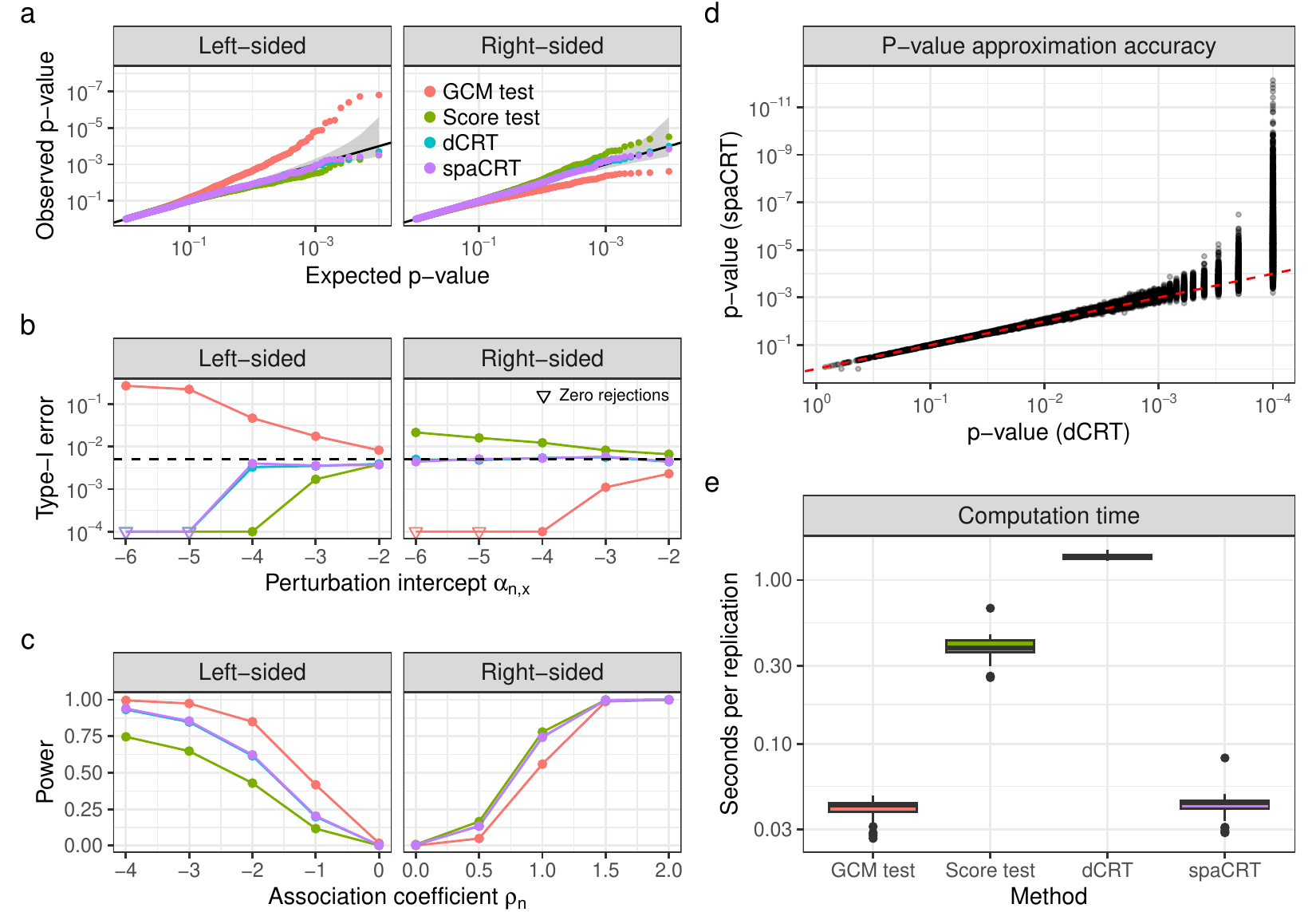}
\caption{Single-cell CRISPR simulation with dispersion $\eta_{n,y}=1$,
expression intercept $\alpha_{n,y}=-5$, sample size $n=5{,}000$, and nominal level $0.005$.
(a) Null $p$-value QQ-plots for $\alpha_{n,x}=-3$.
(b) Type-I error as perturbations become sparser, varying $\alpha_{n,x}$.
(c) Power as a function of $\rho_n$ at $\alpha_{n,x}=-3$.
(d) Paired right-sided \spacrt{} and \dCRT{} $p$-values at $\alpha_{n,x}=-3$
and $\rho_n=1$. (e) Mean computation time per dataset across parameter settings.}
\label{fig:new-simulation}
\end{figure*}

Figure~\ref{fig:new-simulation} reports calibration, power, approximation accuracy, and runtime with $\eta_{n,y}=1$ and $\alpha_{n,y}=-5$. Supplementary Figures~\ref{fig:crispr-sensitivity} and~\ref{fig:crispr-relative-error-sensitivity} examine sensitivity of calibration and power, and relative discrepancies between SPA and Monte Carlo $p$-values, respectively.

\paragraph{spaCRT and dCRT have the most consistent Type-I error control.} As seen in Figure~\ref{fig:new-simulation}(a,b) and Supplementary Figure~\ref{fig:crispr-sensitivity}(a,b), dCRT and spaCRT generally achieve rejection rates close to the nominal level, becoming conservative in very sparse settings. On the other hand, the GCM test tends to be liberal for left-sided tests and conservative for right-sided tests, with the score test following the opposite pattern. 

\paragraph{spaCRT and dCRT are the most powerful among methods controlling Type-I error.} This is seen from Figure~\ref{fig:new-simulation}(c) and Supplementary Figure~\ref{fig:crispr-sensitivity}(c,d), after excluding the GCM test and score test from the left-sided and right-sided comparisons, respectively, due to Type-I error inflation. 

\paragraph{spaCRT accurately approximates dCRT $p$-values.} Figure~\ref{fig:new-simulation}(d) shows that spaCRT $p$-values offer excellent approximations to dCRT $p$-values when the latter are at least $10^{-3}$; below this range, the Monte Carlo accuracy of dCRT $p$-values is insufficient to assess the approximation. This panel underscores the ability of spaCRT to compute very small $p$-values without the Monte Carlo resolution limit, while dCRT $p$-values are at least $1/(M+1)$, where $M$ is the number of resamples. The approximation rate bound~\eqref{eq:headline-bnb-rate} from Theorem~\ref{thm:new-bnb} gives a weaker accuracy guarantee with increasing signal strength, dispersion, and predictor and response sparsities; Supplementary Figure~\ref{fig:crispr-relative-error-sensitivity} shows discrepancies broadly consistent with these qualitative trends, although these discrepancies also include Monte Carlo error in the dCRT $p$-values.

\paragraph{spaCRT substantially accelerates dCRT.} Figure~\ref{fig:new-simulation}(e) shows that the \spacrt{} is approximately as fast as the GCM test, and faster than the score test and dCRT. Ratios of mean dCRT to spaCRT runtimes within each setting give accelerations of 17--52-fold, with a median of 32-fold.

\subsection{Single-cell CRISPR screens: application}
\label{sec:crispr-application}

We apply the \spacrt{} to the single-cell CRISPR screen of \citet{Gasperini2019a}. After preprocessing, the data contain $207{,}324$ cells, $13{,}135$ measured genes, perturbations targeting $6{,}105$ regulatory elements, and six cell-level covariates. Many genes are expressed in about $1\%$ of cells, and individual perturbations are still rarer, exemplifying the challenges motivating this work. We work with perturbation-gene pairs whose association status is known. In particular, we use 51 non-targeting perturbations as negative controls. Pairing each with $3{,}000$ randomly selected genes gives $153{,}000$ candidate null pairs, of which $148{,}570$ remain after \texttt{sceptre}'s quality-control filtering and are used for all four methods. We use 754 perturbations targeting genes as positive controls, pairing each perturbation with its known target gene.

\begin{figure*}[!ht]
\centering
\includegraphics[width=\textwidth]{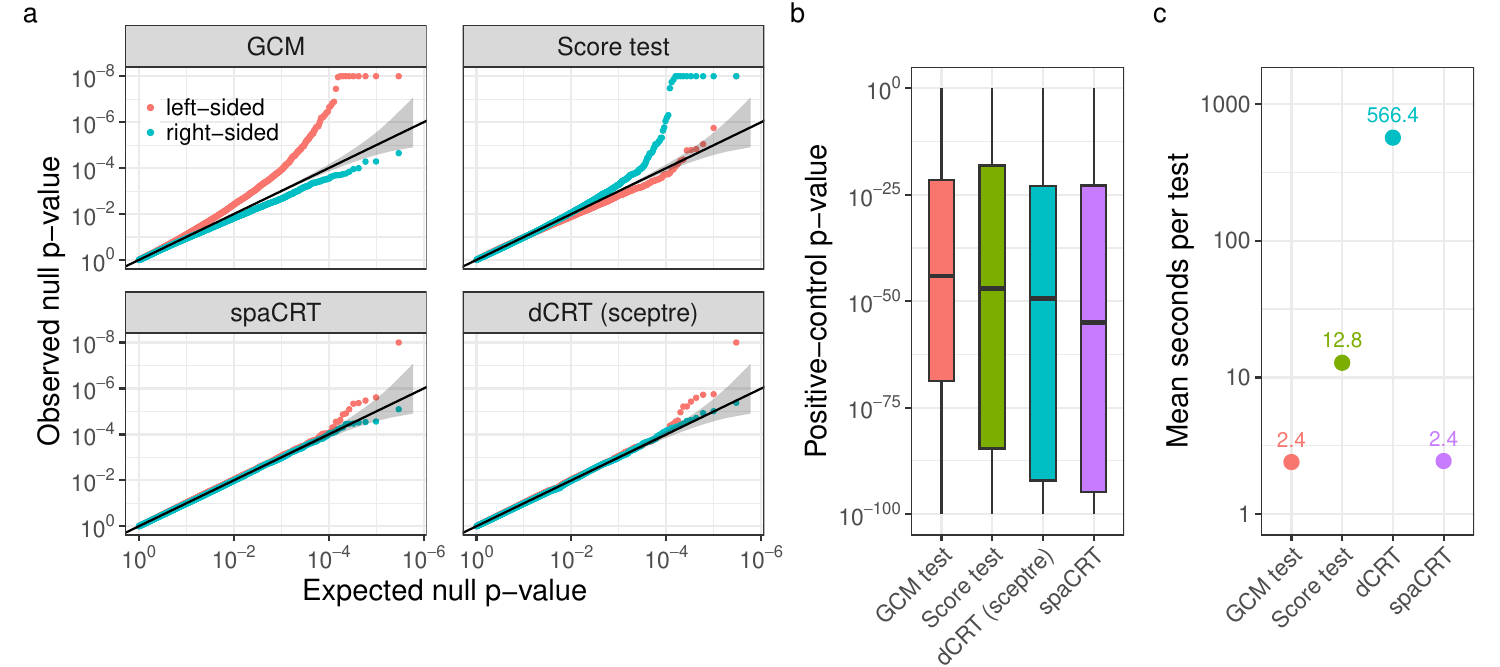}
\caption{Gasperini data analysis. (a) Null $p$-value
QQ-plots for negative-control perturbation--gene pairs. (b) Left-sided
$p$-values for positive-control pairs. (c) Mean computation times per test
over 102 perturbation--gene pairs; labels give times in seconds. Panel (c)
benchmarks direct Monte Carlo dCRT with $100{,}000$ resamples (implemented
in R), rather than dCRT (\texttt{sceptre}), which uses parametric tail
extrapolation and is shown in panels (a) and (b).}
\label{fig:new-real-data}
\end{figure*}

\paragraph{Methods compared.} We compare \spacrt{}, GCM, the negative-binomial score test, and dCRT (Figure~\ref{fig:new-real-data}). Because direct Monte Carlo dCRT is infeasible at full scale, statistical comparisons use the optimized \texttt{sceptre} implementation \citep{Katsevich2020c,Barry2024}. It implements performance-critical steps in C++ and extrapolates a fitted skew-normal tail from a few hundred resampled statistics, a heuristic permitting very small reported $p$-values. Timing comparisons instead use direct Monte Carlo dCRT on a subset of pairs, comparing theoretically supported methods implemented in R. Negative controls use both tails; positive controls use the left tail because CRISPR perturbations are expected to decrease target-gene expression.

\paragraph{spaCRT preserves dCRT's statistical performance.} As in our simulations (Section~\ref{sec:simulation}), spaCRT and dCRT (\texttt{sceptre}) are well-calibrated, while the GCM and score tests show the same direction-dependent miscalibration. The extent of null $p$-value inflation is more striking in the real data analysis than in the simulation, likely because we examine more null $p$-values, pushing further into the tail of the distribution. Power is somewhat more challenging to assess, because all four procedures return very small $p$-values on positive controls, and therefore essentially saturate this strong-signal comparison. Nevertheless, spaCRT does not appear to lose any power compared to the other methods considered.

\paragraph{spaCRT substantially reduces the cost of resampling.} We benchmarked the R-based dCRT implementation on 102 perturbation--gene pairs. We use $M=100{,}000$ resamples, giving roughly $10\%$ relative Monte Carlo standard error for $p$-values near $10^{-3}$, the approximate Benjamini--Hochberg cutoff implied by the original study's discovery yield and multiplicity. The computational advantage of spaCRT over dCRT is even more dramatic on the real data, with an acceleration of approximately 250-fold. For the roughly $85{,}000$ tests in the original analysis, spaCRT reduces the projected computation time from 1.5 CPU-years to 2.4 CPU-days.

\section{Future directions} \label{sec:future-directions}

\paragraph{Quantifying resampling calibration.} Existing validity arguments for plug-in resampling, including those used here, establish fixed-level asymptotic Type-I error control but do not explain its empirical calibration advantage over normality-based tests. Quantifying that advantage requires sharper comparisons between the fitted resampling law and the sampling distribution of the observed statistic. Such results would complement our approximation guarantees and help determine how far into the tail the resulting tests remain calibrated.

\paragraph{Extending the resampling framework.} The independent-sum theory may apply beyond the observation-level resampling procedures studied here. For example, some SPA methods in statistical genetics preserve dependence within families \citep{xuSPAGRMEffectivelyControlling2025}. When family blocks are independent, Theorem~\ref{thm:conditional-lr} could be applied to their contributions, provided its conditions can be verified. A distinct direction concerns procedures that introduce resample-dependent studentization, as in the estimating function bootstrap \citep{huEstimatingFunctionBootstrap2000}. Unlike weights computed from the observed data and held fixed during resampling,
resample-dependent normalization generally requires analysis beyond a scalar additive statistic. Developing suitable self-normalized extensions could broaden the reach of the framework. %

\section*{Declaration of the use of generative AI and AI-assisted technologies}

During the preparation of this work, the authors used OpenAI Codex, Claude Code, and Refine to assist with literature search, language revision, drafting proofs of the lower bound statements in Theorems~\ref{thm:headline-fastSPA} and~\ref{thm:headline-bnb}, and proof checking. After using these tools, the authors reviewed and edited any generated content as necessary and take full responsibility for the content of the publication.

\section*{Acknowledgement}

We thank an editor, associate editor, two referees, and John Kolassa, who provided valuable feedback on earlier versions of this paper. We thank the Wharton research computing team for computing assistance. This work was supported by NSF DMS-2113072 and NSF DMS-2310654.

\section*{Code availability}

The spaCRT R package is available at \url{https://github.com/Katsevich-Lab/spacrt}. Code to reproduce all numerical analyses is available at \url{https://github.com/Katsevich-Lab/spacrt-manuscript}.

\clearpage
\begingroup
\renewcommand{\E}{\mathbb E}
\renewcommand{\Var}{\operatorname{Var}}
\renewcommand{\Cov}{\operatorname{Cov}}
\renewcommand{\R}{\mathbb R}
\renewcommand{\cF}{\mathcal F}
\renewcommand{\cD}{\mathcal D}
\renewcommand{\indicator}{\mathbbm 1}
\renewcommand{\norm}[1]{\left\lVert #1\right\rVert}
\renewcommand{\abs}[1]{\left|#1\right|}
\renewcommand{\diag}{\operatorname{diag}}
\renewcommand{\expit}{\operatorname{expit}}
\renewcommand{\logit}{\operatorname{logit}}
\renewcommand{\LR}{\ensuremath{\mathrm{LR}}}
\renewcommand{\Rob}{\ensuremath{\mathrm R}}
\renewcommand{\RR}{\textnormal{RR}}
\renewcommand{\dCRT}{\textnormal{dCRT}}
\renewcommand{\spacrt}{\textnormal{spaCRT}}
\renewcommand{\fastspa}{\textnormal{fastSPA}}
\renewcommand{\Ber}{\operatorname{Bernoulli}}
\renewcommand{\NB}{\operatorname{NB}}
\renewcommand{\Unif}{\operatorname{Unif}}
\renewcommand{\Phibar}{\overline\Phi}
\renewcommand{\eps}{\varepsilon}
\renewcommand{\RE}{\operatorname{RE}}
\renewcommand{\op}{\mathrm{op}}
\renewcommand{\convp}{\overset{p}{\longrightarrow}}
\renewcommand{\toD}{\xrightarrow{d}}
\renewcommand{\convd}{\xrightarrow{d}}
\renewcommand{\Sx}{S_x}
\renewcommand{\Sy}{S_y}
\renewcommand{\mTwo}{m_2}
\renewcommand{\mThree}{m_3}
\renewcommand{\OP}[1]{O_{#1}}
\renewcommand{\PRR}{\mathcal P^{\mathrm{GLM\text{-}RR}}}
\renewcommand{\PLL}{\mathcal P^{\mathrm{LL}}}
\renewcommand{\PBNB}{\mathcal P^{\mathrm{CRISPR}}}
\renewcommand{\PRRnull}{\mathcal P^{\mathrm{GLM\text{-}RR},0}}
\renewcommand{\PLLnull}{\mathcal P^{\mathrm{LL},0}}
\renewcommand{\PBNBnull}{\mathcal P^{\mathrm{CRISPR},0}}
\renewcommand{\Pnullgen}{\mathcal P^{0}}
\renewcommand{\V}{\operatorname{Var}}
\renewcommand{\Q}{\mathbb Q}
\renewcommand{\Z}{\mathbb Z}
\renewcommand{\N}{\mathbb N}
\renewcommand{\independent}{{\perp\!\!\!\perp}}
\renewcommand{\iidsim}{\stackrel{\mathrm{i.i.d.}}{\sim}}
\renewcommand{\indsim}{\stackrel{\mathrm{ind}}{\sim}}
\renewcommand{\convdp}{\overset {d,p} \longrightarrow}
\renewcommand{\convpp}{\overset {p,p} \longrightarrow}
\renewcommand{\bgam}{\bm\gamma}
\renewcommand{\bgamhat}{\widehat{\bm\gamma}}
\def\T{\top}   
\def\sgn{\operatorname{sgn}}
\renewcommand{\indep}{\perp\!\!\!\perp}
\renewcommand{\obs}{\bm O_{1:n}}
\renewcommand{\Bin}{\operatorname{Bin}}
\renewcommand{\Pois}{\operatorname{Pois}}
\renewcommand{\pn}{p_{\infty,n}^{\mathrm{GLM\text{-}RR}}}
\renewcommand{\ps}{p_{\mathrm{SPA},n}^{\mathrm{GLM\text{-}RR}}}
\renewcommand{\gam}{\bm\gamma}
\renewcommand{\gh}{\widehat{\bm\gamma}_n}
\renewcommand{\gb}{\overline{\bm\gamma}}
\renewcommand{\ab}{\overline{\bm\alpha}}
\renewcommand{\ah}{\widehat{\bm\alpha}_n}
\renewcommand{\exk}{\withtwosubscripts{\widetilde X}}
\allowdisplaybreaks
\setcounter{equation}{0}
\setcounter{figure}{0}
\setcounter{table}{0}
\setcounter{theorem}{0}
\setcounter{proposition}{0}
\setcounter{lemma}{0}
\setcounter{corollary}{0}
\setcounter{assumption}{0}
\setcounter{example}{0}
\setcounter{remark}{0}
\setcounter{definition}{0}
\renewcommand{\theequation}{S\arabic{equation}}
\renewcommand{\thefigure}{S\arabic{figure}}
\renewcommand{\thetable}{S\arabic{table}}
\renewcommand{\thetheorem}{S\arabic{theorem}}
\renewcommand{\theproposition}{S\arabic{proposition}}
\renewcommand{\thelemma}{S\arabic{lemma}}
\renewcommand{\thecorollary}{S\arabic{corollary}}
\renewcommand{\theassumption}{S\arabic{assumption}}
\renewcommand{\theexample}{S\arabic{example}}
\renewcommand{\theremark}{S\arabic{remark}}
\renewcommand{\thedefinition}{S\arabic{definition}}

\renewcommand{\theHequation}{supp.\arabic{equation}}
\renewcommand{\theHfigure}{supp.\arabic{figure}}
\renewcommand{\theHtable}{supp.\arabic{table}}
\expandafter\def\csname theHtheorem\endcsname{supp.\arabic{theorem}}
\expandafter\def\csname theHproposition\endcsname{supp.\arabic{proposition}}
\expandafter\def\csname theHlemma\endcsname{supp.\arabic{lemma}}
\expandafter\def\csname theHcorollary\endcsname{supp.\arabic{corollary}}
\expandafter\def\csname theHassumption\endcsname{supp.\arabic{assumption}}
\expandafter\def\csname theHexample\endcsname{supp.\arabic{example}}
\expandafter\def\csname theHremark\endcsname{supp.\arabic{remark}}
\expandafter\def\csname theHdefinition\endcsname{supp.\arabic{definition}}
\section*{Supplementary material}
\addtocontents{toc}{\protect\setcounter{tocdepth}{2}}
\makeatletter
{\def\@tocrmarg{4em}\def\@pnumwidth{2em}
\@starttoc{toc}}
\makeatother
\clearpage
\appendix
\counterwithout{equation}{section}
\renewcommand{\theequation}{S\arabic{equation}}

\section{Resampling representations of statistical-genetics SPAs}
\label{sec:genetics-resampling-mappings}

The statistical-genetics SPAs in Table~\ref{tab:genetics-resampling-mappings}
span binary, ordinal, count, survival and longitudinal outcomes. They can be
represented through three resampling mechanisms: drawing responses from a
fitted law, drawing residual contributions from an empirical law, or drawing
genotypes from a fitted law. In each case, the reference statistic is a sum
of independent, centered contributions conditional on the observed data,
with all fitted quantities held fixed. The construction below makes this
shared structure explicit; Sections~\ref{sec:genetics-response-resampling}--\ref{sec:genetics-predictor-resampling}
specify the methods within each mechanism.

\begin{samepage}
\paragraph{Common construction and inferential target.}
For each observation $i$, let $W_i$ denote the observed quantity to be resampled,
$\widetilde\Q_i$ its conditional resampling law, and $a_i$ its fixed weight.
A positive fitted factor $\widehat c$ allows a variance adjustment, with
$\widehat c=1$ when no adjustment is used. Set
\begin{equation}\label{eq:genetics-common-construction}
\begin{aligned}
 \widehat m_i&=\widetilde\E(\tilde W_i\mid\obs), &
 s_i(O_i,\widehat\gamma)&=a_i(W_i-\widehat m_i), &
 S_n&=\sum_i s_i(O_i,\widehat\gamma),\\
 \tilde W_i\mid\obs&\indsim\widetilde\Q_i, &
 \tilde s_i&=\widehat c\,a_i(\tilde W_i-\widehat m_i), &
 \widetilde S_n&=\sum_i\tilde s_i.
\end{aligned}
\end{equation}
\end{samepage}
All fitted quantities and weights stay fixed during resampling. The
independent draws and centering give the conditional resampling structure
of Algorithm~\ref{alg:plug-in-resampling}. For a right-tailed test, the
inferential target is the conditional tail probability in~\eqref{eq:infinite-M},
\[
 p_{\infty,n}(\obs)
 =\widetilde\P\bigl(\widetilde S_n\ge S_n\mid\obs\bigr).
\]
These representations identify
the resampling targets approximated by the SPAs; approximation accuracy
and Type-I error control require their respective assumptions in the main
text. Table~\ref{tab:genetics-resampling-mappings} displays the three
mechanisms using the same notation throughout.

\begingroup
\small
\setlength{\tabcolsep}{3pt}
\renewcommand{\arraystretch}{1.1}
\setlength{\LTcapwidth}{\textwidth}
\begin{longtable}{@{}>{\raggedright\arraybackslash}p{0.16\textwidth}
                  >{\raggedright\arraybackslash}p{0.22\textwidth}
                  >{\raggedright\arraybackslash}p{0.23\textwidth}
                  >{\raggedright\arraybackslash}p{0.18\textwidth}
                  >{\raggedright\arraybackslash}p{\dimexpr0.21\textwidth-24pt\relax}@{}}
\caption{Statistical-genetics SPAs grouped by their resampling mechanism.
Each row specifies $W_i$, $\widetilde\Q_i$ and $a_i$ in~\eqref{eq:genetics-common-construction}.
The index $i$ refers to cells in SAIGE-QTL and subjects otherwise.
For TrajGWAS, we present the mean test.
Definitions of the component transformations and weights follow the table.}
\label{tab:genetics-resampling-mappings}\\
\toprule
Method & Authors & Resampled quantity $W_i$ & Resampling law $\widetilde\Q_i$ & Fixed weight $a_i$ \\
\midrule
\endfirsthead
\multicolumn{5}{l}{\textit{Table~\thetable\ (continued)}}\\
\toprule
Method & Authors & Resampled quantity $W_i$ & Resampling law $\widetilde\Q_i$ & Fixed weight $a_i$ \\
\midrule
\endhead
\midrule
\multicolumn{5}{r}{\textit{Continued on next page}}\\
\endfoot
\bottomrule
\endlastfoot
\multicolumn{5}{@{}l}{\textit{Response resampling}}\\*[3pt]
fastSPA
& \citet{Dey2017}
& Binary phenotype $Y_i$
& $\operatorname{Bernoulli}(\widehat\mu_i)$
& $r_{x,i}\equiv x_i-\bm z_i^{\mathsf T}\widehat{\bm\alpha}$\\[3pt]
SAIGE
& \citet{Zhou2018}
& Binary phenotype $Y_i$
& $\operatorname{Bernoulli}(\widehat\mu_i)$
& $r_{x,i}$\\[3pt]
SPAGE
& \citet{biFastAccurateMethod2019}
& Binary phenotype $Y_i$
& $\operatorname{Bernoulli}(\widehat\mu_i)$
& $r_{xE,i}-\lambda r_{x,i}$\\[3pt]
REGENIE
& \citet{mbatchouComputationallyEfficientWholegenome2021}
& Binary phenotype $Y_i$
& $\operatorname{Bernoulli}(\widehat\mu_i)$
& $r_{x,i}$\\[3pt]
fastGWA-GLMM
& \citet{Jiang_2021}
& Binary phenotype $Y_i$
& $\operatorname{Bernoulli}(\widehat\mu_i)$
& $r_{x,i}$\\[3pt]
POLMM
& \citet{biEfficientMixedModel2021}
& $d_i(Y_i)$ from~\eqref{eq:genetics-polmm-coding}
& $\sum_{k=1}^J\widehat\pi_{i,k}\delta_{d_i(k)}$
& $r_{x,i}$\\[3pt]
GATE
& \citet{deyEfficientAccurateFrailty2022}
& Event indicator $Y_i$
& $\operatorname{Poisson}(\widehat\mu_i)$
& $r_{x,i}$\\[3pt]
SAIGE-QTL
& \citet{zhouEfficientAccurateMixed2024}
& Expression count $Y_i$
& $\operatorname{Poisson}(\widehat\mu_i)$
& $r_{x,i}$\\[3pt]
\midrule
\multicolumn{5}{@{}l}{\textit{Empirical-residual resampling}}\\*[3pt]
SPACox
& \citet{biFastAccurateMethod2020}
& $Y_i-\widehat\mu_i$
& $\widetilde\Q_{\mathrm{emp}}$
& $r_{x,i}$\\[3pt]
TrajGWAS
& \citet{Ko_2022}
& $W_i$ from~\eqref{eq:genetics-traj-components}
& $\widetilde\Q_{\mathrm{emp}}$
& $x_i^{\mathrm{norm}}$ from~\eqref{eq:genetics-traj-weight}\\[3pt]
SPARE
& \citet{hofFastAccurateRecurrent2023}
& $Y_i-\widehat\mu_i$
& $\widetilde\Q_{\mathrm{emp}}$
& $\displaystyle\frac{x_i-\bar x}{\|\bm x-\bar x\bm 1\|_2^2}$\\[3pt]
\midrule
\multicolumn{5}{@{}l}{\textit{Predictor resampling}}\\*[3pt]
SPAmix
& \citet{maSPAmixScalableAccurate2025}
& Genotype $X_i$
& $\operatorname{Binomial}(2,\widehat q_i)$
& $Y_i$ residual \\
\end{longtable}
\endgroup

\paragraph{Default weights.}
Unless specified otherwise, the fixed weight is the genotype residual
$a_i=r_{x,i}\equiv x_i-\bm z_i^{\mathsf T}\widehat{\bm\alpha}$,
where $x_i$ is the observed genotype and $\widehat{\bm\alpha}$ is obtained by
variance-weighted regression on covariates $\bm z_i$, including an intercept.
This is the projection in~\eqref{eq:new-glm-projection} of the main text,
with diagonal weights $D_{ii}=\operatorname{Var}_{\widetilde\Q_i}(W)$.
All fits and projections use each method's null model.

\paragraph{Variance adjustment: $\widehat c$.}
If $\widehat V_n$ is the estimated score variance used by the method, the
adjustment is
\begin{equation}\label{eq:genetics-variance-adjustment}
 \widehat c
 =\left\{\frac{\widehat V_n}
 {\sum_i a_i^2\operatorname{Var}_{\widetilde\Q_i}(W)}\right\}^{1/2},
 \qquad
 \widetilde\Var(\widetilde S_n\mid\obs)=\widehat V_n,
\end{equation}
with both variances finite and positive. The factor $\widehat c$ stays fixed during resampling
and rescales only the reference sum. Equivalently, the SPA for the unscaled
sum $\sum_i a_i(\tilde W_i-\widehat m_i)$ is evaluated at $S_n/\widehat c$.
For brevity, we omit method-specific details of the variance adjustment $\widehat c$, referring readers to the original papers.

\subsection{Response resampling}
\label{sec:genetics-response-resampling}

For the Bernoulli rows of Table~\ref{tab:genetics-resampling-mappings},
$W_i=Y_i$ is the binary phenotype and $\widehat\mu_i$ is its fitted mean;
in SAIGE-QTL, $Y_i$ is an expression count and the working law is Poisson.
The following methods use a survival working law, an ordinal transformation,
or interaction weights within the same response-resampling construction.

\paragraph{GATE.}
GATE \citep{deyEfficientAccurateFrailty2022} illustrates how a working
resampling law can represent a survival SPA. An event is the endpoint
chosen for the analysis: the first recorded diagnosis of the disease under
study in GATE's disease analyses, or death in its lifespan analysis. Let
$T_i^\star$ be subject $i$'s event time and $C_i$ the right-censoring time,
measured on the same time scale. The observed survival data are
\[
 T_i=\min\{T_i^\star,C_i\},\qquad
 Y_i=\indicator\{T_i^\star\le C_i\},\qquad W_i=Y_i.
\]
Thus $Y_i=1$ means the endpoint was observed at $T_i$, while $Y_i=0$ means
follow-up ended at $T_i=C_i$ without observing it. This $Y_i$ is the
indicator denoted $\delta_i$ in the GATE paper. The replacement of this
binary component is a Poisson draw whose mean is the fitted cumulative
hazard.
GATE first fits a Gaussian frailty model under the null. Let $\bm u_i$
contain the survival covariates without an intercept. Its null hazard model is
\[
 \lambda_i(s\mid \bm u_i,b_i)
 =\lambda_0(s)\exp\{\bm u_i^{\mathsf T}\bm\beta+b_i\},
 \qquad \bm b\sim\mathcal N(\bm0,\tau\bm\Psi),
\]
where $\bm\Psi$ is the genetic relatedness matrix. With fitted frailty
$\widehat b_i$ and fitted baseline cumulative hazard $\widehat\Lambda_0$,
\begin{equation}\label{eq:genetics-gate-components}
 \widehat\mu_i=\widehat\Lambda_0(T_i)
 \exp\{\bm u_i^{\mathsf T}\widehat{\bm\beta}+\widehat b_i\},
 \qquad S_n^{\mathrm{GATE}}=\sum_i r_{x,i}(Y_i-\widehat\mu_i).
\end{equation}
Here $\bm z_i=(1,\bm u_i^{\mathsf T})^{\mathsf T}$ and
$D_{ii}=\widehat\mu_i$, so $r_{x,i}$ is the
$\widehat\mu_i$-weighted genotype residual, denoted $\widetilde G_i$ by GATE.
The centered cumulant generating function used by its implementation is
that of the independent-Poisson working law
\begin{equation}\label{eq:genetics-gate-working-law}
 \tilde W_i\mid\obs\indsim\operatorname{Poisson}(\widehat\mu_i),
 \qquad \widehat m_i=\widehat\mu_i.
\end{equation}
Thus the observed time enters through $\widehat\mu_i$, which is a
cumulative hazard rather than an event probability. The binary observation
and the potentially larger Poisson replacement have different supports;
the latter identifies the reference law of GATE's implemented SPA.

\paragraph{POLMM.}
POLMM \citep{biEfficientMixedModel2021} incorporates an ordinal phenotype
$Y_i\in\{1,\ldots,J\}$ through the scalar transformation $W_i=d_i(Y_i)$.
It first fits the null proportional-odds logistic mixed model, with the
tested genotype effect set to zero:
\[
 \logit\P_0(Y_i\le k\mid\bm z_i,b_i)
 =\alpha_k-\bm z_i^{\mathsf T}\bm\beta-b_i,
 \qquad k=1,\ldots,J-1,\qquad
 \bm b\sim\mathcal N(\bm0,\tau\bm\Psi).
\]
Here $\alpha_1<\cdots<\alpha_{J-1}$ are cutpoints and $\bm\Psi$ is the
genetic relatedness matrix. POLMM includes an intercept in $\bm z_i$ and
fixes $\alpha_1=0$ for identifiability. Using the fitted cutpoints,
coefficients and subject-specific random effects, define
\begin{equation}\label{eq:genetics-polmm-probabilities}
 \widehat\nu_{i,k}
 =\expit\{\widehat\alpha_k-\bm z_i^{\mathsf T}\widehat{\bm\beta}
             -\widehat b_i\},\quad k=1,\ldots,J-1,
 \qquad
 \widehat\pi_{i,k}=\widehat\nu_{i,k}-\widehat\nu_{i,k-1},
 \quad k=1,\ldots,J,
\end{equation}
where $\expit(t)=(1+e^{-t})^{-1}$,
$\widehat\nu_{i,0}=0$ and $\widehat\nu_{i,J}=1$.
Thus $\widehat\pi_{i,k}$ is the fitted null probability that subject $i$
belongs to category $k$, evaluated at $\widehat b_i$ rather than integrated
over the random-effects distribution. These probabilities and all fitted
quantities are held fixed during resampling. The transformation is
\begin{equation}\label{eq:genetics-polmm-coding}
 d_i(k)=\frac{\widehat\nu_{i,k-1}(1-\widehat\nu_{i,k-1})-
                   \widehat\nu_{i,k}(1-\widehat\nu_{i,k})}{\widehat\pi_{i,k}}
         -\frac{\widehat\nu_{i,J-1}(1-\widehat\nu_{i,J-1})}{\widehat\pi_{i,J}}.
\end{equation}
Resampling the fitted category and applying this transformation gives
$\widetilde\Q_i=\sum_{k=1}^J\widehat\pi_{i,k}\delta_{d_i(k)}$,
where $\delta_w$ is a point mass at $w$.

\paragraph{SPAGE.}
SPAGE \citep{biFastAccurateMethod2019} retains the Bernoulli resampling law
and uses weights for a genotype--environment interaction. If $E_i$ is the
environmental exposure, let $\bm r_{xE}$ be the variance-weighted residual
of $x_iE_i$ on $\bm z_i$, using the same projection as $\bm r_x$. Then
$a_i=r_{xE,i}-\lambda r_{x,i}$, with
$\lambda=(\bm r_{xE}^{\mathsf T}\bm D\bm r_x)/(\bm r_x^{\mathsf T}\bm D\bm r_x)$.

\subsection{Empirical-residual resampling}
\label{sec:genetics-empirical-resampling}

SPACox, TrajGWAS and SPARE draw each $\tilde W_i$ independently with
replacement from the fitted contributions $W_1,\ldots,W_n$. Their common law
and centering are
\[
 \widetilde\Q_{\mathrm{emp}}=\frac1n\sum_{j=1}^n\delta_{W_j},
 \qquad \widehat m_i=\bar W\equiv\frac1n\sum_{j=1}^n W_j.
\]
Here $\delta_w$ denotes a point mass at $w$. The contributions are computed
before resampling; subtracting $\bar W$ centers each empirical draw. SPACox \citep{biFastAccurateMethod2020} and
SPARE \citep{hofFastAccurateRecurrent2023} use
$W_i=Y_i-\widehat\mu_i$, where $Y_i$ is the event indicator in SPACox and
the total event count in SPARE, and $\widehat\mu_i$ is the fitted cumulative
hazard over that subject's at-risk periods. SPACox uses the default weight
$r_{x,i}$, while SPARE uses
$a_i=(x_i-\bar x)/\|\bm x-\bar x\bm1\|_2^2$,
where $\bar x=n^{-1}\sum_{j=1}^n x_j$.

\paragraph{TrajGWAS.}
We illustrate the mean test of TrajGWAS \citep{Ko_2022}, which reduces each
subject's repeated outcomes to a scalar contribution and applies the same
empirical resampling construction. Let $\bm Y_i$ contain the repeated
outcomes, with fitted mean $\widehat{\bm\mu}_i$, and let $\bm V_i^{(0)}$
be the initial covariance estimate used for weighting in the null-model fit.
With $\bm1$ denoting a vector of ones, the contribution is
\begin{equation}\label{eq:genetics-traj-components}
 W_i=\bm1^{\mathsf T}[\bm V_i^{(0)}]^{-1}
 (\bm Y_i-\widehat{\bm\mu}_i).
\end{equation}
The resampling law is $\widetilde\Q_{\mathrm{emp}}$, with $\widehat c=1$.
Here $x_i$ is the genotype dosage after any mean imputation. Define
\[
 \bar W\equiv\frac1n\sum_{j=1}^n W_j,\qquad
 \widehat v\equiv\frac1{n-1}\sum_{j=1}^n(W_j-\bar W)^2.
\]
The fixed weight is the normalized genotype
\begin{equation}\label{eq:genetics-traj-weight}
 a_i=x_i^{\mathrm{norm}}
 \equiv
 \frac{x_i-\bar x}
 {\left\{\widehat v\sum_{j=1}^n(x_j-\bar x)^2\right\}^{1/2}},
\end{equation}
which gives $\widehat v\sum_{i=1}^n(x_i^{\mathrm{norm}})^2=1$.

\subsection{Predictor resampling}
\label{sec:genetics-predictor-resampling}

SPAmix \citep{maSPAmixScalableAccurate2025} uses the genotype $W_i=X_i$
as its resampled quantity and takes
$\widetilde\Q_i=\operatorname{Binomial}(2,\widehat q_i)$,
where $\widehat q_i$ is the fitted allele frequency. The fixed weight $a_i$
is the fitted phenotype-model residual. This places genotype resampling
in the same construction as the response and residual resampling methods
above.

\section{Two-sided tests and numerical comparison with fastSPA}
\label{sec:two-sided-and-fastspa}

\subsection{Generalization to two-sided tests}
\label{sec:two-sided-tests}

The two-sided
resampling target is
\[
 p_{\infty,n}^{(2)}
 =\widetilde\P(|\widetilde S_n|\ge |S_n|\mid\obs)
 =\widetilde\P(\widetilde S_n\ge |S_n|\mid\obs)
  +\widetilde\P(\widetilde S_n\le-|S_n|\mid\obs),\qquad |S_n|>0.
\]
At $S_n=0$ the target is one; the two tails must not be added because
they overlap at any atom at zero. Write $L_{n,+}(a)$ and $L_{n,-}(a)$
for the ordinary upper-tail LR formulas with cutoff $a$ and summands
$\tilde s_i$ and $-\tilde s_i$, respectively, including the conventions
in Section~\ref{sec:lugannani-rice} of the main text. We use
$p_{\mathrm{SPA},n}^{(2)}=L_{n,+}(|S_n|)+L_{n,-}(|S_n|)$ for $S_n\ne0$
and $p_{\mathrm{SPA},n}^{(2)}=1$ at zero.

\paragraph{Two-sided accuracy and null validity.}
Theorem~\ref{thm:conditional-lr} also bounds the relative error of
$p_{\mathrm{SPA},n}^{(2)}$ by $CB_n(1+Z_n)/\sqrt n$ on its stated event.
Indeed, changing the sign of the summands or replacing the cutoff by
$|S_n|$ leaves the moment condition and $V_n,B_n,Z_n,D_n$ unchanged.
For $a=|S_n|>0$, let $P_+=\widetilde\P(\widetilde S_n\ge a\mid\obs)$,
$P_-=\widetilde\P(\widetilde S_n\le-a\mid\obs)$ and
$Q_\pm=L_{n,\pm}(a)$. Both $Q_\pm$ are positive on that event, and
\[
 \frac{P_++P_-}{Q_++Q_-}-1
 =\sum_{\nu\in\{+,-\}}\frac{Q_\nu}{Q_++Q_-}
       \left(\frac{P_\nu}{Q_\nu}-1\right).
\]
The claimed bound follows by taking absolute values; at $S_n=0$ the
relative error is zero. Thus the upper relative-error rates in our
applications also apply to this two-sided LR construction.

Null validity requires a separate sampling argument. If, for a positive
data-dependent scale $\sigma_n$,
$S_n/\sigma_n\convd N(0,1)$ and
$\sup_z|\widetilde\P(\widetilde S_n/\sigma_n\le z\mid\obs)-\Phi(z)|
\convp0$, then continuity of $\Phi$ also controls non-strict upper tails,
so
\[
 p_{\infty,n}^{(2)}=2\Phibar(|S_n|/\sigma_n)+o_{\P_n}(1)
 \convd\Unif(0,1).
\]
Whenever the preceding relative-error bound tends to zero in probability
and its event has probability tending to one, it follows that
$p_{\mathrm{SPA},n}^{(2)}\convd\Unif(0,1)$ as well. The matched-CLT
premises follow from the proofs of Theorems~\ref{thm:glmrr-target-validity}
and~\ref{thm:spacrt-validity}; the required SPA conditions follow from
the proofs of Theorems~\ref{thm:new-glmrr} and~\ref{thm:spacrt-validity},
under their respective null assumptions. These arguments are uniform
over the stated GLM-RR null class.

\subsection{Numerical comparison with the fastSPA implementation}
\label{sec:fastspa-numerical-comparison}

We compare the ordinary Lugannani--Rice approximation studied in the main
text with the fastSPA implementation of \citet{Dey2017} in a simple
GWAS setting. We generate 200 independent datasets with $n=20{,}000$
participants. Independently, genotype $X_i$ follows
$\operatorname{Binomial}(2,0.01)$, age $A_i$ is normal with mean 50 and
standard deviation 10, clipped to $[20,80]$, and sex $B_i$ is
$\operatorname{Bernoulli}(0.5)$. The binary phenotype satisfies
\[
 \logit\P(Y_i=1\mid X_i,A_i,B_i)
 =a+\log(1.3)(A_i-50)/10+\log(1.3)B_i+\log(2)X_i,
\]
where $a\simeq-3.114$ gives 5\% marginal prevalence when the genotype
coefficient is zero. Thus, the plotted setting has a per-allele odds ratio
of two.

For each dataset, we fit the logistic null model with an intercept, age
and sex. All calculations use the same fitted Bernoulli probabilities,
information-weighted genotype residuals and observed score.

We compare
the ordinary Lugannani--Rice formula using the full fitted cumulant
generating function, the \texttt{fastSPA} method in \texttt{SPAtest}
version 3.1.2 with its default settings, and the normal score approximation.
Both SPA calculations use the outward cutoffs defined in
Section~\ref{sec:two-sided-tests}; the normal calculation uses $2\Phibar(|S_n|/\sqrt{V_n})$. The package uses the
Barndorff--Nielsen formula and approximates the contribution of
zero-genotype individuals by a Gaussian CGF; its defaults also include
a normal approximation for standardized scores of magnitude less than two
and numerical fallback rules. The latter normal branch is used in nine
of the 200 datasets. All LR evaluations succeed.

Figure~\ref{fig:fastspa-numerical-comparison} shows close agreement between
the two SPAs, including where the normal approximation differs by several
orders of magnitude. Across all 200 datasets, the maximum
absolute relative difference $|p_{\LR}/p_{\mathrm{fastSPA}}-1|$ is
1.63\%. Thus, their agreement is not merely a consequence of
both reducing to a normal approximation.

\begin{figure}[!htbp]
\centering
\includegraphics[width=0.4226\textwidth]{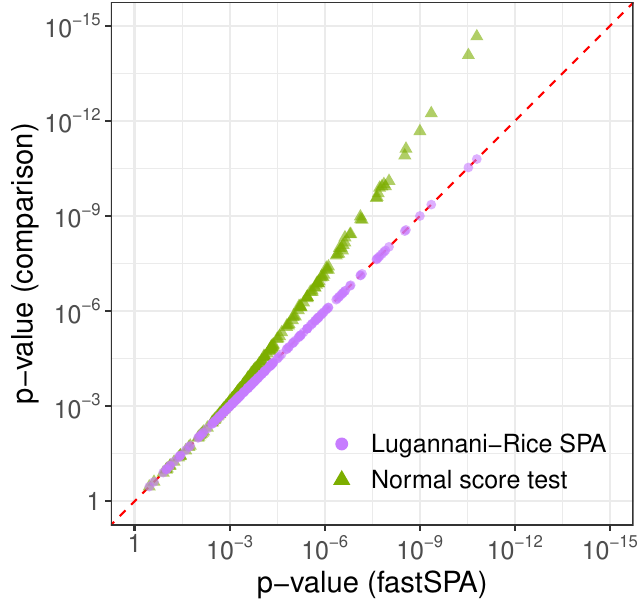}
\caption{Comparison of Lugannani--Rice and normal score-test $p$-values
with fastSPA in 200 simulated GWAS datasets with per-allele odds ratio two.
Both axes use reversed logarithmic scales; the dashed line denotes equality.}
\label{fig:fastspa-numerical-comparison}
\end{figure}

\section{Additional CRISPR simulation results}
\label{sec:crispr-sensitivity}

\paragraph{Left-tail convention.}
The non-strict left-sided resampling target and its plus-one Monte Carlo
version are
\[
 p_{\infty,n}^{\mathrm{left}}
 =\widetilde\P(\widetilde S_n\le S_n\mid\obs),\qquad
 p_{M,n}^{\mathrm{left}}
 =\frac{1+\sum_{m=1}^M\indicator\{\widetilde S_n^{(m)}\le S_n\}}{M+1}.
\]
They are the upper-tail constructions applied to summands $-\tilde s_i$
and cutoff $-S_n$. The left-sided LR approximation uses the same sign
reversal, with CGF $K_n(-t)$; at $S_n=0$ its continuous branch is
$1/2+K_n'''(0)/\{6\sqrt{2\pi n}\,K_n''(0)^{3/2}\}$.
The two continuous LR branches sum to one whenever admissible roots exist,
whereas the two exact non-strict tails sum to
$1+\widetilde\P(\widetilde S_n=S_n\mid\obs)$.

Figure~\ref{fig:crispr-sensitivity} complements the main-text
Figure~\ref{fig:new-simulation} by examining sensitivity to expression
sparsity and negative-binomial dispersion, using the same simulation model
and fitting procedures. In the expression-sparsity sweep, we vary
$\alpha_{n,y}\in\{-6,-5,-4,-3,-2\}$ while fixing
$\alpha_{n,x}=-5$ and $\eta_{n,y}=1$; more negative expression intercepts
produce sparser responses. In the dispersion sweep, we vary
$\eta_{n,y}\in\{0.1,1,20\}$ while fixing
$(\alpha_{n,x},\alpha_{n,y})=(-3,-5)$, matching the main figure's power
setting. Thus, the two sweeps use different perturbation intercepts.
For each setting, we evaluate Type-I error under $\rho_n=0$ and power
under $\rho_n=-2$ for left-sided tests and $\rho_n=1$ for right-sided tests.

\begin{figure}[!ht]
\centering
\includegraphics[width=\textwidth]{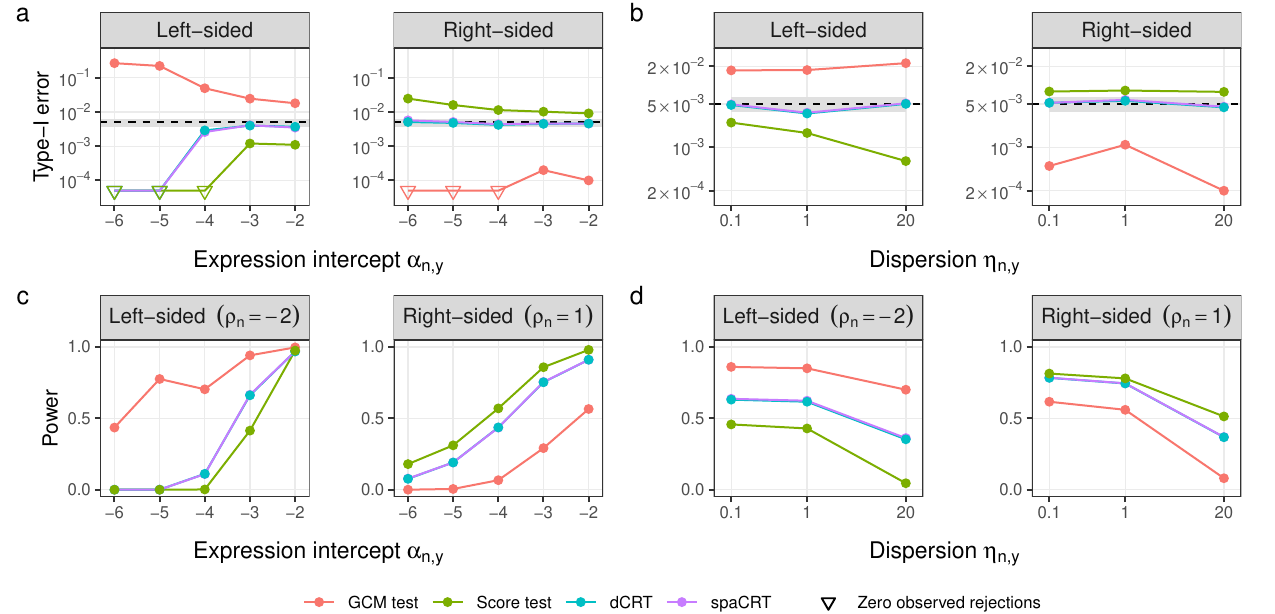}
\caption[CRISPR calibration and power sensitivity]{Sensitivity of Type-I error (top) and power (bottom) in the CRISPR
simulation. Each point uses 10,000 datasets of size $n=5{,}000$ and nominal
level $0.005$; dCRT uses 10,000 resamples per dataset. The left pair of columns
varies the expression intercept $\alpha_{n,y}$, fixing perturbation intercept
$\alpha_{n,x}=-5$ and dispersion $\eta_{n,y}=1$. The right pair varies
$\eta_{n,y}$ on a logarithmic scale, fixing
$(\alpha_{n,x},\alpha_{n,y})=(-3,-5)$ as in the main figure's power panels.
Both covariate slopes are one. Power uses $\rho_n=-2$ for left-sided tests
and $\rho_n=1$ for right-sided tests; Type-I error uses $\rho_n=0$.
The dashed line marks the nominal level; the gray band is the central 95\%
binomial reference range for 10,000 repetitions of an exactly calibrated test.
Open downward triangles denote zero observed rejections, displayed at
$0.5/10{,}000$ on the logarithmic Type-I error axes.
Power comparisons are not size-adjusted and
should be interpreted together with the corresponding Type-I error panels.}
\label{fig:crispr-sensitivity}
\end{figure}

Figure~\ref{fig:crispr-relative-error-sensitivity} examines agreement between
the paired \spacrt{} and finite-resampling \dCRT{} $p$-values on the same
simulated datasets. For each dataset and one-sided direction, we compute
$|p_{\mathrm{SPA},n}^{\dCRT}/p_{M,n}^{\dCRT}-1|$, with $M=10{,}000$.
The boxplots summarize these discrepancies across 10,000 datasets per
setting, without pooling the two directions. Panels (a), (c) and (d)
use all null replicates. Panel (b) retains only pairs with
$p_{M,n}^{\dCRT}\ge 10/M=10^{-3}$, including at $\rho_n=0$, to limit
the effect of Monte Carlo resolution on comparisons of small $p$-values.
Consequently, the retained datasets in panel (b) also differ across signal strengths.
These discrepancies include Monte Carlo error in the finite-resampling
\dCRT{} $p$-values and therefore do not isolate the SPA approximation error.

\begin{figure}[!ht]
\centering
\includegraphics[width=\textwidth]{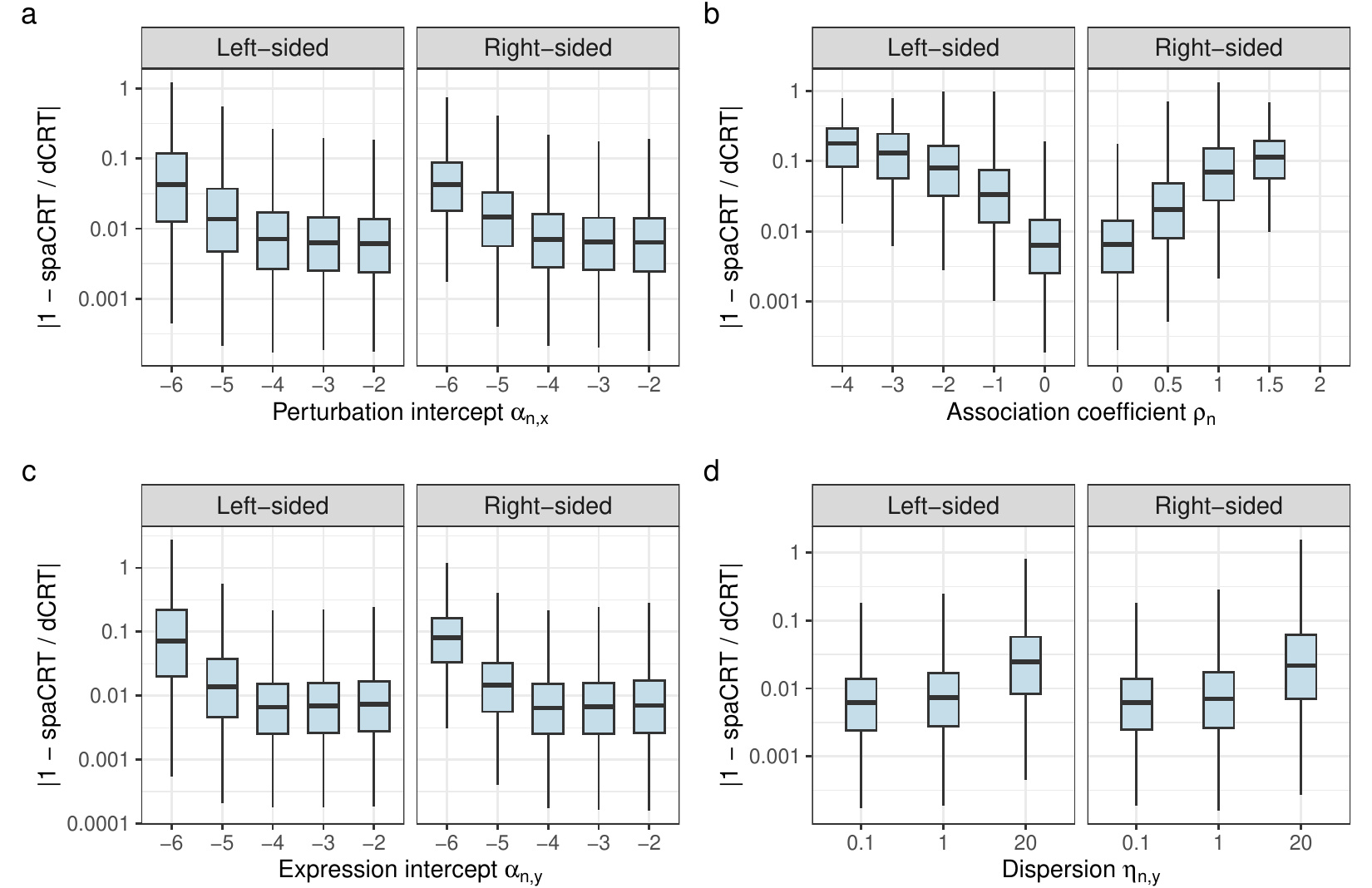}
\caption[CRISPR relative-error sensitivity]{Relative discrepancies between \spacrt{} and finite-resampling \dCRT{} $p$-values
in the CRISPR simulation, with $n=5{,}000$ and both covariate slopes equal
to one. (a) Perturbation intercept $\alpha_{n,x}$ varies from $-6$ to $-2$,
with $\alpha_{n,y}=-5$, $\eta_{n,y}=1$ and $\rho_n=0$.
(b) Association coefficient $\rho_n$ varies over the displayed values,
with $(\alpha_{n,x},\alpha_{n,y})=(-3,-5)$ and $\eta_{n,y}=1$.
Panel (b) includes only datasets with Monte Carlo \dCRT{} $p$-values
at least $10^{-3}$. At $\rho_n=2$, all 10,000 right-sided \dCRT{}
$p$-values fall below this cutoff, so no box is shown.
(c) Expression intercept $\alpha_{n,y}$ varies from $-6$ to $-2$,
with $\alpha_{n,x}=-5$, $\eta_{n,y}=1$ and $\rho_n=0$.
(d) Dispersion $\eta_{n,y}\in\{0.1,1,20\}$ varies with
$(\alpha_{n,x},\alpha_{n,y})=(-5,-2)$ and $\rho_n=0$; this less-sparse
response setting makes the effect of dispersion more visible.
Boxes show medians and interquartile ranges; whiskers extend to the most
extreme observations within 1.5 interquartile ranges of the box, and
outlying points are omitted. All vertical axes are logarithmic;
horizontal positions index the displayed parameter settings.}
\label{fig:crispr-relative-error-sensitivity}
\end{figure}

\newpage
\paragraph{Organization and conventions for the proofs.}
Sections~\ref{sec:uniform-null-diagonal}--\ref{sec:proof-main-universal-results}
establish the general approximation results and their uniform formulations.
Sections~\ref{sec:proof-saige-validity}--\ref{sec:glmrr-supporting-proofs}
contain the response-resampling proofs,
Sections~\ref{sec:proof-main-spacrt-results}--\ref{sec:spacrt-detailed-proofs}
the predictor-resampling proofs, and
Sections~\ref{sec:spacrt-high-dimensional}--\ref{sec:spacrt-logistic-supporting-proofs}
the high-dimensional extension and its supporting proofs.
Unqualified sums over $i$ run from $1$ to $n$; omitted row indices are
proof-local shorthand. Vectors and matrices are bold, whereas scalar
coordinates and moment scales are not. Throughout, $\P_n$ denotes the original data law and $\widetilde \P$ and
$\widetilde\E$ denote resampling conditional on the data $\obs$.
In generic probability lemmas, $\P$ and $\E$ denote the stipulated law;
unindexed probabilities and expectations in application proofs refer to $\P_n$
unless explicitly marked as resampling quantities.
A p-value is set to one if a prerequisite needed to define that p-value
fails. A failed common fit or resampling-law construction affects both
the exact target and its SPA; failure of the admissible saddlepoint alone
affects only the SPA. Failure of a sufficient theorem condition is not a
construction failure. Auxiliary estimator extensions on failure events are
used only to define proof quantities, not to replace these reported p-values.
For all asymptotic ratio statements, define
$p_{\infty,n}/p_{\mathrm{SPA},n}=1$ on
$\{p_{\mathrm{SPA},n}\le0\}$, and define relative error using this extended
ratio; the same convention applies with procedure-specific superscripts.
This extension leaves the reported LR approximation unchanged. Under the
relevant assumptions, the exceptional event has probability tending to zero,
uniformly wherever a uniform conclusion is asserted.

\section{Class-uniform upgrades by diagonalization}
\label{sec:uniform-null-diagonal}

The following sequence-to-class upgrades are used for the random-design
CRISPR class and the binary logistic-lasso class. In this section
$\mathcal P_n(\delta)$ denotes the constant-level version of $\PBNB_n$ or
$\PLL_n$ (or its null slice), with $\delta_n$ replaced by $\delta$.
The fixed-design GLM-RR results below are proved directly with uniform
constants and do not use a random-design filler argument.

\subsection{Statements}
\label{sec:uniform-diagonal-statements}

\begin{lemma}[Diagonal upgrade to class-uniform distributional
validity]
\label{lem:uniform-null-diagonal}
For each sample size $n\ge1$ and each level $\delta>0$, let
$\mathcal P_n(\delta)$ be a (possibly empty) collection of
data-generating processes for the size-$n$ data, and for each $n$ and
each $\P\in\bigcup_{\delta>0}\mathcal P_n(\delta)$ let $\xi_{n,\P}$ be
a real statistic of the size-$n$ data (auxiliary randomization, such
as conditional resampling, is permitted), with distribution function
$G_{n,\P}(t)\equiv \P(\xi_{n,\P}\le t)$.  Let $F$ be a fixed continuous
distribution function on $\R$.  Assume:
\begin{enumerate}[label=\textup{(\alph*)}]
\item \emph{(filler property)} there is an $n_0\ge1$ such that,
if there exists $n_0\le n$, $\delta>0$, such that
$\mathcal P_n(\delta)\neq\emptyset$, then for any $m \geq n$, we have
$\mathcal P_m(\delta)\neq\emptyset$;
\item \emph{(sequence-wise limits)} for every $n_1\ge1$, every
sequence $(\delta_n)_{n\ge n_1}$ with $\delta_n\to\infty$, and every
array $(\P_n)_{n\ge n_1}$ with $\P_n\in\mathcal P_n(\delta_n)$ for all
$n\ge n_1$, one has $G_{n,\P_n}(t)\to F(t)$ for every $t\in\R$, that
is, $\xi_{n,\P_n}\convd F$.
\end{enumerate}
Then, for every sequence $\delta_n\to\infty$,
\[
  \sup_{\P\in\mathcal P_n(\delta_n)}\,
  \sup_{t\in\R}\,
  |G_{n,\P}(t)-F(t)|
  \longrightarrow0,
  \qquad n\to\infty,
\]
with the supremum over an empty class read as zero.
\end{lemma}

\begin{corollary}[Uniform null upgrade for the application classes]
\label{cor:uniform-null-diagonal-use}
Let $\mathcal C_n^0(\delta)$ be the constant-level version of any one
of $\PLLnull_n$ or $\PBNBnull_n$, and put
$\mathcal C_n^0\equiv\mathcal C_n^0(\delta_n)$.  For every $n\ge1$,
$\delta>0$, and $\P\in\mathcal C_n^0(\delta)$, let
$p_{n,\P}$ be a real-valued statistic.  If, for every $n_1\ge1$,
$p_{n,\P_n}\convd\Unif(0,1)$ along every array $(\P_n)_{n\ge n_1}$
satisfying the envelope constraints and the limit form of the
corresponding regime conditions with $\rho_n=0$, then
\[
 \sup_{\P\in\mathcal C_n^0}\sup_{t\in[0,1]}
 \left|\P(p_{n,\P}\le t)-t\right|\longrightarrow0.
\]
The supremum over an empty class is read as zero.
\end{corollary}

\begin{lemma}[Equivalence of the sup form and the sequence-wise form
of $\OP{\mathcal P}$]
\label{lem:op-sup-seq-equivalence}
For each sample size $n\ge1$ and each level $\delta>0$, let
$\mathcal P_n(\delta)$ be a (possibly empty) collection of
data-generating processes for the size-$n$ data, and for each $n$
and each $\P\in\bigcup_{\delta>0}\mathcal P_n(\delta)$ let $X_{n,\P}$
be a real statistic of the size-$n$ data (auxiliary randomization
permitted) and $a_n(\P)>0$ a deterministic rate.  Fix $\eps>0$ and
$C<\infty$, and assume the filler property in Assumption (a) of
Lemma~\ref{lem:uniform-null-diagonal}.
Then the following are equivalent:
\begin{enumerate}[label=\textup{(\alph*)}]
\item \emph{(sup form)} for every sequence $\delta_n\to\infty$,
\[
  \limsup_n\,
  \sup_{\P\in\mathcal P_n(\delta_n)}
  \P\{|X_{n,\P}|>C\,a_n(\P)\}\le\eps,
\]
with the supremum over an empty class read as zero;
\item \emph{(sequence-wise form)} for every $n_1\ge1$, every
sequence $(\delta_n)_{n\ge n_1}$ with $\delta_n\to\infty$, and every
array $(\P_n)_{n\ge n_1}$ with $\P_n\in\mathcal P_n(\delta_n)$ for all
$n\ge n_1$,
\[
  \limsup_n\,
  \P_n\{|X_{n,\P_n}|>C\,a_n(\P_n)\}\le\eps .
\]
\end{enumerate}
\end{lemma}

\begin{corollary}[Class-uniform stochastic-order upgrade]
\label{cor:op-sup-seq-use}
Let $\mathcal C_n(\delta)$ be the constant-level version of any one of
$\PLL_n$ or $\PBNB_n$, put
$\mathcal C_n\equiv\mathcal C_n(\delta_n)$, and write
$\mathcal C\equiv\{\mathcal C_n\}$.  For every $n\ge1$, $\delta>0$,
and $\P\in\mathcal C_n(\delta)$, let $X_{n,\P}$ be real-valued and let
$a_n(\P)>0$ be deterministic.
Suppose that, for every
$\eps>0$, there is a constant $C(\eps)$ depending only on the fixed
envelope constants such that
\[
 \limsup_n \P_n\{|X_{n,\P_n}|>C(\eps)a_n(\P_n)\}\le\eps
\]
along every array $(\P_n)_{n\ge n_1}$, for every $n_1\ge1$, satisfying
the envelope constraints and the limit form of the corresponding
regime conditions.  Then
\[
 X_{n,\P}=\OP{\mathcal C}\{a_n(\P)\}
\]
in the sense of~\eqref{eq:op-def}.
\end{corollary}

\subsection{Proofs}
\label{sec:uniform-diagonal-proofs}

\begin{proof}[of Lemma~\ref{lem:uniform-null-diagonal}]
Suppose the conclusion fails for some sequence $\delta_n\to\infty$:
there are $\eps>0$ and a strictly increasing sequence
$n_1<n_2<\cdots$, which may be taken with $n_1\ge n_0$, such that for
every $k$
\[
  \sup_{\P\in\mathcal P_{n_k}(\delta_{n_k})}
  \sup_{t\in\R}|G_{n_k,\P}(t)-F(t)|>\eps ;
\]
in particular $\mathcal P_{n_k}(\delta_{n_k})\neq\emptyset$, and one
may choose $\Q_k\in\mathcal P_{n_k}(\delta_{n_k})$ with
\begin{equation}
  \sup_{t\in\R}|G_{n_k,\Q_k}(t)-F(t)|>\eps .
  \label{eq:diag-separation}
\end{equation}
Complete the subsequential choice to an array: for $n\ge n_1$ put
$k(n)\equiv\max\{k:n_k\le n\}$ and
$\delta'_n\equiv\delta_{n_{k(n)}}$, set $\P'_{n_k}\equiv \Q_k$, and
for $n\notin\{n_1,n_2,\ldots\}$ choose
$\P'_n\in\mathcal P_n(\delta'_n)$ arbitrarily.  This is possible by
the filler property because
$\mathcal P_{n_{k(n)}}(\delta'_n)\ni \Q_{k(n)}$ is nonempty and
$n_0\le n_{k(n)}\le n$.  Moreover, $\delta'_n\to\infty$ because
$k(n)\to\infty$ and $\delta_{n_k}\to\infty$ as $k\to\infty$.
Hypothesis~(b), applied to the array $(\P'_n)_{n\ge n_1}$, gives
$G_{n,\P'_n}(t)\to F(t)$ for every $t\in\R$.  Since the $G_{n,\P'_n}$
are distribution functions and the limit $F$ is a continuous
distribution function, P\'olya's theorem upgrades this pointwise
convergence to
\[
  \sup_{t\in\R}|G_{n,\P'_n}(t)-F(t)|\longrightarrow0 .
\]
Restricting to $n=n_k$, along which $\P'_{n_k}=\Q_k$,
contradicts~\eqref{eq:diag-separation}.
\end{proof}

\begin{proof}[of Corollary~\ref{cor:uniform-null-diagonal-use}]
Apply Lemma~\ref{lem:uniform-null-diagonal} with $F$ the
$\Unif(0,1)$ distribution function and
$\mathcal P_n(\delta)=\mathcal C_n^0(\delta)$.  Two observations
verify its hypotheses.
First, the filler property~(a) holds with $n_0=55$.  Indeed, the
envelope constraints do not involve $n$, and at a fixed
process each regime quantity is of one of the forms $na(\P)$, $a(\P)$,
$b(\P)/n$, or $na(\P)/\log^4n$ with $a(\P),b(\P)\ge0$, hence monotone in
$n$ in the favourable direction for all $n\ge55$ (the map
$n\mapsto n/\log^4n$ is increasing there).  Thus, if the
constant-level null class is nonempty at $n$, the same null model
specification at $m\ge n$ gives a member of the corresponding class
at $m$; these are exactly the regime forms in the class definitions
accompanying Theorems~\ref{thm:spacrt-logistic-lasso-local-alternative}, and
\ref{thm:spacrt-rate}.  Second, along any array with
$\P_n\in\mathcal P_n(\delta'_n)$ and $\delta'_n\to\infty$, the
memberships recover the limit form of every regime condition, which is that each
divergent quantity is at least $\delta'_n\to\infty$, each vanishing
quantity is at most $1/\delta'_n\to0$, and the alternative-decay
conditions hold trivially because $\rho_n=0$ on the null class---so
the assumed sequence-wise null limit applies to the
array.  The lemma now gives the displayed conclusion for the fixed
main-text sequence $(\delta_n)$.
\end{proof}

\begin{proof}[of Lemma~\ref{lem:op-sup-seq-equivalence}]
(a)$\Rightarrow$(b): extend the level sequence backwards by
$\delta_n=\delta_{n_1}$ for $n<n_1$; each term of the array
probability sequence is bounded by the corresponding supremum, so
its $\limsup$ is at most $\eps$.  The filler property is not used here.

(b)$\Rightarrow$(a): suppose (a) fails for some $\delta_n\to\infty$:
there are $\eta>\eps$ and $n_0\le n_1<n_2<\cdots$ with, for every
$k$, some $\Q_k\in\mathcal P_{n_k}(\delta_{n_k})$ satisfying
\begin{equation}
  \Q_k\{|X_{n_k,\Q_k}|>C\,a_{n_k}(\Q_k)\}>\eta .
  \label{eq:op-diag-separation}
\end{equation}
Complete the subsequential choice to an array: for $n\ge n_1$ put
$k(n)\equiv\max\{k:n_k\le n\}$ and $\delta'_n\equiv\delta_{n_{k(n)}}$, set
$\P'_{n}\equiv \Q_k$ for $n=n_k$, and for $n\notin\{n_1,n_2,\dots\}$ choose
$\P'_n\in\mathcal P_n(\delta'_n)$ arbitrarily, which is possible by
the filler property because
$\mathcal P_{n_{k(n)}}(\delta'_n)\ni \Q_{k(n)}$ is nonempty and
$n_0\le n_{k(n)}\le n$.  Then $\delta'_n\to\infty$, so
hypothesis~(b), applied to the array $(\P'_n)_{n\ge n_1}$ with the
levels $(\delta'_n)$, gives
$\limsup_n\P'_n\{|X_{n,\P'_n}|>C\,a_n(\P'_n)\}\le\eps<\eta$;
restricting to $n=n_k$
contradicts~\eqref{eq:op-diag-separation}.
\end{proof}

\begin{proof}[of Corollary~\ref{cor:op-sup-seq-use}]
Apply Lemma~\ref{lem:op-sup-seq-equivalence} with
$\mathcal P_n(\delta)=\mathcal C_n(\delta)$.  Its filler property
holds with $n_0=55$: given
$\P\in\mathcal P_n(\delta)$, its null modification
$\P^0$---set $\rho=0$, all else unchanged---satisfies the envelope
constraints, makes every $\rho$-regime quantity zero, and leaves the
remaining regime quantities unchanged apart from their dependence on
$n$, each being monotone in the favourable direction for $n\ge55$ (those
quantities being of the forms $na(\P)$, $a(\P)$, $b(\P)/n$, or
$na(\P)/\log^4n$), so $\P^0\in\mathcal P_m(\delta)$ for
$m\ge n\ge55$.  Along any array with
$\P_n\in\mathcal P_n(\delta'_n)$ and $\delta'_n\to\infty$, the
memberships recover the limit form of every regime condition in the
class definitions accompanying Theorems~\ref{thm:spacrt-logistic-lasso-local-alternative}, and
\ref{thm:spacrt-rate}, so the
assumed sequence-wise bound applies to the array.  Finally, apply the
assumed sequence-wise bound at level $\eps/2$.  The lemma gives
$\limsup_n\sup_{\P}(\cdot)\le\eps/2<\eps$, hence an onset
$N(\eps)$ beyond which the supremum is at most $\eps$, which is
exactly~\eqref{eq:op-def} with the envelope-built constant
$C(\eps/2)$.
\end{proof}

\section{Proofs of the results in Section~\ref{sec:universal-theory}}
\label{sec:proof-main-universal-results}

\subsection{Finite-row approximation lemmas}
\label{sec:heterogeneous-translator}

This subsection collects the foundational lemmas used to prove
Theorem~\ref{thm:conditional-lr}. The following direct consequence of
Theorem~2, Corollaries~3--5, and equation~(88)
of~\citet[pp.~652--656]{sakhanenkoBerryEsseenTypeEstimates1991} is used without proof.

\begin{lemma}[Theorem 2, Corollaries 3-5 in~\citet{sakhanenkoBerryEsseenTypeEstimates1991}]\label{lem:sakhanenko-finite-row}
Let $X_1,\ldots,X_m$ be independent and centered, and put
$T\equiv\sum_iX_i$, $A(s)\equiv\log\E e^{sT}$, and
$V^2\equiv\sum_i\E X_i^2\in(0,\infty)$. For $a\in\R$, define
$\mathcal L_{\mathrm S}(a)\equiv\sum_i\E\{|X_i|^3\max(e^{aX_i},1)\}$.
Fix $t\ne0$, set $a_t\equiv2t/V^2$ and
$\Lambda\equiv\mathcal L_{\mathrm S}(a_t)$, and suppose
\begin{align*}
  16|t|\Lambda\le V^4.
\end{align*}
Then $A$ is three times continuously differentiable and strictly convex on the segment joining $0$ and $a_t$, and $A'(s)=t$ has a unique solution $\widehat s$ on that segment. Moreover,
\begin{align}\label{eq:sakhanenko-cited-root}
  \sgn(\widehat s)=\sgn(t),\qquad|\widehat s|\le\frac{2|t|}{V^2},\qquad |A''(\widehat s)^2-V^4|\le8|t|\Lambda,\qquad\sup_{0\le q\le1}|A'''(q\widehat s)|\le4\Lambda.
\end{align}
Under the tilted law
$\frac{d\P_{\widehat s}}{d\P}\equiv
\exp\{\widehat sT-A(\widehat s)\}$, put
\[
 L^*(\widehat s)\equiv A''(\widehat s)^{-3/2}
 \sum_i\E_{\widehat s}|X_i-\E_{\widehat s}X_i|^3,
 \quad
 u\equiv\widehat s\sqrt{A''(\widehat s)},
 \quad
 r\equiv\sgn(t)[2\{\widehat st-A(\widehat s)\}]^{1/2}.
\]
Then
\begin{equation}
 L^*(\widehat s)\le\frac{7\Lambda}{V^3},
 \label{eq:sakhanenko-cited-lyapunov}
\end{equation}
and, for some $|\theta_+|,|\theta_-|\le1$,
\begin{align}
 t>0:\quad \P(T\ge t)
 &=\exp\{(u^2-r^2)/2\}
   \{\Phibar(u)+4\theta_+L^*(\widehat s)\phi(u)\},
 \label{eq:sakhanenko-cited-upper}\\
 t<0:\quad \P(T<t)
 &=\exp\{(u^2-r^2)/2\}
   \{\Phi(u)+4\theta_-L^*(\widehat s)\phi(u)\}.
 \label{eq:sakhanenko-cited-lower}
\end{align}
Thus the source pairs a non-strict upper tail with a strict lower tail.
\end{lemma}

We first show that Robinson's formula follows from
Lemma~\ref{lem:sakhanenko-finite-row}.

\begin{lemma}[Bernstein--Sakhanenko reduction and Robinson approximation]\label{lem:bernstein-sakhanenko}
Let $X_1,\ldots,X_m$ be independent and centered, and suppose, with
$v_i\equiv\E X_i^2$ and $b_i\ge0$, that
\[
 \E|X_i|^k\le\frac{k!}{2}b_i^{k-2}v_i,
 \qquad k\ge3.
\]
Put
\[
 T\equiv\sum_iX_i,\quad A(s)\equiv\log\E e^{sT},\quad
 V^2\equiv\sum_i v_i,\qquad H\equiv\sum_i b_iv_i,\qquad
 b^\star\equiv\max_i b_i.
\]
There are universal constants $c_{\mathrm R},C_{\mathrm R}>0$ for which
the following statements hold whenever $V>0$.
\begin{enumerate}[label=\textnormal{(\alph*)},leftmargin=2.2em,itemsep=5pt,topsep=5pt]
\item \label{part:bernstein-domain-center}
\textnormal{\textbf{Domain and center.}}
$H,b^\star>0$, and $A$ is analytic and strictly convex on $|s|b^\star<1$.  With $\eta\equiv H/V^3$,
\begin{align}\label{eq:finite-row-center}
  \left|\P(T\ge0)-\frac12\right|\le C_{\mathrm R}\eta,
  \qquad |A'''(0)|\le3H.
\end{align}

\item \label{part:bernstein-saddlepoint-control}
\textnormal{\textbf{Nonzero-cutoff saddlepoint control.}}
Fix $t\ne0$ and set
\[
 a_t\equiv\frac{2t}{V^2},\qquad
 Z\equiv\frac{|t|}{V},\qquad
 D\equiv\frac{b^\star|t|}{V^2},\qquad
 \epsilon\equiv(1+Z)\eta.
\]
Set $\Lambda\equiv\mathcal L_{\mathrm S}(a_t)$.
If $D\le c_{\mathrm R}$ and $\epsilon\le c_{\mathrm R}$, then
\begin{align}\label{eq:sakhanenko-verification}
  \Lambda\le48H,\qquad16|t|\Lambda/V^4\le1,
\end{align}
$A'(s)=t$ has a unique solution $\widehat s$ on $|s|b^\star<1$, with
$\sgn(\widehat s)=\sgn(t)$ and
\begin{align}\label{eq:finite-row-root}
  b^\star|\widehat s|\le2D,\qquad\frac12V^4\le A''(\widehat s)^2\le\frac32V^4,\qquad\sup_{0\le q\le1}|A'''(q\widehat s)|\le4\Lambda.
\end{align}
Define
\begin{align}\label{eq:r-u}
  u\equiv\widehat s\sqrt{A''(\widehat s)},\qquad r\equiv\sgn(t)[2\{\widehat st-A(\widehat s)\}]^{1/2}.
\end{align}
Then $|u|\le C_{\mathrm R}Z$.

\item \label{part:robinson-approximation}
\textnormal{\textbf{Robinson approximation.}}
Set $R(0)\equiv1/2$.  Under the conditions and notation of part~\ref{part:bernstein-saddlepoint-control}, define the nonzero branches of Robinson's tilted-normal functional~\citep{Robinson1982}; together, these conventions are
\begin{align}\label{eq:robinson}
 R(t)\equiv
 \begin{cases}
  e^{(u^2-r^2)/2}\Phibar(u),&t>0,\\
  1-e^{(u^2-r^2)/2}\Phi(u),&t<0,\\
  1/2,&t=0.
 \end{cases}
\end{align}
For $t\ne0$, the following bounds hold:
\begin{align}
 t>0:\quad&
 \left|\frac{\P(T\ge t)}{R(t)}-1\right|\le C_{\mathrm R}\epsilon,
 \label{eq:finite-row-positive}\\
 t<0:\quad&
 |\P(T\ge t)-R(t)|\le C_{\mathrm R}\eta.
 \label{eq:finite-row-negative}
\end{align}
The probability in \eqref{eq:finite-row-negative} is obtained by complementing the strict lower tail $\P(T<t)$, so atoms are retained.
\end{enumerate}
\end{lemma}

\begin{proof}
\noindent\emph{Part~\ref{part:bernstein-domain-center}: domain and center.}\quad
For each $i$, if $b_i=0$, the $k=3$ bound makes $X_i=0$ almost surely. If $b_i>0$ and $|s|b_i<1$, the moment series and $\E|X_i|\le\sqrt{v_i}$ give
\[
 \E e^{|s||X_i|}
 \le1+|s|\sqrt{v_i}+\frac{s^2v_i}{2(1-|s|b_i)}<\infty.
\]
Thus $A$ is analytic on $|s|b^\star<1$.  Because $V>0$, some $X_i$ is nonconstant; the preceding observation forces its $b_i$ to be positive. Writing $\Var_s$ for variance under exponential tilting by $s$, tilting
preserves this summand's nondegeneracy, so
$A''(s)=\sum_i\Var_s(X_i)>0$ throughout this interval.  The same summand also gives $H,b^\star>0$. Recall the definition of $\mathcal{L}_{S}(a)\equiv \sum_i\E\{|X_i|^3\max(e^{aX_i},1)\}$. Then for $|a|b^\star<1$,
\begin{align}\label{eq:bernstein-sakhanenko-bridge}
  \mathcal L_{\mathrm S}(a)\le\frac{3H}{(1-|a|b^\star)^4},\qquad |A'''(0)|\le\mathcal L_{\mathrm S}(0)\le3H.
\end{align}
Indeed, the zero-scale summands contribute nothing, while for $b_i>0$,
\[
\begin{aligned}
  \E\{|X_i|^3\max(e^{aX_i},1)\}\le \E\{|X_i|^3e^{|a||X_i|}\}\le\frac{b_iv_i}{2}\sum_{j\ge0}(j+1)(j+2)(j+3)(|a|b_i)^j=\frac{3b_iv_i}{(1-|a|b_i)^4}.
\end{aligned}
\]
Summation proves the first bound in~\eqref{eq:bernstein-sakhanenko-bridge}. Since the summands are centered, $A'''(0)=\sum_i\E X_i^3$, which proves the second bound. Finally, the ordinary Berry--Esseen inequality, evaluated at the left limit of zero, gives
\[
 \left|\P(T<0)-\frac12\right|
 \le C\frac{\sum_i\E|X_i|^3}{V^3}
 \le C\eta.
\]
Complementing $\P(T<0)$ proves the probability bound in~\eqref{eq:finite-row-center}, including any atom at zero.  This proves
part~\ref{part:bernstein-domain-center}.

\par\smallskip\noindent
\emph{Part~\ref{part:bernstein-saddlepoint-control}: nonzero-cutoff
saddlepoint control.}\quad Choose $c_{\mathrm R}\le\min(1/4,1/768)$. Since $|a_t|b^\star=2D\le1/2$, the first bound in~\eqref{eq:bernstein-sakhanenko-bridge} gives
\[
 \Lambda=\mathcal{L}_{S}(a_t)\le\frac{3H}{(1-2D)^4}\le48H,
 \qquad
 \frac{16|t|\Lambda}{V^4}
 \le768Z\eta\le768\epsilon\le1.
\]
This is \eqref{eq:sakhanenko-verification}.  Moreover, the segment joining
$0$ and $a_t$ lies inside $|s|b^\star<1$ because $|a_t|b^\star\le1/2$.
Thus Lemma~\ref{lem:sakhanenko-finite-row} applies.  Its root localization gives
$\sgn(\widehat s)=\sgn(t)$ and
$b^\star|\widehat s|\le2D$.  Moreover,
\[
 |A''(\widehat s)^2-V^4|
 \le8|t|\Lambda\le\frac12V^4,
 \qquad
 \sup_{0\le q\le1}|A'''(q\widehat s)|\le4\Lambda,
\]
which proves all the bounds in \eqref{eq:finite-row-root}. Part~\ref{part:bernstein-domain-center} makes $A'$ strictly increasing on the whole interval $|s|b^\star<1$, so the root supplied on the segment joining $0$ and $a_t$ is unique throughout that interval. Since $A(0)=A'(0)=0$ and $\widehat s\ne0$, strict convexity also gives $\widehat st-A(\widehat s)>0$, so $r$ is real. Finally,
\[
 |u|=|\widehat s|\sqrt{A''(\widehat s)}
 \le\frac{2|t|}{V^2}\left(\frac32V^4\right)^{1/4}
 \le CZ.
\]
This proves part~\ref{part:bernstein-saddlepoint-control}.

\par\smallskip\noindent
\emph{Part~\ref{part:robinson-approximation}: Robinson approximation.}\quad
The value $R(0)=1/2$ is the stated center convention, and
part~\ref{part:bernstein-domain-center} gives its center error.  For
$t\ne0$, retain the conditions and notation of
part~\ref{part:bernstein-saddlepoint-control}.  Equations
\eqref{eq:sakhanenko-cited-lyapunov} and
\eqref{eq:sakhanenko-verification} give
\[
 L^*(\widehat s)\le\frac{7\Lambda}{V^3}\le336\eta.
\]
If $t>0$, then $u>0$, and
\eqref{eq:sakhanenko-cited-upper} together with the definition of $R(t)$
gives
\[
 \left|\frac{\P(T\ge t)}{R(t)}-1\right|
 \le4L^*(\widehat s)\frac{\phi(u)}{\Phibar(u)}
 \le C L^*(\widehat s)(1+u)
 \le C(1+Z)\eta=C\epsilon,
\]
where the second inequality is Mills' ratio and the third uses the bound on
$u$ from part~\ref{part:bernstein-saddlepoint-control}.  This proves
\eqref{eq:finite-row-positive}.

If $t<0$, complementing the strict lower-tail identity~\eqref{eq:sakhanenko-cited-lower} gives
\[
\begin{aligned}
 \P(T\ge t)-R(t)
 &=-4\theta_-L^*(\widehat s)e^{(u^2-r^2)/2}\phi(u)\\
 &=-4\theta_-L^*(\widehat s)\phi(r).
\end{aligned}
\]
Consequently $|\P(T\ge t)-R(t)|\le C\eta$, proving~\eqref{eq:finite-row-negative}. Because the argument complements $\P(T<t)$ exactly, any atom at $t$ remains in the non-strict upper tail. After enlarging $C_{\mathrm R}$ to absorb the universal constants above, all three parts follow.
\end{proof}

\begin{lemma}[Robinson--Lugannani--Rice transfer]
\label{lem:robinson-lr-transfer}
Under the assumptions and notation of Lemma~\ref{lem:bernstein-sakhanenko}, there are universal constants $c_{\mathrm L},C_{\mathrm L}>0$, with $c_{\mathrm L}\le c_{\mathrm R}$, such that the following statements hold. Recall that $\operatorname{LR}(0)\equiv\frac12-\frac{A'''(0)}{6\sqrt{2\pi}\,V^3}$. Then
\begin{equation}
 |\operatorname{LR}(0)-R(0)|\le C_{\mathrm L}\eta.
 \label{eq:finite-row-lr-center}
\end{equation}
For $t\ne0$, suppose $D\le c_{\mathrm L}$ and
$\epsilon\le c_{\mathrm L}$, and use $\widehat s,u,r,R(t)$ from
Lemma~\ref{lem:bernstein-sakhanenko}.  Recall the LR formula in~\eqref{eq:new-lr} as $\operatorname{LR}(t)\equiv\Phibar(r)+\phi(r)\left(\frac1u-\frac1r\right)$. Then recalling the definition of $u$ and $r$ in~\eqref{eq:r-u}, we have $|u-r|\le C_{\mathrm L}\eta u^2$, and
\begin{align}
  t>0:\quad
  &\label{eq:finite-row-lr-positive}
  \left|\frac{\operatorname{LR}(t)}{R(t)}-1\right|\le C_{\mathrm L}\eta,\\
  t<0:\quad
  &\label{eq:finite-row-lr-negative}
  |\operatorname{LR}(t)-R(t)|\le C_{\mathrm L}\eta.
\end{align}
\end{lemma}

\begin{proof}
Put $\kappa_3\equiv A'''(0)$.  As $t\to0$, write $t=A'(s)$; strict convexity gives $s\to0$, and Taylor expansion gives $r=Vs\left\{1+\frac{\kappa_3}{3V^2}s+o(s)\right\}$ and $u=Vs\left\{1+\frac{\kappa_3}{2V^2}s+o(s)\right\}$. Consequently, $\frac1u-\frac1r=-\frac{\kappa_3}{6V^3}+o(1)$, while $\Phibar(r)\to1/2$ and $\phi(r)\to(2\pi)^{-1/2}$. Substitution therefore verifies the displayed continuous branch. Since $R(0)=1/2$ by~\eqref{eq:robinson} and $|\kappa_3|\le3H$ by~\eqref{eq:finite-row-center}, it also proves~\eqref{eq:finite-row-lr-center}. For $t\ne0$, the segmentwise third-derivative bound in~\eqref{eq:finite-row-root} gives
\begin{align}\label{eq:finite-row-u-r}
 r^2=2\widehat s^{\,2}\int_0^1qA''(q\widehat s)\,dq,\qquad |u^2-r^2|\le\frac43\Lambda|\widehat s|^3.
\end{align}
Since $u^2=\widehat s^{\,2}A''(\widehat s)$ and
$A''(\widehat s)\ge V^2/\sqrt2$, the localization bound
$|\widehat s|\le2|t|/V^2$ in~\eqref{eq:sakhanenko-cited-root}
and~\eqref{eq:sakhanenko-verification} give
\[
 \frac{|u^2-r^2|}{u^2}
 \le \frac{4\Lambda|\widehat s|}{3A''(\widehat s)}
 \le \frac{8\sqrt2}{3}\frac{|t|\Lambda}{V^4}
 \le \frac{\sqrt2}{6}.
\]
Thus $|u|$ and $|r|$ are comparable.  Moreover, both are nonzero and
$\sgn(u)=\sgn(r)=\sgn(t)$, so $|u+r|=|u|+|r|\ge|u|$.  Factoring
$u^2-r^2=(u-r)(u+r)$, using the second bound in
\eqref{eq:finite-row-u-r}, and then using
$A''(\widehat s)\ge V^2/\sqrt2$, $\Lambda\le48H$, and
$\eta=H/V^3$, we obtain
\begin{align*}
  |u-r|=\frac{|u^2-r^2|}{|u+r|}\le\frac{4\Lambda|\widehat s|^3}{3|u|}=\frac{4\Lambda}{3A''(\widehat s)^{3/2}}u^2\le \frac{4}{3}\,2^{3/4}\frac{\Lambda}{V^3}u^2\le 64\,2^{3/4}\eta u^2.
\end{align*}
For $x>0$, put $m(x)\equiv\Phibar(x)/\phi(x)$ and $g(x)\equiv m(x)-x^{-1}$. Standard Mills bounds and the integral representations of $m,g'$ give $m(x)\asymp(1+x)^{-1}$ and $|g'(x)|\le\frac{C}{x^2(1+x^2)}$. By the mean-value theorem, there is some $\xi$ between $|r|$ and $|u|$. Their comparability gives $\xi\asymp|u|$, and their common sign gives
$\big||r|-|u|\big|=|r-u|$. The derivative bound above and the bound on $|r-u|$ therefore give
\[
 \begin{aligned}
 |g(|r|)-g(|u|)|
 &=|g'(\xi)|\,\big||r|-|u|\big|
 \le \frac{C|r-u|}{u^2(1+u^2)}
 \le \frac{C\eta}{1+u^2}.
 \end{aligned}
\]
If $t>0$, then $u,r>0$.  The identity
$e^{(u^2-r^2)/2}\phi(u)=\phi(r)$ and the definitions of $m$ and $g$ yield $R(t)=\phi(r)m(u)=\phi(r)\left\{g(u)+\frac1u\right\}$ and $\operatorname{LR}(t)=\phi(r)\left\{m(r)+\frac1u-\frac1r\right\}=\phi(r)\left\{g(r)+\frac1u\right\}$. Consequently, $R(t)>0$ and
\[
 \begin{aligned}
 \left|\frac{\operatorname{LR}(t)}{R(t)}-1\right|
 &=\frac{|g(r)-g(u)|}{m(u)}
 \le C\eta\frac{1+u}{1+u^2}
 \le C\eta,
 \end{aligned}
\]
where the penultimate inequality uses $m(u)\asymp(1+u)^{-1}$.  This
proves \eqref{eq:finite-row-lr-positive}.

If $t<0$, put $x\equiv-r>0$ and $y\equiv-u>0$.  Normal symmetry and the same exponential identity give $R(t)=1-\phi(x)m(y)=1-\phi(x)\left\{g(y)+\frac1y\right\}$ and $\operatorname{LR}(t)=1-\phi(x)m(x)+\phi(x)\left(\frac1x-\frac1y\right)=1-\phi(x)\left\{g(x)+\frac1y\right\}$. It follows that
\[
 \begin{aligned}
 |\operatorname{LR}(t)-R(t)|
 &=\phi(x)|g(x)-g(y)|
 \le \phi(0)\frac{C\eta}{1+u^2}
 \le C\eta,
 \end{aligned}
\]
which proves \eqref{eq:finite-row-lr-negative}, after enlarging
$C_{\mathrm L}$ if necessary.  Thus division by $R(t)$ is needed only for
the positive-cutoff relative bound.
\end{proof}

\subsection{Proof of Theorem~\ref{thm:conditional-lr}}
\label{sec:proof-main-one}

\begin{proof}
We apply the finite-row lemmas conditionally on the data. Write
$\mathcal F_n=\sigma(\obs)$,
$T_n=\sum_i\tilde s_i$, and $t_n=S_n$ for the resampled sum and observed
cutoff in~\eqref{eq:infinite-M}. Throughout this proof, all moments are
conditional on $\mathcal F_n$. Use the following proof-local notation
$v_{in}=\widetilde\E(\tilde s_i^2\mid\mathcal F_n)$ and
$\sigma_n^2=\sum_i v_{in}$, so that $\sigma_n$ here is the standard deviation,
with total variance $\sigma_n^2$; the main text denotes that total variance by $V_n$.
Put $b_{in}=b_i(\obs)$,
$b_n^\star=\max_i b_{in}$,
$H_n=\sum_i b_{in}v_{in}$, and
$A_n(s)=\log\widetilde\E\{e^{sT_n}\mid\mathcal F_n\}=nK_n(s)$.
On $\{\sigma_n>0\}$, set
\[
 \eta_n=\frac{H_n}{\sigma_n^3}
       =\frac{B_n(\obs)}{\sqrt n},
 \qquad
 Z_n=\frac{|t_n|}{\sigma_n},\qquad
 D_n=\frac{b_n^\star|t_n|}{\sigma_n^2},\qquad
 \epsilon_n=(1+Z_n)\eta_n.
\]

\noindent\textbf{Reduction to absolute moments.}\quad
For even $k$, the signed-moment condition~\eqref{eq:HB-main} already
bounds the absolute moment. For odd $k\ge3$, conditional
Cauchy--Schwarz gives
\begin{align*}
 \widetilde\E(|\tilde s_i|^k\mid\mathcal F_n)
 &\le
 \{\widetilde\E(\tilde s_i^{k-1}\mid\mathcal F_n)
    \widetilde\E(\tilde s_i^{k+1}\mid\mathcal F_n)\}^{1/2}\\
 &\le \frac{k!}{2}b_{in}^{k-2}v_{in}
       \sqrt{\frac{k+1}{k}}
 \le \frac{k!}{2}(c_{\mathrm B}b_{in})^{k-2}v_{in},
 \qquad c_{\mathrm B}=\frac2{\sqrt3}.
\end{align*}
For $k=3$, the lower neighboring moment is $v_{in}$ itself.
Thus the absolute-moment hypotheses of
Lemmas~\ref{lem:bernstein-sakhanenko} and
\ref{lem:robinson-lr-transfer} hold with
$\bar b_{in}=c_{\mathrm B}b_{in}$.
This replacement multiplies $H_n$, $\eta_n$, $D_n$, and $\epsilon_n$
by $c_{\mathrm B}$ and leaves $\sigma_n$ and $Z_n$ unchanged.
Let $\mathcal E_n$ denote the event~\eqref{eq:scale-adaptive-lr-conditions}.
Choose $c_\star$ so that $c_{\mathrm B}c_\star$ is below the universal
constants in both lemmas and $c_\star<1/32$.
The lemmas then apply pathwise on $\mathcal E_n$ with the rescaled
Bernstein scales. Absorbing the fixed factor $c_{\mathrm B}$ into
universal constants, we express the following bounds in the original
coordinates $\eta_n$, $D_n$, and $\epsilon_n$.

\par\noindent
\textbf{Robinson companion.}\quad
For $t_n\ne0$ on $\mathcal E_n$,
Lemma~\ref{lem:bernstein-sakhanenko} gives
$b_n^\star|\widehat s_n|\le2D_n<1/16$, hence the root interval and
positive tilted curvature.  Since $A_n=nK_n$, \eqref{eq:r-u} gives
$r=r_n$ and $u=u_n$; put $R_n\equiv R(t_n)$ as in
\eqref{eq:robinson}; also
$\operatorname{LR}(t_n)=p_{\mathrm{SPA},n}(\obs)$ by \eqref{eq:new-lr}.

For $t_n>0$, \eqref{eq:finite-row-positive} and
\eqref{eq:finite-row-lr-positive} give
\[
 \widetilde \P(T_n\ge t_n\mid\cF_n)/R_n=1+O(\epsilon_n),
 \qquad \operatorname{LR}(t_n)/R_n=1+O(\eta_n).
\]
The second ratio is at least $1/2$ by our choice of $c_\star$, so their
quotient proves \eqref{eq:master-bound-main}.  If $t_n<0$,
\eqref{eq:finite-row-center} and monotonicity give
\[
 \widetilde \P(T_n\ge t_n\mid\cF_n)\ge
 \widetilde \P(T_n\ge0\mid\cF_n)\ge\frac12-C\eta_n.
\]
By \eqref{eq:finite-row-negative} and
\eqref{eq:finite-row-lr-negative}, both $R_n$ and
$\operatorname{LR}(t_n)$ differ from this exact tail by $O(\eta_n)$.  All three
are therefore bounded below by a universal positive constant, so division
proves \eqref{eq:master-bound-main}.  Exact complementation retains any atom
at $t_n$.

If $t_n=0$, part~\ref{part:bernstein-domain-center} of
Lemma~\ref{lem:bernstein-sakhanenko} makes zero the unique root.
Because
$A_n=nK_n$, we have
$A_n'''(0)=nK_n'''(0)$ and
$\sigma_n^2=A_n''(0)=nK_n''(0)$.  Thus \eqref{eq:new-lr-zero} and the derivative
bound $|A_n'''(0)|\le3H_n$ from the signed $k=3$ condition give
\[
 \begin{aligned}
 \left|\operatorname{LR}(0)-\frac12\right|
 &=\frac{|K_n'''(0)|}
 {6\sqrt{2\pi n}\,K_n''(0)^{3/2}}
 =\frac{|A_n'''(0)|}{6\sqrt{2\pi}\,\sigma_n^3}\\
 &\le\frac{3H_n}{6\sqrt{2\pi}\,\sigma_n^3}
 =\frac{\eta_n}{2\sqrt{2\pi}}.
 \end{aligned}
\]
The probability bound in \eqref{eq:finite-row-center} is
\[
 \left|\widetilde \P(T_n\ge0\mid\cF_n)-\frac12\right|
 \le C_{\mathrm R}\eta_n.
\]
Thus both tails are $1/2+O(\eta_n)$.  Here $Z_n=0$, so
$\eta_n=\epsilon_n\le c_\star$, and our choice of $c_\star$ makes
$\operatorname{LR}(0)\ge1/4$.  Consequently,
\[
 \begin{aligned}
 \left|
 \frac{\widetilde \P(T_n\ge0\mid\cF_n)}{\operatorname{LR}(0)}-1
 \right|
 &=\frac{\left|\widetilde \P(T_n\ge0\mid\cF_n)-\operatorname{LR}(0)\right|}
 {\operatorname{LR}(0)}\\
 &\le4\left(C_{\mathrm R}+\frac1{2\sqrt{2\pi}}\right)\eta_n
 \le C\eta_n=C\epsilon_n.
 \end{aligned}
\]

\par\noindent
\textbf{Measurability.}\quad
Conditional kernel integration makes the total CGF $A_n$ and its derivatives
jointly measurable on the interior $J_n$ of its finite domain. On the event
that an admissible root exists, strict convexity and continuity identify it as
\[
 \widehat s_n=\inf\{q\in J_n:q\text{ rational},\ A_n'(q)\ge t_n\}.
\]
The countable representation is measurable and agrees with the root
constructed above on $\mathcal E_n$; at $t_n=0$ the root is zero.
The admissibility conditions require a finite interior root with positive
second derivative. Where they fail, assign the SPA value one, as stipulated
in the main text. Where they hold outside $\mathcal E_n$, retain the actual
LR value and its continuous center extension; the theorem asserts no bound
there. Thus no truncation on the complement of the sufficient theorem event
is imposed. This establishes a measurable version of the stated functional.
\end{proof}

\subsection{Two Bernstein certificates}
\label{sec:bounded-support-translators}
For later use, an independent centered row satisfies the conditional
absolute-moment Bernstein property (CB) with envelope $\bar b_n$ if
\begin{equation}\label{eq:cb-property}
 \widetilde\E(|W_{in}|^k\mid\mathcal F_n)
 \le\frac{k!}{2}\bar b_n^{k-2}
       \widetilde\E(W_{in}^2\mid\mathcal F_n),\qquad k\ge3.
\end{equation}
The same definition with ordinary conditional expectation applies to a row
under the original sampling law. The resampling version implies the signed-moment assumption of Theorem~\ref{thm:conditional-lr}.
The envelope $\bar b_n$ is distinct from the standardized coordinate $B_n$.
\begin{corollary}[Bernstein certificates]\label{cor:translators-main}
Use the proof-local standard deviation $\sigma_n$, so that
$\sigma_n^2=\sum_i v_{in}$, and put $H_n=\sum_i b_{in}v_{in}$.
\begin{enumerate}[label=\textnormal{(\alph*)}]
\item\label{part:translator-common-envelope}
If $b_{in}\le\bar b_n$, then
\[
 \frac{B_n}{\sqrt n}=\frac{H_n}{\sigma_n^3}
 \le\frac{\bar b_n}{\sigma_n},\qquad
 D_n\le\frac{\bar b_n|S_n|}{\sigma_n^2}.
\]
\item\label{part:translator-bounded-support}
If $W_{in}=a_{in}(\widetilde X_i-\mu_{in})$, where
$\mu_{in}=\widetilde\E(\widetilde X_i\mid\mathcal F_n)$, the conditional support
diameter of $\widetilde X_i$ is at most $D_0$, its conditional variance
is $\nu_{in}$, and $a_{in}$ is data-measurable, one may take
\[
 b_{in}=D_0|a_{in}|,\quad
 \sigma_n^2=\sum_i a_{in}^2\nu_{in},\quad
 H_n=D_0\sum_i|a_{in}|^3\nu_{in}.
\]
\end{enumerate}
\end{corollary}
\begin{proof}
For (a), $H_n\le\bar b_n\sigma_n^2$. For (b), the mean belongs to the
convex hull of the support, so $|\widetilde X_i-\mu_{in}|\le D_0$.
Consequently the kth absolute moment is at most
$(D_0|a_{in}|)^{k-2}a_{in}^2\nu_{in}$; summing the variances and their
Bernstein-weighted counterparts gives the formulas.
\end{proof}

\section{Proofs of the results in Section~\ref{sec:response-resampling}}\label{sec:proof-saige-validity}

This section states the supporting lemmas and proves
Theorems~\ref{thm:glmrr-target-validity}--\ref{thm:saige-rate}.
Detailed proofs of the supporting lemmas are collected in
Supplementary Section~\ref{sec:glmrr-supporting-proofs}.

\subsection{Proof notation and common ingredients}
For every $n$, fix numbers $x_i$ and vectors
$\bm z_i=(1,\bm u_i^\T)^\T\in\R^d$, $i=1,\ldots,n$.
The row index $n$ on these quantities is suppressed; $d$ is fixed.
Only the responses are random:
\begin{equation}\label{fd:eq:model}
 Y_i\ \text{independently}\sim f_Y(\,\cdot\,;\theta_i^\rho),\qquad
 \theta_i^\rho=\bm z_i^\T\gam_n+\rho_nx_i,\qquad
 f_Y(y;\theta)=\exp\{y\theta-b_Y(\theta)\}h_Y(y).
\end{equation}
The family is the one in Example~\ref{ex:fastspa}. Write $v_Y=b_Y''$, and let $\P_n$ be the product response
law in \eqref{fd:eq:model}. Let $\P_n^0$ be the product law obtained by setting
$\rho_n=0$ while holding $\gam_n$ and the design fixed. Testing concerns
$H_0:\rho_n=0$. Tildes on probabilities and expectations refer to resampling conditional
on the observed responses. Put $\tau_n=v_Y(\alpha_{n,y})$.

Define
\begin{align}\label{fd:eq:oracle}
 \theta_i^0=\bm z_i^\T\gam_n,\quad
 \mu_i^0=b_Y'(\theta_i^0),\quad
 A_i^0=v_Y(\theta_i^0),\quad
 \bm\alpha_n^0=\left(\sum_i A_i^0\bm z_i\bm z_i^\T\right)^{-1}\sum_i A_i^0\bm z_ix_i,\quad
 m_i^0=x_i-\bm z_i^\T\bm\alpha_n^0.
\end{align}
The total baseline-null efficient information is
\begin{equation}\label{fd:eq:J}
 I_{n,x}=\V_{\P_n^0}\left\{\sum_i m_i^0(Y_i-\mu_i^0)\right\}=\sum_i A_i^0(m_i^0)^2=\min_{\bm a\in\R^d}\sum_i A_i^0(x_i-\bm z_i^\T\bm a)^2.
\end{equation}
Thus $I_{n,x}/n$ is average information in the fixed design. It is not
the variance of a generic observation, since the observations need not be
identically distributed. Under an alternative, $I_{n,x}$ is baseline-null
information, not actual alternative information or information evaluated
at the restricted pseudo-true parameter.

Fit the restricted model by maximizing
\[
 \ell_n(\bm\eta)=\frac1n\sum_i
 \{Y_i\bm z_i^\T\bm\eta-b_Y(\bm z_i^\T\bm\eta)\}.
\]
Let its maximizer be $\gh$, and define
\begin{align*}
 \widehat\theta_i=\bm z_i^\T\gh,\quad
 \widehat\mu_i=b_Y'(\widehat\theta_i),\quad
 \widehat A_i=v_Y(\widehat\theta_i),\quad
 \ah=\left(\sum_i\widehat A_i\bm z_i\bm z_i^\T\right)^{-1}\sum_i\widehat A_i\bm z_ix_i,\quad
 \widehat m_i=x_i-\bm z_i^\T\ah.
\end{align*}
Further, define $S_n=\sum_i\widehat m_i(Y_i-\widehat\mu_i)$ and $\widehat I_{n,x}=\sum_i\widehat A_i\widehat m_i^2$. The response-resampling law and target are
\begin{equation}\label{fd:eq:target}
 \widetilde Y_i\mid\bm Y_{1:n}\ \text{independently}\sim
 f_Y(\,\cdot\,;\widehat\theta_i),\qquad
 W_i=\widehat m_i(\widetilde Y_i-\widehat\mu_i),\qquad
 \pn=\widetilde \P\left(\sum_iW_i\ge S_n\right).
\end{equation}
The non-strict inequality is essential to the sharpness result.
For brevity we will use the total CGF $H_n=nK_n$:
\begin{align*}
  H_n(t)=\sum_i\left[b_Y(\widehat\theta_i+t\widehat m_i)-b_Y(\widehat\theta_i)-t\widehat m_i\widehat\mu_i\right].
\end{align*}

\paragraph{Uniformity.} Throughout this section constants depend only on $(K,\lambda,d)$ and fixed neighborhood radii; their values may change between displays. Statements are along any sequence in $\PRR_n$ under Assumption~\ref{assu:fastspa-regularity}. All displayed stochastic orders and failure probabilities are uniform over this class.

\subsubsection{Variance comparison and Bernstein moments}
\paragraph{Taylor's integral formula controls relative changes in variance.}
Since $v_Y(\theta)>0$, the function $\log v_Y$ is well defined.
The fundamental theorem of calculus, equivalently Taylor's formula in
integral form applied to $\log v_Y$, gives the exact identity
\[
 \log\frac{v_Y(\theta+s)}{v_Y(\theta)}
 =s\int_0^1\frac{v_Y'(\theta+us)}{v_Y(\theta+us)}\,du,
 \qquad \theta,s\in\R.
\]
Assumption~\ref{assu:fastspa-regularity} bounds the absolute value of the
integrand by $K$, so
\[
 \left|\log\frac{v_Y(\theta+s)}{v_Y(\theta)}\right|
 \le |s|\int_0^1
       \left|\frac{v_Y'(\theta+us)}{v_Y(\theta+us)}\right|\,du
 \le K|s|.
\]
Exponentiating gives
\begin{equation}\label{fd:eq:logvar}
 e^{-K|s|}\le \frac{v_Y(\theta+s)}{v_Y(\theta)}\le e^{K|s|}.
\end{equation}
No Taylor remainder is discarded, and the shift $s$ need not tend to zero.

\paragraph{The oracle variances are uniformly comparable to the intercept variance.}
Write $\gam_n=(\alpha_{n,y},\bm\beta_{n,y}^\T)^\T$, where
$\bm\beta_{n,y}$ contains the nonintercept coefficients. Since
$\bm z_i=(1,\bm u_i^\T)^\T$,
\[
 \theta_i^0=\bm z_i^\T\gam_n
 =\alpha_{n,y}+\bm u_i^\T\bm\beta_{n,y}.
\]
The bounds $\|\bm u_i\|\le K$ and $\|\bm\beta_{n,y}\|\le K$ in
Assumption~\ref{assu:fastspa-regularity} and Cauchy--Schwarz imply
\[
 |\theta_i^0-\alpha_{n,y}|
 =|\bm u_i^\T\bm\beta_{n,y}|
 \le\|\bm u_i\|\,\|\bm\beta_{n,y}\|\le K^2.
\]
Taking $\theta=\alpha_{n,y}$ and $s=\theta_i^0-\alpha_{n,y}$ in
\eqref{fd:eq:logvar}, and recalling
$A_i^0=v_Y(\theta_i^0)$ and $\tau_n=v_Y(\alpha_{n,y})$, therefore gives
\[
 e^{-K^3}\le\frac{A_i^0}{\tau_n}\le e^{K^3},
 \qquad
 e^{-K^3}\tau_n\le A_i^0\le e^{K^3}\tau_n.
\]
Thus $c\tau_n\le A_i^0\le C\tau_n$ with constants depending only on
$K$, uniformly in $i,n$, including when $\tau_n\to0$.

\paragraph{Diverging efficient information forces the total variance scale to diverge.}
Evaluating the minimum in \eqref{fd:eq:J} at $\bm a=0$ and using
$|x_i|\le K$ gives
\[
 I_{n,x}=\min_{\bm a}\sum_i A_i^0(x_i-\bm z_i^\T\bm a)^2
 \le\sum_i A_i^0x_i^2
 \le K^2e^{K^3}n\tau_n.
\]
Consequently, whenever $I_{n,x}\ge\delta_n\to\infty$,
\[
 n\tau_n\ge\frac{\delta_n}{K^2e^{K^3}}\longrightarrow\infty
\]
uniformly over the stated class.

\begin{lemma}[A variance-sensitive moment certificate]\label{fd:lem:mom}
If $Y\sim f_Y(\cdot;\theta)$, $v=v_Y(\theta)\le V_0$, and~\eqref{fd:eq:logvar} holds, then a constant $b_0$, depending only on $K,V_0$, satisfies $\E_\theta|Y-b_Y'(\theta)|^k\le \frac{k!}{2}b_0^{k-2}v, k\ge3$. Multiplication by a scalar $a$ gives a certificate $b_0|a|$ with variance $a^2v$.
\end{lemma}

\begin{lemma}[Comparison of weighted residual information]\label{fd:lem:wls}
For positive weights $w_i,w_i'$ satisfying
$(1-\epsilon)w_i\le w_i'\le(1+\epsilon)w_i$, $0\le\epsilon<1$,
\[
 (1-\epsilon)\min_{\bm a}\sum_iw_i(x_i-\bm z_i^\T\bm a)^2
 \le \min_{\bm a}\sum_iw_i'(x_i-\bm z_i^\T\bm a)^2
 \le (1+\epsilon)\min_{\bm a}\sum_iw_i(x_i-\bm z_i^\T\bm a)^2.
\]
\end{lemma}

\subsubsection{Restricted fit and deterministic pseudo-null parameter}
Define $\mu_i^\rho=b_Y'(\theta_i^\rho)$ and $\bar\ell_n(\bm\eta)
 =\frac1n\sum_i\{\mu_i^\rho\bm z_i^\T\bm\eta-b_Y(\bm z_i^\T\bm\eta)\}.$
Its maximizer $\gb_Y$ is the pseudo-null parameter for this deterministic
design, not a population projection over random covariates.

\begin{lemma}[Fit stability]\label{fd:lem:fit}
The restricted MLE $\gh$, when it exists, is obtained by imposing
$\rho=0$ in the fitted model \eqref{fd:eq:model}, that is, fixing the
coefficient of $x_i$ at zero:
\[
 \gh\in\operatorname*{arg\,max}_{\bm\eta\in\R^d}
 \frac1n\sum_i
 \{Y_i\bm z_i^\T\bm\eta-b_Y(\bm z_i^\T\bm\eta)\}.
\]
All nuisance coefficients, including the intercept, are estimated over
$\R^d$. The same restricted fit is used when the data are generated
under an alternative $\rho_n\ne0$.
For all sufficiently large $n$, $\gb_Y$ exists uniquely and
$\|\gb_Y-\gam_n\|\le C|\rho_n|$.
The restricted MLE exists uniquely with probability tending to one, and
\begin{equation}\label{fd:eq:fitrate}
 \|\gh-\gb_Y\|=O_{\P_n}((n\tau_n)^{-1/2}),\qquad
 \max_i|\widehat\theta_i-\theta_i^0|
 =O_{\P_n}((n\tau_n)^{-1/2}+|\rho_n|).
\end{equation}
Under the null $\gb_Y=\gam_n$ exactly.
\end{lemma}

\subsubsection{Oracle variances and residual multipliers}
Set $\bar\theta_i=\bm z_i^\T\gb_Y$,
$\bar\mu_i=b_Y'(\bar\theta_i)$ and $\bar A_i=v_Y(\bar\theta_i)$.
Define $\ab_Y,\bar m_i,\bar I_{n,x}$ by replacing $A_i^0$ in
\eqref{fd:eq:oracle}--\eqref{fd:eq:J} by $\bar A_i$.
The barred objects are deterministic.

\begin{lemma}[Residual and variance bounds]\label{fd:lem:variance}
There exist deterministic constants $C,B>0$, depending only on the fixed
class bounds, and events $\mathcal E_n$ satisfying
$\sup_{\P_n\in\PRR_n}\P_n(\mathcal E_n^c)\to0$, on which the
restricted MLE and fitted projection exist, with the following properties.
All bounds are uniform over the stated class.
\begin{enumerate}[label=\textnormal{(\alph*)},leftmargin=2.2em,itemsep=5pt,topsep=5pt]
\item \textnormal{\textbf{The oracle coefficients and residuals are uniformly bounded.}}
For every $n$,
\[
 \|\bm\alpha_n^0\|+\max_{1\le i\le n}|m_i^0|\le C.
\]
\item \textnormal{\textbf{The pseudo-null and fitted coefficients and residuals are uniformly bounded.}}
For all sufficiently large $n$, the deterministic pseudo-null quantities
satisfy
\[
 \|\ab_Y\|+\max_{1\le i\le n}|\bar m_i|\le C.
\]
On $\mathcal E_n$, the fitted quantities satisfy
\[
 \|\ah\|+\max_{1\le i\le n}|\widehat m_i|\le C.
\]
\item \textnormal{\textbf{The pseudo-null and fitted information have the stated relative errors.}}
The fitted ratio is considered on $\mathcal E_n$:
\[
 \frac{\bar I_{n,x}}{I_{n,x}}=1+O(|\rho_n|),\qquad
 \frac{\widehat I_{n,x}}{\bar I_{n,x}}
 =1+O_{\P_n}((n\tau_n)^{-1/2}).
\]
\item \textnormal{\textbf{The resampling summands share a deterministic Bernstein scale.}}
On $\mathcal E_n$, the variables $W_i$ in \eqref{fd:eq:target} satisfy,
conditionally on the observed responses,
\[
 \widetilde\E W_i=0,\qquad
 \widetilde\E W_i^2=\widehat m_i^2\widehat A_i,
\]
and, simultaneously for $i=1,\ldots,n$ and every integer $k\ge3$,
\[
 \widetilde\E|W_i|^k
 \le\frac{k!}{2}B^{k-2}\widetilde\E W_i^2.
\]
The same deterministic $B$ applies to all summands and moment orders.
\end{enumerate}
\end{lemma}

\subsubsection{Score transfer and alternative drift}
\begin{lemma}[Score expansion]\label{fd:lem:score}
Put $ \bar U_n=\sum_i\bar m_i(Y_i-\bar\mu_i)$, $N_n=\sum_i\bar m_i(Y_i-\mu_i^\rho)$ and $\Delta_n^{\mathrm{drift}}=\sum_i\bar m_i(\mu_i^\rho-\bar\mu_i)$. Then uniformly over $\PRR_n$,
\begin{equation}\label{fd:eq:scorebounds}
 S_n-\bar U_n=O_{\P_n}(1),\qquad
 N_n=O_{\P_n}(\sqrt{I_{n,x}}),\qquad
 |\Delta_n^{\mathrm{drift}}|\le C|\rho_n|I_{n,x}.
\end{equation}
In particular $S_n=O_{\P_n}(\sqrt{I_{n,x}}+|\rho_n|I_{n,x})$.
Under the null, $\Delta_n^{\mathrm{drift}}=0$ and
$S_n-\sum_i m_i^0(Y_i-\mu_i^0)=O_{\P_n}(1)$.
\end{lemma}

\subsection{Proof of Theorem~\ref{thm:glmrr-target-validity}: null validity}
\begin{proof}[of Theorem~\ref{thm:glmrr-target-validity}]
Under the null define $U_n=\sum_i m_i^0(Y_i-\mu_i^0)$.
The summands are independent and centered, with variance sum $I_{n,x}$.
Lemma \ref{fd:lem:mom} and bounded $m_i^0$ give
$\sum_i\E|m_i^0(Y_i-\mu_i^0)|^3\le CI_{n,x}$.
The Berry--Esseen inequality therefore yields
\begin{equation}\label{fd:eq:BE}
 \sup_z|\P_n(U_n/\sqrt{I_{n,x}}\le z)-\Phi(z)|\le C/\sqrt{I_{n,x}}.
\end{equation}
Lemmas \ref{fd:lem:variance} and \ref{fd:lem:score} imply
$S_n/\sqrt{\widehat I_{n,x}}-U_n/\sqrt{I_{n,x}}=o_{\P_n}(1)$ uniformly.
Conditionally on the responses, another application of Berry--Esseen gives
\[
 \sup_z\left|\widetilde \P\left(\sum_iW_i/\sqrt{\widehat I_{n,x}}\le z\right)-\Phi(z)\right|
 \le C/\sqrt{\widehat I_{n,x}}
\]
on the common localization event. The same bound applies to the survival
function with a non-strict inequality, by taking left limits and using
continuity of $\Phi$. Thus
\[
 \pn=\Phibar(S_n/\sqrt{\widehat I_{n,x}})+o_{\P_n}(1)
     =\Phibar(U_n/\sqrt{I_{n,x}})+o_{\P_n}(1).
\]
For completeness, this implies uniformity of the distribution functions,
not just convergence at fixed levels. If $W_n=\Phibar(U_n/\sqrt{I_{n,x}})$,
then \eqref{fd:eq:BE} bounds its distance from the uniform distribution by
$C/\sqrt{I_{n,x}}$. For any fixed $\epsilon>0$,
\[
 \sup_{a\in[0,1]}|\P_n(\pn\le a)-a|
 \le C/\sqrt{I_{n,x}}+\epsilon+\P_n(|\pn-W_n|>\epsilon).
\]
Take the supremum over configurations, then $n\to\infty$, then
$\epsilon\downarrow0$. The failure-event convention contributes only its
uniformly vanishing probability.
\end{proof}

\subsection{Proof of Theorem~\ref{thm:new-glmrr}: SPA accuracy}
Theorem~\ref{thm:conditional-lr} has the following common-envelope consequence: if independent centered resampling variables have total variance $V>0$ and obey $\widetilde\E|W_i|^k\le(k!/2)\bar b^{k-2}\widetilde\E W_i^2, k\ge3$, then, whenever $\bar b/\sqrt V+\bar b|S|/V\le c$ for a sufficiently small universal $c$, an admissible saddlepoint exists, the ordinary LR value is positive, and
\begin{equation}\label{fd:eq:input}
 \left|\frac{\widetilde \P(\sum_iW_i\ge S)}{p_{\rm LR}(S)}-1\right|
 \le C\bar b\left(V^{-1/2}+\frac{|S|}{V}\right).
\end{equation}
This follows directly by taking all Bernstein scales equal to $\bar b$ in
the main theorem. Absolute moments imply its signed-moment assumption.
Here $V$ is total variance.

\begin{proof}[of Theorem~\ref{thm:new-glmrr}]
Existence of the restricted MLE and projection follows from
Lemmas \ref{fd:lem:fit} and \ref{fd:lem:variance}. On their common event,
choose $b_0$ from Lemma~\ref{fd:lem:mom} using a common deterministic upper
bound on the fitted variances $\widehat A_i$. Applied conditionally to
$W_i=\widehat m_i(\widetilde Y_i-\widehat\mu_i)$, that lemma gives the
Bernstein scale $b_0|\widehat m_i|$ for each summand. We may therefore choose
the common Bernstein envelope
\[
 \bar b\equiv b_0\max_{1\le i\le n}|\widehat m_i|\le C,
\]
where the deterministic bound follows from Lemma~\ref{fd:lem:variance}.
Take this $\bar b$ and $V=\widehat I_{n,x}$ in \eqref{fd:eq:input}.
The same lemmas and \eqref{fd:eq:scorebounds} give
\[
 \widehat I_{n,x}/I_{n,x}\convp1,\qquad
 \widehat I_{n,x}^{-1/2}+\frac{|S_n|}{\widehat I_{n,x}}
 =O_{\P_n}(I_{n,x}^{-1/2}+|\rho_n|)=o_{\P_n}(1).
\]
The finite-sample theorem therefore applies with probability tending to
one, proving saddlepoint existence and the rate in \eqref{eq:glmrr-quantitative-main}.
Strict convexity of the CGF when $\widehat I_{n,x}>0$ gives uniqueness.
All stochastic bounds and event probabilities were uniform in
$(K,\lambda,d)$ and the divergence envelope $\delta_n$, proving
the asserted uniform rate.

Under the null the relative error tends to zero uniformly in probability.
Since $0\le\pn\le1$, on the event
$\{\ps>0,\ |\pn/\ps-1|\le1/2\}$ we have $\ps\le2$ and
$|\ps-\pn|\le2|\pn/\ps-1|$.
The positivity conclusion above and relative-error convergence make
the probability of this event tend to one uniformly.
Thus the two reported values differ by $o_{\P_n}(1)$ uniformly.
The distribution-function sandwich in the proof of
Theorem \ref{thm:glmrr-target-validity} transfers its uniform null conclusion to $\ps$.
This proves only fixed-level asymptotic validity, not shrinking-level
Type-I error control or improved finite-sample calibration over a normal test.
\end{proof}

\subsection{Proof of Theorem~\ref{thm:saige-rate}: genotype specialization and sharpness}
\label{sec:proof-saige-rate}

\subsubsection{Upper rate and null validity}

\begin{proof}[of the rate and validity assertions in Theorem~\ref{thm:saige-rate}]
For the Bernoulli logistic model,
\begin{equation}\label{fd:eq:gwas-logistic-variance}
 v_Y(\theta)=\expit(\theta)\{1-\expit(\theta)\}
 =\frac{e^\theta}{(1+e^\theta)^2},
 \qquad
 \frac{d}{d\theta}\log v_Y(\theta)=1-2\expit(\theta).
\end{equation}
Because
$\bm z_i^\T\bm\gamma_n=\alpha_{n,y}+\bm u_i^\T\bm\beta_{n,y}$ and
$|\bm u_i^\T\bm\beta_{n,y}|\le K^2$, integrating the derivative in
\eqref{fd:eq:gwas-logistic-variance} gives
\begin{equation}\label{fd:eq:gwas-log-variance-ratio}
 \left|\log\frac{v_Y(\bm z_i^\T\bm\gamma_n)}
 {v_Y(\alpha_{n,y})}\right|
 =\left|\int_{\alpha_{n,y}}^{\bm z_i^\T\bm\gamma_n}
       \{1-2\expit(s)\}\,ds\right|
 \le |\bm u_i^\T\bm\beta_{n,y}|\le K^2.
\end{equation}
Exponentiating \eqref{fd:eq:gwas-log-variance-ratio} and multiplying by
$v_Y(\alpha_{n,y})>0$ gives
\begin{equation}\label{fd:eq:gwas-baseline-variance-comparison}
 e^{-K^2}v_Y(\alpha_{n,y})
 \le v_Y(\bm z_i^\T\bm\gamma_n)
 \le e^{K^2}v_Y(\alpha_{n,y}).
\end{equation}
By the definition $\pi_{n,y}=\expit(\alpha_{n,y})$,
\begin{equation}\label{fd:eq:gwas-prevalence-complement}
 v_Y(\alpha_{n,y})=\pi_{n,y}(1-\pi_{n,y}),
 \qquad
 1-\pi_{n,y}=\frac{1}{1+e^{\alpha_{n,y}}}
 \ge\frac{1}{1+e^K}=1-\expit(K)>0.
\end{equation}
The lower bound in \eqref{fd:eq:gwas-prevalence-complement} uses
$\alpha_{n,y}\le K$ in \eqref{eq:genotype-regularity}. Combining
\eqref{fd:eq:gwas-prevalence-complement} with $1-\pi_{n,y}\le1$ gives
\begin{equation}\label{fd:eq:gwas-baseline-variance-bounds}
 \{1-\expit(K)\}\pi_{n,y}
 \le v_Y(\alpha_{n,y})\le\pi_{n,y}.
\end{equation}
Substituting \eqref{fd:eq:gwas-baseline-variance-bounds} into
\eqref{fd:eq:gwas-baseline-variance-comparison} yields
\begin{equation}\label{fd:eq:gwas-weight-comparison}
 e^{-K^2}\{1-\expit(K)\}\pi_{n,y}
 \le v_Y(\bm z_i^\T\bm\gamma_n)
 \le e^{K^2}\pi_{n,y},
 \qquad i=1,\ldots,n.
\end{equation}

Since the nuisance design includes an intercept, the ordinary least-squares
residual sum of squares is
\begin{equation}\label{fd:eq:gwas-ols-identity}
 \min_{\bm a}\sum_i(x_i-\bm z_i^\top\bm a)^2
 =(1-R_{n,x}^2)\sum_i(x_i-2\pi_{n,x})^2.
\end{equation}
The Hardy--Weinberg condition \eqref{eq:genotype-hwe} gives
\begin{equation}\label{fd:eq:gwas-hwe-variance-bounds}
 2n(1-\eta_n)\pi_{n,x}(1-\pi_{n,x})
 \le \sum_i(x_i-2\pi_{n,x})^2
 \le2n(1+\eta_n)\pi_{n,x}(1-\pi_{n,x}).
\end{equation}
For all sufficiently large $n$, $\eta_n\le1/2$; using
$1/2\le1-\pi_{n,x}<1$ in \eqref{fd:eq:gwas-hwe-variance-bounds} gives
\begin{equation}\label{fd:eq:gwas-genotype-variance-bounds}
 \frac12n\pi_{n,x}
 \le \sum_i(x_i-2\pi_{n,x})^2
 \le3n\pi_{n,x}.
\end{equation}
Substituting \eqref{fd:eq:gwas-genotype-variance-bounds} and
$K^{-1}\le1-R_{n,x}^2\le1$ into \eqref{fd:eq:gwas-ols-identity} gives
\begin{equation}\label{fd:eq:gwas-residual-variance-bounds}
 \frac{n\pi_{n,x}}{2K}
 \le \min_{\bm a}\sum_i(x_i-\bm z_i^\T\bm a)^2
 \le3n\pi_{n,x}.
\end{equation}
The variational expression in \eqref{fd:eq:J} and
\eqref{fd:eq:gwas-weight-comparison} now give
\begin{equation}\label{fd:eq:gwas-weighted-residual-comparison}
\begin{aligned}
 I_{n,x}
 &=\min_{\bm a}\sum_i v_Y(\bm z_i^\T\bm\gamma_n)
       (x_i-\bm z_i^\T\bm a)^2\\
 &\ge e^{-K^2}\{1-\expit(K)\}\pi_{n,y}
       \min_{\bm a}\sum_i(x_i-\bm z_i^\T\bm a)^2,\\
 I_{n,x}
 &\le e^{K^2}\pi_{n,y}
       \min_{\bm a}\sum_i(x_i-\bm z_i^\T\bm a)^2.
\end{aligned}
\end{equation}
Substituting \eqref{fd:eq:gwas-residual-variance-bounds} into
\eqref{fd:eq:gwas-weighted-residual-comparison} gives
\begin{equation}\label{fd:eq:gwas-information-comparison}
 \frac{e^{-K^2}\{1-\expit(K)\}}{2K}
 n\pi_{n,x}\pi_{n,y}
 \le I_{n,x}\le
 3e^{K^2}n\pi_{n,x}\pi_{n,y}.
\end{equation}

It remains to verify the assumptions of the general GLM-RR results. The
design and coefficient requirements in
Assumption~\ref{assu:fastspa-regularity}(i) follow explicitly from
\[
 \max_i|x_i|\le2\le K,
 \qquad \max_i\|\bm u_i\|\le K,
 \qquad \|\bm\gamma_n\|_{\mathrm{nonint}}
 =\|\bm\beta_{n,y}\|\le K,
 \qquad
 \lambda_{\min}\!\left(\frac1n\sum_i\bm z_i\bm z_i^\T\right)\ge\lambda.
\]
For Assumption~\ref{assu:fastspa-regularity}(ii), the natural-parameter
space is $\R$, $v_Y(\theta)>0$, $v_Y(\alpha_{n,y})\le1/4\le K$, and
\[
 \sup_{\theta\in\R}\left|\frac{v_Y'(\theta)}{v_Y(\theta)}\right|
 =\sup_{\theta\in\R}|1-2\expit(\theta)|\le1\le K.
\]
Thus Assumption~\ref{assu:fastspa-regularity} holds uniformly. Combining
\eqref{fd:eq:gwas-information-comparison} with \eqref{eq:genotype-class}
gives, for every law in $\mathcal P_n^{\mathrm{GWAS}}$,
\begin{equation}\label{fd:eq:gwas-class-envelope}
 I_{n,x}\ge
 \frac{e^{-K^2}\{1-\expit(K)\}}{2K}\,\delta_n,
 \qquad
 |\rho_n|\le\delta_n^{-1}
 \le\left[
 \frac{e^{-K^2}\{1-\expit(K)\}}{2K}\,\delta_n
 \right]^{-1}.
\end{equation}
Hence this class is contained in the GLM-RR class with the divergence
envelope in \eqref{fd:eq:gwas-class-envelope}. Theorems
\ref{thm:glmrr-target-validity} and~\ref{thm:new-glmrr} consequently give
the asserted existence and null-validity conclusions and, explicitly,
\[
 \frac{\pn}{\ps}-1
 =O_{\mathcal P^{\mathrm{GWAS}}}
 \{(n\pi_{n,x}\pi_{n,y})^{-1/2}+|\rho_n|\}.
\]
This is \eqref{eq:genotype-rate} and completes the upper-bound part of the
proof.
\end{proof}

\subsubsection{Auxiliary Poisson and LR approximations}

The next two elementary expansions are used for the fixed-design sharpness result in Theorem~\ref{thm:saige-rate}. Their proofs are in Supplementary Section~\ref{sec:glmrr-supporting-proofs}.
\begin{lemma}[Symmetric Poisson difference]\label{fd:lem:skellam}
Let $A,B$ be independent $\Pois(L/2)$ and $D=A-B$, with $L\to\infty$.
For each fixed $M$, uniformly over integers $k$ with $|k|\le M\sqrt L$,
\[
 \P(D\ge k)
 =\Phibar(k/\sqrt L)+\frac{\phi(k/\sqrt L)}{2\sqrt L}
   +o(L^{-1/2}).
\]
\end{lemma}

\begin{lemma}[Central LR approximation for a symmetric binomial difference]
\label{fd:lem:symLR}
Let
\[
 H(t)=m\log(1-p+pe^t)+m\log(1-p+pe^{-t}),\qquad
 \Lambda=2mp(1-p)\to\infty,\qquad 0<p\le1/4.
\]
Uniformly over $|s|/\sqrt\Lambda\le M$, its ordinary LR upper-tail
functional satisfies
\[
 p_{\rm LR}(s)=\Phibar(s/\sqrt\Lambda)+O_M(\Lambda^{-1}).
\]
\end{lemma}

\subsubsection{Sharpness}

\begin{proof}[of the sharpness assertion in Theorem~\ref{thm:saige-rate}]
For $n\ge8$, put $m_n=\lfloor n/4\rfloor$ and choose the deterministic
genotype vector
\begin{equation}\label{fd:eq:gwas-witness-genotypes}
 x_i=\begin{cases}
 0,&1\le i\le m_n,\\
 2,&m_n<i\le2m_n,\\
 1,&2m_n<i\le n.
 \end{cases}
\end{equation}
Here $x_i$ is the allele count for individual $i$; the genotypes are fixed,
not sampled. Only the responses are random. Take the intercept-only null
response model
\[
 Y_i\mathrel{\mathrm{i.i.d.}}\sim\Ber(\pi_{n,y}),
 \qquad \pi_{n,y}=n^{-2/3},
 \qquad \rho_n=0.
\]
We first verify that this sequence belongs to the class in the theorem.
By \eqref{fd:eq:gwas-witness-genotypes}, its empirical minor-allele
frequency is
\[
 \pi_{n,x}=\frac1{2n}\sum_i x_i
 =\frac{2m_n+(n-2m_n)}{2n}=\frac12.
\]
Also, $R_{n,x}^2=0$. Furthermore,
\[
 \frac1n\sum_i(x_i-2\pi_{n,x})^2=\frac{2m_n}{n},
 \qquad
 \left|\frac{2m_n/n}{2\pi_{n,x}(1-\pi_{n,x})}-1\right|
 =\left|\frac{4m_n}{n}-1\right|\le\frac4n.
\]
Thus \eqref{eq:genotype-hwe} holds with $\eta_n=4/n$. With $K=2$,
$\lambda=1$, and $d=1$, the nuisance design is $z_i=1$, the nonintercept
coefficient vector is empty, and
\[
 \max_i|x_i|=2,\qquad
 \alpha_{n,y}=\logit(n^{-2/3})\le0<2,
 \qquad
 \frac1n\sum_i z_i^2=1.
\]
Thus all the remaining design and coefficient conditions hold. The null
information and the information scale in the theorem satisfy
\begin{equation}\label{fd:eq:gwas-witness-information}
 I_{n,x}=2m_n\pi_{n,y}(1-\pi_{n,y})\sim\frac12n^{1/3},
 \qquad
 n\pi_{n,x}\pi_{n,y}=\frac12n^{1/3}.
\end{equation}
The choice $\delta_n=n^{1/4}/4$ satisfies
$n\pi_{n,x}\pi_{n,y}\ge\delta_n$ for all sufficiently large $n$, while
$|\rho_n|=0\le\delta_n^{-1}$. Hence the constructed law belongs to
$\mathcal P_n^{\mathrm{GWAS},0}$ for every sufficiently large $n$.

\paragraph{Step 1: Reduce the score to a symmetric difference of binomial counts.}
The fitted response probability is
$\widehat\pi_{n,y}=\overline Y_n$. Since
$n\pi_{n,y}=n^{1/3}\to\infty$,
\begin{equation}\label{fd:eq:gwas-witness-prevalence-moments}
 \E_{\P_n}(\widehat\pi_{n,y})=\pi_{n,y},
 \qquad
 \frac{\V_{\P_n}(\widehat\pi_{n,y})}{\pi_{n,y}^2}
 =\frac{1-\pi_{n,y}}{n\pi_{n,y}}\longrightarrow0.
\end{equation}
Applying Chebyshev's inequality to
\eqref{fd:eq:gwas-witness-prevalence-moments} gives
\begin{equation}\label{fd:eq:gwas-witness-fit}
 \widehat\pi_{n,y}/\pi_{n,y}\convp1,
 \qquad
 \P_n(0<\widehat\pi_{n,y}<1/4)\longrightarrow1.
\end{equation}
On the event $0<\widehat\pi_{n,y}<1/4$ in
\eqref{fd:eq:gwas-witness-fit}, all fitted variance weights are equal
and positive. The intercept-only fitted projection is therefore
\begin{equation}\label{fd:eq:gwas-witness-projection}
 \widehat\alpha
 =\frac{\sum_i\widehat\pi_{n,y}(1-\widehat\pi_{n,y})x_i}
        {\sum_i\widehat\pi_{n,y}(1-\widehat\pi_{n,y})}
 =\frac{2m_n+(n-2m_n)}{n}=1.
\end{equation}
Thus $\widehat m_i=x_i-1$ equals $-1$ for the first $m_n$ individuals,
$1$ for the next $m_n$, and $0$ for the remaining individuals.
Let $N_{2n}=\sum_{i:x_i=2}Y_i$ and $N_{0n}=\sum_{i:x_i=0}Y_i$.
Since $\sum_i(x_i-1)=0$, substituting
\eqref{fd:eq:gwas-witness-projection} into the score gives
\begin{equation}\label{fd:eq:gwas-witness-score}
 S_n=\sum_i(x_i-1)(Y_i-\widehat\pi_{n,y})
     =\sum_i(x_i-1)Y_i=N_{2n}-N_{0n}.
\end{equation}
Conditionally on the responses, the fitted Bernoulli responses are
independent. Hence the resampled score and its total CGF are
\[
 \widetilde S_n=\widetilde N_{2n}-\widetilde N_{0n},\qquad
 \widetilde N_{2n},\widetilde N_{0n}
 \mathrel{\stackrel{\mathrm{ind}}\sim}
 \Bin(m_n,\widehat\pi_{n,y}).
\]
\begin{equation}\label{fd:eq:gwas-witness-cgf}
 H_n(t)=m_n\log(1-\widehat\pi_{n,y}+\widehat\pi_{n,y}e^t)
       +m_n\log(1-\widehat\pi_{n,y}+\widehat\pi_{n,y}e^{-t}).
\end{equation}
All subsequent calculations involving the fitted variance are on
$0<\widehat\pi_{n,y}<1/4$. Write
$\widehat I_{n,x}=2m_n\widehat\pi_{n,y}(1-\widehat\pi_{n,y})$ and
$z_n=S_n/\sqrt{\widehat I_{n,x}}$.
For the following moment and convergence calculations, extend $S_n$ by
$N_{2n}-N_{0n}$ and set $\widehat I_{n,x}=1$ on the complement of this
event. These auxiliary extensions do not change the defined $p$-values
or the limiting results, since the complement has probability tending
to zero by \eqref{fd:eq:gwas-witness-fit}. Under the data-generating law,
\begin{equation}\label{fd:eq:gwas-witness-score-moments}
 \E_{\P_n}(S_n)=0,
 \qquad
 \V_{\P_n}(S_n)=2m_n\pi_{n,y}(1-\pi_{n,y})=I_{n,x}\longrightarrow\infty.
\end{equation}
By \eqref{fd:eq:gwas-witness-score},
$S_n=\sum_i(x_i-1)(Y_i-\pi_{n,y})$ is a sum of independent centered
variables of absolute value at most one. Thus, for every $\epsilon>0$,
once $\epsilon\sqrt{I_{n,x}}>1$,
\begin{equation}\label{fd:eq:gwas-witness-lindeberg}
 \frac{1}{I_{n,x}}\sum_i
 \E_{\P_n}\!\left[
 (x_i-1)^2(Y_i-\pi_{n,y})^2
 \mathbf{1}\{|(x_i-1)(Y_i-\pi_{n,y})|>\epsilon\sqrt{I_{n,x}}\}
 \right]=0.
\end{equation}
The Lindeberg--Feller theorem, applied using
\eqref{fd:eq:gwas-witness-score-moments} and
\eqref{fd:eq:gwas-witness-lindeberg}, yields
$S_n/\sqrt{I_{n,x}}\convd Z\sim N(0,1)$.
Moreover, \eqref{fd:eq:gwas-witness-fit} and
\eqref{fd:eq:gwas-witness-information} give
\begin{equation}\label{fd:eq:gwas-witness-standardization}
 \frac{\widehat I_{n,x}}{I_{n,x}}
 =\frac{\widehat\pi_{n,y}}{\pi_{n,y}}
  \frac{1-\widehat\pi_{n,y}}{1-\pi_{n,y}}\convp1,
 \qquad
 \frac{n\pi_{n,x}\pi_{n,y}}{\widehat I_{n,x}}\convp1.
\end{equation}
Consequently, Slutsky's theorem gives
\begin{equation}\label{fd:eq:gwas-witness-clt}
 z_n=\frac{S_n}{\sqrt{I_{n,x}}}
       \sqrt{\frac{I_{n,x}}{\widehat I_{n,x}}}
 \convd Z\sim N(0,1).
\end{equation}

\paragraph{Step 2: Obtain the expansion of the exact resampling tail.}
Lemma~\ref{lem:le-cam-elementary}, applied to each fitted binomial
count, and the total-variation inequality for product measures give
\begin{equation}\label{fd:eq:gwas-witness-poisson-error}
 \begin{aligned}
 d_{\mathrm{TV}}\!\left(
   \Bin(m_n,\widehat\pi_{n,y})^{\otimes2},
   \Pois(m_n\widehat\pi_{n,y})^{\otimes2}\right)
 &\le4m_n\widehat\pi_{n,y}^{\,2}\\
 &=O_{\P_n}(n^{-1/3})=o_{\P_n}(I_{n,x}^{-1/2}).
 \end{aligned}
\end{equation}
Here $\otimes2$ denotes the joint law of two independent copies.
Taking their difference cannot increase total variation, so
\eqref{fd:eq:gwas-witness-poisson-error} also bounds the error in every
upper-tail probability of $\widetilde S_n$.
Put $L=2m_n\widehat\pi_{n,y}$. Equations
\eqref{fd:eq:gwas-witness-fit} and \eqref{fd:eq:gwas-witness-information}
give
\begin{equation}\label{fd:eq:gwas-witness-poisson-scale}
 \frac{L}{I_{n,x}}
 =\frac{\widehat\pi_{n,y}}{\pi_{n,y}(1-\pi_{n,y})}\convp1,
 \qquad L\longrightarrow\infty\quad\text{in probability}.
\end{equation}
After conditioning, $L$ and the integer cutoff $S_n$ are fixed.
On $|z_n|\le M$, the cutoff is in the range of
Lemma~\ref{fd:lem:skellam}: the identity
$\widehat I_{n,x}=L(1-\widehat\pi_{n,y})$ gives
\begin{equation}\label{fd:eq:gwas-witness-argument-comparison}
 \frac{S_n}{\sqrt L}=z_n\sqrt{1-\widehat\pi_{n,y}},
 \qquad
 \left|\frac{S_n}{\sqrt L}-z_n\right|
 \le M\widehat\pi_{n,y}.
\end{equation}
In particular, $|S_n|/\sqrt L\le M$.
The deterministic remainder bound in
\eqref{fd:eq:skellam-summed-error}--\eqref{fd:eq:skellam-trapezoid}
is $C_M\sqrt{\log L}/L$ for all sufficiently large real $L$.
Thus \eqref{fd:eq:gwas-witness-poisson-error} and
\eqref{fd:eq:gwas-witness-poisson-scale} imply, on $|z_n|\le M$,
\begin{equation}\label{fd:eq:gwas-witness-poisson-tail}
 \begin{aligned}
 \left|\pn-\Phibar(S_n/\sqrt L)
          -\frac{\phi(S_n/\sqrt L)}{2\sqrt L}\right|
 &\le4m_n\widehat\pi_{n,y}^{\,2}
       +C_M\frac{\sqrt{\log L}}{L}\\
 &=O_{\P_n}(n^{-1/3}\sqrt{\log n})
   =o_{\P_n}(I_{n,x}^{-1/2}).
 \end{aligned}
\end{equation}
The normal density is bounded, so the mean-value theorem and
\eqref{fd:eq:gwas-witness-argument-comparison}, together with
\eqref{fd:eq:gwas-witness-fit}, give, uniformly on $|z_n|\le M$,
\begin{equation}\label{fd:eq:gwas-witness-normal-tail-error}
 \left|\Phibar(S_n/\sqrt L)-\Phibar(z_n)\right|
 =O_{\P_n}(\pi_{n,y})
 =o_{\P_n}(I_{n,x}^{-1/2}).
\end{equation}
Using \eqref{fd:eq:gwas-witness-argument-comparison} and
$L^{-1/2}=(1-\widehat\pi_{n,y})^{1/2}\widehat I_{n,x}^{-1/2}$,
boundedness and Lipschitz continuity of $\phi$ give
\begin{equation}\label{fd:eq:gwas-witness-boundary-error}
 \left|
 \frac{\phi(S_n/\sqrt L)}{2\sqrt L}
 -\frac{\phi(z_n)}{2\sqrt{\widehat I_{n,x}}}
 \right|
 =O_{\P_n}(\pi_{n,y}I_{n,x}^{-1/2})
 =o_{\P_n}(I_{n,x}^{-1/2}).
\end{equation}
Combining \eqref{fd:eq:gwas-witness-poisson-tail} with
\eqref{fd:eq:gwas-witness-normal-tail-error} and
\eqref{fd:eq:gwas-witness-boundary-error} gives
\begin{equation}\label{fd:eq:gwas-witness-exact-tail}
 \pn=\Phibar(z_n)+\frac{\phi(z_n)}{2\sqrt{\widehat I_{n,x}}}
       +o_{\P_n}(I_{n,x}^{-1/2}).
\end{equation}

\begin{samepage}
\paragraph{Step 3: Obtain the expansion of the LR formula.}
For the CGF \eqref{fd:eq:gwas-witness-cgf}, Lemma~\ref{fd:lem:symLR} applies
with $m=m_n$, $p=\widehat\pi_{n,y}$, $s=S_n$, and
$\Lambda=\widehat I_{n,x}\to\infty$ in probability by
\eqref{fd:eq:gwas-witness-standardization}.
Its proof constructs a unique admissible saddlepoint on $|z_n|\le M$
for all sufficiently large $\widehat I_{n,x}$, and gives the bound
\par
\end{samepage}
\begin{equation}\label{fd:eq:gwas-witness-spa-tail}
 |\ps-\Phibar(z_n)|\le\frac{C_M}{\widehat I_{n,x}}
 =O_{\P_n}(I_{n,x}^{-1}).
\end{equation}
This bound includes $S_n=0$: symmetry in
\eqref{fd:eq:gwas-witness-cgf} gives $H_n'''(0)=0$, so
\eqref{eq:new-lr-zero} gives $\ps=1/2$ there.
Also, on $|z_n|\le M$, \eqref{fd:eq:gwas-witness-spa-tail} gives
\[
 \ps\ge\Phibar(M)-C_M/\widehat I_{n,x}\ge\Phibar(M)/2>0
\]
for sufficiently large $\widehat I_{n,x}$.

\paragraph{Step 4: Subtract the expansions to obtain the limiting error.}
Combining \eqref{fd:eq:gwas-witness-spa-tail} with
\eqref{fd:eq:gwas-witness-exact-tail} yields, uniformly on $|z_n|\le M$,
\begin{equation}\label{fd:eq:gwas-witness-tail-difference}
 \pn-\ps
 =\frac{\phi(z_n)}{2\sqrt{\widehat I_{n,x}}}
  +o_{\P_n}(I_{n,x}^{-1/2}),
 \qquad
 \ps=\Phibar(z_n)+o_{\P_n}(1).
\end{equation}

To remove the compact restriction, \eqref{fd:eq:gwas-witness-clt} gives,
for every $M>0$,
\begin{equation}\label{fd:eq:gwas-witness-tightness}
 \limsup_{n\to\infty}\P_n(|z_n|>M)\le\P(|Z|\ge M)\longrightarrow0
 \qquad(M\to\infty).
\end{equation}
For every $\epsilon>0$ and fixed $M>0$,
\begin{equation}\label{fd:eq:gwas-witness-truncation}
 \begin{aligned}
 &\P_n\!\left(
 \sqrt{I_{n,x}}\left|\pn-\ps
       -\frac{\phi(z_n)}{2\sqrt{\widehat I_{n,x}}}\right|>\epsilon
 \right)\\
 &\quad\le \P_n(|z_n|>M)
 +\P_n\!\left(
 \sqrt{I_{n,x}}\left|\pn-\ps
       -\frac{\phi(z_n)}{2\sqrt{\widehat I_{n,x}}}\right|>\epsilon,\
 |z_n|\le M\right).
 \end{aligned}
\end{equation}
The second probability in \eqref{fd:eq:gwas-witness-truncation}
tends to zero by \eqref{fd:eq:gwas-witness-tail-difference}.
Taking the limit superior as $n\to\infty$ and then letting $M\to\infty$
in \eqref{fd:eq:gwas-witness-tightness} therefore gives
\[
 \pn-\ps-\frac{\phi(z_n)}{2\sqrt{\widehat I_{n,x}}}
 =o_{\P_n}(I_{n,x}^{-1/2}).
\]
For the denominator, the same event decomposition gives directly
\[
 \P_n(|\ps-\Phibar(z_n)|>\epsilon)
 \le \P_n(|z_n|>M)
 +\P_n(|\ps-\Phibar(z_n)|>\epsilon,\ |z_n|\le M).
\]
Using \eqref{fd:eq:gwas-witness-tail-difference} and
\eqref{fd:eq:gwas-witness-tightness}, first let $n\to\infty$ and then
$M\to\infty$ to obtain $\ps-\Phibar(z_n)=o_{\P_n}(1)$.
Thus both conclusions of \eqref{fd:eq:gwas-witness-tail-difference}
hold without the compact restriction. The same truncation argument
shows that the admissible saddlepoint exists and $\ps>0$ with probability
tending to one. Combining
\eqref{fd:eq:gwas-witness-tail-difference} with
\eqref{fd:eq:gwas-witness-standardization} and
\eqref{fd:eq:gwas-witness-clt} gives
\[
 \sqrt{n\pi_{n,x}\pi_{n,y}}\left(\frac{\pn}{\ps}-1\right)
 =\sqrt{\frac{n\pi_{n,x}\pi_{n,y}}{\widehat I_{n,x}}}
   \frac{\phi(z_n)}{2\ps}+o_{\P_n}(1).
\]
Here $1/\ps=O_{\P_n}(1)$ because
$\ps\convd\Phibar(Z)>0$ almost surely. The continuous mapping theorem
therefore gives the limiting relative error
\[
 \sqrt{n\pi_{n,x}\pi_{n,y}}\left(\frac{\pn}{\ps}-1\right)
 \convd\frac{\phi(Z)}{2\Phibar(Z)}.
\]
The limit is finite and strictly positive almost surely. Hence, for every
$\epsilon>0$, there exist $0<c_\epsilon<C_\epsilon<\infty$ such that
\[
 \liminf_{n\to\infty}\P_n\!\left(
 c_\epsilon\le
 \sqrt{n\pi_{n,x}\pi_{n,y}}\left|\frac{\pn}{\ps}-1\right|
 \le C_\epsilon\right)\ge1-\epsilon.
\]
This is the two-sided stochastic-order assertion
\[
 \left|\frac{\pn}{\ps}-1\right|
 \asymp_{\P_n}(n\pi_{n,x}\pi_{n,y})^{-1/2},
\]
in \eqref{eq:genotype-sharpness}. The first-order discrepancy is the
half-atom boundary correction for the inclusive lattice tail, which is
absent from ordinary LR. This witness establishes
the claimed rate without asserting sharpness under a vanishing minor-allele
frequency.
\end{proof}
\section{Detailed proofs of the supporting lemmas for
Theorems~\ref{thm:glmrr-target-validity}--\ref{thm:saige-rate}}
\label{sec:glmrr-supporting-proofs}

The notation and uniformity conventions in these proofs are inherited
from Supplementary Section~\ref{sec:proof-saige-validity}.
Except for the localization lemma below, the lemma statements appear in
that section.

\subsection{A localization lemma for concave criteria}

\begin{lemma}[Localization of a concave maximizer]\label{fd:lem:localization}
Let $L:\R^d\to\R$ be twice continuously differentiable and concave.
Suppose that, for some $\bm\eta_0\in\R^d$ and $r,\kappa>0$,
\[
 -\nabla^2L(\bm\eta)\succeq\kappa\bm I_d
 \quad\text{whenever }\|\bm\eta-\bm\eta_0\|\le r,
 \qquad
 \|\nabla L(\bm\eta_0)\|<\frac{\kappa r}{2}.
\]
Then $L$ has a global maximizer $\widehat{\bm\eta}$ satisfying
\[
 \|\widehat{\bm\eta}-\bm\eta_0\|<r,\qquad
 \nabla L(\widehat{\bm\eta})=0,\qquad
 \|\widehat{\bm\eta}-\bm\eta_0\|
 \le\frac{\|\nabla L(\bm\eta_0)\|}{\kappa}.
\]
If $L$ is strictly concave on $\R^d$, this maximizer is unique.
\end{lemma}

\begin{proof}
For $\|\bm\delta\|\le r$, Taylor's integral formula and the curvature
bound give
\[
 L(\bm\eta_0+\bm\delta)-L(\bm\eta_0)
 \le\nabla L(\bm\eta_0)^\T\bm\delta
       -\frac{\kappa}{2}\|\bm\delta\|^2.
\]
On the boundary $\|\bm\delta\|=r$, the right-hand side is at most
$r\|\nabla L(\bm\eta_0)\|-\kappa r^2/2<0$.
By continuity, $L$ attains a maximum on the closed ball, and the strict
boundary inequality places any such maximizer $\widehat{\bm\eta}$ in its
interior. Thus $\nabla L(\widehat{\bm\eta})=0$. Concavity then gives,
for every $\bm\eta\in\R^d$,
\[
 L(\bm\eta)\le L(\widehat{\bm\eta})
 +\nabla L(\widehat{\bm\eta})^\T(\bm\eta-\widehat{\bm\eta})
 =L(\widehat{\bm\eta}),
\]
so this is a global maximizer.
Put $\bm\Delta=\widehat{\bm\eta}-\bm\eta_0$.
The segment from $\bm\eta_0$ to $\widehat{\bm\eta}$ lies in the ball,
and hence
\[
 \nabla L(\bm\eta_0)^\T\bm\Delta
 =-\int_0^1\bm\Delta^\T
       \nabla^2L(\bm\eta_0+t\bm\Delta)\bm\Delta\,dt
 \ge\kappa\|\bm\Delta\|^2.
\]
Cauchy--Schwarz and division by $\kappa\|\bm\Delta\|$ give the distance
bound when $\bm\Delta\ne0$; it is immediate when $\bm\Delta=0$.
Strict concavity gives uniqueness under the additional hypothesis.
\end{proof}

\subsection{Proof of Lemma~\ref{fd:lem:mom}}

\begin{proof}
Put $R=Y-b_Y'(\theta)$. Its log MGF is
$g(t)=b_Y(\theta+t)-b_Y(\theta)-tb_Y'(\theta)$.
Taylor's integral formula and \eqref{fd:eq:logvar} give
\begin{equation}\label{fd:eq:moment-cgf-bound}
 0\le g(t)
 =t^2\int_0^1(1-s)v_Y(\theta+st)\,ds
 \le \frac{vt^2}{2}e^{K|t|}.
\end{equation}
Taking $t=\pm1$ in \eqref{fd:eq:moment-cgf-bound} and using
$e^a-1\le ae^a$ for $a\ge0$, with $v\le V_0$, yields
\begin{equation}\label{fd:eq:moment-cosh-bound}
 \E_\theta\cosh(R)-1
 =\frac{e^{g(1)}+e^{g(-1)}}2-1
 \le \frac{e^K}{2}\exp(V_0e^K/2)\,v
 \le Cv,
\end{equation}
where $C\ge1$ depends only on $K,V_0$.
The nonnegative power series of $\cosh$ and
\eqref{fd:eq:moment-cosh-bound} imply
\begin{equation}\label{fd:eq:moment-even-bound}
 \sum_{j=1}^\infty\frac{\E_\theta|R|^{2j}}{(2j)!}
 =\E_\theta\cosh(R)-1\le Cv,
 \qquad
 \E_\theta|R|^{2j}\le C(2j)!\,v\quad(j\ge1).
\end{equation}
For odd $k\ge3$, Cauchy--Schwarz and
\eqref{fd:eq:moment-even-bound} give
\begin{equation}\label{fd:eq:moment-odd-bound}
 \E_\theta|R|^k
 \le\{\E_\theta|R|^{k-1}\E_\theta|R|^{k+1}\}^{1/2}
 \le C\sqrt{(k-1)!(k+1)!}\,v
 \le\sqrt2 Ck!\,v.
\end{equation}
Choose $b_0=\max\{1,2\sqrt2 C\}$. Since
$b_0^{k-2}\ge2\sqrt2 C$ for $k\ge3$,
\eqref{fd:eq:moment-even-bound} and \eqref{fd:eq:moment-odd-bound} yield
\begin{equation}\label{fd:eq:moment-certificate}
 \E_\theta|R|^k\le\frac{k!}{2}b_0^{k-2}v,\qquad k\ge3.
\end{equation}
Finally, multiplying \eqref{fd:eq:moment-certificate} by $|a|^k$ gives
\[
 \E_\theta|aR|^k
 \le\frac{k!}{2}(b_0|a|)^{k-2}a^2v,
 \qquad \V_\theta(aR)=a^2v,
\]
which proves the scaled certificate as well.
\end{proof}

\subsection{Proof of Lemma~\ref{fd:lem:wls}}

\begin{proof}
Multiplying the weight bounds by the nonnegative squared residuals and
summing gives, for every $\bm a$,
\begin{equation}\label{fd:eq:wls-pointwise-comparison}
 (1-\epsilon)\sum_iw_i(x_i-\bm z_i^\T\bm a)^2
 \le \sum_iw_i'(x_i-\bm z_i^\T\bm a)^2
 \le (1+\epsilon)\sum_iw_i(x_i-\bm z_i^\T\bm a)^2.
\end{equation}
For the lower bound, the first inequality in
\eqref{fd:eq:wls-pointwise-comparison} implies, for every $\bm a$,
\[
 \sum_iw_i'(x_i-\bm z_i^\T\bm a)^2
 \ge (1-\epsilon)\sum_iw_i(x_i-\bm z_i^\T\bm a)^2
 \ge (1-\epsilon)\min_{\bm b}\sum_iw_i(x_i-\bm z_i^\T\bm b)^2.
\]
Minimizing the left-hand side over $\bm a$ gives
\begin{equation}\label{fd:eq:wls-lower-comparison}
 \min_{\bm a}\sum_iw_i'(x_i-\bm z_i^\T\bm a)^2
 \ge (1-\epsilon)\min_{\bm a}\sum_iw_i(x_i-\bm z_i^\T\bm a)^2.
\end{equation}
For the upper bound, taking infima in the last inequality of
\eqref{fd:eq:wls-pointwise-comparison} gives
\begin{equation}\label{fd:eq:wls-upper-comparison}
 \min_{\bm a}\sum_iw_i'(x_i-\bm z_i^\T\bm a)^2
 \le (1+\epsilon)\min_{\bm a}\sum_iw_i(x_i-\bm z_i^\T\bm a)^2.
\end{equation}
Weighted least-squares minima are attained. Thus
\eqref{fd:eq:wls-lower-comparison} and \eqref{fd:eq:wls-upper-comparison}
give the claimed two-sided bound.
\end{proof}

\subsection{Proof of Lemma~\ref{fd:lem:fit}}

\begin{proof}
We prove the two nuisance-parameter bounds by successive applications of
Lemma~\ref{fd:lem:localization}: first to the deterministic criterion
$\bar\ell_n$ centered at $\gam_n$, and then to the random criterion
$\ell_n$ centered at $\gb_Y$.
Fix $r>0$.
Both criteria share Hessian
$-n^{-1}\sum_i v_Y(\bm z_i^\T\bm\eta)\bm z_i\bm z_i^\T$.
For $\|\bm\eta-\gam_n\|\le2r$, the covariate and coefficient bounds give
\[
 |\bm z_i^\T\bm\eta-\alpha_{n,y}|
 \le K^2+2r\sqrt{1+K^2}.
\]
Consequently, \eqref{fd:eq:logvar} gives
\[
 e^{-K(K^2+2r\sqrt{1+K^2})}\tau_n
 \le v_Y(\bm z_i^\T\bm\eta)
 \le e^{K(K^2+2r\sqrt{1+K^2})}\tau_n.
\]
Combining these bounds with the deterministic Gram bound yields
\begin{equation}\label{fd:eq:fit-curvature}
 -\nabla^2\ell_n(\bm\eta)
 =-\nabla^2\bar\ell_n(\bm\eta)
 =\frac1n\sum_i v_Y(\bm z_i^\T\bm\eta)\bm z_i\bm z_i^\T
 \succeq c\tau_n\bm I_d
 \qquad(\|\bm\eta-\gam_n\|\le2r),
\end{equation}
where $c>0$ depends only on the fixed envelopes and $r$.
The Hessian is strictly negative definite everywhere because the weights
are positive and the design spans $\R^d$. Thus both criteria are strictly
concave, and each maximizer is unique whenever it exists.

\paragraph{The first application bounds the deterministic pseudo-null displacement.}
We prove that $\gb_Y$ exists uniquely and
$\|\gb_Y-\gam_n\|\le C|\rho_n|$.
Apply Lemma~\ref{fd:lem:localization} with radius $r$ and
\[
 L=\bar\ell_n,\qquad \bm\eta_0=\gam_n,\qquad \kappa=c\tau_n.
\]
The curvature bound \eqref{fd:eq:fit-curvature} applies on this ball. To verify the gradient
condition, differentiate the deterministic criterion at $\gam_n$:
\[
 \nabla\bar\ell_n(\gam_n)
 =\frac1n\sum_i\bm z_i(\mu_i^\rho-\mu_i^0).
\]
Since $\theta_i^\rho=\theta_i^0+\rho_nx_i$ and $b_Y''=v_Y$,
Taylor's integral formula gives the exact identity
\[
 \mu_i^\rho-\mu_i^0
 =b_Y'(\theta_i^\rho)-b_Y'(\theta_i^0)
 =\rho_nx_i\int_0^1v_Y(\theta_i^0+t\rho_nx_i)\,dt.
\]
No remainder is discarded. We next bound the variance along this entire
integration segment. By \eqref{fd:eq:logvar}, the bound
$A_i^0\le e^{K^3}\tau_n$ established in Supplementary
Section~\ref{sec:proof-saige-validity}, and $|x_i|\le K$, we have,
uniformly in $i$ and $t\in[0,1]$,
\[
 v_Y(\theta_i^0+t\rho_nx_i)
 \le A_i^0e^{K|\rho_nx_i|}
 \le e^{K^3+K^2}\tau_n
\]
for all sufficiently large $n$, since $|\rho_n|\le1$ eventually.
Consequently, the triangle inequality and
$\|\bm z_i\|\le\sqrt{1+K^2}$ yield
\begin{align*}
 \|\nabla\bar\ell_n(\gam_n)\|
 &\le\frac{|\rho_n|}{n}\sum_i\|\bm z_i\|\,|x_i|
       \int_0^1v_Y(\theta_i^0+t\rho_nx_i)\,dt\\
 &\le K\sqrt{1+K^2}\,e^{K^3+K^2}\tau_n|\rho_n|
 \le C\tau_n|\rho_n|.
\end{align*}
Here $C$ depends only on $K$, including when $\tau_n\to0$.
Hence the gradient condition holds for all sufficiently large $n$, since
\[
 \frac{\|\nabla\bar\ell_n(\gam_n)\|}{c\tau_n}
 \le\frac Cc|\rho_n|<\frac r2
\]
by $\rho_n\to0$. Thus the lemma gives existence of the deterministic
maximizer $\gb_Y$ and the bound
\begin{equation}\label{fd:eq:fit-population-distance}
 \|\gb_Y-\gam_n\|
 \le\frac{\|\nabla\bar\ell_n(\gam_n)\|}{c\tau_n}
 \le\frac Cc|\rho_n|.
\end{equation}
Strict concavity gives uniqueness.

\paragraph{The second application bounds the restricted MLE error around the pseudo-null parameter.}
We now prove that $\gh$ exists uniquely with probability tending to one
and $\|\gh-\gb_Y\|=O_{\P_n}((n\tau_n)^{-1/2})$.
Apply Lemma~\ref{fd:lem:localization} with radius $r$ and
\[
 L=\ell_n,\qquad \bm\eta_0=\gb_Y,\qquad \kappa=c\tau_n.
\]
We first check the curvature bound at this new center. The first
application gives $\|\gb_Y-\gam_n\|\to0$, so for all sufficiently
large $n$ and every $\bm\eta$ with $\|\bm\eta-\gb_Y\|\le r$,
\[
 \|\bm\eta-\gam_n\|
 \le\|\bm\eta-\gb_Y\|+\|\gb_Y-\gam_n\|\le2r.
\]
Hence \eqref{fd:eq:fit-curvature} also holds on the ball of radius $r$
about $\gb_Y$.

It remains to check the gradient condition at this center.
The random score at this deterministic center is centered because
$\gb_Y$ maximizes $\bar\ell_n$:
\[
 \E_{\P_n}\nabla\ell_n(\gb_Y)
 =\nabla\bar\ell_n(\gb_Y)=0,\qquad
 \E_{\P_n}\|\nabla\ell_n(\gb_Y)\|^2
 =\frac1{n^2}\sum_i\|\bm z_i\|^2v_Y(\theta_i^\rho)
 \le C\tau_n/n.
\]
Consequently its norm is $O_{\P_n}(\sqrt{\tau_n/n})$ by Chebyshev, and
\[
 \frac{\|\nabla\ell_n(\gb_Y)\|}{c\tau_n}
 =O_{\P_n}((n\tau_n)^{-1/2})=o_{\P_n}(1).
\]
In particular, the gradient condition holds with probability tending to
one. More explicitly,
\[
 \P_n\left\{\|\nabla\ell_n(\gb_Y)\|\ge\frac{c\tau_n r}{2}\right\}
 \le\frac{4\E_{\P_n}\|\nabla\ell_n(\gb_Y)\|^2}{c^2\tau_n^2r^2}
 \le\frac{C}{n\tau_n r^2}\longrightarrow0.
\]

On the complementary event, Lemma~\ref{fd:lem:localization} and strict
concavity give existence and uniqueness of the restricted MLE $\gh$,
with
\[
 \|\gh-\gb_Y\|
 \le\frac{\|\nabla\ell_n(\gb_Y)\|}{c\tau_n}
 =O_{\P_n}((n\tau_n)^{-1/2}).
\]
This is the first assertion of \eqref{fd:eq:fitrate}.

\paragraph{The remaining assertions follow by combining the two bounds.}
The natural-parameter bound follows from bounded $\bm z_i$ and
\[
 \max_i|\widehat\theta_i-\theta_i^0|
 \le\max_i\|\bm z_i\|
       \bigl(\|\gh-\gb_Y\|+\|\gb_Y-\gam_n\|\bigr)
 =O_{\P_n}((n\tau_n)^{-1/2}+|\rho_n|).
\]
Under the null, $\rho_n=0$ and the first bound gives
$\gb_Y=\gam_n$ exactly; equivalently,
$\nabla\bar\ell_n(\gam_n)=0$.
The lemma is applied separately for each $n$, so the curvature
$\kappa=c\tau_n$ may tend to zero. Its gradient condition concerns the
ratio of the gradient norm to this curvature: $\rho_n\to0$ controls
that ratio in the first application, and $n\tau_n\to\infty$ controls
it in the second.
The score tail and failure probabilities are uniform in the stated envelopes.
\end{proof}

\subsection{Proof of Lemma~\ref{fd:lem:variance}}

\begin{proof}
We prove parts (a)--(d) in order, treating the deterministic and fitted
claims in parts (b) and (c) separately.

\paragraph{Part (a): the oracle coefficients and residuals are uniformly bounded.}
The null weights lie between $c\tau_n$ and $C\tau_n$.
Define the normalized weighted Gram matrix and cross moment
\[
 \bm G_n^0=\frac1{n\tau_n}\sum_i A_i^0\bm z_i\bm z_i^\T,
 \qquad
 \bm q_n^0=\frac1{n\tau_n}\sum_i A_i^0\bm z_ix_i.
\]
The deterministic Gram bound gives
\[
 \bm G_n^0\succeq\frac cn\sum_i\bm z_i\bm z_i^\T
 \succeq c\lambda\bm I_d,
\]
while bounded $x_i,\bm z_i$ and $A_i^0/\tau_n$ give
$\|\bm q_n^0\|\le C$. Consequently,
\[
 \|\bm\alpha_n^0\|
 =\|(\bm G_n^0)^{-1}\bm q_n^0\|
 \le\frac{\|\bm q_n^0\|}{c\lambda}\le C,
 \qquad
 \max_i|m_i^0|
 \le\max_i|x_i|+\max_i\|\bm z_i\|\,\|\bm\alpha_n^0\|\le C.
\]
The normalization cancels the common scale $\tau_n$ from the regression
coefficient, so these bounds remain valid when $\tau_n\to0$.

\begin{samepage}
\paragraph{Part (b): the deterministic pseudo-null coefficients and residuals are uniformly bounded.}
By \eqref{fd:eq:logvar},
$|\log v_Y(\theta')-\log v_Y(\theta)|\le K|\theta'-\theta|$.
Combining this inequality with bounded $\bm z_i$ and
the deterministic bound in Lemma~\ref{fd:lem:fit} gives
\[
 \max_i|\log(\bar A_i/A_i^0)|
 \le C\|\gb_Y-\gam_n\|\le C|\rho_n|.
\]
\end{samepage}
Since $\rho_n\to0$, this maximum is at most $1$ for all sufficiently
large $n$. Exponentiating and using $c\tau_n\le A_i^0\le C\tau_n$
show that $\bar A_i$ lies between fixed positive multiples of $\tau_n$.
The normalized Gram-matrix and cross-moment argument in part (a), with
$A_i^0$ replaced by $\bar A_i$, therefore gives
\[
 \|\ab_Y\|+\max_i|\bar m_i|\le C.
\]

\paragraph{Part (b): the fitted coefficients and residuals are uniformly bounded on $\mathcal E_n$.}
The same log-variance inequality and the stochastic bound in
Lemma~\ref{fd:lem:fit} give
\[
 \max_i|\log(\widehat A_i/\bar A_i)|
 \le C\|\gh-\gb_Y\|
 =O_{\P_n}((n\tau_n)^{-1/2}).
\]
For all sufficiently large $n$, let $\mathcal E_n$ be the event that
the restricted MLE exists uniquely, $\|\gh-\gb_Y\|\le1$, and this
maximum is at most $1$.
Lemma~\ref{fd:lem:fit} and $n\tau_n\to\infty$ imply
$\sup_{\P_n\in\PRR_n}\P_n(\mathcal E_n^c)\to0$.
On $\mathcal E_n$, exponentiating and using the preceding deterministic
bound on $\bar A_i$ show that $\widehat A_i$ also lies between fixed
positive multiples of $\tau_n$. Applying the argument in part (a) with
$A_i^0$ replaced by $\widehat A_i$ proves that the fitted Gram matrix
is invertible and
\[
 \|\ah\|+\max_i|\widehat m_i|\le C.
\]
Combining the three projection bounds gives, on $\mathcal E_n$,
\begin{equation}\label{fd:eq:variance-residual-bounds}
 \begin{aligned}
 \|\bm\alpha_n^0\|\vee\|\ab_Y\|\vee\|\ah\|\le C,\qquad \max_i(|m_i^0|\vee|\bar m_i|\vee|\widehat m_i|)\le C.
 \end{aligned}
\end{equation}
The oracle and pseudo-null bounds are deterministic; only the fitted
bounds require $\mathcal E_n$. The same events will be used in parts
(c) and (d).

\paragraph{Part (c): the pseudo-null information has relative error $O(|\rho_n|)$.}
We first establish a weight comparison that applies to both information
ratios.
For positive weights $\bm w=(w_i)_{i=1}^n$, put
\[
 J(\bm w)=\min_{\bm a}\sum_iw_i(x_i-\bm z_i^\T\bm a)^2.
\]
If $\max_i|\log(w_i'/w_i)|\le\delta$, then for every $\bm a$,
\[
 e^{-\delta}\sum_iw_i(x_i-\bm z_i^\T\bm a)^2
 \le\sum_iw_i'(x_i-\bm z_i^\T\bm a)^2
 \le e^{\delta}\sum_iw_i(x_i-\bm z_i^\T\bm a)^2.
\]
As in Lemma~\ref{fd:lem:wls}, taking minima preserves these inequalities:
\[
 e^{-\delta}J(\bm w)\le J(\bm w')\le e^{\delta}J(\bm w).
\]
This comparison holds even though the minimizing coefficients may differ.
Apply it with $\bm w=\bm A^0$ and $\bm w'=\bar{\bm A}$, for which
$J(\bm A^0)=I_{n,x}$ and $J(\bar{\bm A})=\bar I_{n,x}$.
The deterministic log-weight bound in part (b) and
$e^{\pm\delta}=1+O(\delta)$ as $\delta\to0$ give
\[
 \bar I_{n,x}/I_{n,x}=1+O(|\rho_n|).
\]
The denominator is positive because $I_{n,x}>0$, and the comparison
also gives $\bar I_{n,x}>0$.

\paragraph{Part (c): the fitted information has relative error $O_{\P_n}((n\tau_n)^{-1/2})$.}
On $\mathcal E_n$, apply the same weight comparison with
$\bm w=\bar{\bm A}$ and $\bm w'=\widehat{\bm A}$, so that
$J(\bar{\bm A})=\bar I_{n,x}$ and
$J(\widehat{\bm A})=\widehat I_{n,x}$.
The stochastic log-weight bound in part (b), together with the same
expansion of $e^{\pm\delta}$, yields
\[
 \widehat I_{n,x}/\bar I_{n,x}
 =1+O_{\P_n}((n\tau_n)^{-1/2}).
\]
The comparison applies to the fitted weights for each realized data set
on $\mathcal E_n$; its denominator is positive by the preceding paragraph.
In particular, the two pointwise information comparisons used below are
\begin{equation}\label{fd:eq:variance-information-ratios}
 \begin{aligned}
 e^{-C|\rho_n|}&\le\bar I_{n,x}/I_{n,x}\le e^{C|\rho_n|},\\
 e^{-C\|\gh-\gb_Y\|}
 &\le\widehat I_{n,x}/\bar I_{n,x}\le e^{C\|\gh-\gb_Y\|}.
 \end{aligned}
\end{equation}

\begin{samepage}
\paragraph{Part (d): the resampling summands share a deterministic Bernstein scale.}
On $\mathcal E_n$, the fitted variances satisfy
$\widehat A_i\le C\tau_n\le CK$, and
$\max_i|\widehat m_i|\le C_m$ for a deterministic constant $C_m$.
Conditionally on the observed responses, the fitted quantities are fixed
and $\widetilde Y_i\sim f_Y(\,\cdot\,;\widehat\theta_i)$.
Thus $W_i=\widehat m_i(\widetilde Y_i-\widehat\mu_i)$ from
\eqref{fd:eq:target} satisfies
\[
 \widetilde\E W_i=0,\qquad
 \widetilde\E W_i^2=\widehat m_i^2\widehat A_i.
\]
\end{samepage}
Apply Lemma~\ref{fd:lem:mom} with the common variance upper bound
$V_0=CK$. It supplies one deterministic $b_0$, depending only on $K,V_0$,
for all the centered resampled responses. Multiplication by
$\widehat m_i$ then gives, with $B=b_0C_m$ and $k\ge3$,
\begin{align*}
 \widetilde\E|W_i|^k\le\frac{k!}{2}(b_0|\widehat m_i|)^{k-2}\widehat m_i^2\widehat A_i\le\frac{k!}{2}B^{k-2}\widetilde\E W_i^2.
\end{align*}
Thus $B$ is a common deterministic Bernstein scale on $\mathcal E_n$.
All bounds and failure probabilities are
uniform under the stated conventions.
\end{proof}

\subsection{Proof of Lemma~\ref{fd:lem:score}}

\begin{proof}
We prove the three bounds in \eqref{fd:eq:scorebounds} in order, then
derive the overall score bound and the assertions under the null.

\paragraph{The score transfer error satisfies $S_n-\bar U_n=O_{\P_n}(1)$.}
Work on the events in Lemmas~\ref{fd:lem:fit} and \ref{fd:lem:variance}.
The restricted score equation and the pseudo-null weighted projection give
\begin{equation}\label{fd:eq:score-orthogonality}
 \sum_i\bm z_i(Y_i-\widehat\mu_i)=0,\qquad
 \sum_i\bar A_i\bar m_i\bm z_i=0.
\end{equation}
Using the first identity in \eqref{fd:eq:score-orthogonality},
\begin{equation}\label{fd:eq:score-transfer-identity}
 S_n=\sum_i x_i(Y_i-\widehat\mu_i)
     =\sum_i\bar m_i(Y_i-\widehat\mu_i),
 \qquad
 S_n-\bar U_n=-\sum_i\bar m_i(\widehat\mu_i-\bar\mu_i).
\end{equation}
Put $\bm\Delta=\gh-\gb_Y$. Taylor's formula gives
\begin{equation}\label{fd:eq:score-mean-remainder}
 \begin{aligned}
 \widehat\mu_i-\bar\mu_i&=\bar A_i\bm z_i^\T\bm\Delta+r_i,\\
 r_i&=(\bm z_i^\T\bm\Delta)^2
       \int_0^1(1-s)v_Y'(\bar\theta_i+s\bm z_i^\T\bm\Delta)\,ds,\\
 |r_i|&\le \frac K2\bar A_i e^{K|\bm z_i^\T\bm\Delta|}
                 (\bm z_i^\T\bm\Delta)^2
 \le C\bar A_i(\bm z_i^\T\bm\Delta)^2.
 \end{aligned}
\end{equation}
The last inequality uses \eqref{fd:eq:logvar} and $\|\bm\Delta\|\le1$.
Substituting \eqref{fd:eq:score-mean-remainder} into
\eqref{fd:eq:score-transfer-identity} and using the second identity
in \eqref{fd:eq:score-orthogonality} yields
\begin{equation}\label{fd:eq:score-transfer-bound}
 \begin{aligned}
 S_n-\bar U_n
 &=-\left(\sum_i\bar A_i\bar m_i\bm z_i\right)^\T\bm\Delta
       -\sum_i\bar m_i r_i=-\sum_i\bar m_i r_i,\\
 |S_n-\bar U_n|
 &\le C\sum_i|\bar m_i|\bar A_i(\bm z_i^\T\bm\Delta)^2
 \le Cn\tau_n\|\bm\Delta\|^2=O_{\P_n}(1),
 \end{aligned}
\end{equation}
by \eqref{fd:eq:variance-residual-bounds} and \eqref{fd:eq:fitrate}.

\paragraph{The centered term satisfies $N_n=O_{\P_n}(\sqrt{I_{n,x}})$.}
Equations \eqref{fd:eq:model} and \eqref{fd:eq:logvar} give
\begin{align}\label{fd:eq:score-noise-variance}
  \E_{\P_n}N_n=0,\qquad \V_{\P_n}(N_n)=\sum_i\bar m_i^2v_Y(\theta_i^\rho)\le C\sum_i\bar m_i^2\bar A_i=C\bar I_{n,x}\le CI_{n,x}.
\end{align}
Here $|\theta_i^\rho-\bar\theta_i|\le K|\rho_n|+\sqrt{1+K^2}\|\gb_Y-\gam_n\|\le C|\rho_n|$ by \eqref{fd:eq:fit-population-distance}, and the last comparison uses
\eqref{fd:eq:variance-information-ratios}.
Chebyshev's inequality applied to \eqref{fd:eq:score-noise-variance} gives
\[
 \P_n(|N_n|>t\sqrt{I_{n,x}})\le C/t^2\quad(t>0),
 \qquad N_n=O_{\P_n}(\sqrt{I_{n,x}}).
\]

\paragraph{The deterministic drift satisfies $|\Delta_n^{\mathrm{drift}}|\le C|\rho_n|I_{n,x}$.}
The pseudo-null score equation $\nabla\bar\ell_n(\gb_Y)=0$ implies
$\sum_i\bm z_i(\mu_i^\rho-\bar\mu_i)=0$.
When $\rho_n\ne0$, put
\begin{equation}\label{fd:eq:score-drift-weights}
 \bm a_\rho=\frac{\gb_Y-\gam_n}{\rho_n},\qquad
 w_i=\int_0^1v_Y\{\bar\theta_i+s(\theta_i^\rho-\bar\theta_i)\}\,ds.
\end{equation}
The integral form of the mean-value theorem states that, for a
continuously differentiable function $f$,
\[
 f(b)-f(a)=(b-a)\int_0^1f'\{a+s(b-a)\}\,ds.
\]
Applying this identity with $f=b_Y'$, $a=\bar\theta_i$ and
$b=\theta_i^\rho$, and using $v_Y=b_Y''$, gives
\[
 \begin{aligned}
 \mu_i^\rho-\bar\mu_i
 &=b_Y'(\theta_i^\rho)-b_Y'(\bar\theta_i)\\
 &=(\theta_i^\rho-\bar\theta_i)
   \int_0^1v_Y\{\bar\theta_i+s(\theta_i^\rho-\bar\theta_i)\}\,ds
   =(\theta_i^\rho-\bar\theta_i)w_i.
 \end{aligned}
\]
By the definitions of the natural parameters and $\bm a_\rho$,
\[
 \theta_i^\rho-\bar\theta_i
 =\bm z_i^\T(\gam_n-\gb_Y)+\rho_nx_i
 =\rho_n(x_i-\bm z_i^\T\bm a_\rho).
\]
Substitution gives the first identity below. Inserting it into the
pseudo-null score equation $\sum_i\bm z_i(\mu_i^\rho-\bar\mu_i)=0$
and dividing by $\rho_n\ne0$ gives the second:
\begin{equation}\label{fd:eq:score-drift-normal-equation}
 \mu_i^\rho-\bar\mu_i
 =\rho_n w_i(x_i-\bm z_i^\T\bm a_\rho),\qquad
 \sum_iw_i\bm z_i(x_i-\bm z_i^\T\bm a_\rho)=0.
\end{equation}
The second identity characterizes the weighted least-squares minimizer.
Using both identities in \eqref{fd:eq:score-drift-normal-equation},
\begin{equation}\label{fd:eq:drift}
 \begin{aligned}
 \Delta_n^{\mathrm{drift}}
 &=\sum_i x_i(\mu_i^\rho-\bar\mu_i)
 =\rho_n\sum_iw_ix_i(x_i-\bm z_i^\T\bm a_\rho)\\
 &=\rho_n\sum_iw_i(x_i-\bm z_i^\T\bm a_\rho)^2
 =\rho_n\min_{\bm a}\sum_iw_i(x_i-\bm z_i^\T\bm a)^2.
 \end{aligned}
\end{equation}
The first equality uses $\bar m_i=x_i-\bm z_i^\T\ab_Y$ and the
pseudo-null score equation. Recall from \eqref{fd:eq:score-drift-weights} that
\[
 w_i=\int_0^1v_Y\{\bar\theta_i+s(\theta_i^\rho-\bar\theta_i)\}\,ds.
\]
Thus $w_i$ is the average of the response variance function $v_Y=b_Y''$
along the segment from the pseudo-null natural parameter $\bar\theta_i$
to the true natural parameter $\theta_i^\rho$.
Applying \eqref{fd:eq:logvar} along this segment gives
\begin{equation}\label{fd:eq:score-drift-weight-comparison}
 e^{-C|\rho_n|}\bar A_i\le w_i\le e^{C|\rho_n|}\bar A_i.
\end{equation}
Substituting \eqref{fd:eq:score-drift-weight-comparison} into
\eqref{fd:eq:drift}, with
\eqref{fd:eq:variance-information-ratios}, proves
\begin{equation}\label{fd:eq:score-drift-bound}
 |\Delta_n^{\mathrm{drift}}|
 \le|\rho_n|e^{C|\rho_n|}\bar I_{n,x}
 \le C|\rho_n|I_{n,x}.
\end{equation}
\newpage
\paragraph{The fitted score satisfies $S_n=O_{\P_n}(\sqrt{I_{n,x}}+|\rho_n|I_{n,x})$.}
Combining \eqref{fd:eq:score-transfer-bound},
\eqref{fd:eq:score-noise-variance}, and \eqref{fd:eq:score-drift-bound} gives
\[
 S_n=(S_n-\bar U_n)+N_n+\Delta_n^{\mathrm{drift}}
 =O_{\P_n}(1+\sqrt{I_{n,x}}+|\rho_n|I_{n,x})
 =O_{\P_n}(\sqrt{I_{n,x}}+|\rho_n|I_{n,x}),
\]
since $I_{n,x}\to\infty$.

\paragraph{Under the null, $\Delta_n^{\mathrm{drift}}=0$ and $S_n-\sum_i m_i^0(Y_i-\mu_i^0)=O_{\P_n}(1)$.}
Under the null, Lemma~\ref{fd:lem:fit} gives
$\gb_Y=\gam_n$, hence $\bar m_i=m_i^0$ and $\bar\mu_i=\mu_i^\rho=\mu_i^0$.
Recalling the definition of the pseudo-null score, we therefore have
\[
 \bar U_n=\sum_i\bar m_i(Y_i-\bar\mu_i)
         =\sum_i m_i^0(Y_i-\mu_i^0).
\]
The score-transfer bound \eqref{fd:eq:score-transfer-bound} states that
$S_n-\bar U_n=O_{\P_n}(1)$. Substituting the displayed expression for
$\bar U_n$, and using $\mu_i^\rho=\bar\mu_i$ in the definition of the drift,
gives
\[
 \Delta_n^{\mathrm{drift}}=0,\qquad
 S_n-\sum_i m_i^0(Y_i-\mu_i^0)=S_n-\bar U_n=O_{\P_n}(1).
\]
\end{proof}

\subsection{Proof of Lemma~\ref{fd:lem:skellam}}

\begin{proof}
Independence of $A,B$ gives
\[
 \E e^{itD}
 =\exp\{(L/2)(e^{it}-1)+(L/2)(e^{-it}-1)\}
 =e^{L(\cos t-1)}.
\]
Fourier inversion for the integer-valued law and the Gaussian density gives
\begin{equation}\label{fd:eq:skellam-inversion}
 \P(D=k)=\frac1{2\pi}\int_{-\pi}^{\pi}e^{-ikt}e^{L(\cos t-1)}\,dt,
 \qquad
 \frac{\phi(k/\sqrt L)}{\sqrt L}
 =\frac1{2\pi}\int_{\R}e^{-ikt}e^{-Lt^2/2}\,dt.
\end{equation}
Choose fixed $t_0\in(0,1)$ sufficiently small. For $|t|\le t_0$,
\[
 |\cos t-1+t^2/2|\le Ct^4,\qquad
 \cos t-1\le-ct^2,\qquad
 |e^{L(\cos t-1)}-e^{-Lt^2/2}|
 \le CLt^4e^{-cLt^2}.
\]
For $t_0\le|t|\le\pi$, $1-\cos t\ge1-\cos t_0>0$.
Since $|e^{-ikt}|=1$, subtracting the integrals in
\eqref{fd:eq:skellam-inversion} yields the uniform bound
\begin{equation}\label{fd:eq:skellam-local-error}
 \sup_{k\in\mathbb Z}
 \left|\P(D=k)-\frac{\phi(k/\sqrt L)}{\sqrt L}\right|
 \le CL\int_{\R}t^4e^{-cLt^2}\,dt+Ce^{-cL}
 \le CL^{-3/2}.
\end{equation}
In particular,
\[
 \P(D=k)=L^{-1/2}\phi(k/\sqrt L)+O(L^{-3/2}),
 \qquad k\in\mathbb Z.
\]

To sum the local errors, take $T_L=\lfloor\sqrt{16L\log L}\rfloor$.
The MGF is $\E e^{tD}=\exp\{L(\cosh t-1)\}$. With
$t=T_L/(2L)\to0$, Chernoff's inequality and $\cosh t-1\le t^2$ give
\begin{equation}\label{fd:eq:skellam-truncation-tails}
 \P(|D|>T_L)
 \le2\exp\{-tT_L+L(\cosh t-1)\}
 \le2e^{-T_L^2/(4L)}=O(L^{-2}).
\end{equation}
The omitted Gaussian lattice tail satisfies
\begin{equation}\label{fd:eq:skellam-gaussian-tail}
 \frac1{\sqrt L}\sum_{j>T_L}\phi(j/\sqrt L)
 \le\int_{T_L/\sqrt L}^{\infty}\phi(u)\,du=O(L^{-2}).
\end{equation}
For $|k|\le M\sqrt L$, the number of terms from $k$ through $T_L$
is at most $T_L+M\sqrt L+1$. Thus \eqref{fd:eq:skellam-local-error} gives
\begin{equation}\label{fd:eq:skellam-summed-error}
 \begin{aligned}
 \left|\sum_{j=k}^{T_L}\P(D=j)
       -\frac1{\sqrt L}\sum_{j=k}^{T_L}\phi(j/\sqrt L)\right|
 &\le C(T_L+M\sqrt L+1)L^{-3/2}\\
 &=O_M(\sqrt{\log L}/L)=o(L^{-1/2}).
 \end{aligned}
\end{equation}
Combining \eqref{fd:eq:skellam-truncation-tails},
\eqref{fd:eq:skellam-gaussian-tail}, and
\eqref{fd:eq:skellam-summed-error} gives
\begin{equation}\label{fd:eq:skellam-tail-sum}
 \P(D\ge k)
 =\frac1{\sqrt L}\sum_{j=k}^{\infty}\phi(j/\sqrt L)
   +o(L^{-1/2}),
 \qquad |k|\le M\sqrt L.
\end{equation}
The trapezoidal rule, using
$\int_{\R}|\phi''(u)|\,du<\infty$, gives
\begin{equation}\label{fd:eq:skellam-trapezoid}
 \left|
 \frac1{\sqrt L}\sum_{j=k}^{\infty}\phi(j/\sqrt L)
 -\int_{k/\sqrt L}^{\infty}\phi(u)\,du
 -\frac{\phi(k/\sqrt L)}{2\sqrt L}
 \right|
 \le\frac C L\int_{\R}|\phi''(u)|\,du.
\end{equation}
Substituting \eqref{fd:eq:skellam-trapezoid} into
\eqref{fd:eq:skellam-tail-sum} proves, uniformly for $|k|\le M\sqrt L$,
\[
 \P(D\ge k)
 =\Phibar(k/\sqrt L)+\frac{\phi(k/\sqrt L)}{2\sqrt L}
   +o(L^{-1/2}).
\]
\end{proof}

\subsection{Proof of Lemma~\ref{fd:lem:symLR}}

\begin{proof}
The function $H$ is even, with
$H(0)=H'(0)=H'''(0)=0$ and $H''(0)=\Lambda$.
For the tilted Bernoulli probability
$p_t=pe^t/(1-p+pe^t)$, differentiation gives
\begin{equation}\label{fd:eq:symmetric-lr-derivatives}
 \begin{aligned}
 H''(t)&=m\{p_t(1-p_t)+p_{-t}(1-p_{-t})\}>0,\\
 H^{(4)}(t)
 &=m\sum_{\varepsilon\in\{-1,1\}}
 p_{\varepsilon t}(1-p_{\varepsilon t})
       \{1-6p_{\varepsilon t}(1-p_{\varepsilon t})\}.
 \end{aligned}
\end{equation}
For $|t|\le1$ and $p\le1/4$, $p_{\pm t}\le e p/(1-p)\le4ep/3$.
Since $mp=\Lambda/\{2(1-p)\}\le2\Lambda/3$,
\eqref{fd:eq:symmetric-lr-derivatives} implies
$\sup_{|t|\le1}|H^{(4)}(t)|\le C\Lambda$.
Taylor's formula therefore gives
\begin{equation}\label{fd:eq:symmetric-lr-taylor}
 \begin{aligned}
 |H(t)-\Lambda t^2/2|&\le C\Lambda t^4,\\
 |H'(t)-\Lambda t|&\le C\Lambda|t|^3,\qquad
 |H''(t)-\Lambda|\le C\Lambda t^2.
 \end{aligned}
\end{equation}
Choose fixed $t_0>0$ sufficiently small that
$H''(t)\ge\Lambda/2$ for $|t|\le t_0$.
For $|s|\le M\sqrt\Lambda$ and large $\Lambda$, integrating this bound gives
\[
 H'\!\left(\frac{2(M+1)}{\sqrt\Lambda}\right)
 \ge(M+1)\sqrt\Lambda>s,\qquad
 H'\!\left(-\frac{2(M+1)}{\sqrt\Lambda}\right)
 \le-(M+1)\sqrt\Lambda<s.
\]
Strict monotonicity in \eqref{fd:eq:symmetric-lr-derivatives} gives a
unique solution $t$ to $H'(t)=s$ between these two endpoints.
Writing $z=s/\sqrt\Lambda$, the curvature bound also gives
$|t|\le2|s|/\Lambda=2|z|/\sqrt\Lambda$. Substitution into
\eqref{fd:eq:symmetric-lr-taylor} yields
\begin{equation}\label{fd:eq:symmetric-lr-saddlepoint}
 \left|t-\frac{z}{\sqrt\Lambda}\right|
 =\frac{|H'(t)-\Lambda t|}{\Lambda}
 \le C|t|^3\le C_M|z|^3\Lambda^{-3/2}.
\end{equation}
For $s\ne0$, the LR quantities are
$r=\operatorname{sgn}(t)\sqrt{2\{ts-H(t)\}}$ and
$u=t\sqrt{H''(t)}$. Using $s=H'(t)$ in
\eqref{fd:eq:symmetric-lr-taylor} gives
\begin{equation}\label{fd:eq:symmetric-lr-energy}
 2\{ts-H(t)\}=\Lambda t^2+O(\Lambda t^4),\qquad
 H''(t)=\Lambda\{1+O(t^2)\}.
\end{equation}
Taking square roots in \eqref{fd:eq:symmetric-lr-energy} and substituting
\eqref{fd:eq:symmetric-lr-saddlepoint} gives, uniformly for $0<|z|\le M$,
\begin{equation}\label{fd:eq:symmetric-lr-ru}
 r=\sqrt\Lambda\,t\{1+O(t^2)\}
   =z+O_M(|z|^3/\Lambda),\qquad
 u=\sqrt\Lambda\,t\{1+O(t^2)\}
   =z+O_M(|z|^3/\Lambda).
\end{equation}
For large $\Lambda$, \eqref{fd:eq:symmetric-lr-ru} implies
$|r|,|u|\ge|z|/2$ and hence
\begin{equation}\label{fd:eq:symmetric-lr-reciprocals}
 \left|\frac1u-\frac1r\right|
 =\frac{|r-u|}{|ru|}
 \le\frac{C_M|z|^3/\Lambda}{|z|^2/4}
 \le C_M|z|/\Lambda.
\end{equation}
Substituting \eqref{fd:eq:symmetric-lr-ru} and
\eqref{fd:eq:symmetric-lr-reciprocals} into \eqref{eq:new-lr} yields
\[
 \left|p_{\rm LR}(s)-\Phibar(z)\right|
 \le\frac1{\sqrt{2\pi}}
       \left(|r-z|+\left|\frac1u-\frac1r\right|\right)
 \le\frac{C_M}{\Lambda}(|z|^3+|z|)
 \le\frac{C_M}{\Lambda}.
\]
As $z\to0$, \eqref{fd:eq:symmetric-lr-ru} gives $r\to0$ and
\eqref{fd:eq:symmetric-lr-reciprocals} gives $1/u-1/r\to0$.
Thus the continuous value at $s=0$ is $p_{\rm LR}(0)=1/2=\Phibar(0)$.
Including this value proves
\[
 \sup_{|s|/\sqrt\Lambda\le M}
 \left|p_{\rm LR}(s)-\Phibar(s/\sqrt\Lambda)\right|
 \le C_M\Lambda^{-1}.
\]
\end{proof}

\section{Proofs of the results in Section~\ref{sec:predictor-resampling}}
\label{sec:proof-main-spacrt-results}

\subsection{Notation for Supplementary Sections~\ref{sec:proof-main-spacrt-results}
and~\ref{sec:spacrt-detailed-proofs}}
\label{sec:spacrt-notation}

The notation fixed here is used throughout this section and Supplementary Section~\ref{sec:spacrt-detailed-proofs}. The fitted response residuals are $\widehat e_{y,i}\equiv Y_i-\widehat\mu_{y,i}$; write $e_n^\star\equiv\max_{1\le i\le n}|\widehat e_{y,i}|$. For $j\in\{x,y\}$, let $\mu^0_{j,i}$, $e^0_{j,i}$, and $v^0_{j,i}$ denote the oracle conditional means, residuals, and variances.  Put $\sigma_{n,\dCRT}^2\equiv\frac1n\sum_i v^0_{x,i}v^0_{y,i}$. With $\Delta_{j,i}\equiv\widehat\mu_{j,i}-\mu^0_{j,i}$, define
\[
 E_{n,j}^2\equiv\frac1n\sum_i\Delta_{j,i}^2,
 \qquad
 (E'_{n,x})^2\equiv\frac1n\sum_i v^0_{y,i}\Delta_{x,i}^2,
 \qquad
 (E'_{n,y})^2\equiv\frac1n\sum_i v^0_{x,i}\Delta_{y,i}^2,
\]
and $(\widehat E'_{n,y})^2\equiv\frac1n\sum_i\widehat v_{x,i}\Delta_{y,i}^2$ and $V_{n,\dCRT}^2\equiv\frac1n\sum_i\widehat v_{x,i}\widehat e_{y,i}^2=A_{2n}$.

Write $\mathcal Z_n\equiv\sigma\{\rz{i}:1\leq i\leq n\}$, $\mathcal X_n\equiv\sigma\{(\ex{i},\rz{i}):1\leq i\leq n\}$ and $\mathcal Y_n\equiv\sigma\{(\ey{i},\rz{i}):1\leq i\leq n\}$. Let $\mathcal F_n\equiv\mathcal X_n\vee\mathcal Y_n=\sigma\{(\ex{i},\ey{i},\rz{i}):1\leq i\leq n\}$ denote the $\sigma$-field generated by the observed data. In particular, $\widehat\mu_{x,i}$, $\widehat v_{x,i}$, $\widehat e_{y,i}$, and $S_n$ are $\mathcal F_n$-measurable, and, conditionally on $\mathcal F_n$, the resampled $\widetilde X_i$ are independent with mean $\widehat\mu_{x,i}$ and variance $\widehat v_{x,i}$.

\subsection{Proof of Corollary~\ref{thm:spacrt-approximation}: the observable conditional approximation}
\label{sec:proof-spacrt-approximation}

\begin{proof}[of Corollary~\ref{thm:spacrt-approximation}]
Condition on the observed data and put $W_{in}\equiv\widehat e_{y,i}(\widetilde X_i-\widehat\mu_{x,i})$ and $t_n\equiv S_n$. By part~\ref{part:translator-bounded-support} of Corollary~\ref{cor:translators-main}, the heterogeneous Bernstein scales may be chosen as $b_{in}=D_0|\widehat e_{y,i}|$.  Consequently,
\[
  \sigma_n^2=nA_{2n},\qquad
  H_n=nD_0A_{3n},\qquad
  b_n^\star=D_0e_n^\star.
\]
Substitution into \eqref{eq:three-coordinates} gives the $B_n$, $Z_n$, and $D_n$ displayed in Corollary~\ref{thm:spacrt-approximation}. Applying Theorem~\ref{thm:conditional-lr} gives the claimed conclusions.
\end{proof}

\subsection{Proof of Theorem~\ref{thm:spacrt-rate}: sparse Ber--NB spaCRT}
\label{sec:proof-spacrt-rate}
\mbox{}\par

The shared GLM tools are stated in Supplementary Section~\ref{sec:regression} and
proved in Supplementary Section~\ref{sec:saige-rate-supporting-proofs}.  The
Ber--NB-specific supporting lemmas are stated below and proved in
Supplementary Section~\ref{sec:spacrt-rate-supporting-proofs}.

\subsubsection{Shared GLM and logistic-regression tools}
\label{sec:regression}

Let $\bm Z\in\R^d$ include an intercept, with first coordinate equal to one almost surely.  The
following standing design condition is used in the random-design predictor-resampling arguments:
\begin{equation}\label{eq:design}
 \norm{\bm Z}_2\le B_z\quad\text{a.s.},
 \qquad
 \lambda_{\min}\{\E(\bm Z\bm Z^{\T})\}
 \ge\lambda_z,
 \qquad 0<\lambda_z\le1.
\end{equation}

\begin{lemma}[Sparsity scales implied by bounded GLM coefficients]
\label{lem:glm-scales}
Write $\bm Z=(1,\bm U^{\T})^{\T}$, suppose
$\|\bm U\|_2\le B_u$ almost surely (so that $\|\bm Z\|_2^2\le1+B_u^2$), and let
$\|\bm\beta\|_2\le B_\beta$.

For a Bernoulli logistic model
\[
 p(x,\bm u)\equiv\expit(\alpha+\rho x+\bm\beta^{\T}\bm u),
 \qquad x\in\{0,1\},
\]
define $S\equiv\expit(\alpha)$.  If $\alpha\le B_\alpha$ and
$|\rho|\le B_\rho$, then there are fixed constants
$0<c_\mu<C_\mu<\infty$ such that
\begin{equation}\label{eq:logistic-scale-comparison}
 c_\mu S\le p(x,\bm U)\le C_\mu S,
 \qquad 1-p(x,\bm U)\ge c_\mu
\end{equation}
almost surely, uniformly in $x$.

For a negative-binomial log-mean model
\[
 m(x,\bm u)\equiv\exp(\alpha+\rho x+\bm\beta^{\T}\bm u),
\]
define $S\equiv e^\alpha$.  Under the same bounds,
\begin{equation}\label{eq:nb-scale-comparison}
 c_\mu S\le m(x,\bm U)\le C_\mu S
\end{equation}
almost surely, uniformly in $x$.  The constants depend only on
$B_u,B_\beta,B_\alpha$, and $B_\rho$.  In particular, when the logistic
intercept is bounded above, $\expit(\alpha)\asymp e^\alpha$, so this
 probability-scale definition is equivalent up to fixed constants to using
 the exponentiated intercept.
\end{lemma}

The next deterministic lemma gives existence, uniqueness, and localization
of the MLE from a small gradient and a curvature lower bound.

\begin{lemma}[Deterministic generalized-self-concordant bound]
\label{lem:det-sc}
Let $L_n:\R^d\to\R$ be convex and three times continuously differentiable on
an open neighborhood of $\overline B(\bm\theta_0,r_0)$ for some $r_0>0$.
Suppose
that for all $\bm\theta\in\overline B(\bm\theta_0,r_0)$ and all $\bm a,\bm v\in\R^d$,
\begin{equation}\label{eq:gsc}
 |D^3L_n(\bm\theta)[\bm a,\bm v,\bm v]|
 \le B_z\norm{\bm a}_2D^2L_n(\bm\theta)[\bm v,\bm v].
\end{equation}
Let $\bm{\mathcal H}_n\equiv\nabla^2L_n(\bm\theta_0)$ and $\bm g_n\equiv\nabla L_n(\bm\theta_0)$.  If
$\bm{\mathcal H}_n\succeq \underline h \bm I_d$ and
\[
 R\equiv 4\norm{\bm g_n}_2/\underline h
 \le\min\{r_0,(2B_z)^{-1}\},
\]
then $L_n$ has a unique global minimizer $\widehat{\bm\theta}$, and it satisfies
$\norm{\widehat{\bm\theta}-\bm\theta_0}_2\le R$.
\end{lemma}

\paragraph{Logistic regression concentration.}

\begin{lemma}[Logistic MLE under the null and bounded alternatives]
\label{lem:logistic-mle}\label{lem:logistic-null-local}
Suppose $(X_i,Y_i,\bm Z_i)_{i=1}^n$ are i.i.d., the design satisfies
\eqref{eq:design}, and
\[
 X_i\mid \bm Z_i\sim\operatorname{Bernoulli}(q_i),\qquad
 Y_i\mid X_i,\bm Z_i\sim\operatorname{Bernoulli}
 \{\expit(\bm\theta_0^{\T}\bm Z_i+\rho_nX_i)\}.
\]
Let $m_i=\expit(\bm\theta_0^{\T}\bm Z_i)$ and suppose, for
$0<S_y\le1$ and $0\le S_x\le1$, that
\begin{equation}\label{eq:logistic-means}
 c_mS_y\le m_i\le C_mS_y,\qquad m_i\le1-c_m,
 \qquad 0\le q_i\le C_qS_x,
 \qquad |\rho_n|\le H_\rho.
\end{equation}
Put $L\equiv\log(C_0d/\delta)$, where $0<\delta\le1/4$.  If
\begin{equation}\label{eq:logistic-n}
 n\ge C\max\left\{\frac{d^2L}{\lambda_z^2},
 \frac{dL}{S_y\lambda_z^2}\right\},
\end{equation}
and
\begin{equation}\label{eq:logistic-local-rho-small}
 S_x|\rho_n|\le c_0
\end{equation}
holds for a sufficiently small constant $c_0>0$, then, with probability at
least $1-\delta$, the restricted logistic MLE $\widehat{\bm\theta}$ fitted with
$\rho=0$ exists, is unique, and, with
$\widehat m_i\equiv\expit(\widehat{\bm\theta}^{\T}\bm Z_i)$, satisfies
\begin{equation}\label{eq:logistic-theta-error}
 \|\widehat{\bm\theta}-\bm\theta_0\|_2
 \le\frac{C}{\lambda_z}
 \left\{\sqrt{\frac{dL}{nS_y}}+S_x|\rho_n|\right\},
\end{equation}
and
\begin{equation}\label{eq:logistic-error}
 \max_i\left|\frac{\widehat m_i}{m_i}-1\right|
 \le\frac{C}{\lambda_z}
 \left\{\sqrt{\frac{dL}{nS_y}}+S_x|\rho_n|\right\}.
\end{equation}
Here $C$ and the admissible $c_0$ may depend on $H_\rho$ and the fixed
design and comparability constants.  On this event
$\sum_i\bm Z_i(Y_i-\widehat m_i)=0$.

The correctly specified null model is included by taking
$X_i\equiv0$, $S_x=0$, and $\rho_n=0$.  Then the right-hand sides of
\eqref{eq:logistic-theta-error} and \eqref{eq:logistic-error} reduce to
$C\lambda_z^{-1}\sqrt{dL/(nS_y)}$.
\end{lemma}

\subsubsection{Negative-binomial regression tools}

We use the mean--dispersion parameterization
\[
 Y\sim\operatorname{NB}(\mu,\eta),
 \qquad
 \Var(Y)=\mu+\eta\mu^2,
\]
with the Poisson law at $\eta=0$.  Put $A_\mu\equiv1+\eta\mu$ and
$R\equiv Y-\mu$.

\begin{lemma}[NB cumulants, moments, and exponential tails]
\label{lem:nb-facts}
Suppose $0<\mu\le M_0$ and $\eta\ge0$.  For every fixed integer $k\ge2$
there is a constant $C_k>0$ depending only on $k$ and $M_0$, and there are
a constant $C>0$ depending only on $M_0$ and an absolute constant $c>0$,
such that
\begin{align}
 \kappa_k(Y)&\le (k-1)!\,\mu A_\mu^{k-1},
 \label{eq:nb-cumulant}\\
 \E|R|^{2k}&\le C_k\mu A_\mu^{2k-1},
 \label{eq:nb-even-moment}\\
 \E|R|^3&\le C\mu A_\mu^2,
 \label{eq:nb-abs3}\\
 \P\{|R|\ge t\}&\le2\exp\left[-c\min\left
 \{\frac{t^2}{\mu A_\mu},\frac{t}{A_\mu}\right\}\right].
 \label{eq:nb-tail}
\end{align}
Moreover, for each fixed integer $0\le k\le6$ and sufficiently small fixed
$c>0$,
\begin{equation}\label{eq:nb-weighted-moment}
 \E\left[e^{c|R|/A_\mu}|R|^k\right]
 \le C_k\mu A_\mu^{k-1},
\end{equation}
where the right-hand side for $k=0$ is interpreted as a fixed constant.
\end{lemma}

\begin{lemma}[Known-dispersion NB MLE under the null and bounded alternatives]
\label{lem:nb-mle}\label{lem:nb-null-local}
Suppose $(X_i,Y_i,\bm Z_i)_{i=1}^n$ are i.i.d., the design satisfies
\eqref{eq:design}, $\eta_y\ge0$ is known, and
\[
 X_i\mid \bm Z_i\sim\operatorname{Bernoulli}(q_i),\qquad
 Y_i\mid X_i,\bm Z_i\sim\operatorname{NB}
 \{m_i e^{\rho_nX_i},\eta_y\},
 \qquad m_i\equiv e^{\bm\theta_0^{\T}\bm Z_i}.
\]
Suppose, for $0\le S_x\le1$, that
\begin{equation}\label{eq:nb-means}
 c_yS_y\le m_i\le C_yS_y\le M_0,
 \qquad 0\le q_i\le C_qS_x,
 \qquad |\rho_n|\le H_\rho.
\end{equation}
Put $L\equiv\log(C_0d/\delta)$.  If
\begin{equation}\label{eq:nb-mle-n}
 n\ge C\frac{d^2(1+\eta_yS_y)L}{S_y\lambda_z^2},
\end{equation}
and
\begin{equation}\label{eq:nb-local-smallness}
 S_x|\rho_n|\le c_\star
\end{equation}
holds for a sufficiently small constant $c_\star>0$, then, with probability at
least $1-\delta$, the restricted known-dispersion NB MLE
$\widehat{\bm\theta}_y$ fitted with $\rho=0$ exists, is unique, and, with
$\widehat m_i=\exp(\widehat{\bm\theta}_y^{\T}\bm Z_i)$, satisfies
\begin{equation}\label{eq:nb-theta-error}
 \|\widehat{\bm\theta}_y-\bm\theta_0\|_2
 \le\frac{C}{\lambda_z}
 \left\{\sqrt{\frac{(1+\eta_yS_y)dL}{S_yn}}+S_x|\rho_n|\right\},
\end{equation}
and
\begin{equation}\label{eq:nb-mle-error}
 \max_i\left|\frac{\widehat m_i}{m_i}-1\right|
 \le\frac{C}{\lambda_z}
 \left\{\sqrt{\frac{(1+\eta_yS_y)dL}{S_yn}}+S_x|\rho_n|\right\}.
\end{equation}
Here $C$ and the admissible $c_\star$ may depend on $H_\rho$ and the fixed
design and comparability constants.

The correctly specified model is included by taking $X_i\equiv0$,
$S_x=0$, and $\rho_n=0$.  Then the right-hand sides of
\eqref{eq:nb-theta-error} and \eqref{eq:nb-mle-error} reduce to
$C\lambda_z^{-1}\sqrt{(1+\eta_yS_y)dL/(S_yn)}$.
\end{lemma}

\begin{lemma}[Empirical restricted-NB residual moments]
\label{lem:nb-empirical}
Suppose the triples $(X_i,Y_i,\bm Z_i)_{i=1}^n$ are independent and
\[
 X_i\mid \bm Z_i\sim\operatorname{Bernoulli}(q_i),\qquad
 Y_i\mid X_i,\bm Z_i\sim\operatorname{NB}\{m_i e^{\rho_nX_i},\eta_y\},
\]
where $q_i\asymp S_x$, $m_i\asymp S_y\le M_0$, and
$|\rho_n|\le H_\rho$.  For an arbitrary estimator
$\widehat m_i$, define the fit event
\begin{equation}\label{eq:nb-fit-event}
 \mathcal F_{n,\mathrm{NB}}
 \equiv\left\{\max_i|\widehat m_i/m_i-1|\le c_0\right\}
\end{equation}
for a sufficiently small fixed $c_0$, and put
$a_i\equiv Y_i-\widehat m_i$, $L\equiv\log(C_0/\delta)$, and
$H\equiv\log(C_0n/\delta)$.  If
\begin{equation}\label{eq:nb-moment-n}
 n\ge C\frac{(1+\eta_yS_y)H^3L}{S_y},
\end{equation}
then there exists an event $\mathcal G_{n,\mathrm{NB}}$, with
$\mathbb P(\mathcal G_{n,\mathrm{NB}})\ge1-\delta$, such that on
$\mathcal F_{n,\mathrm{NB}}\cap\mathcal G_{n,\mathrm{NB}}$,
\begin{align}\label{eq:nb-empirical-events}
 cS_y(1+\eta_yS_y)
 &\le\frac1n\sum_i a_i^2\le CS_y(1+\eta_yS_y),\\
 \frac1n\sum_i|a_i|^3
 &\le CS_y(1+\eta_yS_y)^2,
 \qquad
 \max_i|a_i|\le C(1+\eta_yS_y)H.\notag
\end{align}
The constants may depend on $H_\rho$.
\end{lemma}

\subsubsection{Bernoulli--negative-binomial spaCRT verification lemmas}
\label{sec:bnb}

This module uses the main-text sparsity and dispersion notation and
proof-local full coefficient vectors $\bm\theta_x,\bm\theta_y$.
The theory treats dispersion as known. The simulations in
Section~\ref{sec:simulation} estimate dispersion and use a Gaussian covariate;
they evaluate the practical procedure beyond the known-dispersion and bounded-design
theorem assumptions.
The resampled predictor $X_i^*$ is $\widetilde X_i$; the fitted means
$\widehat\mu_{ix},\widehat\mu_{iy}$ are
$\widehat\mu_{x,i},\widehat\mu_{y,i}$; the response residual $a_i$ is
$\widehat e_{y,i}$; and the single envelope constant $B$ of
\eqref{eq:crispr-regularity-conditions} may be taken for each of
$B_u,B_\beta,B_\alpha,H_\rho$ below, with $\lambda=\lambda_z$.  In the
predictor-resampling arguments, $m_i$ denotes the null response mean, as in
Lemma~\ref{lem:logistic-mle} and Lemma~\ref{lem:nb-empirical}; it is
unrelated to the oracle residual multiplier $m_i$ of the GLM-RR proofs
(Supplementary Section~\ref{sec:proof-saige-validity}).

\paragraph{Parametric model and null fit.}
Assume that $(X_i,Y_i,\bm Z_i)_{i=1}^n$ are i.i.d., with
$\bm Z_i=(1,\bm U_i^{\T})^{\T}\in\R^d$ for a dimension $d$ that is
fixed throughout this module, and
\begin{align*}
 X_i\mid \bm Z_i
 &\sim\operatorname{Bernoulli}(q_i),
 &q_i&\equiv\expit(\alpha_{n,x}+\bm\beta_{n,x}^{\T}\bm U_i),
 \\
 Y_i\mid X_i,\bm Z_i
 &\sim\operatorname{NB}(m_i e^{\rho_nX_i},\eta_{n,y}),
 &m_i&=\exp(\alpha_{n,y}+\bm\beta_{n,y}^{\T}\bm U_i).
\end{align*}
Write, for $\bm z=(1,\bm u^{\T})^{\T}$,
\[
 q(\bm z)\equiv\expit(\alpha_{n,x}+\bm\beta_{n,x}^{\T}\bm u),
 \qquad
 m(\bm z)\equiv\exp(\alpha_{n,y}+\bm\beta_{n,y}^{\T}\bm u),
\]
so that $q_i=q(\bm Z_i)$ and $m_i=m(\bm Z_i)$, and let
$\bm\theta_x\equiv(\alpha_{n,x},\bm\beta_{n,x}^{\T})^{\T}$ and
$\bm\theta_y\equiv(\alpha_{n,y},\bm\beta_{n,y}^{\T})^{\T}$.
The following conditions are standing assumptions for this entire
module: the design condition \eqref{eq:design} holds, and, for both the
$X$-model and the $Y$-model, the coefficient bounds of
Lemma~\ref{lem:glm-scales} hold, namely
\begin{equation*}
 \|\bm U_i\|_2\le B_u\ \text{a.s.},\qquad
 \|\bm\beta_{n,x}\|_2\vee\|\bm\beta_{n,y}\|_2\le B_\beta,\qquad
 \alpha_{n,x}\vee\alpha_{n,y}\le B_\alpha,\qquad
 |\rho_n|\le H_\rho.
\end{equation*}
In the main-text notation,
\begin{equation*}
 \pi_{n,x}=\expit(\alpha_{n,x}),\qquad \mu_{n,y}=e^{\alpha_{n,y}}.
\end{equation*}
Lemma~\ref{lem:glm-scales}, applied to each model with $B_\rho=H_\rho$,
gives
\begin{equation}\label{eq:bnb-sparsity}
 q_i\asymp \pi_{n,x},
 \qquad m_i e^{\rho_nX_i}\asymp \mu_{n,y},
 \qquad 1-q_i\ge c.
\end{equation}
The conditional-independence null is equivalent to $\rho_n=0$.

Let $\widehat{\bm\theta}_x$ be the logistic MLE for $\bm X$ on $\bm Z$ and
let $\widehat{\bm\theta}_y$ be the known-dispersion NB MLE in the
restricted model $\rho=0$; if either fails to exist, $p_{\infty,n}^{\dCRT}$
and $p_{\mathrm{SPA},n}^{\dCRT}$ are defined to be $1$ by the prerequisite-failure
convention stated above. All statements below concern events on which
both MLEs exist and are unique.  Set
\[
 \widehat\mu_{ix}\equiv\expit(\widehat{\bm\theta}_x^{\T}\bm Z_i),
 \qquad
 \widehat\mu_{iy}\equiv e^{\widehat{\bm\theta}_y^{\T}\bm Z_i},
 \qquad
 a_i\equiv Y_i-\widehat\mu_{iy}.
\]
The observed statistic is
\begin{equation}\label{eq:bnb-statistic}
 T_n^{\mathrm{dCRT}}\equiv w_n
 =\frac1n\sum_{i=1}^n(X_i-\widehat\mu_{ix})a_i.
\end{equation}
Let $\mathcal D_n\equiv\sigma(X_{1:n},Y_{1:n},\bm Z_{1:n})$.  Conditional on
$\mathcal D_n$, draw independent
$X_i^*\sim\operatorname{Bernoulli}(\widehat\mu_{ix})$ and define
\begin{equation*}
 W_{in}\equiv(X_i^*-\widehat\mu_{ix})a_i,
 \qquad \overline W_n\equiv n^{-1}\sum_iW_{in}.
\end{equation*}
Here $\widetilde \P$ denotes the conditional resampling law of the starred draws
given $\mathcal D_n$; the conditioning bar is retained for emphasis.  The
exact infinite-resampling plug-in reference p-value
under the fitted law is
\begin{equation}\label{eq:pdcrt}
 p_{\infty,n}^{\dCRT}
 \equiv\widetilde \P(\overline W_n\ge T_n^{\mathrm{dCRT}}\mid\mathcal D_n).
\end{equation}
This is the main-text target $p_{\infty,n}^{\dCRT}$, with the generic conditioning
$\sigma$-field $\cF_n$ instantiated as $\mathcal D_n$.
The conditional average CGF is
\begin{equation}\label{eq:bnb-cgf}
 K_n(s)\equiv\frac1n\sum_{i=1}^n
 \left[-s\widehat\mu_{ix}a_i+
 \log\{1-\widehat\mu_{ix}+\widehat\mu_{ix}e^{sa_i}\}\right].
\end{equation}
In this module $p_{\mathrm{SPA},n}^{\dCRT}$ is the ordinary Lugannani--Rice
functional for this average CGF and cutoff, with the admissibility and
continuous-center conventions in the main text.

Under a conditional tilt $s$, write $\widetilde\E_s$ for expectation under
the tilted law; the Bernoulli success probability becomes
\begin{equation}\label{eq:bnb-qs}
 \mu_{ix,s}=\frac{\widehat\mu_{ix}e^{sa_i}}
 {1-\widehat\mu_{ix}+\widehat\mu_{ix}e^{sa_i}},
\end{equation}
so
\begin{align}
 K_n''(s)&=\frac1n\sum_i a_i^2\mu_{ix,s}(1-\mu_{ix,s}),
 \label{eq:bnb-K2-exact}\\
 \frac1n\sum_i\widetilde\E_s|W_{in}-\widetilde\E_sW_{in}|^3
 &=\frac1n\sum_i|a_i|^3\mu_{ix,s}(1-\mu_{ix,s})
 \{\mu_{ix,s}^2+(1-\mu_{ix,s})^2\}.
 \label{eq:bnb-K3-exact}
\end{align}

For later use, write
\[
 \kappa_n(s)\equiv
 \frac{n^{-1}\sum_i
   \widetilde\E_s|W_{in}-\widetilde\E_sW_{in}|^3}
 {\{K_n''(s)\}^{3/2}},
\]
the tilted Lyapunov ratio associated with
\eqref{eq:bnb-K2-exact}--\eqref{eq:bnb-K3-exact}.

\paragraph{Cutoff and conditional moments.}

\begin{lemma}[Ber--NB cutoff under the null and bounded alternatives]
\label{lem:bnb-cutoff}
Let $L\equiv\log(C_0/\delta)$, and let $\mathcal F_{n,\mathrm{bnb}}$ be the
intersection of the conclusion events in Lemma~\ref{lem:logistic-mle}
and Lemma~\ref{lem:nb-mle}.  If
\begin{equation}\label{eq:bnb-cutoff-n}
 n\ge C\frac{(1+\eta_{n,y}\mu_{n,y})L}{\pi_{n,x}\mu_{n,y}},
\end{equation}
then there exists an event $\mathcal G_{n,\mathrm{bnb}}^{\mathrm{cut}}$ with
$\mathbb P(\mathcal G_{n,\mathrm{bnb}}^{\mathrm{cut}})\ge1-C\delta$ such
that, on
$\mathcal F_{n,\mathrm{bnb}}\cap
 \mathcal G_{n,\mathrm{bnb}}^{\mathrm{cut}}$,
\begin{equation}\label{eq:bnb-cutoff-bound}
 |T_n^{\mathrm{dCRT}}|
 \le C\left\{\sqrt{\frac{\pi_{n,x}\mu_{n,y}(1+\eta_{n,y}\mu_{n,y})L}{n}}
 +\pi_{n,x}\mu_{n,y}|\rho_n|\right\}.
\end{equation}
The constant may depend on $H_\rho$.
\end{lemma}

\begin{lemma}[Ber--NB conditional variance and tilted third moment]
\label{lem:bnb-moments}
Let $L\equiv\log(C_0d/\delta)$ and $H\equiv\log(C_0n/\delta)$, and assume both the
burn-in \eqref{eq:nb-moment-n} of Lemma~\ref{lem:nb-empirical}, which
guarantees the moment bounds \eqref{eq:nb-empirical-events} on
$\mathcal F_{n,\mathrm{NB}}\cap\mathcal G_{n,\mathrm{NB}}$, and
\begin{equation}\label{eq:bnb-moments-n}
 n\ge C\frac{dL}{\pi_{n,x}\lambda_z^2},
\end{equation}
with $C$ large enough that the relative-error bound in
\eqref{eq:logistic-error} is at most a fixed small constant.  On
$\mathcal F_{n,\mathrm{bnb}}\cap
\mathcal F_{n,\mathrm{NB}}\cap\mathcal G_{n,\mathrm{NB}}$ from
Lemma~\ref{lem:nb-empirical}, for every
\begin{equation}\label{eq:bnb-s0}
 |s|\le s_{0n}:=\frac{c}{(1+\eta_{n,y}\mu_{n,y})H},
\end{equation}
we have
\begin{align}
 c\pi_{n,x}\mu_{n,y}(1+\eta_{n,y}\mu_{n,y})&\le K_n''(s)\le
 C\pi_{n,x}\mu_{n,y}(1+\eta_{n,y}\mu_{n,y}),\label{eq:bnb-K2-bound}\\
 \frac1n\sum_i\widetilde\E_s|W_{in}-\widetilde\E_sW_{in}|^3
 &\le C\pi_{n,x}\mu_{n,y}(1+\eta_{n,y}\mu_{n,y})^2,\label{eq:bnb-K3-bound}\\
 \sup_{|s|\le s_{0n}}\kappa_n(s)
 &\le C\sqrt{\frac{1+\eta_{n,y}\mu_{n,y}}{\pi_{n,x}\mu_{n,y}}}.
 \label{eq:bnb-kappa}
\end{align}
\end{lemma}

\subsubsection{Upper rate and null validity}
\label{sec:bnb-upper-validity}

The sequence-wise bounds below inherit envelope-built constants from the
verification chain: every constant depends only on the fixed envelope,
design, and dimension constants.  Dispersion enters through
$A_{n,y}=1+\eta_{n,y}\mu_{n,y}$.  The upgrade from the sequence-wise bound
to the sup form~\eqref{eq:op-def} is
Corollary~\ref{cor:op-sup-seq-use}.

\begin{proof}[Upper rate in Theorem~\ref{thm:spacrt-rate}]
The argument is sequence-wise, along an arbitrary array satisfying
\eqref{eq:crispr-regularity-conditions} together with the limit form of the
regime conditions,
\begin{equation}\label{eq:bnb-proof-regime}
 \frac{n\pi_{n,x}\mu_{n,y}}
 {(1+\eta_{n,y}\mu_{n,y})\log^4n}\to\infty,
 \qquad |\rho_n|\log n\to0.
\end{equation}
Every array with $\P_n\in\PBNB_n$ satisfies these conditions by
\eqref{eq:crispr-class}, since $\delta_n\to\infty$.

Fix $\eps\in(0,1/4]$.  Apply Lemmas~\ref{lem:logistic-mle},
\ref{lem:nb-mle}, \ref{lem:nb-empirical}, \ref{lem:bnb-cutoff}, and
\ref{lem:bnb-moments} with failure levels chosen as fixed multiples of
$\eps$.  After absorbing those multiples into $C_0$, write
$L=\log(C_0d/\eps)$ and $H=\log(C_0n/\eps)$.
Since $\pi_{n,x}\le1$, $\mu_{n,y}\le e^B$, and $A_{n,y}\ge1$,
\[
 \min\left\{n,n\pi_{n,x},\frac{n\mu_{n,y}}{A_{n,y}}\right\}
 \ge c\frac{n\pi_{n,x}\mu_{n,y}}{A_{n,y}},
 \qquad
 \frac{n\pi_{n,x}\mu_{n,y}}{A_{n,y}H^3L}\longrightarrow\infty.
\]
The last limit follows from \eqref{eq:bnb-proof-regime}, since
$L$ is fixed and $H=O(\log n)$ for fixed $\eps$.
These bounds imply, for all sufficiently large $n$, the
sample-size conditions~\eqref{eq:logistic-n}, \eqref{eq:nb-mle-n},
\eqref{eq:nb-moment-n}, \eqref{eq:bnb-cutoff-n},
and~\eqref{eq:bnb-moments-n}. Also,
$\pi_{n,x}|\rho_n|\le|\rho_n|\to0$ verifies the local smallness
conditions~\eqref{eq:logistic-local-rho-small}
and~\eqref{eq:nb-local-smallness}, and the NB relative fit error tends to
zero, as required by \eqref{eq:nb-fit-event}.
A union bound therefore gives an event
of probability at least $1-\eps$ on which both restricted maximum
likelihood estimators exist uniquely and all the bounds below hold.

Put
\[
 A_2\equiv nA_{2n}\equiv\sum_{i=1}^n\widehat v_{x,i}\widehat e_{y,i}^{\,2},
 \qquad
 A_3\equiv nA_{3n}\equiv\sum_{i=1}^n\widehat v_{x,i}|\widehat e_{y,i}|^{3},
\]
where $\widehat v_{x,i}=\widehat\mu_{ix}(1-\widehat\mu_{ix})$ and
$\widehat e_{y,i}=Y_i-\widehat\mu_{iy}=a_i$, and let
$t_n=nT_n^{\mathrm{dCRT}}$ be the sum-scale cutoff, so that
$A_2=nK_n''(0)$ for the conditional CGF \eqref{eq:bnb-cgf}.  Equations
\eqref{eq:bnb-K2-bound}, \eqref{eq:bnb-K3-bound},
\eqref{eq:nb-empirical-events}, and \eqref{eq:bnb-cutoff-bound} give the
explicit bounds
\[
 cn\pi_{n,x}\mu_{n,y}A_{n,y}\le A_2
 \le Cn\pi_{n,x}\mu_{n,y}A_{n,y},
 \qquad
 A_3\le Cn\pi_{n,x}\mu_{n,y}A_{n,y}^2,
\]
\[
 e_n^\star\le CA_{n,y}H,
 \qquad
 |t_n|\le C\{\sqrt{n\pi_{n,x}\mu_{n,y}A_{n,y}L}
 +n\pi_{n,x}\mu_{n,y}|\rho_n|\}.
\]
In particular, $A_2>0$, so the conditional score variance is positive.

For the conditional Bernoulli row ($D_0=1$),
part~\ref{part:translator-bounded-support} of
Corollary~\ref{cor:translators-main} supplies a certificate for which
$b_{in}=|\widehat e_{y,i}|$, $\sigma_n^2=A_2$, $H_n=A_3$, and
$b_n^\star=e_n^\star$.  The preceding bounds therefore close all three
saddlepoint coordinates explicitly:
\[
 \frac{B_n}{\sqrt n}
 =\frac{A_3}{A_2^{3/2}}
 \le C\sqrt{\frac{A_{n,y}}{n\pi_{n,x}\mu_{n,y}}},
 \qquad
 Z_n=\frac{|t_n|}{\sqrt{A_2}}
 \le C\left\{\sqrt L+|\rho_n|
       \sqrt{\frac{n\pi_{n,x}\mu_{n,y}}{A_{n,y}}}\right\},
\]
\[
 D_n=\frac{e_n^\star|t_n|}{A_2}
 \le CH\left\{\sqrt{\frac{A_{n,y}L}{n\pi_{n,x}\mu_{n,y}}}
 +|\rho_n|\right\},
 \qquad
 \frac{B_n(1+Z_n)}{\sqrt n}
 \le C\left\{\sqrt{\frac{A_{n,y}L}{n\pi_{n,x}\mu_{n,y}}}
 +|\rho_n|\right\}.
\]
By \eqref{eq:bnb-proof-regime}, these bounds give, for fixed $\eps$,
\[
 D_n\le C_\eps\left\{
 \log n\sqrt{\frac{A_{n,y}}{n\pi_{n,x}\mu_{n,y}}}
 +|\rho_n|\log n\right\}=o(1),\quad
 \frac{B_n(1+Z_n)}{\sqrt n}
 \le C_\eps\left\{\sqrt{\frac{A_{n,y}}{n\pi_{n,x}\mu_{n,y}}}
 +|\rho_n|\right\}=o(1).
\]
Together with $A_2>0$, these are
\eqref{eq:scale-adaptive-lr-conditions} for sufficiently large $n$.
Since \eqref{eq:pdcrt} is the conditional non-strict upper tail of
$\sum_iW_{in}$ at $t_n$ and equals $p_{\infty,n}^{\dCRT}$ here,
Theorem~\ref{thm:conditional-lr} gives the unique saddlepoint and
\[
 \left|\frac{p_{\infty,n}^{\dCRT}}{p_{\mathrm{SPA},n}^{\dCRT}}-1\right|
 \le C\left\{\sqrt{\frac{A_{n,y}L}{n\pi_{n,x}\mu_{n,y}}}
              +|\rho_n|\right\}.
\]
Hence, with probability at
least $1-\eps$ and for all sufficiently large $n$,
\[
 \left|\frac{p_{\infty,n}^{\dCRT}}{p_{\mathrm{SPA},n}^{\dCRT}}-1\right|
 \le C\sqrt{\log(C_0d/\eps)}
 \left\{\sqrt{\frac{1+\eta_{n,y}\mu_{n,y}}{n\pi_{n,x}\mu_{n,y}}}
       +|\rho_n|\right\}.
\]
This fixed-$\eps$ bound proves
\[
 \frac{p_{\infty,n}^{\dCRT}}{p_{\mathrm{SPA},n}^{\dCRT}}-1
 =O_{\P_n}\!\left(
 \sqrt{\frac{1+\eta_{n,y}\mu_{n,y}}{n\pi_{n,x}\mu_{n,y}}}
 +|\rho_n|\right),
\]
and also shows that both
restricted maximum likelihood estimators, the conditional variance, and
the saddlepoint exist with probability tending to one.  Because the array
was arbitrary and all constants depend only on the fixed envelopes,
Corollary~\ref{cor:op-sup-seq-use}, applied to the relative error and
to the indicator of the complement of the existence event with
deterministic rate $n^{-1}$, gives the uniform conclusions over
$\PBNB_n$.  This completes the upper-rate
argument.
\end{proof}

\begin{corollary}[Null validity of sparse Bernoulli--negative-binomial spaCRT]
\label{cor:bnb-null-validity}
Let $\P_n\in\PBNBnull_n$ for each $n$.  Then
\[
  p_{\infty,n}^{\dCRT}\convd\Unif(0,1),
  \qquad
  p_{\mathrm{SPA},n}^{\dCRT}\convd\Unif(0,1).
\]
Moreover, both limits are uniform over the null models in the sense
of~\eqref{eq:bnb-null-uniformity}.  Explicitly, at $\rho_n=0$ the
$\rho$-conditions hold trivially, so $\PBNBnull_n$ consists of
the null ($\rho_n=0$) data-generating processes---covariate law,
coefficients $(\alpha_{n,x},\bm\beta_{n,x},\alpha_{n,y},\bm\beta_{n,y})$,
and dispersion $\eta_{n,y}\ge0$---satisfying
\eqref{eq:crispr-regularity-conditions} with these $(B,\lambda,d)$ and
\[
 \frac{n\pi_{n,x}\mu_{n,y}}{1+\eta_{n,y}\mu_{n,y}}
 \ge\delta_n\log^4n.
\]
Then, for
$p\in\{p_{\infty,n}^{\dCRT},p_{\mathrm{SPA},n}^{\dCRT}\}$,
\[
  \sup_{\P\in\PBNBnull_n}
  \sup_{t\in[0,1]}
  |\P(p\le t)-t|
  \longrightarrow0 .
\]
\end{corollary}

\begin{proof}
Under the null, apply the correctly specified null specializations of
Lemmas~\ref{lem:logistic-mle} and~\ref{lem:nb-mle}, in which the
auxiliary alternative variables in those lemmas are identically zero.
With fixed dimension and design eigenvalue, the
bounds~\eqref{eq:logistic-error} and~\eqref{eq:nb-mle-error} imply
\[
 \max_i\left|\frac{\widehat\mu_{x,i}}{\mu^0_{x,i}}-1\right|
 =O_{\P_n}\{(n\pi_{n,x})^{-1/2}\},
 \qquad
 \max_i\left|\frac{\widehat\mu_{y,i}}{\mu^0_{y,i}}-1\right|
 =O_{\P_n}\left\{\sqrt{\frac{A_{n,y}}{n\mu_{n,y}}}\right\}.
\]
Because the true means are uniformly comparable to $\pi_{n,x}$ and
$\mu_{n,y}$ by~\eqref{eq:bnb-sparsity}, these bounds give
\[
 E_{n,x}=O_{\P_n}\sqrt{\frac{\pi_{n,x}}n},
 \qquad
 E_{n,y}=O_{\P_n}\sqrt{\frac{\mu_{n,y}A_{n,y}}n}.
\]
Moreover $v^0_{x,i}\asymp\pi_{n,x}$,
$v^0_{y,i}\asymp\mu_{n,y}A_{n,y}$, and hence
$\sigma_{n,\dCRT}^2\asymp\pi_{n,x}\mu_{n,y}A_{n,y}$.  It follows that
\[
 \frac{\sqrt nE_{n,x}E_{n,y}}{\sigma_{n,\dCRT}}
 =O_{\P_n}(n^{-1/2})=o_{\P_n}(1).
\]
The weighted-error definitions in
Supplementary Section~\ref{sec:spacrt-notation} then give
\[
 (E'_{n,x})^2
 =O_{\P_n}\left(\frac{\pi_{n,x}\mu_{n,y}A_{n,y}}n\right)
 =O_{\P_n}\left(\frac{\sigma_{n,\dCRT}^2}n\right),
\]
\[
 (E'_{n,y})^2
 =O_{\P_n}\left(\frac{\pi_{n,x}\mu_{n,y}A_{n,y}}n\right)
 =O_{\P_n}\left(\frac{\sigma_{n,\dCRT}^2}n\right).
\]
The predictor mean bound and $v^0_{x,i}\asymp\pi_{n,x}$ also imply
\[
 \max_i\left|\frac{\widehat v_{x,i}}{v^0_{x,i}}-1\right|
 =O_{\P_n}\{(n\pi_{n,x})^{-1/2}\}=o_{\P_n}(1),
 \qquad
 \max_i\widehat v_{x,i}=O_{\P_n}(\pi_{n,x}),
\]
where the regime condition implies $n\pi_{n,x}\to\infty$.  Consequently,
\[
 (\widehat E'_{n,y})^2
 =O_{\P_n}\left(\frac{\pi_{n,x}\mu_{n,y}A_{n,y}}n\right)
 =O_{\P_n}\left(\frac{\sigma_{n,\dCRT}^2}n\right).
\]
The three explicit bounds therefore give
\[
 E'_{n,x}+E'_{n,y}+\widehat E'_{n,y}
 =O_{\P_n}(\sigma_{n,\dCRT}/\sqrt n)
 =o_{\P_n}(\sigma_{n,\dCRT}).
\]
They also complete the fitted-variance calibration, since
\[
 \left|\frac1n\sum_i(\widehat v_{x,i}-v^0_{x,i})v^0_{y,i}\right|
 \le
 \max_i\left|\frac{\widehat v_{x,i}}{v^0_{x,i}}-1\right|
 \frac1n\sum_i v^0_{x,i}v^0_{y,i}
 =o_{\P_n}(\sigma_{n,\dCRT}^2).
\]

The true and fitted Bernoulli predictor residuals are bounded by one, so
the Bernstein scales in Assumption~\ref{assu:spacrt-regularity}(ii),
stated in Subsubsection~\ref{sec:spacrt-validity-statement}, may be taken
as
\[
 b^0_{n,x}=1,
 \qquad
 \widehat b_{n,x}=1.
\]
For the true response residuals $R_i\equiv Y_i-\mu^0_{y,i}$, the moment form of the
conditional Bernstein bound required by
Assumption~\ref{assu:spacrt-regularity}(ii) follows from
Lemma~\ref{lem:nb-facts}: the weighted-moment bound
\eqref{eq:nb-weighted-moment} with $k=2$, applied conditionally on
$\bm Z_i$, and the elementary inequality
$x^{k-2}\le(k-2)!\,(A_{n,y}/c)^{k-2}e^{cx/A_{n,y}}$ for $x\ge0$ give,
for every integer $k\ge3$,
\[
\begin{aligned}
 \E(|R_i|^k\mid \bm Z_i)
 &=\E(R_i^2|R_i|^{k-2}\mid \bm Z_i)\\
 &\le(k-2)!\left(\frac{A_{n,y}}{c}\right)^{k-2}
   \E\left(R_i^2e^{c|R_i|/A_{n,y}}\ \middle|\ \bm Z_i\right)\\
 &\le\frac{k!}{2}\left(\frac{A_{n,y}}{c}\right)^{k-2}
   C\mu_{n,y}A_{n,y},
\end{aligned}
\]
using $(k-2)!\le k!/2$, $\mu^0_{y,i}\asymp\mu_{n,y}$, and
$1+\eta_{n,y}\mu^0_{y,i}\asymp A_{n,y}$. Since the exact conditional
variance is
$v^0_{y,i}=\mu^0_{y,i}(1+\eta_{n,y}\mu^0_{y,i})
\asymp\mu_{n,y}A_{n,y}$, a fixed enlargement of the scale absorbs the
comparison constants and gives, for every $k\ge3$,
\[
 \E(|R_i|^k\mid\bm Z_i)
 \le\frac{k!}{2}(C_bA_{n,y})^{k-2}v^0_{y,i}.
\]
Thus the true response residuals satisfy the required conditional
Bernstein bound with $b^0_{n,y}=C_bA_{n,y}$. The smallness condition in
clause~(ii) therefore
ends with the explicit bound
\[
 \frac{b^0_{n,y}(b^0_{n,x}+\widehat b_{n,x})}
 {\sqrt n\,\sigma_{n,\dCRT}}
 =O\left\{\frac{A_{n,y}}
 {\sqrt{n\pi_{n,x}\mu_{n,y}A_{n,y}}}\right\}
 =O\!\left(\sqrt{\frac{1+\eta_{n,y}\mu_{n,y}}
 {n\pi_{n,x}\mu_{n,y}}}\right)=o(1).
\]
The maximal-residual bound in~\eqref{eq:nb-empirical-events} gives
$e_n^\star=\max_i|\widehat e_{y,i}|=O_{\P_n}(A_{n,y}\log n)$, while
the conditional-variance bound~\eqref{eq:bnb-K2-bound} at $s=0$ gives
$V_{n,\dCRT}^2=\Theta_{\P_n}(\pi_{n,x}\mu_{n,y}A_{n,y})$.  Hence
\[
 \frac{\widehat b_{n,x}e_n^\star}
 {\sqrt nV_{n,\dCRT}}
 =O_{\P_n}\!\left(\log n\sqrt{\frac{1+\eta_{n,y}\mu_{n,y}}
 {n\pi_{n,x}\mu_{n,y}}}\right)
 =o_{\P_n}(1).
\]
Finally, the existence conclusions of Lemmas~\ref{lem:logistic-mle}
and~\ref{lem:nb-mle}, applied above, ensure that both nuisance fits
are finite with probability tending to one. On this event, the fitted
Bernoulli probabilities define the conditional resampling law, verifying
Assumption~\ref{assu:spacrt-regularity}(iv).
All clauses of Assumption~\ref{assu:spacrt-regularity} are now verified,
so Theorem~\ref{thm:spacrt-validity} proves asymptotic uniformity of the
exact fitted dCRT target and of its saddlepoint approximation; the
verification covers every array satisfying
\eqref{eq:crispr-regularity-conditions} with
\[
 \frac{n\pi_{n,x}\mu_{n,y}}
 {(1+\eta_{n,y}\mu_{n,y})\log^4n}\to\infty,
 \qquad \rho_n=0.
\]
The uniform null clause now follows from
Corollary~\ref{cor:uniform-null-diagonal-use}.
\end{proof}

\subsubsection{Auxiliary Poisson approximations}

The following Poisson approximations are used in the sharpness argument
for the sparse Bernoulli--negative-binomial result in Theorem~\ref{thm:spacrt-rate}.

\begin{lemma}[Bernoulli-to-Poisson couplings]\label{lem:le-cam-elementary}
Let $\xi_1,\ldots,\xi_N$ be independent Bernoulli variables with
success probabilities $0<p_i\le1/2$, put $B\equiv\sum_i\xi_i$ and
$\lambda\equiv\sum_i p_i$, and let $Z\sim\operatorname{Poisson}(\lambda)$.  Then
\[
 d_{\mathrm{TV}}\{\mathcal L(B),\mathcal L(Z)\}
 \le 2\sum_{i=1}^Np_i^2.
\]
\end{lemma}

\begin{lemma}[Uniform lattice expansions for finite Poisson jump arrays]
\label{lem:poisson-jump-edgeworth}
Let $L\to\infty$ and fix an integer $J\ge1$.  Let
$d_1,\ldots,d_J$ be fixed integers with greatest common divisor one, and
fix constants $h,c,C>0$.  For each $L$, let
$N_{1L},\ldots,N_{JL}$ be independent Poisson variables with means
$\mu_{1L},\ldots,\mu_{JL}\in[cL,CL]$.  Define the centered variable
\[
 A_L\equiv h\sum_{j=1}^Jd_j(N_{jL}-\mu_{jL}),
\]
and write
\[
 \sigma_L^2\equiv\Var(A_L),\qquad
 \kappa_{3,L}\equiv\operatorname{cum}_3(A_L).
\]
For every fixed $M<\infty$, uniformly over the means in the stated ranges
and over support points $a_L$ for which
$z_L\equiv a_L/\sigma_L$ satisfies $|z_L|\le M$,
\begin{align}
 \P(A_L=a_L)
 &=\frac{h}{\sigma_L}\phi(z_L)+O_M(L^{-1}),
 \label{eq:poisson-jump-local}\\
 \P(A_L>a_L)+\frac12\P(A_L=a_L)
 &=\Phibar(z_L)
   +\frac{\kappa_{3,L}}{6\sigma_L^3}
    (z_L^2-1)\phi(z_L)
   +O_M(L^{-1}).
 \label{eq:poisson-jump-midtail}
\end{align}
The constants in the remainders depend only on
$(J,d_1,\ldots,d_J,h,c,C,M)$.
\end{lemma}

\begin{lemma}[Poisson local mass, midpoint tail, and central tail]
\label{lem:poisson-central-tail}\label{lem:poisson-midtail-lr}
Let $P_\lambda\sim\operatorname{Poisson}(\lambda)$, let $a$ be an integer,
put $z\equiv(a-\lambda)/\sqrt\lambda$, and define
\[
 Q_\lambda(a)\equiv\P(P_\lambda>a)+\frac12\P(P_\lambda=a).
\]
For every fixed $M<\infty$, uniformly over $|z|\le M$ as
$\lambda\to\infty$,
\begin{align}
 \P(P_\lambda=a)
 &=\frac{\phi(z)}{\sqrt\lambda}+O_M(\lambda^{-1}),
 \label{eq:poisson-local-compact}\\
 Q_\lambda(a)
 &=\Phibar(z)+\frac{\phi(z)(z^2-1)}{6\sqrt\lambda}
 +O_M(\lambda^{-1}).
 \label{eq:poisson-midtail-compact}
\end{align}
In particular, for integer $k\to\infty$ and
$Z_k\sim\operatorname{Poisson}(k)$,
\begin{align}
 \P(Z_k=k)&=\frac1{\sqrt{2\pi k}}+O(k^{-3/2}),
 \label{eq:poisson-mass}\\
 \P(Z_k\ge k)&=\frac12+\frac1{3\sqrt{2\pi k}}+O(k^{-1}).
 \label{eq:poisson-tail}
\end{align}
\end{lemma}

\begin{lemma}[Central LR expansion for a Poisson-like CGF]
\label{lem:poisson-like-lr}
Let $K_\lambda$ be a centered CGF with cumulants
$\kappa_j=K_\lambda^{(j)}(0)$.  Fix $M<\infty$.  Suppose, uniformly over a
possibly random collection of such CGFs,
\begin{equation}\label{eq:poisson-like-cumulant-assumptions}
 \kappa_2=\lambda+O(\lambda^{-1}),\qquad
 \kappa_3=\lambda+O(\lambda^{-1}),
\end{equation}
and, for $j=4,5$ and $|s|\le C_M\lambda^{-1/2}$ with $C_M\ge M+2$,
\begin{equation}\label{eq:poisson-like-derivative-assumptions}
 |K_\lambda^{(j)}(s)|\le C_M\lambda.
\end{equation}
If the centered cutoff satisfies
\begin{equation}\label{eq:poisson-like-cutoff-assumption}
 w=z\sqrt\lambda+O(\lambda^{-1/2}),\qquad |z|\le M,
\end{equation}
then
\begin{equation}\label{eq:poisson-like-lr}
 p_{\LR}(w)=\Phibar(z)
 +\frac{\phi(z)(z^2-1)}{6\sqrt\lambda}
 +O_M(\lambda^{-1}).
\end{equation}
The assertion remains valid at $z=0$ under the continuous convention
\eqref{eq:new-lr-zero}.
The conclusion remains valid with
$O_{\P_n},o_{\P_n}$ hypotheses and conclusions uniformly on events whose probability
tends to one.
\end{lemma}

Comparing \eqref{eq:poisson-midtail-compact} and
\eqref{eq:poisson-like-lr}, LR reproduces the midpoint Poisson tail up to
$O_M(\lambda^{-1})$. The fitted boundary fraction is analyzed below.

\subsubsection{Sharpness}

Throughout this subsection the exact conditional tail is
\eqref{eq:pdcrt}, and $p_{\mathrm{SPA},n}^{\dCRT}$ is its ordinary LR approximation.

\begin{proposition}[A covariate phase makes the Ber--NB saddlepoint rate tight]
\label{prop:bnb-phase-obstruction}
Let $U$ be uniform on $\{-1,0,1\}$, let $\bm Z=(1,U)^{\T}$, and fix
$\beta\ne0$.  Consider the correctly specified null sequence
\[
 X\mid U\sim\operatorname{Bernoulli}(n^{-2/5}),\qquad
 Y\mid U\sim\operatorname{Poisson}\{n^{-2/5}e^{\beta U}\},
 \qquad X\perp Y\mid U.
\]
Fit the logistic model for $X\mid \bm Z$ and the restricted Poisson log-link
model for $Y\mid \bm Z$.  This sequence has
\[
 \pi_{n,x}=\mu_{n,y}=n^{-2/5},\qquad
 \rho_n=\eta_{n,y}=0,\qquad
 n\pi_{n,x}\mu_{n,y}=n^{1/5}.
\]
It satisfies \eqref{eq:crispr-regularity-conditions} and lies in $\PBNBnull_n$ whenever
$\delta_n\le n^{1/5}/\log^4n$ eventually.

Put $f(u)\equiv e^{\beta u}$, let $\Pi f$ be its
$L^2\{\mathcal L(U)\}$ projection onto $\operatorname{span}\{1,u\}$, and
define
\[
 r\equiv f-\Pi f,\qquad
 \tau\equiv\frac{\{\E r(U)^2\}^{1/2}}
 {\{\E f(U)^2\}^{1/2}}.
\]
Then $\tau>0$. For independent standard normals $G_1,G_2$,
\begin{equation}\label{eq:bnb-phase-limit}
 n^{1/10}\left(\frac{p_{\infty,n}^{\dCRT}}{p_{\mathrm{SPA},n}^{\dCRT}}-1\right)
 \convd
 -\frac{\{1/2-\Phi(\tau G_2)\}\phi(G_1)}
 {\{\E f(U)\}^{1/2}\Phibar(G_1)},
\end{equation}
The limit is finite and nonzero almost surely.  Consequently,
\begin{equation}\label{eq:bnb-phase-asymp-p}
 \left|\frac{p_{\infty,n}^{\dCRT}}{p_{\mathrm{SPA},n}^{\dCRT}}-1\right|
 =\Theta_{\P_n}(n^{-1/10})
 =\Theta_{\P_n}\{(n\pi_{n,x}\mu_{n,y})^{-1/2}\}.
\end{equation}
Thus the exponent in the uniform saddlepoint bound
\eqref{eq:headline-bnb-rate} is tight in probability even when $\eta_{n,y}=0$.
\end{proposition}

\begin{proof}
All calculations below are under the sampling law unless a tilde is shown on the probability or expectation.
All statistics below are written explicitly at the sum scale.  The variables
$G_1,G_2$ are the independent standard normals in the proposition statement.
For brevity in the proof, write $q\equiv n^{-2/5}$.
For all sufficiently large $n$, this construction satisfies
\eqref{eq:crispr-regularity-conditions} with
$B=\max\{1,|\beta|\}$, $\lambda=2/3$, and $d=2$.
For example, $\delta_n=n^{1/10}/\log^4n$ tends to infinity and satisfies
$\delta_n\log^4n=n^{1/10}\le n\pi_{n,x}\mu_{n,y}=n^{1/5}$.
Also, the projection in the proposition has the explicit residual
\[
 r(u)=\frac{\cosh\beta-1}{3}(3u^2-2),\qquad
 \E r(U)^2=\frac{2(\cosh\beta-1)^2}{9}>0,
\]
so $\tau>0$ and the phase standardizations below are well defined.

The proof has four stages.  First, we localize the nuisance fits and identify
the lattice count $L_n^*$ and fitted correction $\Delta_n^*$.  Second, we
expand $p_{\infty,n}^{\dCRT}$ and retain the random fraction $\theta_n$ of its
boundary atom.  Third, we expand the LR formula $p_{\mathrm{SPA},n}^{\dCRT}$ around the
midpoint count tail $Q_n$.  Finally, we subtract the expansions and apply the
joint limit of the count and phase coordinates.

\paragraph{Step 1: Reduce the problem to a lattice count perturbed by the fitted correction.}
Let $\bm\theta_{x0}\equiv(\logit q,0)^{\T}$ and
$\bm\theta_{y0}\equiv(\log q,\beta)^{\T}$ be the true coefficients.  We first
localize the fitted probabilities, then separate the integer count that
governs the reference tail from the fitted correction that decides the
boundary atom.

Both nuisance models are correctly specified, fixed-dimensional GLMs with
bounded design.  In particular,
\[
 \|\bm Z\|_2\le\sqrt2,
 \qquad
 \lambda_{\min}\{\E(\bm Z\bm Z^{\T})\}=\frac23,
\]
and their true conditional means satisfy
\[
 \expit(\bm\theta_{x0}^{\T}\bm Z_i)=q,
 \qquad
 e^{-|\beta|}q
 \le \exp(\bm\theta_{y0}^{\T}\bm Z_i)
 =qf(U_i)\le e^{|\beta|}q.
\]
Thus the design and mean-comparability constants are uniform in $n$.

Because the true intercepts drift, we use Lemmas~\ref{lem:logistic-mle}
and~\ref{lem:nb-mle} to establish existence, uniqueness, and the coefficient
rates together.  Their concavity argument, supplied by
Lemma~\ref{lem:det-sc}, does not require prior consistency of the maximizers.
Fix $\delta\in(0,1/4]$, set $d=2$, and put
$L_\delta\equiv\log(C_0d/\delta)$.

For the logistic fit of $X$ on $\bm Z$, apply the correctly specified null case
of Lemma~\ref{lem:logistic-mle} with auxiliary alternative variable zero,
$S_y=q=n^{-2/5}$.  Condition~\eqref{eq:logistic-means} holds with
fixed constants.  The sample-size requirement~\eqref{eq:logistic-n} is of
order $L_\delta n^{2/5}$ and therefore holds for all large $n$.
The lemma also supplies the restricted score equation
\[
 \sum_{i=1}^n \bm Z_i(X_i-\widehat\mu_{ix})=0.
\]

For the outcome fit, apply the correctly specified null case of
Lemma~\ref{lem:nb-mle} with auxiliary alternative variable zero,
$\eta_{n,y}=0$, and $\mu_{n,y}=q$.  Condition~\eqref{eq:nb-means} holds
with fixed constants, and the sample-size requirement
\eqref{eq:nb-mle-n} is again of order $L_\delta n^{2/5}$.

Combining the two lemmas, for all large $n$, with probability at least
$1-2\delta$, both restricted MLEs exist uniquely and
\[
 \|\widehat{\bm\theta}_x-\bm\theta_{x0}\|_2
 \le C\sqrt{\frac{L_\delta}{nq}}
 =C\sqrt{L_\delta}\,n^{-3/10},
 \qquad
 \|\widehat{\bm\theta}_y-\bm\theta_{y0}\|_2
 \le C\sqrt{\frac{L_\delta}{nq}}
 =C\sqrt{L_\delta}\,n^{-3/10}.
\]
Since $\delta$ was arbitrary, this localization ends with the explicit
probability and rate conclusions
\[
 \P(\text{both restricted MLEs exist and are unique})\longrightarrow1.
\]
Moreover,
\[
 \widehat{\bm\theta}_x-\bm\theta_{x0}=O_{\P_n}(n^{-3/10}),\qquad
 \widehat{\bm\theta}_y-\bm\theta_{y0}=O_{\P_n}(n^{-3/10}).
\]

Expanding the logistic and log links uniformly over the observed support
points gives
\[
\begin{aligned}
 \widehat\mu_{ix}
 &=q+q(1-q)\bm Z_i^{\T}(\widehat{\bm\theta}_x-\bm\theta_{x0})
   +O_{\P_n}(q\|\widehat{\bm\theta}_x-\bm\theta_{x0}\|_2^2)
 =q+q\bm Z_i^{\T}(\widehat{\bm\theta}_x-\bm\theta_{x0})
   +O_{\P_n}(qn^{-3/5}),\\
 \widehat\mu_{iy}
 &=qf(U_i)\exp\{\bm Z_i^{\T}(\widehat{\bm\theta}_y-\bm\theta_{y0})\}
 =qf(U_i)\{1+\bm Z_i^{\T}(\widehat{\bm\theta}_y-\bm\theta_{y0})
   +O_{\P_n}(n^{-3/5})\},
\end{aligned}
\]
and therefore
\begin{equation}\label{eq:phase-fit-rates}
 \max_i\left|\frac{\widehat\mu_{ix}}q-1\right|
 +\max_i\left|\frac{\widehat\mu_{iy}}{qf(U_i)}-1\right|
 =O_{\P_n}(n^{-3/10}).
\end{equation}
Thus the MLE lemmas provide existence and uniqueness, while
\eqref{eq:phase-fit-rates} records the fitted-mean approximation used below.

We pass from the fitted probabilities to the conditional reference tail.
The purpose of the next decomposition is to expose the integer-valued count
whose boundary atom will determine the leading discreteness error.
Centering cancels exactly from the dCRT tail event.  Indeed, both sides
of the conditional event in \eqref{eq:pdcrt} are $1/n$-averages sharing
the common term $-n^{-1}\sum_{i=1}^n\widehat\mu_{ix}(Y_i-\widehat\mu_{iy})$;
multiplying through by $n$ and adding
$\sum_{i=1}^n \widehat\mu_{ix}(Y_i-\widehat\mu_{iy})$ to both sides leaves
\begin{equation}\label{eq:phase-tail-event}
 p_{\infty,n}^{\dCRT}
 =\widetilde \P\left\{
 \sum_{i=1}^n X_i^*(Y_i-\widehat\mu_{iy})
 \ge \sum_{i=1}^n X_i(Y_i-\widehat\mu_{iy})\right\}.
\end{equation}
Define
\[
 L_n^*\equiv\sum_{i=1}^n X_i^*\indicator(Y_i=1),\quad
 L_n\equiv\sum_{i=1}^n X_i\indicator(Y_i=1),\quad
 \Delta_n^*\equiv\sum_{i=1}^n X_i^*\widehat\mu_{iy},\quad
 \Delta_n\equiv\sum_{i=1}^n X_i\widehat\mu_{iy},
\]
and
\[
 R_n\equiv\Delta_n-\widetilde\E\Delta_n^*
 =\sum_{i=1}^n (X_i-\widehat\mu_{ix})\widehat\mu_{iy}.
\]
(The symbol $\Delta_n$ is local to this proof; $D_n$ is reserved
throughout the manuscript for the localization coordinate of the main
theorems.)

We first analyze $R_n$, because this fitted correction will decide what
fraction of the equality layer $L_n^*=L_n$ is included in the exact tail.
The link expansions just obtained give, uniformly over the observed support
points,
\[
 \widehat\mu_{ix}
 =q+q\bm Z_i^{\T}(\widehat{\bm\theta}_x-\bm\theta_{x0})+O_{\P_n}(qn^{-3/5}),
 \qquad
 \widehat\mu_{iy}
 =qf(U_i)\{1+\bm Z_i^{\T}(\widehat{\bm\theta}_y-\bm\theta_{y0})
   +O_{\P_n}(n^{-3/5})\}.
\]
Since $R_n=\sum_{i=1}^n (X_i-\widehat\mu_{ix})\widehat\mu_{iy}$, the first-order
outcome fit decomposes it as
\[
\begin{aligned}
 R_n
 &=q\sum_{i=1}^n f(U_i)(X_i-\widehat\mu_{ix})
   +q(\widehat{\bm\theta}_y-\bm\theta_{y0})^{\T}
     \sum_{i=1}^n f(U_i)\bm Z_i(X_i-\widehat\mu_{ix})
   +O_{\P_n}\left(qn^{-3/5}\sum_{i=1}^n |X_i-\widehat\mu_{ix}|\right)\\
 &=
   q\sum_{i=1}^n f(U_i)(X_i-\widehat\mu_{ix})
   +q(\widehat{\bm\theta}_y-\bm\theta_{y0})^{\T}
     \sum_{i=1}^n f(U_i)\bm Z_i(X_i-\widehat\mu_{ix})
   +O_{\P_n}(n^{-2/5}),
\end{aligned}
\]
because $\sum_{i=1}^n |X_i-\widehat\mu_{ix}|=O_{\P_n}(nq)=O_{\P_n}(n^{3/5})$ and
$q n^{-3/5}=n^{-1}$; the bound on $\sum_{i=1}^n |X_i-\widehat\mu_{ix}|$ follows from
$\sum_{i=1}^n X_i=O_{\P_n}(nq)$ by Markov's inequality and
$\sum_{i=1}^n \widehat\mu_{ix}=O_{\P_n}(nq)$ by \eqref{eq:phase-fit-rates}.  The
logistic score equations annihilate the affine part of $f=\Pi f+r$:
\[
\begin{aligned}
 q\sum_{i=1}^n f(U_i)(X_i-\widehat\mu_{ix})
 &=q\sum_{i=1}^n \{\Pi f(U_i)+r(U_i)\}(X_i-\widehat\mu_{ix})\\
 &\overset{(a)}{=}
   q\sum_{i=1}^n r(U_i)(X_i-\widehat\mu_{ix})\\
 &=q\sum_{i=1}^n r(U_i)(X_i-q)
   -q\sum_{i=1}^n r(U_i)(\widehat\mu_{ix}-q).
\end{aligned}
\]
where in (a), the logistic score equation
$\sum_{i=1}^n \bm Z_i(X_i-\widehat\mu_{ix})=0$ annihilates the affine function $\Pi f$.
The second term in the last line is negligible at the $n^{-1/10}$ scale:
\[
\begin{aligned}
 q\sum_{i=1}^n r(U_i)(\widehat\mu_{ix}-q)
 &=q\sum_{i=1}^n r(U_i)
   \{q\bm Z_i^{\T}(\widehat{\bm\theta}_x-\bm\theta_{x0})+O_{\P_n}(qn^{-3/5})\}\\
 &=q^2(\widehat{\bm\theta}_x-\bm\theta_{x0})^{\T}
   \sum_{i=1}^n r(U_i)\bm Z_i
   +O_{\P_n}(q^2n^{2/5})\\
 &\overset{(a)}{=}
   q^2O_{\P_n}(n^{-3/10})O_{\P_n}(\sqrt n)+O_{\P_n}(n^{-2/5})
 \\&=O_{\P_n}(n^{-3/5})+O_{\P_n}(n^{-2/5})
 =O_{\P_n}(n^{-2/5}).
\end{aligned}
\]
where in (a), $\E\{r(U)\bm Z\}=0$ gives
$\sum_{i=1}^n r(U_i)\bm Z_i=O_{\P_n}(\sqrt n)$ by Chebyshev's inequality.
To see that the remaining outcome-fit term has the same smaller order,
first bound the vector that multiplies
$\widehat{\bm\theta}_y-\bm\theta_{y0}$:
\[
 \sum_{i=1}^n f(U_i)\bm Z_i(X_i-\widehat\mu_{ix})
 =\sum_{i=1}^n f(U_i)\bm Z_i(X_i-q)
   -\sum_{i=1}^n f(U_i)\bm Z_i(\widehat\mu_{ix}-q).
\]
The first term has norm $O_{\P_n}(\sqrt{nq})$, while the second has norm
$O_{\P_n}(nq\|\widehat{\bm\theta}_x-\bm\theta_{x0}\|_2)+O_{\P_n}(1)$.  Thus
\[
 \left\|\sum_{i=1}^n f(U_i)\bm Z_i(X_i-\widehat\mu_{ix})\right\|_2
 \le O_{\P_n}(\sqrt{nq})+O_{\P_n}(nq\|\widehat{\bm\theta}_x-\bm\theta_{x0}\|_2)+O_{\P_n}(1)
 =O_{\P_n}(n^{3/10}).
\]
By the Cauchy--Schwarz inequality and
$\|\widehat{\bm\theta}_y-\bm\theta_{y0}\|_2=O_{\P_n}(n^{-3/10})$, the corresponding term
in the expansion of $R_n$ satisfies
\[
 \left|q(\widehat{\bm\theta}_y-\bm\theta_{y0})^{\T}
 \sum_{i=1}^n f(U_i)\bm Z_i(X_i-\widehat\mu_{ix})\right|
 \le q\|\widehat{\bm\theta}_y-\bm\theta_{y0}\|_2
 \left\|\sum_{i=1}^n f(U_i)\bm Z_i(X_i-\widehat\mu_{ix})\right\|_2
 =O_{\P_n}(n^{-2/5}).
\]
Finally,
\[
 \frac1{nq^3}
 \Var\left(q\sum_{i=1}^n r(U_i)(X_i-q)\ \middle|\ U_{1:n}\right)
 =(1-q)\frac1n\sum_{i=1}^n r(U_i)^2
 \convp\E r(U)^2,
\]
where the displayed convergence in probability is the law of large numbers.
Since
$nq^3=n^{-1/5}$, the displayed remainders are $o_{\P_n}(n^{-1/10})$, and
\begin{equation}\label{eq:phase-observed-expansion}
 R_n=q\sum_{i=1}^n r(U_i)(X_i-q)+o_{\P_n}(n^{-1/10}).
\end{equation}
The summands in the leading term are independent and bounded by a constant
times $q$, and
\[
 \max_{1\le i\le n}
 \frac{|q r(U_i)(X_i-q)|}{\{nq^3\E r(U)^2\}^{1/2}}
 =O(n^{-3/10}).
\]
Thus Lindeberg's theorem gives
\begin{equation}\label{eq:phase-observed-clt}
 \frac{R_n}{\{nq^3\E r(U)^2\}^{1/2}}\convd G_2,
 \qquad nq^3=n^{-1/5}.
\end{equation}

We next turn to the lattice coordinate $L_n$.  Its conditional mean and
central limit theorem will determine the size of the boundary atom
$\widetilde \P(L_n^*=L_n)$ and the smooth midpoint tail.  Write
\[
 I_i\equiv\indicator(Y_i=1),\qquad
 \lambda_n\equiv\sum_{i:Y_i=1}\widehat\mu_{ix}.
\]
The mean count is localized by
\[
\begin{aligned}
 \P(I_i=1\mid U_i)
 &=qf(U_i)e^{-qf(U_i)}
 =qf(U_i)+O(q^2),\\
 \E\left(q\sum_{i=1}^n I_i\ \middle|\ U_{1:n}\right)
 &=q^2\sum_{i=1}^n f(U_i)+O(nq^3)\\
 &=n^{1/5}\E f(U)+q^2\sum_{i=1}^n \{f(U_i)-\E f(U)\}+O(n^{-1/5})
 =n^{1/5}\E f(U)+O_{\P_n}(n^{-1/5}),\\
 \Var\left(q\sum_{i=1}^n I_i\ \middle|\ U_{1:n}\right)
 &\le Cnq^3=O(n^{-1/5}).
\end{aligned}
\]
where $n^{-1}\sum_{i=1}^n f(U_i)$ converges to $\E f(U)$ in probability by
the law of large numbers, and the final stochastic fluctuation follows from
Chebyshev's inequality applied conditionally on $U_{1:n}$.  Thus
$q\sum_{i=1}^n I_i=n^{1/5}\E f(U)+O_{\P_n}(n^{-1/10})$.  The fitted mean $\lambda_n$
differs from this count mean only by the fitted logistic perturbation:
\[
\begin{aligned}
 \lambda_n-q\sum_{i=1}^n I_i
 &=\sum_{i=1}^n I_i(\widehat\mu_{ix}-q)
 =\sum_{i=1}^n I_i
   \{q\bm Z_i^{\T}(\widehat{\bm\theta}_x-\bm\theta_{x0})+O_{\P_n}(qn^{-3/5})\}\\
 &=q(\widehat{\bm\theta}_x-\bm\theta_{x0})^{\T}\sum_{i=1}^n I_i\bm Z_i
   +O_{\P_n}(qn^{-3/5}\sum_{i=1}^n I_i)\\
 &=n^{-2/5}O_{\P_n}(n^{-3/10})O_{\P_n}(n^{3/5})+O_{\P_n}(n^{-2/5})
 =O_{\P_n}(n^{-1/10}),
\end{aligned}
\]
because $\sum_{i=1}^n I_i\bm Z_i=O_{\P_n}(nq)=O_{\P_n}(n^{3/5})$ by Markov's inequality.

The count coordinate must be standardized carefully: $\lambda_n$ depends
on $X$ through the logistic fit, so it is not $(Y,U)$-measurable,
and a conditional CLT given $(Y,U)$ may only use $(Y,U)$-measurable
centerings and scalings.  We therefore standardize by
\[
 s_n^2\equiv q(1-q)\sum_{i=1}^n I_i,
\]
which is $(Y,U)$-measurable.  Conditional on $(Y,U)$, under the null the
variables $X_1,\ldots,X_n$ remain independent
$\operatorname{Bernoulli}(q)$, and the integer variable
$L_n=\sum_{i=1}^n X_iI_i$ satisfies
\[
 \E(L_n\mid Y,U)=q\sum_{i=1}^n I_i,
 \qquad
 \Var(L_n\mid Y,U)=s_n^2=n^{1/5}\E f(U)+o_{\P_n}(n^{1/5}).
\]
The summands are bounded and $s_n^2=\Theta_{\P_n}(n^{1/5})\to\infty$, so the
conditional Lindeberg condition is automatic and the conditional CLT
gives
\[
 \widetilde W_n\equiv\frac{L_n-q\sum_{i=1}^n I_i}{s_n}
 \convd G_1,
\]
where sampling-law convergence follows by taking expectations of the
bounded conditional characteristic functions.  The $X$-dependent standardization is now
reattached by Slutsky's theorem.  Since
$\lambda_n-q\sum_{i=1}^nI_i=O_{\P_n}(n^{-1/10})$,
$q^2\sum_{i=1}^nI_i=O_{\P_n}(n^{-1/5})$, and $s_n^2=\Theta_{\P_n}(n^{1/5})$,
\[
 \lambda_n-s_n^2
 =\left(\lambda_n-q\sum_{i=1}^nI_i\right)+q^2\sum_{i=1}^nI_i
 =O_{\P_n}(n^{-1/10}),
 \qquad
 \frac{s_n}{\sqrt{\lambda_n}}=1+O_{\P_n}(n^{-3/10}),
\]
and therefore
\begin{equation}\label{eq:phase-count-clt}
\begin{aligned}
 \lambda_n
 &=n^{1/5}\E f(U)+O_{\P_n}(n^{-1/10}),
 \qquad
 \frac{\lambda_n}{n^{1/5}}\convp\E f(U),\\
 W_n
 &\equiv\frac{L_n-\lambda_n}{\sqrt{\lambda_n}}
 =\widetilde W_n+o_{\P_n}(1)\convd G_1.
\end{aligned}
\end{equation}
To combine the two marginal CLTs, condition on $(Y,U)$.  Both
$\widetilde W_n$ and the leading phase term
$q\sum_{i=1}^n r(U_i)(X_i-q)$ are then sums of independent centered
variables with $(Y,U)$-measurable coefficients.  Their conditional
covariance satisfies
\[
 \left|\Cov\left(
 \frac{q\sum_{i=1}^n r(U_i)(X_i-q)}{\{nq^3\E r(U)^2\}^{1/2}},
 \widetilde W_n
 \ \middle|\ Y,U\right)\right|
 \le
 \frac{Cq^2\sum_{i=1}^n I_i}
      {\{nq^3\E r(U)^2\}^{1/2}s_n}.
\]
The numerator has the explicit order
\[
 q^2\sum_{i=1}^n I_i
 =O_{\P_n}(q^2nq)
 =O_{\P_n}(nq^3)
 =O_{\P_n}(n^{-1/5}).
\]
For the denominator, $nq^3=n^{-1/5}$ and
$s_n^2=n^{1/5}\E f(U)+o_{\P_n}(n^{1/5})$ by the variance calculation above.
Therefore
\[
 \{nq^3\E r(U)^2\}^{1/2}s_n
 =\{nq^3\E r(U)^2s_n^2\}^{1/2}
 \convp
 \{\E r(U)^2\E f(U)\}^{1/2}>0.
\]
Combining the numerator and denominator bounds gives
\[
 \left|\Cov\left(
 \frac{q\sum_{i=1}^n r(U_i)(X_i-q)}{\{nq^3\E r(U)^2\}^{1/2}},
 \widetilde W_n
 \ \middle|\ Y,U\right)\right|
 =O_{\P_n}(n^{-1/5})\convp0.
\]
Hence the two limiting Gaussian coordinates are independent.  The bounded
summands satisfy the conditional bivariate Lindeberg condition by the same
estimates used for the two marginal CLTs.  The conditional
Lindeberg--Feller theorem and the Cram\'er--Wold device therefore give a
standard bivariate normal conditional limit.  Taking expectations of the
bounded conditional characteristic functions transfers this limit to the
sampling law.

It remains to replace the two leading coordinates by $W_n$ and $R_n$.
Equations~\eqref{eq:phase-count-clt} and
\eqref{eq:phase-observed-expansion}, together with
$\{nq^3\E r(U)^2\}^{1/2}\asymp n^{-1/10}$, give the explicit bounds
\[
 W_n-\widetilde W_n=o_{\P_n}(1),\qquad
 \frac{R_n-q\sum_{i=1}^n r(U_i)(X_i-q)}
 {\{nq^3\E r(U)^2\}^{1/2}}=o_{\P_n}(1).
\]
Slutsky's theorem now gives
\begin{equation}\label{eq:phase-joint-clt}
 \left(
 W_n,\frac{R_n}{\{nq^3\E r(U)^2\}^{1/2}}
 \right)\convd (G_1,G_2).
\end{equation}
Thus the preceding analysis has produced the two coordinates needed for the
comparison.
The count coordinate $W_n$ locates the observed lattice cutoff, while the
phase coordinate $R_n$ will determine how much of the equality atom enters
the exact tail.  We next convert these two coordinates into an
expansion of $p_{\infty,n}^{\dCRT}$.

\paragraph{Step 2: Obtain the expansion of the exact dCRT tail.}
Under the conditional law $\widetilde \P$, the count $L_n^*$ is integer-valued and
$L_n$ is the observed cutoff for its upper tail.  By the boundary atom
we mean the entire conditional probability mass at this cutoff,
\[
 \widetilde \P(L_n^*=L_n\mid X,Y,U).
\]
Thus the boundary atom is an atom of the lattice coordinate $L_n^*$, not
an atom of the complete fitted statistic in \eqref{eq:phase-tail-event}.
The event comparison below shows that the fitted correction cannot change
the tail decision when $L_n^*\ne L_n$, except on an event of conditional
probability $O_{\P_n}(n^{-1/5})$, whereas on $L_n^*=L_n$ the decision is determined
by $\Delta_n^*\le \Delta_n$.
Consequently, the relevant fraction of the boundary atom is
\[
 \widetilde \P(\Delta_n^*\le \Delta_n\mid L_n^*=L_n,X,Y,U).
\]
The Poisson approximation at the end of this step shows in
\eqref{eq:phase-local-mass} that the boundary atom itself has order
$\lambda_n^{-1/2}=O_{\P_n}(n^{-1/10})$.

We return to the resampling law and study $\Delta_n^*$.  Its conditional variance
is the scale on which $R_n$ randomizes the displayed fraction of the boundary
atom.
Conditionally on the data,
\begin{equation}\label{eq:phase-bootstrap-variance}
 \frac{v_n}{nq^3}\equiv\frac{\widetilde\Var(\Delta_n^*)}{nq^3}
 =\frac{1}{nq^3}\sum_{i=1}^n \widehat\mu_{iy}^2
   \widehat\mu_{ix}(1-\widehat\mu_{ix})
 \convp\E f(U)^2.
\end{equation}
The convergence follows by inserting the uniform fit expansions:
$\widehat\mu_{iy}^2\widehat\mu_{ix}(1-\widehat\mu_{ix})
=q^3f(U_i)^2\{1+o_{\P_n}(1)\}$ uniformly over $i$, and the empirical average of
$f(U_i)^2$ converges by the law of large numbers because $U$ has fixed finite
support.  The total conditional third absolute moment is $O_{\P_n}(nq^4)$, so the
Berry--Esseen ratio for $\Delta_n^*$ is $O_{\P_n}(nq^4/v_n^{3/2})=O_{\P_n}(n^{-3/10})$.

We will use this normal approximation after conditioning on the boundary
event $L_n^*=L_n$.  The part of $\Delta_n^*$ coming from the boundary indices
has negligible size on the scale $\sqrt{v_n}$.  Decompose
\[
 \Delta_n^*=\sum_{i:Y_i=1}X_i^*\widehat\mu_{iy}
      +\sum_{i:Y_i\ne1}X_i^*\widehat\mu_{iy}.
\]
Indices with $Y_i=1$ contribute at most
$\sum_{i:Y_i=1}Cq^3=O_{\P_n}(n^{-3/5})$ to
\eqref{eq:phase-bootstrap-variance}, whereas $v_n=\Theta_{\P_n}(n^{-1/5})$.  The
second term is independent of $L_n^*$, and its variance differs from $v_n$
by $O_{\P_n}(n^{-3/5})$. Its third absolute moment sum is bounded by
$O_{\P_n}(nq^4)$. Berry--Esseen and the variance comparison therefore give
\[
 \sup_x\left|\widetilde\P\left\{
 \frac{\sum_{i:Y_i\ne1}\widehat\mu_{iy}(X_i^*-\widehat\mu_{ix})}
      {\sqrt{v_n}}\le x\right\}-\Phi(x)\right|
 =O_{\P_n}(n^{-3/10}+n^{-2/5}).
\]
By \eqref{eq:phase-count-clt}, $L_n=\lambda_n+O_{\P_n}(\sqrt{\lambda_n})
=O_{\P_n}(n^{1/5})$. Conditioning on $L_n^*=L_n$ gives
\[
 0\le\sum_{i:Y_i=1}X_i^*\widehat\mu_{iy}
 \le L_n\max_i\widehat\mu_{iy}=O_{\P_n}(n^{-1/5})
 =o_{\P_n}(\sqrt{v_n}),
\]
Its resampling mean is at most $Cq\lambda_n=O_{\P_n}(n^{-1/5})$.
The conditional boundary event has positive probability: on the fit
event, all fitted Bernoulli probabilities are in $(0,1)$ and
$0\le L_n\le\sum_i I_i$.
Independence of the two index groups and boundedness of the normal density
now give the explicit conditional approximation
\begin{equation}\label{eq:phase-boundary-normal-error}
 \begin{aligned}
 \sup_x\left|\widetilde \P\left\{
 \frac{\Delta_n^*-\widetilde\E\Delta_n^*}{\sqrt{v_n}}\le x
 \ \middle|\ L_n^*=L_n,X,Y,U\right\}-\Phi(x)\right|
 &\le O_{\P_n}(n^{-3/10}+n^{-2/5})
       +\frac{Cq(L_n+\lambda_n)}{\sqrt{v_n}}\\
 &=O_{\P_n}(n^{-1/10})=o_{\P_n}(1).
 \end{aligned}
\end{equation}
Substituting $\Delta_n=\widetilde\E\Delta_n^*+R_n$ into
\eqref{eq:phase-boundary-normal-error} gives the boundary fraction
\begin{align}
 \theta_n
 &\equiv\widetilde \P(\Delta_n^*\le \Delta_n\mid L_n^*=L_n,X,Y,U)\nonumber\\
 &=\Phi\left(\frac{R_n}{\sqrt{v_n}}\right)+o_{\P_n}(1)
 =\Phi\left(
   \tau\frac{R_n}{\{nq^3\E r(U)^2\}^{1/2}}
  \right)+o_{\P_n}(1),
 \label{eq:phase-boundary-fraction}
\end{align}
where the final equality uses \eqref{eq:phase-bootstrap-variance} and
tightness in \eqref{eq:phase-observed-clt}. Combining
\eqref{eq:phase-boundary-fraction} with \eqref{eq:phase-joint-clt} gives
\[
 \theta_n\convd \Phi(\tau G_2).
\]
Since $\tau>0$, $\Phi(\tau G_2)$ is not the constant $1/2$; indeed it equals
$1/2$ only on the event $G_2=0$, which has probability zero.  The fitted
correction therefore selects a random asymptotic fraction of the equality
atom rather than the midpoint fraction $1/2$.

We translate this boundary fraction back to the original tail event.
The goal is to show that, away from $L_n^*=L_n$, the fitted correction is too
small to change the sign of the lattice difference; on the equality layer,
the sign is exactly the event $\Delta_n^*\le \Delta_n$.
Uniformly in $u$, $\P(Y_i\ge2\mid U_i=u)\le Cq^2$, so
$\E\,\#\{i:Y_i\ge2\}\le Cnq^2$ and, by Markov's inequality,
$\#\{i:Y_i\ge2\}=O_{\P_n}(nq^2)$.  For the observed side, $X\perp Y\mid U$
with $\E(X_i\mid U_i)=q$ gives
\[
 \E\sum_{i=1}^n X_i\indicator(Y_i\ge2)\le Cnq^3=O(n^{-1/5}),
\]
so Markov's inequality yields a sampling event of probability
$1-O(n^{-1/5})$ on which no observed selected index has $Y_i\ge2$.  For
the resampled side, an unconditional expectation bound is not available,
because $\widehat\mu_{ix}$ is controlled only on the fit event of
\eqref{eq:phase-fit-rates}; we argue on that event instead.  On the
event of \eqref{eq:phase-fit-rates}, whose probability tends to one,
$\max_i\widehat\mu_{ix}\le Cq$, so
\[
 \widetilde\E\sum_{i=1}^n X_i^*\indicator(Y_i\ge2)
 =\sum_{i=1}^n \widehat\mu_{ix}\indicator(Y_i\ge2)
 \le Cq\,\#\{i:Y_i\ge2\}
 =O_{\P_n}(nq^3)=O_{\P_n}(n^{-1/5}),
\]
and conditional Markov's inequality yields a conditional event of
probability $1-O_{\P_n}(n^{-1/5})$ on which no resampled selected index has
$Y_i\ge2$.  On the intersection of those two good events,
\[
 \sum_{i=1}^n X_i^*(Y_i-\widehat\mu_{iy})
   -\sum_{i=1}^n X_i(Y_i-\widehat\mu_{iy})
 =(L_n^*-\Delta_n^*)-(L_n-\Delta_n)=(L_n^*-L_n)-(\Delta_n^*-\Delta_n).
\]
The fitted correction cannot cross an adjacent integer except on a small
conditional-probability event:
\[
 \widetilde \P(|\Delta_n^*-\Delta_n|>1/2\mid X,Y,U)
 \le 4\,\widetilde\E\{(\Delta_n^*-\Delta_n)^2\mid X,Y,U\}=4\{v_n+R_n^2\}
 \overset{(a)}{=}O_{\P_n}(n^{-1/5}).
\]
where in (a), \eqref{eq:phase-bootstrap-variance} gives
$v_n=O_{\P_n}(n^{-1/5})$ and \eqref{eq:phase-observed-expansion} gives
$R_n^2=O_{\P_n}(n^{-1/5})$.
On the complement of this event,
\[
\begin{array}{lll}
 L_n^*>L_n
 &\Longrightarrow&
 (L_n^*-L_n)-(\Delta_n^*-\Delta_n)\ge 1-\frac12>0,\\[2mm]
 L_n^*<L_n
 &\Longrightarrow&
 (L_n^*-L_n)-(\Delta_n^*-\Delta_n)\le -1+\frac12<0,\\[2mm]
 L_n^*=L_n
 &\Longrightarrow&
 \eqref{eq:phase-tail-event}\ \text{holds exactly when }\Delta_n^*\le \Delta_n .
\end{array}
\]
Thus, except for a conditional event of probability $O_{\P_n}(n^{-1/5})$, the
indicator of the exact tail event is
\[
 \indicator\{(L_n^*-L_n)-(\Delta_n^*-\Delta_n)\ge0\}
 =\indicator\{L_n^*>L_n\}
  +\indicator\{L_n^*=L_n\}\indicator\{\Delta_n^*\le \Delta_n\}.
\]
Taking $\widetilde \P$ and using the definition of $\theta_n$ gives
\begin{equation}\label{eq:phase-exact-decomp}
\begin{aligned}
 p_{\infty,n}^{\dCRT}
 &=\widetilde \P(L_n^*>L_n)
   +\widetilde \P(\Delta_n^*\le \Delta_n,\ L_n^*=L_n)
   +O_{\P_n}(n^{-1/5})\\
 &=\widetilde \P(L_n^*>L_n)
   +\theta_n\widetilde \P(L_n^*=L_n)+O_{\P_n}(n^{-1/5}).
\end{aligned}
\end{equation}
The remainder in \eqref{eq:phase-exact-decomp} collects exactly two
exceptional cases.  The Markov bounds
\[
 \E\sum_{i=1}^nX_i\indicator(Y_i\ge2)=O(n^{-1/5}),
 \qquad
 \widetilde\E\sum_{i=1}^nX_i^*\indicator(Y_i\ge2)=O_{\P_n}(n^{-1/5})
 \ \text{on the fit event of \eqref{eq:phase-fit-rates}},
\]
control selected observed or resampled indices with $Y_i\ge2$, while
\[
 \widetilde \P(|\Delta_n^*-\Delta_n|>1/2\mid X,Y,U)=O_{\P_n}(n^{-1/5})
\]
controls a fitted correction large enough to cross an adjacent integer
lattice point.

The remaining task is to approximate the two purely count-based terms in
\eqref{eq:phase-exact-decomp}.  This is where the usual midpoint treatment of
the boundary atom enters.
Given the data, $L_n^*$ is Poisson--binomial with mean $\lambda_n$.  Lemma
\ref{lem:le-cam-elementary} compares its conditional law with the Poisson law
of mean $\lambda_n$.  More precisely, define
\[
 \begin{aligned}
 d_{\mathrm{TV}}
 &\equiv
 \sup_{B\subseteq\mathbb Z_+}
 \left|\widetilde \P(L_n^*\in B)-\operatorname{Pois}(\lambda_n)(B)\right|\\
 &=\frac12\sum_{\ell=0}^{\infty}
 \left|\widetilde \P(L_n^*=\ell)
       -e^{-\lambda_n}\frac{\lambda_n^\ell}{\ell!}\right|.
 \end{aligned}
\]
On the fit event of \eqref{eq:phase-fit-rates}, whose probability tends
to one, $\max_i\widehat\mu_{ix}\le2q\le1/2$ for all large $n$, so the
hypothesis of Lemma \ref{lem:le-cam-elementary} holds and the lemma
controls this distance by
\[
 d_{\mathrm{TV}}
 \le 2\sum_{i:Y_i=1}\widehat\mu_{ix}^2
 \le 2\max_i\widehat\mu_{ix}\sum_{i:Y_i=1}\widehat\mu_{ix}
 =2q\{1+o_{\P_n}(1)\}\lambda_n
 =O_{\P_n}(n^{-1/5}).
\]
The equality with $2q\{1+o_{\P_n}(1)\}\lambda_n$ uses
\eqref{eq:phase-fit-rates}, and the last equality uses $q=n^{-2/5}$ and
$\lambda_n=O_{\P_n}(n^{1/5})$.  Applying Lemma
\ref{lem:poisson-midtail-lr} to that Poisson variable gives the following
bounds uniformly on $|W_n|\le M$.  This substitution is valid for the
random mean $\lambda_n$ because the lemma's remainders are deterministic
and uniform in $\lambda$, while
$\lambda_n/n^{1/5}\convp\E f(U)>0$ by \eqref{eq:phase-count-clt}:
\begin{align}
 \widetilde \P(L_n^*=L_n)
 &=\frac{\phi(W_n)}{\sqrt{\lambda_n}}+O_{\P_n}(n^{-1/5}),
 \label{eq:phase-local-mass}\\
 Q_n\equiv\widetilde \P(L_n^*>L_n)+\frac12\widetilde \P(L_n^*=L_n)
 &=\Phibar(W_n)
 +\frac{\phi(W_n)(W_n^2-1)}{6\sqrt{\lambda_n}}
 +O_{\P_n}(n^{-1/5}).
 \label{eq:phase-midtail}
\end{align}
The exact decomposition \eqref{eq:phase-exact-decomp} can now be written as
\begin{equation}\label{eq:phase-dcrt-expansion}
\begin{aligned}
 p_{\infty,n}^{\dCRT}
 &=Q_n+
   \left(\theta_n-\frac12\right)\widetilde \P(L_n^*=L_n)
   +O_{\P_n}(n^{-1/5})\\
 &=Q_n+
   \frac{(\theta_n-\frac12)\phi(W_n)}{\sqrt{\lambda_n}}
   +o_{\P_n}(n^{-1/10}).
\end{aligned}
\end{equation}
Here the second line uses \eqref{eq:phase-local-mass},
$0\le\theta_n\le1$, and $n^{-1/5}=o(n^{-1/10})$.  This is the required
expansion of the exact dCRT tail: relative to the midpoint tail $Q_n$, it
contains the random fraction $\theta_n-1/2$ of the boundary atom.  Together
with \eqref{eq:phase-midtail}, it gives, on every fixed compact $W_n$ event,
\[
 p_{\infty,n}^{\dCRT}-\Phibar(W_n)=O_{\P_n}(n^{-1/10})=o_{\P_n}(1).
\]
The truncation error is explicit: by \eqref{eq:phase-count-clt},
\[
 \limsup_{n\to\infty}\P(|W_n|>M)\le\P(|G_1|\ge M)\longrightarrow0
 \qquad(M\to\infty).
\]
Therefore the compact relation extends to
\[
 p_{\infty,n}^{\dCRT}=\Phibar(W_n)+o_{\P_n}(1)
 \convd\Phibar(G_1).
\]

\paragraph{Step 3: Obtain the expansion of the LR formula.}
We compare the LR formula with $Q_n$ using
Lemma~\ref{lem:poisson-like-lr}.  Let $K_n^{\rm av}$ be the average CGF in
\eqref{eq:bnb-cgf}, and work with the sum-scale pair
\[
 K_n\equiv nK_n^{\rm av},\qquad
 t_n\equiv nw_n
 =\sum_{i=1}^n(X_i-\widehat\mu_{ix})(Y_i-\widehat\mu_{iy}).
\]
Here $K_n$ is the conditional CGF of
$\sum_i(X_i^*-\widehat\mu_{ix})(Y_i-\widehat\mu_{iy})$.
The saddlepoint equation is
$K_n'(\widehat s_n)=t_n$, and the corresponding $r_n,u_n$ are those in
\eqref{eq:r-u}.  Thus Lemma~\ref{lem:poisson-like-lr} applies once its
cumulant conditions are verified.

The first two required cumulants are exactly
\begin{equation}\label{eq:phase-cumulant-identities}
 \begin{aligned}
 K_n''(0)&=\sum_i\widehat\mu_{ix}(1-\widehat\mu_{ix})
                      (Y_i-\widehat\mu_{iy})^2,\\
 K_n'''(0)&=\sum_i\widehat\mu_{ix}(1-\widehat\mu_{ix})
                    (1-2\widehat\mu_{ix})(Y_i-\widehat\mu_{iy})^3.
 \end{aligned}
\end{equation}
For the Poisson responses, $\P(\max_iY_i>2)\le Cnq^3=O(n^{-1/5})$.
Work on $\max_iY_i\le2$ and the fit event of
\eqref{eq:phase-fit-rates}. Then $|Y_i-\widehat\mu_{iy}|\le3$ for large $n$.
On $|s|\le C_M\lambda_n^{-1/2}$, the tilted Bernoulli probabilities
are at most $C_M\widehat\mu_{ix}$. Their cumulant polynomials give,
for $m=2,\ldots,5$,
\[
 \left|\frac{d^m}{ds^m}
 \log\{1-\widehat\mu_{ix}
       +\widehat\mu_{ix}e^{s(Y_i-\widehat\mu_{iy})}\}\right|
 \le C_M\widehat\mu_{ix}|Y_i-\widehat\mu_{iy}|^m .
\]
The three outcome layers satisfy, for $m=2,\ldots,5$,
\begin{equation}\label{eq:phase-layer-moment-bounds}
\begin{aligned}
 \sum_{i:Y_i=1}\widehat\mu_{ix}(1-\widehat\mu_{iy})^m
 &=\lambda_n+O_{\P_n}(q\lambda_n)
   =\lambda_n+O_{\P_n}(n^{-1/5}),\\
 \sum_{i:Y_i=0}\widehat\mu_{ix}|\widehat\mu_{iy}|^m
 &\le Cnq^{m+1}=O_{\P_n}(n^{-1/5})\qquad(m\ge2),\\
 \sum_{i:Y_i\ge2}\widehat\mu_{ix}|Y_i-\widehat\mu_{iy}|^m
 &=O_{\P_n}(nq^3)=O_{\P_n}(n^{-1/5}),
\end{aligned}
\end{equation}
The last bound uses $\#\{i:Y_i\ge2\}=O_{\P_n}(nq^2)$.
Since $\max_i\widehat\mu_{ix}=O_{\P_n}(q)$, the factors
$1-\widehat\mu_{ix}$ and $1-2\widehat\mu_{ix}$ in
\eqref{eq:phase-cumulant-identities} change the leading layer by at most
$O_{\P_n}(q\lambda_n)=O_{\P_n}(n^{-1/5})$.
Thus \eqref{eq:phase-layer-moment-bounds} gives
\begin{equation}\label{eq:phase-lr-cumulant-bounds}
 \begin{aligned}
 K_n''(0)&=\lambda_n+O_{\P_n}(n^{-1/5}),\qquad
 K_n'''(0)=\lambda_n+O_{\P_n}(n^{-1/5}),\\
 \sup_{|s|\le C_M\lambda_n^{-1/2}}|K_n^{(m)}(s)|
 &=O_{\P_n}(\lambda_n),\qquad m=4,5.
 \end{aligned}
\end{equation}

The centered observed cutoff for this CGF is obtained from the same layer
decomposition:
\[
 \sum_{i=1}^n (X_i-\widehat\mu_{ix})(Y_i-\widehat\mu_{iy})
 =\sum_{i=1}^n (X_i-\widehat\mu_{ix})Y_i
   -\sum_{i=1}^n (X_i-\widehat\mu_{ix})\widehat\mu_{iy}.
\]
The first term differs from $L_n-\lambda_n$ only through indices with
$Y_i\ge2$, and the second term is $R_n$.
On $\max_iY_i\le2$, the remaining contribution is bounded by
\[
 \left|\sum_{i:Y_i\ge2}(X_i-\widehat\mu_{ix})Y_i\right|
 \le2\sum_iX_i\indicator(Y_i\ge2)+Cq\,\#\{i:Y_i\ge2\}
 =O_{\P_n}(n^{-1/5}).
\]
The first term is zero outside an event of probability $O(n^{-1/5})$,
as proved in Step 2; the second has the displayed stochastic order.
Using \eqref{eq:phase-observed-clt} and \eqref{eq:phase-count-clt}, we obtain
\begin{equation}\label{eq:phase-lr-cutoff-bound}
 t_n=(L_n-\lambda_n)-R_n+O_{\P_n}(n^{-1/5})
 =W_n\sqrt{\lambda_n}+O_{\P_n}(\lambda_n^{-1/2}).
\end{equation}
Equations \eqref{eq:phase-lr-cumulant-bounds} and
\eqref{eq:phase-lr-cutoff-bound}, with
$n^{-1/5}=\Theta_{\P_n}(\lambda_n^{-1})$, verify
\eqref{eq:poisson-like-cumulant-assumptions}--\eqref{eq:poisson-like-cutoff-assumption}
for $\lambda=\lambda_n$ and $z=W_n$ on $|W_n|\le M$.
Lemma~\ref{lem:poisson-like-lr} therefore constructs an admissible
saddlepoint and gives, including the continuous value at $t_n=0$,
\[
 p_{\mathrm{SPA},n}^{\dCRT}
 =\Phibar(W_n)+\frac{\phi(W_n)(W_n^2-1)}{6\sqrt{\lambda_n}}
   +O_{\P_n}(n^{-1/5}).
\]
Comparison with \eqref{eq:phase-midtail} proves, on $|W_n|\le M$,
\begin{equation}\label{eq:phase-saddlepoint-midtail}
 p_{\mathrm{SPA},n}^{\dCRT}=Q_n+O_{\P_n}(n^{-1/5}).
\end{equation}
Thus LR has the leading midpoint tail but no first-order random
boundary-fraction term.

\paragraph{Step 4: Subtract the expansions to obtain the limiting error.}
The midpoint terms cancel, leaving the random boundary fraction.
Subtracting \eqref{eq:phase-dcrt-expansion} from
\eqref{eq:phase-saddlepoint-midtail} gives, on every fixed $|W_n|\le M$,
\[
 p_{\mathrm{SPA},n}^{\dCRT}-p_{\infty,n}^{\dCRT}
 =\frac{\{1/2-\theta_n\}\phi(W_n)}
 {\sqrt{\lambda_n}}
 +o_{\P_n}(n^{-1/10}).
\]
To remove this restriction at the required scale, split the error event
over $|W_n|\le M$ and its complement. For each $\epsilon>0$,
\[
 \begin{aligned}
 &\limsup_{n\to\infty}\P_n\!\left(
 \left|\sqrt{\lambda_n}
 (p_{\mathrm{SPA},n}^{\dCRT}-p_{\infty,n}^{\dCRT})
 -(1/2-\theta_n)\phi(W_n)\right|>\epsilon\right)\\
 &\qquad\le\limsup_{n\to\infty}\P_n(|W_n|>M)
 \le\P(|G_1|\ge M)\longrightarrow0\qquad(M\to\infty),
 \end{aligned}
\]
where \eqref{eq:phase-count-clt} gives both
$\sqrt{\lambda_n}=\Theta_{\P_n}(n^{1/10})$ and tightness of $W_n$.
Consequently, without a compact restriction,
\begin{equation}\label{eq:phase-global-tail-difference}
 \sqrt{\lambda_n}(p_{\mathrm{SPA},n}^{\dCRT}-p_{\infty,n}^{\dCRT})
 =(1/2-\theta_n)\phi(W_n)+o_{\P_n}(1).
\end{equation}
Finally, $p_{\infty,n}^{\dCRT}\convd\Phibar(G_1)>0$ almost surely,
and \eqref{eq:phase-global-tail-difference} gives
\[
 p_{\mathrm{SPA},n}^{\dCRT}-p_{\infty,n}^{\dCRT}=O_{\P_n}(n^{-1/10})=o_{\P_n}(1),
 \qquad
 p_{\mathrm{SPA},n}^{\dCRT}\convd\Phibar(G_1).
\]
In particular, $p_{\mathrm{SPA},n}^{\dCRT}>0$ with probability tending to
one and $1/p_{\mathrm{SPA},n}^{\dCRT}=O_{\P_n}(1)$.
Dividing \eqref{eq:phase-global-tail-difference} by this denominator gives
\[
 \sqrt{\lambda_n}\left(\frac{p_{\infty,n}^{\dCRT}}{p_{\mathrm{SPA},n}^{\dCRT}}-1\right)
 =-\frac{\{1/2-\theta_n\}\phi(W_n)}{p_{\mathrm{SPA},n}^{\dCRT}}
 +o_{\P_n}(1).
\]
More explicitly, \eqref{eq:phase-boundary-fraction},
\eqref{eq:phase-joint-clt}, and \eqref{eq:phase-global-tail-difference}
give the joint convergence
\[
 (W_n,\theta_n,p_{\mathrm{SPA},n}^{\dCRT})
 \convd
 \left(G_1,\Phi(\tau G_2),\Phibar(G_1)\right).
\]
Because $\Phibar(G_1)>0$ almost surely, the continuous mapping theorem
applied to the preceding ratio expansion gives
\[
 \sqrt{\lambda_n}\left(\frac{p_{\infty,n}^{\dCRT}}{p_{\mathrm{SPA},n}^{\dCRT}}-1\right)
 \convd
 -\frac{\{1/2-\Phi(\tau G_2)\}\phi(G_1)}
 {\Phibar(G_1)},
\]
Since $\lambda_n/n^{1/5}\convp\E f(U)>0$, another application of
Slutsky's theorem gives
\[
 n^{1/10}\left(\frac{p_{\infty,n}^{\dCRT}}{p_{\mathrm{SPA},n}^{\dCRT}}-1\right)
 \convd
 -\frac{\{1/2-\Phi(\tau G_2)\}\phi(G_1)}
 {\{\E f(U)\}^{1/2}\Phibar(G_1)},
\]
which is \eqref{eq:bnb-phase-limit}.  Since $\tau>0$, conditional on
$G_1$ the phase term is non-atomic with a nonzero coefficient, so the
limit is finite and nonzero almost surely.  The proof therefore ends with
the explicit order comparison
\[
 \left|\frac{p_{\infty,n}^{\dCRT}}{p_{\mathrm{SPA},n}^{\dCRT}}-1\right|
 =\Theta_{\P_n}(\lambda_n^{-1/2})
 =\Theta_{\P_n}(n^{-1/10})
 =\Theta_{\P_n}\{(n\pi_{n,x}\mu_{n,y})^{-1/2}\},
\]
which is \eqref{eq:bnb-phase-asymp-p}.
Thus the discrepancy is exactly of order $\lambda_n^{-1/2}$: the exact dCRT
uses the random fraction $\theta_n$ of the boundary atom, while the
ordinary LR formula uses its midpoint fraction at first order.
\end{proof}

\begin{remark}[Boundary phase]\label{rem:phase-scope}
Unlike the fixed lattice of the intercept-only GWAS witness, the
predictor-resampling witness has fitted weights that determine a random
fraction of a boundary atom. A tailored correction would have to account
for that fitted phase, not merely a fixed lattice span.
\end{remark}

\subsubsection{Completion of Theorem~\ref{thm:spacrt-rate}}

\begin{proof}[of Theorem~\ref{thm:spacrt-rate}]
Supplementary Section~\ref{sec:bnb-upper-validity} proves the upper rate uniformly over
$\PBNB_n$.  Proposition~\ref{prop:bnb-phase-obstruction} gives a null
Poisson-subcase sequence with $\eta_{n,y}=0$, and
\eqref{eq:bnb-phase-asymp-p} proves
\[
 \left|\frac{p_{\infty,n}^{\dCRT}}{p_{\mathrm{SPA},n}^{\dCRT}}-1\right|
 =\Theta_{\P_n}\{(n\pi_{n,x}\mu_{n,y})^{-1/2}\}.
\]
Corollary~\ref{cor:bnb-null-validity} supplies the uniform null validity of
the exact target and its SPA. Together with existence in the upper-rate
argument and this sharpness construction, it proves all conclusions.
\end{proof}

\subsection{A scale-adaptive validity theorem for the fitted dCRT target and the spaCRT}
\label{sec:proof-spacrt-validity}
\mbox{}\par

In this section, we will present a scale-adaptive theorem,
Theorem~\ref{thm:spacrt-validity}.  All oracle residuals, error
functionals, variance scales, and sigma-fields below are as defined in
Supplementary Section~\ref{sec:spacrt-notation}.

\subsubsection{Intrinsic conditions and the validity theorem}\label{sec:spacrt-validity-statement}

\begin{assumption}[Intrinsic spaCRT validity conditions]\label{assu:spacrt-regularity}
Under $H_0:X_i\indep Y_i\mid \bm Z_i$, the two nuisance fits use only $(\bm X,\bm Z)$ and $(\bm Y,\bm Z)$, respectively, and:
\begin{enumerate}[label=\textup{(\roman*)}]
\item $\P(\sigma_{n,\dCRT}>0)\to1$ and
\[
  \frac{\sqrt nE_{n,x}E_{n,y}}{\sigma_{n,\dCRT}}\convp0,
  \qquad
  \frac{E'_{n,x}+E'_{n,y}+\widehat E'_{n,y}}
       {\sigma_{n,\dCRT}}\convp0,
\]
with intrinsic fitted-variance calibration $\frac1n\sum_i(\widehat v_{x,i}-v^0_{x,i})v^0_{y,i}=o_{\P_n}(\sigma_{n,\dCRT}^2)$;
\item there exist finite nonnegative $\mathcal Z_n$-measurable scales
$b^0_{n,x}$ and $b^0_{n,y}$ and a finite nonnegative
$\mathcal X_n$-measurable scale $\widehat b_{n,x}$ such that, almost
surely, for every integer $k\geq3$ and every $1\leq i\leq n$, the true
residuals satisfy
\[
  \E(|e^0_{x,i}|^k\mid\rz{i})
  \leq\frac{k!}{2}(b^0_{n,x})^{k-2}v^0_{x,i},
  \qquad
  \E(|e^0_{y,i}|^k\mid\rz{i})
  \leq\frac{k!}{2}(b^0_{n,y})^{k-2}v^0_{y,i},
\]
the fitted predictor resampling residuals satisfy
\[
  \widetilde\E(|\widetilde X_i-\widehat\mu_{x,i}|^k\mid\mathcal F_n)
  \leq\frac{k!}{2}(\widehat b_{n,x})^{k-2}\widehat v_{x,i},
\]
and
\[
  \frac{b^0_{n,y}(b^0_{n,x}+\widehat b_{n,x})}
       {\sqrt n\,\sigma_{n,\dCRT}}\convp0;
\]
\item with $V_{n,\dCRT}^2=A_{2n}$, we assume $\P(V_{n,\dCRT}>0)\to1$ and $\frac{\widehat b_{n,x}e_n^\star}{\sqrt nV_{n,\dCRT}}\convp0$;
\item the nuisance fits and the conditional resampling-law construction
succeed with probability tending to one.
\end{enumerate}
\end{assumption}

\begin{theorem}[Scale-adaptive spaCRT validity]
\label{thm:spacrt-validity}
Under $H_0:X_i\indep Y_i\mid \bm Z_i$ and
Assumption~\ref{assu:spacrt-regularity},
\[
  p_{\infty,n}^{\dCRT}\convd\Unif(0,1).
\]
Moreover,
\[
  \frac{p_{\infty,n}^{\dCRT}}{p_{\mathrm{SPA},n}^{\dCRT}}\convp1,
  \qquad
  p_{\mathrm{SPA},n}^{\dCRT}\convd\Unif(0,1).
\]
\end{theorem}

All ratios involving $\sigma_{n,\dCRT}$ or $V_{n,\dCRT}$ are defined
arbitrarily on the events where their denominators vanish.  Those events
have probability tending to zero by
Assumption~\ref{assu:spacrt-regularity}.

\subsubsection{Preliminary moment and stability lemmas}

The supporting lemmas are stated at the point where they enter the argument; all of their proofs are deferred to Supplementary Section~\ref{sec:spacrt-supporting-proofs}. The first lemma records two elementary consequences of conditional Bernstein control used in the sampling and resampling arguments.

\begin{lemma}[Conditional Bernstein consequences]
\label{lem:spacrt-cb-consequences}
Let $\xi$ be conditionally centered given a $\sigma$-field
$\mathcal G$, with conditional variance $v$, and suppose that, for some
finite nonnegative $\mathcal G$-measurable $b$ and every integer
$k\geq3$,
\[
  \E(|\xi|^k\mid\mathcal G)
  \leq
  \frac{k!}{2}b^{k-2}v.
\]
Then
\begin{equation}
  \E(|\xi|^3\mid\mathcal G)\leq3bv,
  \qquad
  v^{1/2}\leq3b.
  \label{eq:spacrt-cb-third-variance}
\end{equation}
If $a_1,\ldots,a_n$ are $\mathcal G$-measurable and, conditionally on
$\mathcal G$, the variables $\xi_i$ are independent, centered, and
satisfy the preceding bound with variances $v_i$ and a common scale
$b_n$, then
$W_i\equiv a_i\xi_i$ satisfy the conditional Bernstein property (CB), defined
in~\eqref{eq:cb-property}, with
\begin{equation}
  \bar b_n=b_n\max_{1\leq i\leq n}|a_i|.
  \label{eq:spacrt-cb-multiplier-scale}
\end{equation}
\end{lemma}

Under the null, define the oracle dCRT score
\begin{equation}
  U_{n,\dCRT}
  \equiv
  \sum_{i=1}^n e^0_{x,i}e^0_{y,i}.
  \label{eq:spacrt-oracle-score}
\end{equation}
The next lemma transfers its intrinsic central limit theorem to the
observed score without requiring sample splitting.

\begin{lemma}[Oracle CLT and in-sample score transfer]
\label{lem:spacrt-oracle-score-transfer}
Under $H_0:\ex{}\independent\ey{}\mid\rz{}$ and
Assumption~\ref{assu:spacrt-regularity},
\begin{equation}
  \frac{U_{n,\dCRT}}{\sqrt n\sigma_{n,\dCRT}}
  \convd N(0,1),
  \qquad
  \frac{S_n-U_{n,\dCRT}}{\sqrt n\sigma_{n,\dCRT}}
  \convp0.
  \label{eq:spacrt-oracle-score-transfer}
\end{equation}
\end{lemma}

The fitted variance contains the observed squared response residuals.
The following lemma shows that the intrinsic moment and nuisance-rate
conditions are exactly sufficient to replace them by their conditional
variances.

\begin{lemma}[Intrinsic fitted-variance matching]
\label{lem:spacrt-variance-matching}
Under $H_0:\ex{}\independent\ey{}\mid\rz{}$, suppose that the two
nuisance fits are respectively $\mathcal X_n$- and
$\mathcal Y_n$-measurable, and that clauses~\textup{(i)}--\textup{(ii)}
of Assumption~\ref{assu:spacrt-regularity} hold.  Then
\begin{equation}
  \frac{V_{n,\dCRT}^2}{\sigma_{n,\dCRT}^2}
  \convp1.
  \label{eq:spacrt-variance-matching}
\end{equation}
\end{lemma}

Finally, the no-dominance condition yields a conditional central limit
theorem for the fitted resampling law itself.

\begin{lemma}[Conditional CB resampling CLT]
\label{lem:spacrt-resampling-clt}
Under Assumption~\ref{assu:spacrt-regularity}, conditionally on
$\mathcal F_n$ let
\[
  W_{in}
  \equiv
  \widehat e_{y,i}(\widetilde X_i-\widehat\mu_{x,i}).
\]
Then
\begin{equation}
  \sup_{t\in\R}
  \left|
    \widetilde \P\left(
      \left.
      \frac{\sum_{i=1}^nW_{in}}{\sqrt nV_{n,\dCRT}}
      \leq t
      \ \right|\ \mathcal F_n
    \right)
    -\Phi(t)
  \right|
  \convp0.
  \label{eq:spacrt-conditional-resampling-clt}
\end{equation}
The same convergence holds for the corresponding conditional survival
functions with the weak inequality in the upper tail.
\end{lemma}

\subsubsection{Proof of the first conclusion of Theorem~\ref{thm:spacrt-validity}}

\begin{proof}[of the first conclusion of
Theorem~\ref{thm:spacrt-validity}]
Lemmas~\ref{lem:spacrt-oracle-score-transfer} and
\ref{lem:spacrt-variance-matching}, together with Slutsky's theorem,
give
\begin{equation}
  \frac{S_n}{\sqrt nV_{n,\dCRT}}
  \convd N(0,1).
  \label{eq:spacrt-feasible-score-clt}
\end{equation}
In particular, $|S_n|/(\sqrt nV_{n,\dCRT})=O_{\P_n}(1)$ under the null.

Let
\[
  \overline F_n^*(t)
  \equiv
  \widetilde \P\left(
    \left.
    \frac{\sum_{i=1}^nW_{in}}{\sqrt nV_{n,\dCRT}}
    \geq t
    \ \right|\ \mathcal F_n
  \right).
\]
Lemma~\ref{lem:spacrt-resampling-clt} implies
\[
  \sup_{t\in\R}
  |\overline F_n^*(t)-\{1-\Phi(t)\}|
  \convp0.
\]
On the successful-construction event, whose probability tends to one by
Assumption~\ref{assu:spacrt-regularity}(iv), the conditional-tail
definition~\eqref{eq:infinite-M} with the
dCRT summands~\eqref{eq:dcrt-resampling-summand} gives
\[
  p_{\infty,n}^{\dCRT}
  =
  \overline F_n^*\left(\frac{S_n}{\sqrt nV_{n,\dCRT}}\right).
\]
The failure convention therefore does not affect the following
approximation in probability:
\begin{equation*}
  p_{\infty,n}^{\dCRT}
  =
  1-\Phi\left(\frac{S_n}{\sqrt nV_{n,\dCRT}}\right)
  +o_{\P_n}(1).
\end{equation*}
Equation~\eqref{eq:spacrt-feasible-score-clt} and the continuous mapping
theorem now give
$p_{\infty,n}^{\dCRT}\convd\mathrm{Unif}(0,1)$.
\end{proof}

\subsubsection{Proof of the second conclusion of Theorem~\ref{thm:spacrt-validity}}

The second conclusion of Theorem~\ref{thm:spacrt-validity} is an
application of Theorem~\ref{thm:conditional-lr} to the conditional
resampling array $\{W_{in}\}$: the proof below verifies the three
scale-adaptive conditions~\eqref{eq:scale-adaptive-lr-conditions} of that
theorem at the cutoff $nw_n=S_n$, and it is its relative-error conclusion
that transfers validity from $p_{\infty,n}^{\dCRT}$ to $p_{\mathrm{SPA},n}^{\dCRT}$.

The effective-local cutoff condition used to transfer validity from the
exact fitted target to its saddlepoint approximation is
\begin{equation}
  \frac{|S_n|}{\sqrt n\,V_{n,\dCRT}}=O_{\P_n}(1).
  \label{eq:spacrt-intrinsic-local-alternative}
\end{equation}
Under the hypotheses of Theorem~\ref{thm:spacrt-validity} this condition
is automatic: it follows from the feasible-score central limit
theorem~\eqref{eq:spacrt-feasible-score-clt} established above.  It is
displayed separately because the transfer step uses only the
resampling-side conditions in
Assumption~\ref{assu:spacrt-regularity}(ii)--(iii) together with
\eqref{eq:spacrt-intrinsic-local-alternative} itself.

\begin{proof}[of the second conclusion of
Theorem~\ref{thm:spacrt-validity}]
Conditionally on $\mathcal F_n$, the variables
\[
  W_{in}
  \equiv
  \widehat e_{y,i}(\widetilde X_i-\widehat\mu_{x,i})
\]
are independent, centered, and have total conditional variance
\[
  \sum_{i=1}^n\widetilde\V(W_{in}\mid\mathcal F_n)
  =
  \sum_{i=1}^n
  \widehat v_{x,i}\widehat e_{y,i}^2
  =
  nV_{n,\dCRT}^2.
\]
Lemma~\ref{lem:spacrt-cb-consequences} verifies (CB) with
\[
  \bar b_n
  =
  \widehat b_{n,x}
  \max_{1\leq i\leq n}|\widehat e_{y,i}|.
\]
Assumption~\ref{assu:spacrt-regularity}(iii) gives
\[
  \P(V_{n,\dCRT}>0)\longrightarrow1,
  \qquad
  \frac{\bar b_n}{\sqrt nV_{n,\dCRT}}\convp0.
\]
Condition~\eqref{eq:spacrt-intrinsic-local-alternative} holds under
the null by \eqref{eq:spacrt-feasible-score-clt}, so
\[
  \frac{\bar b_n|S_n|}{nV_{n,\dCRT}^2}
  =
  \frac{\bar b_n}{\sqrt nV_{n,\dCRT}}
  \frac{|S_n|}{\sqrt nV_{n,\dCRT}}
  \convp0.
\]
By part~\ref{part:translator-common-envelope} of
Corollary~\ref{cor:translators-main}, the common envelope $\bar b_n$ gives
$\frac{B_n}{\sqrt n}\leq \bar b_n/\sigma_n$ and $D_n\leq \bar b_n|t_n|/\sigma_n^2$; here
$\sigma_n^2=nV_{n,\dCRT}^2$ and $t_n=S_n$, so that
$Z_n=|S_n|/(\sqrt nV_{n,\dCRT})$.  Hence
\[
  \frac{B_n(1+Z_n)}{\sqrt n}
  \leq
  \frac{\bar b_n}{\sqrt nV_{n,\dCRT}}
  +\frac{\bar b_n|S_n|}{nV_{n,\dCRT}^2}
  \convp0,
  \qquad
  D_n
  \leq
  \frac{\bar b_n|S_n|}{nV_{n,\dCRT}^2}
  \convp0,
\]
and all three conditions in
\eqref{eq:scale-adaptive-lr-conditions} hold with probability tending to
one, at the cutoff $nw_n=S_n$.
Theorem~\ref{thm:conditional-lr} gives a unique saddlepoint in the
interior of the effective domain, positive tilted variance, and
\[
  p_{\infty,n}^{\dCRT}
  =
  p_{\mathrm{SPA},n}^{\dCRT}\{1+o_{\P_n}(1)\}.
\]
This saddlepoint is the unique finite solution of the saddlepoint
equation~\eqref{eq:saddlepoint-equation} used in evaluating the Lugannani--Rice approximation
$p_{\mathrm{SPA},n}^{\dCRT}$ defined in Section~\ref{alg:spacrt}.  Combining the
relative-error conclusion with
$p_{\infty,n}^{\dCRT}\convd\mathrm{Unif}(0,1)$ proves
$p_{\mathrm{SPA},n}^{\dCRT}\convd\mathrm{Unif}(0,1)$ and completes the proof.
\end{proof}

\section{Detailed proofs of the supporting lemmas for
Theorem~\ref{thm:spacrt-rate}}
\label{sec:spacrt-rate-supporting-proofs}
\label{sec:saige-rate-supporting-proofs}

The notation in these supporting proofs is inherited from the
corresponding lemma statements in
Supplementary Section~\ref{sec:proof-main-spacrt-results}, in particular the Ber--NB
setup in Supplementary Section~\ref{sec:bnb}.  The shared design condition remains
\eqref{eq:design}.

\subsection{Proof of Lemma~\ref{lem:glm-scales}}

\begin{proof}
Let $M\equiv B_uB_\beta+B_\rho$.  For the logistic model, the odds satisfy
\[
 e^{-M}\frac{S}{1-S}
 \le \frac{p(x,\bm U)}{1-p(x,\bm U)}
 \le e^M\frac{S}{1-S}.
\]
Because $\alpha\le B_\alpha$, the baseline probability
$S=\expit(\alpha)$ is uniformly bounded away from one.  Solving the two odds
inequalities for $p(x,\bm U)$ proves the two-sided comparison with $S$ and the
uniform lower bound on $1-p(x,\bm U)$ in
\eqref{eq:logistic-scale-comparison}.  For the negative-binomial mean,
\[
 e^{-M}S\le m(x,\bm U)\le e^MS,
\]
which proves \eqref{eq:nb-scale-comparison}.  Finally,
$\expit(\alpha)=e^\alpha/(1+e^\alpha)$ and the upper bound on $\alpha$
makes the denominator uniformly bounded above and below.
\end{proof}

\subsection{Proof of Lemma~\ref{lem:det-sc}}

\begin{proof}
If $\bm g_n=0$, take $\widehat{\bm\theta}=\bm\theta_0$; global convexity and the positive
definite Hessian near $\bm\theta_0$ give existence and uniqueness.  Otherwise
$R>0$.  Fix $\bm\Delta\ne0$ with $\|\bm\Delta\|_2\le R$, put
$\bm v\equiv\bm\Delta/\|\bm\Delta\|_2$, and set
$f(t)\equiv D^2L_n(\bm\theta_0+t\bm\Delta)[\bm v,\bm v]$.  By \eqref{eq:gsc},
$|f'(t)|\le B_z\|\bm\Delta\|_2f(t)$ wherever $f(t)>0$; since
$f(0)\ge\underline h>0$, Gr\"onwall's inequality keeps
$f(t)\ge f(0)e^{-B_zt\|\bm\Delta\|_2}>0$ for $0\le t\le1$, so $\log f$ is
well defined there and
\[
 \abs{\frac{d}{dt}\log D^2L_n(\bm\theta_0+t\bm\Delta)[\bm v,\bm v]}
 \le B_z\|\bm\Delta\|_2.
\]
Thus, for $0\le t\le1$,
\[
 D^2L_n(\bm\theta_0+t\bm\Delta)[\bm\Delta,\bm\Delta]
 \ge e^{-B_z t\|\bm\Delta\|_2}\bm\Delta^{\T}\bm{\mathcal H}_n\bm\Delta.
\]
For $\bm\Delta$ with $\norm{\bm\Delta}_2=R$,
\begin{align*}
 \langle\nabla L_n(\bm\theta_0+\bm\Delta),\bm\Delta\rangle
 &=\langle \bm g_n,\bm\Delta\rangle
 +\int_0^1D^2L_n(\bm\theta_0+t\bm\Delta)[\bm\Delta,\bm\Delta]dt\\
 &\ge-\norm{\bm g_n}_2R+e^{-B_zR}\underline h R^2
 \ge-\norm{\bm g_n}_2R+\tfrac12\underline h R^2>0.
\end{align*}
Continuity gives a minimizer over the compact radius-$R$ ball, and the last
display shows that it cannot lie on the boundary.  It is therefore interior
to the ball, where its gradient vanishes.  Global
convexity makes it a global minimizer.  If
there were a second global minimizer, convexity would make $L_n$ constant on
the segment joining the two.  But \eqref{eq:gsc} holds for every pair of
directions, so the Gr\"onwall argument above, applied with an arbitrary
fixed unit vector $\bm w$ in place of the radial direction $\bm v$, gives
\[
 D^2L_n(\bm\theta_0+\bm\Delta')[\bm w,\bm w]
 \ge e^{-B_z\|\bm\Delta'\|_2}\,\bm w^{\T}\bm{\mathcal H}_n\bm w
 \ge e^{-B_zR}\underline h>0
\]
for every $\bm\theta_0+\bm\Delta'$ in the open ball, so $L_n$ cannot be constant
on any nondegenerate segment meeting the interior of the ball.  Thus the
minimizer is unique.
\end{proof}

\subsection{Proof of Lemma~\ref{lem:logistic-mle}}

\begin{proof}
At $\bm\theta_0$, the restricted average negative log-likelihood has
\[
 \bm g_n=\frac1n\sum_i\bm Z_i(m_i-Y_i),
 \qquad
 \bm{\mathcal H}_n=\frac1n\sum_im_i(1-m_i)\bm Z_i\bm Z_i^{\T}.
\]
The separate bounds in \eqref{eq:logistic-means} imply
\[
 c_m^2S_y\le m_i(1-m_i)\le C_mS_y.
\]

For the score, write $\overline m_i\equiv\E(Y_i\mid \bm Z_i)$.  If
$p_{i1}\equiv\expit(\bm\theta_0^{\T}\bm Z_i+\rho_n)$, then
\[
 \overline m_i=m_i+q_i(p_{i1}-m_i).
\]
Along the segment between the two logits, the derivative of the expit map is
$p(1-p)\le C_{H_\rho}S_y$ by the odds comparison
$p/(1-p)=e^t m_i/(1-m_i)$.  Consequently,
\begin{equation}\label{eq:logistic-local-bias}
 \norm{\E(\bm g_n\mid \bm Z_{1:n})}_2
 =\norm{\frac1n\sum_i\bm Z_i(m_i-\overline m_i)}_2
 \le \frac{B_z}{n}\sum_i|\overline m_i-m_i|
 \le CS_yS_x|\rho_n|.
\end{equation}
Moreover, $\overline m_i\asymp S_y$.  Conditional on $\bm Z_i$, the centered
score coordinates $Z_{ij}(\overline m_i-Y_i)$ are bounded and have variance
at most $CS_y$.  Bernstein's inequality and a union bound therefore give
\[
 \norm{\bm g_n-\E(\bm g_n\mid \bm Z_{1:n})}_2
 \le C\sqrt d\left\{\sqrt{\frac{S_yL}{n}}+\frac{L}{n}\right\}
 \le C\sqrt{\frac{dS_yL}{n}},
\]
where the last inequality follows from \eqref{eq:logistic-n}.  Combining
this display with \eqref{eq:logistic-local-bias} closes the score argument:
\begin{equation}\label{eq:logistic-score}
 \norm{\bm g_n}_2
 \le C\left\{\sqrt{\frac{dS_yL}{n}}+S_yS_x|\rho_n|\right\}.
\end{equation}

For the Hessian, its population counterpart satisfies
\[
 \bm{\mathcal H}:=\E\bm{\mathcal H}_n
 \succeq c_HS_y\E(\bm Z\bm Z^{\T})
 \succeq c_HS_y\lambda_z\bm I_d.
\]
Each centered Hessian entry is bounded by $CS_y$ and has variance at most
$CS_y^2$.  Hence Bernstein's inequality over the $d^2$ entries and
$\norm{\bm A}_{\op}\le d\max_{j,k}|A_{jk}|$ give
\[
 \norm{\bm{\mathcal H}_n-\bm{\mathcal H}}_{\op}
 \le CdS_y\left\{\sqrt{\frac{L}{n}}+\frac{L}{n}\right\}
 \le\frac{c_H}{2}S_y\lambda_z.
\]
Thus the Hessian argument ends with
\begin{equation*}
 \bm{\mathcal H}_n\succeq\frac{c_H}{2}S_y\lambda_z\bm I_d.
\end{equation*}

For the scalar logistic loss, $|\ell'''|\le\ell''$, so \eqref{eq:gsc}
holds on all of $\R^d$.  Applying Lemma~\ref{lem:det-sc} with
$\underline h=cS_y\lambda_z$ and using \eqref{eq:logistic-score} gives
\[
 R=\frac{4\norm{\bm g_n}_2}{\underline h}
 \le\frac{C}{\lambda_z}
 \left\{\sqrt{\frac{dL}{nS_y}}+S_x|\rho_n|\right\}.
\]
Equations \eqref{eq:logistic-n} and \eqref{eq:logistic-local-rho-small}
make $R\le(2B_z)^{-1}$ after adjusting their constants.
Lemma~\ref{lem:det-sc} therefore gives
\begin{equation*}
 \norm{\widehat{\bm\theta}-\bm\theta_0}_2
 \le R\le\frac{C}{\lambda_z}
 \left\{\sqrt{\frac{dL}{nS_y}}+S_x|\rho_n|\right\}.
\end{equation*}
Finally, for $h_i\equiv \bm Z_i^{\T}(\widehat{\bm\theta}-\bm\theta_0)$,
\[
 \frac{\widehat m_i}{m_i}
 =\frac{e^{h_i}}{1-m_i+m_ie^{h_i}},
 \qquad \max_i|h_i|\le B_zR\le\tfrac12.
\]
Thus the ratio differs from one by at most $C|h_i|$, and
\[
 \max_i\left|\frac{\widehat m_i}{m_i}-1\right|
 \le CB_zR
 \le\frac{C}{\lambda_z}
 \left\{\sqrt{\frac{dL}{nS_y}}+S_x|\rho_n|\right\}.
\]
This proves \eqref{eq:logistic-theta-error} and
\eqref{eq:logistic-error}, and stationarity at $\widehat{\bm\theta}$ gives the
score identity $\sum_i\bm Z_i(Y_i-\widehat m_i)=0$ stated in
Lemma~\ref{lem:logistic-mle}.
\end{proof}

\subsection{Proof of Lemma~\ref{lem:nb-facts}}

\begin{proof}
For $\eta>0$,
\[
 \E e^{tY}=\{1-\eta\mu(e^t-1)\}^{-1/\eta}.
\]
Writing $q\equiv\eta\mu/(1+\eta\mu)$ and expanding the logarithm gives
\[
 \kappa_k(Y)=\frac1\eta\sum_{\ell\ge1}q^\ell\ell^{k-1}.
\]
The elementary bound
$\sum_{\ell\ge1}q^\ell\ell^{k-1}
 \le(k-1)!q(1-q)^{-k}$ proves \eqref{eq:nb-cumulant}; the Poisson case is
the limit $\eta\downarrow0$.

Every centered moment is a Bell polynomial in
$\kappa_2,\ldots,\kappa_m$.  A monomial of total order $m$ containing $b$
cumulant factors is bounded by $C_m\mu^bA_\mu^{m-b}$.  Since
$\mu\le M_0$ and $A_\mu\ge1$, this is at most
$C_{m,M_0}\mu A_\mu^{m-1}$.  Taking $m=2k$ proves
\eqref{eq:nb-even-moment}.  In particular,
\[
 \E R^2=\mu A_\mu,
 \qquad
 \E R^4\le C\mu A_\mu^3,
\]
so Cauchy--Schwarz gives \eqref{eq:nb-abs3}.

For $|t|<1/A_\mu$, \eqref{eq:nb-cumulant} implies
\[
 \log\E e^{tR}
 \le\sum_{k\ge2}\frac{\mu A_\mu^{k-1}|t|^k}{k}
 \le\frac{\mu A_\mu t^2}{2(1-A_\mu|t|)}.
\]
The Chernoff argument for a sub-gamma variable proves \eqref{eq:nb-tail}.

Finally, fix $t=c/A_\mu$ with $c>0$ sufficiently small.  Exponential
tilting leaves the NB family invariant, with tilted mean
\[
 \mu_t=\frac{\mu e^t}{1-\eta\mu(e^t-1)}.
\]
Uniformly over $\mu\le M_0$ and $\eta\ge0$,
\[
 \E e^{tY}\le C,
 \qquad \mu_t\le C\mu,
 \qquad 1+\eta\mu_t\le CA_\mu.
\]
Let $\kappa_j^{(t)}$ be the cumulants under the tilted law.  The cumulant
calculation above, applied at $(\mu_t,\eta)$, together with
$\kappa_1^{(t)}=\mu_t\le C\mu$, gives, for every $j\ge1$,
\[
 \kappa_j^{(t)}\le C_j\mu A_\mu^{j-1}.
\]
The moment--cumulant formula writes $\E_tY^k$ as a finite sum over
partitions of $\{1,\ldots,k\}$.  A term corresponding to a partition with
$m\ge1$ blocks is a product of $m$ tilted cumulants and is bounded by
\[
 C_k\mu^mA_\mu^{k-m}
 \le C_{k,M_0}\mu A_\mu^{k-1}.
\]
Therefore, for $1\le k\le6$,
\[
 \E[e^{tY}Y^k]
 =\E[e^{tY}]\E_tY^k
 \le C_k\mu A_\mu^{k-1}.
\]
Since $|R|\le Y+\mu$ and
$\mu^k\le M_0^{k-1}\mu A_\mu^{k-1}$,
\[
 \E\{e^{c|R|/A_\mu}|R|^k\}
 \le C\E\{e^{tY}(Y^k+\mu^k)\}
 \le C_k\mu A_\mu^{k-1}.
\]
This proves \eqref{eq:nb-weighted-moment} for $1\le k\le6$; the displayed
bound $\E e^{tY}\le C$ gives the separately interpreted case $k=0$.
\end{proof}

\subsection{Proof of Lemma~\ref{lem:nb-mle}}

\begin{proof}
\[
 \widetilde m_i\equiv m_ie^{\rho_nX_i},\qquad
 \overline m_i\equiv\E(Y_i\mid \bm Z_i)
 =m_i\{1+q_i(e^{\rho_n}-1)\}.
\]
Let $A_i\equiv1+\eta_ym_i$ and $A_y\equiv1+\eta_yS_y$.  The mean
comparisons~\eqref{eq:nb-means}
and the bound $|\rho_n|\le H_\rho$ give
\[
 \widetilde m_i\asymp\overline m_i\asymp m_i\asymp S_y,
 \qquad
 1+\eta_y\widetilde m_i\asymp1+\eta_y\overline m_i\asymp A_i\asymp A_y.
\]

For $\eta_y>0$, the individual restricted negative log-likelihood is
\[
 \ell_i(\xi)\equiv-Y_i\xi+(Y_i+\eta_y^{-1})\log(1+\eta_ye^\xi).
\]
Its derivatives satisfy
\[
 \ell_i'(\xi)=\frac{e^\xi-Y_i}{1+\eta_ye^\xi},\qquad
 \ell_i''(\xi)=\frac{e^\xi(1+\eta_yY_i)}{(1+\eta_ye^\xi)^2},\qquad
 |\ell_i'''(\xi)|\le\ell_i''(\xi).
\]
Thus \eqref{eq:gsc} holds after composition with the bounded design.  At
$\bm\theta_0$, the score and Hessian are
\[
 \bm g_n\equiv\frac1n\sum_i\frac{m_i-Y_i}{A_i}\bm Z_i,
 \qquad
 \bm{\mathcal H}_n\equiv\frac1n\sum_i
 \frac{m_i(1+\eta_yY_i)}{A_i^2}\bm Z_i\bm Z_i^{\T}.
\]

For the score bias conditional on $\mathcal Z_n$, the bound
$|e^{\rho_n}-1|\le e^{H_\rho}|\rho_n|$ gives
\begin{equation}\label{eq:nb-local-score-bias}
 \norm{\E(\bm g_n\mid\mathcal Z_n)}_2
 \le \frac{B_z}{n}\sum_i\frac{|\overline m_i-m_i|}{A_i}
 \le C\frac{S_xS_y|\rho_n|}{A_y}.
\end{equation}
To center the score, use
\[
 Y_i-\overline m_i
 =(Y_i-\widetilde m_i)+(\widetilde m_i-\overline m_i).
\]
Conditionally on $(X_i,\bm Z_i)$, Lemma~\ref{lem:nb-facts} and
\eqref{eq:nb-weighted-moment} give the first term, after multiplication by
$Z_{ij}/A_i$, a Bernstein variance proxy $CS_y/A_y$ and a fixed scale.
The Bernoulli mixture shift is bounded and has variance at most the same
order because
\[
 q_i\left\{\frac{m_i|e^{\rho_n}-1|}{A_i}\right\}^2
 \le C\frac{S_y}{A_y}.
\]
Apply Bernstein's inequality conditionally on $\mathcal Z_n$ and take a
union bound over the two centered terms and the $d$ coordinates. The
conditional failure bound is uniform in the design, so integrating it gives
the same failure bound under the original data law. Thus
\[
 \norm{\bm g_n-\E(\bm g_n\mid\mathcal Z_n)}_2
 \le C\sqrt d\left\{\sqrt{\frac{S_yL}{A_yn}}+\frac{L}{n}\right\}
 \le C\sqrt{\frac{dS_yL}{A_yn}},
\]
where \eqref{eq:nb-mle-n} gives the last inequality.
Combining this with \eqref{eq:nb-local-score-bias} by the triangle inequality
closes the score argument:
\begin{equation}\label{eq:nb-local-score}
 \norm{\bm g_n}_2
 \le C\left\{\sqrt{\frac{dS_yL}{A_yn}}
 +\frac{S_xS_y|\rho_n|}{A_y}\right\}.
\end{equation}

For the Hessian, its population counterpart satisfies
\[
 \bm{\mathcal H}\equiv\E\bm{\mathcal H}_n
 =\E\left\{\frac{m_i(1+\eta_y\overline m_i)}{A_i^2}
 \bm Z_i\bm Z_i^{\T}\right\}
 \succeq c_H\frac{S_y}{A_y}\lambda_z\bm I_d.
\]
Here the lower bound uses the design condition~\eqref{eq:design} and
the mean comparisons~\eqref{eq:nb-means}.
Decompose its fluctuation as
\[
\begin{split}
 \bm{\mathcal H}_n-\bm{\mathcal H}
 &=\underbrace{\frac1n\sum_i
 \frac{m_i\eta_y(Y_i-\overline m_i)}{A_i^2}
 \bm Z_i\bm Z_i^{\T}}_{\bm\Delta_{\mathrm{out}}}+\underbrace{\left[
 \frac1n\sum_i\frac{m_i(1+\eta_y\overline m_i)}{A_i^2}
 \bm Z_i\bm Z_i^{\T}-\bm{\mathcal H}\right]}_{\bm\Delta_{\mathrm{des}}}.
\end{split}
\]
For $\bm\Delta_{\mathrm{out}}$, the same conditional-NB and mixture
decomposition used for the score gives an entrywise Bernstein variance proxy
$CS_y/A_y$ and a fixed scale.  Explicitly, the mixture contribution obeys
\[
 q_i\left\{\frac{m_i^2\eta_y|e^{\rho_n}-1|}{A_i^2}\right\}^2
 \le C\left(\frac{S_y}{A_y}\right)^2
 \le C\frac{S_y}{A_y}.
\]
Thus
\[
 \max_{j,k}|(\bm\Delta_{\mathrm{out}})_{jk}|
 \le C\left\{\sqrt{\frac{S_yL}{A_yn}}+\frac{L}{n}\right\}.
\]
For $\bm\Delta_{\mathrm{des}}$, the centered entry summands are bounded by
$CS_y/A_y$ and have variance at most $C(S_y/A_y)^2$, so
\[
 \max_{j,k}|(\bm\Delta_{\mathrm{des}})_{jk}|
 \le C\frac{S_y}{A_y}
 \left\{\sqrt{\frac{L}{n}}+\frac{L}{n}\right\}.
\]
Recombining the two named terms gives
\[
\begin{split}
 \norm{\bm{\mathcal H}_n-\bm{\mathcal H}}_{\op}
 &\le\norm{\bm\Delta_{\mathrm{out}}}_{\op}
     +\norm{\bm\Delta_{\mathrm{des}}}_{\op}\\
 &\le Cd\left[
 \sqrt{\frac{S_yL}{A_yn}}+\frac{L}{n}
 +\frac{S_y}{A_y}\left\{\sqrt{\frac{L}{n}}+\frac{L}{n}\right\}\right]\le\frac{c_H}{2}\frac{S_y}{A_y}\lambda_z,
\end{split}
\]
where the last line follows from \eqref{eq:nb-mle-n}.  Hence the Hessian
argument ends with
\begin{equation}\label{eq:nb-combined-curvature}
 \bm{\mathcal H}_n\succeq\frac{c_H}{2}\frac{S_y}{A_y}\lambda_z\bm I_d.
\end{equation}

Lemma~\ref{lem:det-sc}, \eqref{eq:nb-local-score}, and
\eqref{eq:nb-combined-curvature} give the localization radius
\[
 R\le\frac{C}{\lambda_z}
 \left\{\sqrt{\frac{A_ydL}{S_yn}}+S_x|\rho_n|\right\}.
\]
Equations \eqref{eq:nb-mle-n} and \eqref{eq:nb-local-smallness} make
$R\le(2B_z)^{-1}$.  Therefore the MLE exists, is unique, and
\begin{equation}\label{eq:nb-parameter-combined}
 \norm{\widehat{\bm\theta}_y-\bm\theta_0}_2
 \le R\le\frac{C}{\lambda_z}
 \left\{\sqrt{\frac{A_ydL}{S_yn}}+S_x|\rho_n|\right\}.
\end{equation}
Finally,
\[
 \frac{\widehat m_i}{m_i}
 =\exp\{\bm Z_i^{\T}(\widehat{\bm\theta}_y-\bm\theta_0)\},
 \qquad
 \max_i|\bm Z_i^{\T}(\widehat{\bm\theta}_y-\bm\theta_0)|\le B_zR\le\tfrac12,
\]
so
\[
 \max_i\left|\frac{\widehat m_i}{m_i}-1\right|
 \le CB_zR
 \le\frac{C}{\lambda_z}
 \left\{\sqrt{\frac{A_ydL}{S_yn}}+S_x|\rho_n|\right\}.
\]
Together with \eqref{eq:nb-parameter-combined}, this proves
\eqref{eq:nb-theta-error} and \eqref{eq:nb-mle-error}.  When $\eta_y=0$,
the score, Hessian, and all displayed bounds are their continuous limits
with $A_y=1$.
\end{proof}

\subsection{Proof of Lemma~\ref{lem:nb-empirical}}

\begin{proof}
Put $A_y\equiv1+\eta_yS_y$ and introduce the conditional mean
$\widetilde m_i\equiv m_ie^{\rho_nX_i}$.  We first establish bounds for the raw
null residual
\[
 R_i\equiv Y_i-m_i
 =(Y_i-\widetilde m_i)+(\widetilde m_i-m_i).
\]
By Lemma~\ref{lem:glm-scales},
\[
 \widetilde m_i\asymp S_y,\qquad
 1+\eta_y\widetilde m_i\asymp A_y,\qquad
 |\widetilde m_i-m_i|\le CS_y|\rho_n|\le CS_y.
\]
Thus the conditional NB fluctuation $Y_i-\widetilde m_i$ has the scale in
Lemma~\ref{lem:nb-facts}, and the second term in $R_i$ is a bounded shift.
The conditional moment bounds~\eqref{eq:nb-even-moment}
and~\eqref{eq:nb-abs3}, followed by averaging over
$(X_i,\bm Z_i)$, give
\begin{align}
 cS_yA_y&\le \E R_i^2\le CS_yA_y,
 &\E|R_i|^3&\le CS_yA_y^2,\label{eq:nb-raw-low-moments}\\
 \E R_i^4&\le CS_yA_y^3,
 &\E|R_i|^6&\le CS_yA_y^5.\label{eq:nb-raw-high-moments}
\end{align}
For clarity, the lower bound in \eqref{eq:nb-raw-low-moments} follows from
\[
 \E R_i^2
 \ge \E\{\Var(Y_i\mid X_i,\bm Z_i)\}
 \ge cS_yA_y.
\]
This completes the population-moment step.

We next control the largest raw residual.  The conditional tail
bound~\eqref{eq:nb-tail}, together with the bounded shift above, implies,
after adjusting constants, that for every $t>0$,
\begin{equation}\label{eq:nb-raw-tail}
 \P(|R_i|\ge t)
 \le 2\exp\left[-c\min\left\{
 \frac{t^2}{S_yA_y},\frac{t}{A_y}\right\}\right].
\end{equation}
Let
\[
 T\equiv C_TA_yH.
\]
Since $H=\log(C_0n/\delta)$, choosing $C_T$ sufficiently large and applying
a union bound in \eqref{eq:nb-raw-tail} gives
\begin{equation}\label{eq:nb-raw-maximum}
 \P\left(\max_{1\le i\le n}|R_i|>T\right)\le\frac{\delta}{4}.
\end{equation}
Hence the maximum-residual part ends with
$\max_i|R_i|\le C A_yH$ outside an event of probability $\delta/4$.

It remains to concentrate the empirical second and third moments.  For
$j\in\{2,3\}$ define the truncated variables
\[
 U_{ij}\equiv|R_i|^j\indicator(|R_i|\le T).
\]
The weighted-moment bound \eqref{eq:nb-weighted-moment}, applied conditionally
on $(X_i,\bm Z_i)$ and then combined with the bounded shift, gives
\[
 \E\left\{e^{c|R_i|/A_y}|R_i|^j\right\}
 \le C_jS_yA_y^{j-1}.
\]
Consequently, the truncation bias is explicitly bounded by
\begin{equation}\label{eq:nb-truncation-bias}
 0\le \E|R_i|^j-\E U_{ij}
 \le C_jS_yA_y^{j-1}e^{-cC_TH},
 \qquad j=2,3.
\end{equation}
After increasing $C_T$, \eqref{eq:nb-truncation-bias} and
\eqref{eq:nb-raw-low-moments} imply
\[
 cS_yA_y\le \E U_{i2}\le CS_yA_y,
 \qquad
 \E U_{i3}\le CS_yA_y^2.
\]

For the truncated second moments, \eqref{eq:nb-raw-high-moments} and the
definition of $T$ give
\[
 \Var(U_{i2})\le CS_yA_y^3,\qquad
 0\le U_{i2}\le C A_y^2H^2.
\]
Bernstein's inequality therefore yields, with failure probability at most
$\delta/4$,
\[
 \left|\frac1n\sum_i\{U_{i2}-\E U_{i2}\}\right|
 \le C\left\{
 \sqrt{\frac{S_yA_y^3L}{n}}+\frac{A_y^2H^2L}{n}
 \right\}.
\]
The sample-size condition \eqref{eq:nb-moment-n} makes the right-hand side at
most a sufficiently small constant times $S_yA_y$.  Combining this with the
bounds for $\E U_{i2}$ gives
\begin{equation}\label{eq:nb-raw-second-empirical}
 cS_yA_y\le\frac1n\sum_iU_{i2}\le CS_yA_y.
\end{equation}

Likewise, for the truncated third moments,
\[
 \Var(U_{i3})\le CS_yA_y^5,\qquad
 0\le U_{i3}\le C A_y^3H^3.
\]
Another Bernstein inequality gives, with failure probability at most
$\delta/4$,
\[
 \left|\frac1n\sum_i\{U_{i3}-\E U_{i3}\}\right|
 \le C\left\{
 \sqrt{\frac{S_yA_y^5L}{n}}+\frac{A_y^3H^3L}{n}
 \right\}
 \le CS_yA_y^2,
\]
where the final inequality is again \eqref{eq:nb-moment-n}.  Hence
\begin{equation}\label{eq:nb-raw-third-empirical}
 \frac1n\sum_iU_{i3}\le CS_yA_y^2.
\end{equation}

On $\{\max_i|R_i|\le T\}$, whose complement has probability at most
$\delta/4$ by \eqref{eq:nb-raw-maximum}, $U_{i2}=R_i^2$ and
$U_{i3}=|R_i|^3$ for every $i$.  Let $\mathcal G_{n,\mathrm{NB}}$ be the
intersection of $\{\max_i|R_i|\le T\}$ and the two Bernstein events.  The union bound
gives $\P(\mathcal G_{n,\mathrm{NB}})\ge1-3\delta/4\ge1-\delta$, and
\eqref{eq:nb-raw-maximum}, \eqref{eq:nb-raw-second-empirical}, and
\eqref{eq:nb-raw-third-empirical} give, on this event,
\begin{equation}\label{eq:nb-raw-empirical-summary}
 \max_i|R_i|\le CA_yH,\qquad
 cS_yA_y\le\frac1n\sum_iR_i^2\le CS_yA_y,\qquad
 \frac1n\sum_i|R_i|^3\le CS_yA_y^2.
\end{equation}
This completes all three raw-residual bounds.

Finally, we transfer \eqref{eq:nb-raw-empirical-summary} to the fitted
residuals.  Write $d_i\equiv\widehat m_i-m_i$, so that $a_i=R_i-d_i$.  On
$\mathcal F_{n,\mathrm{NB}}$,
\[
 |d_i|\le c_0m_i\le Cc_0S_y.
\]
The maximum satisfies
\[
 \max_i|a_i|\le\max_i|R_i|+\max_i|d_i|\le CA_yH.
\]
For the second moment, using
$|R_i-d_i|^2\ge R_i^2/2-d_i^2$ and
$|R_i-d_i|^2\le2R_i^2+2d_i^2$ gives
\[
 \frac12\frac1n\sum_iR_i^2-Cc_0^2S_y^2
 \le\frac1n\sum_i a_i^2
 \le2\frac1n\sum_iR_i^2+Cc_0^2S_y^2.
\]
Because $S_y^2\le M_0S_yA_y$, choosing $c_0$ sufficiently small and using
\eqref{eq:nb-raw-empirical-summary} yields
\[
 cS_yA_y\le\frac1n\sum_i a_i^2\le CS_yA_y.
\]
For the third moment, $|R_i-d_i|^3\le4(|R_i|^3+|d_i|^3)$ gives
\[
 \frac1n\sum_i|a_i|^3
 \le4\frac1n\sum_i|R_i|^3+Cc_0^3S_y^3
 \le CS_yA_y^2,
\]
where $S_y^3\le M_0^2S_yA_y^2$.  Recalling $A_y=1+\eta_yS_y$, these are
exactly the three bounds in \eqref{eq:nb-empirical-events}.
\end{proof}

\subsection{Proof of Lemma~\ref{lem:bnb-cutoff}}

\begin{proof}
Write
\[
 \varepsilon_{xi}\equiv X_i-q_i,\qquad
 \varepsilon_{yi}\equiv Y_i-m_i,\qquad
 d_{xi}\equiv\widehat\mu_{ix}-q_i,\qquad
 d_{yi}\equiv\widehat\mu_{iy}-m_i.
\]
Substitution into the Ber--NB statistic~\eqref{eq:bnb-statistic} gives
\begin{equation}\label{eq:bnb-cutoff-decomp}
 T_n^{\mathrm{dCRT}}
 =\frac1n\sum_i\varepsilon_{xi}\varepsilon_{yi}
 -\frac1n\sum_id_{xi}\varepsilon_{yi}
 -\frac1n\sum_i\varepsilon_{xi}d_{yi}
 +\frac1n\sum_id_{xi}d_{yi}.
\end{equation}
Conditional on $\bm Z_i$,
\[
 \mathbb E(\varepsilon_{xi}\varepsilon_{yi}\mid \bm Z_i)
 =q_i(1-q_i)m_i(e^{\rho_n}-1),
\]
so, since $|e^{\rho_n}-1|\le e^{H_\rho}|\rho_n|$,
\begin{equation}\label{eq:bnb-local-mean}
 \left|\mathbb E(\varepsilon_{xi}\varepsilon_{yi})\right|
 \le C\pi_{n,x}\mu_{n,y}|\rho_n|.
\end{equation}
Moreover,
\[
 \mathbb E(\varepsilon_{xi}^2\varepsilon_{yi}^2\mid \bm Z_i)
 \le C\pi_{n,x}\mu_{n,y}(1+\eta_{n,y}\mu_{n,y}).
\]
Indeed, conditioning further on $X_i$ reduces the calculation to the
second moment of an NB variable whose mean $m_ie^{\rho_nX_i}$ is
comparable to $\mu_{n,y}$ by \eqref{eq:bnb-sparsity}, recentered by the
deterministic amount $m_i(e^{\rho_nX_i}-1)$, which is bounded in
magnitude by $C\mu_{n,y}|\rho_n|\le CM_0$.
Put $G_i=\varepsilon_{xi}\varepsilon_{yi}$ and
$A_{n,y}=1+\eta_{n,y}\mu_{n,y}$.  The weighted-moment bound
\eqref{eq:nb-weighted-moment}, conditioning first on $(X_i,\bm Z_i)$ and then
centering over $(X_i,\bm Z_i)$, gives
\[
 \E\left[(G_i-\E G_i)^2
 \exp\{c|G_i-\E G_i|/A_{n,y}\}\right]
 \le C\pi_{n,x}\mu_{n,y}A_{n,y}.
\]
Expanding the exponential gives the Bernstein moment condition with
variance proxy $C\pi_{n,x}\mu_{n,y}A_{n,y}$ and scale $CA_{n,y}$.  Bernstein's inequality and
\eqref{eq:bnb-cutoff-n} give
\[
 \left|\frac1n\sum_i\varepsilon_{xi}\varepsilon_{yi}
 -\mathbb E(\varepsilon_{xi}\varepsilon_{yi})\right|
 \le C\sqrt{\frac{\pi_{n,x}\mu_{n,y}(1+\eta_{n,y}\mu_{n,y})L}{n}}.
\]
Together with \eqref{eq:bnb-local-mean}, the first term in
\eqref{eq:bnb-cutoff-decomp} therefore satisfies the explicit bound
\begin{equation}\label{eq:bnb-cutoff-leading-term}
 \left|\frac1n\sum_i\varepsilon_{xi}\varepsilon_{yi}\right|
 \le C\left\{\sqrt{\frac{\pi_{n,x}\mu_{n,y}(1+\eta_{n,y}\mu_{n,y})L}{n}}
 +\pi_{n,x}\mu_{n,y}|\rho_n|\right\}.
\end{equation}

The MLE bounds~\eqref{eq:logistic-error}
and~\eqref{eq:nb-mle-error} give
\[
 d_x^*:=\max_i|d_{xi}|\le C\sqrt{\frac{\pi_{n,x}L}{n}},
 \qquad
 d_y^*:=\max_i|d_{yi}|\le
 C\left\{\sqrt{\frac{\mu_{n,y}(1+\eta_{n,y}\mu_{n,y})L}{n}}+\mu_{n,y}\pi_{n,x}|\rho_n|\right\},
\]
where fixed $d$ and $\lambda_z$ are absorbed into $C$: the second bound
multiplies the relative-error conclusion
\eqref{eq:nb-mle-error} of Lemma~\ref{lem:nb-mle} by
$m_i\asymp \mu_{n,y}$.  Nonnegativity of
$Y_i$, the bounded-alternative mean
comparison \eqref{eq:bnb-sparsity}, and Lemma~\ref{lem:nb-facts} give
\[
 \E|\varepsilon_{yi}|\le C\mu_{n,y},
 \qquad
 \Var(|\varepsilon_{yi}|)\le C\mu_{n,y}A_{n,y}.
\]
More precisely, for
$D_i\equiv|\varepsilon_{yi}|-\E|\varepsilon_{yi}|$, the weighted-moment bound
\eqref{eq:nb-weighted-moment} and the bounded deterministic recentering
$|m_i(e^{\rho_nX_i}-1)|\le CM_0$
give
\[
 \E\{D_i^2\exp(c|D_i|/A_{n,y})\}\le C\mu_{n,y}A_{n,y}.
\]
Thus $D_i$ satisfies the Bernstein moment condition with variance proxy
$C\mu_{n,y}A_{n,y}$ and scale $CA_{n,y}$.  The resulting variance-sensitive Bernstein
inequality and
\eqref{eq:bnb-cutoff-n} therefore imply
\[
 \frac1n\sum_i|\varepsilon_{yi}|
 \le C\left\{\mu_{n,y}+\sqrt{\frac{\mu_{n,y}A_{n,y}L}{n}}+\frac{A_{n,y}L}{n}\right\}
 \le C\sqrt{\mu_{n,y}A_{n,y}}
\]
with failure probability at most $C\delta$.  This direct absolute-moment
bound does not invoke the stronger residual-moment burn-in.
Moreover, $\E|X_i-q_i|=2q_i(1-q_i)\asymp \pi_{n,x}$, and Bernstein's
inequality, using~\eqref{eq:bnb-cutoff-n}, gives
\[
 \frac1n\sum_i|\varepsilon_{xi}|\le C\pi_{n,x}
\]
with probability at least $1-C\delta$.
The remaining three terms in
\eqref{eq:bnb-cutoff-decomp} are now bounded on the intersection of the
fit and concentration events.  Under \eqref{eq:bnb-cutoff-n},
$L/n\le \pi_{n,x}\mu_{n,y}/\{C(1+\eta_{n,y}\mu_{n,y})\}\le1$ after enlarging $C$, and
$\pi_{n,x}\le1$.  First,
\[
 \left|\frac1n\sum_id_{xi}\varepsilon_{yi}\right|
 \le d_x^*\cdot\frac1n\sum_i|\varepsilon_{yi}|
 \le C\sqrt{\frac{\pi_{n,x}L}{n}}\sqrt{\mu_{n,y}(1+\eta_{n,y}\mu_{n,y})}
 =C\sqrt{\frac{\pi_{n,x}\mu_{n,y}(1+\eta_{n,y}\mu_{n,y})L}{n}}.
\]
Second,
\[
\begin{aligned}
 \left|\frac1n\sum_i\varepsilon_{xi}d_{yi}\right|
 &\le d_y^*\cdot\frac1n\sum_i|\varepsilon_{xi}|\\
 &\le C\pi_{n,x}\left\{\sqrt{\frac{\mu_{n,y}(1+\eta_{n,y}\mu_{n,y})L}{n}}
 +\mu_{n,y}\pi_{n,x}|\rho_n|\right\}\\
 &\le C\left\{\sqrt{\frac{\pi_{n,x}\mu_{n,y}(1+\eta_{n,y}\mu_{n,y})L}{n}}
 +\pi_{n,x}\mu_{n,y}|\rho_n|\right\},
\end{aligned}
\]
using $\pi_{n,x}\le\sqrt{\pi_{n,x}}\le1$ in the first summand and $\pi_{n,x}\le1$ in the
second.  Third,
\[
\begin{aligned}
 \left|\frac1n\sum_id_{xi}d_{yi}\right|
 &\le d_x^*d_y^*\\
 &\le C\sqrt{\pi_{n,x}\mu_{n,y}(1+\eta_{n,y}\mu_{n,y})}\,\frac{L}{n}
 +C\pi_{n,x}\mu_{n,y}|\rho_n|\sqrt{\frac{\pi_{n,x}L}{n}}\\
 &\le C\left\{\sqrt{\frac{\pi_{n,x}\mu_{n,y}(1+\eta_{n,y}\mu_{n,y})L}{n}}
 +\pi_{n,x}\mu_{n,y}|\rho_n|\right\},
\end{aligned}
\]
because $L/n\le\sqrt{L/n}$ and $\sqrt{\pi_{n,x}L/n}\le1$.

Let $\mathcal G_{n,\mathrm{bnb}}^{\mathrm{cut}}$ be the intersection of
the three concentration events used above: the centered-product event,
the event controlling $n^{-1}\sum_i|\varepsilon_{yi}|$, and the event
controlling $n^{-1}\sum_i|\varepsilon_{xi}|$.  A union bound gives
\[
 \mathbb P(\mathcal G_{n,\mathrm{bnb}}^{\mathrm{cut}})
 \ge1-C\delta.
\]
On $\mathcal F_{n,\mathrm{bnb}}\cap
\mathcal G_{n,\mathrm{bnb}}^{\mathrm{cut}}$, substitute
\eqref{eq:bnb-cutoff-leading-term} and the three displayed fitted-error
bounds into \eqref{eq:bnb-cutoff-decomp}.  This explicitly closes the
decomposition with
\[
 |T_n^{\mathrm{dCRT}}|
 \le C\left\{\sqrt{\frac{\pi_{n,x}\mu_{n,y}(1+\eta_{n,y}\mu_{n,y})L}{n}}
 +\pi_{n,x}\mu_{n,y}|\rho_n|\right\},
\]
which is \eqref{eq:bnb-cutoff-bound}.
\end{proof}

\subsection{Proof of Lemma~\ref{lem:bnb-moments}}

\begin{proof}
On $\mathcal F_{n,\mathrm{bnb}}$, the relative-error bound
\eqref{eq:logistic-error} is at most a fixed small constant by
\eqref{eq:bnb-moments-n}, so $\widehat\mu_{ix}\asymp q_i\asymp \pi_{n,x}$ and
$1-\widehat\mu_{ix}\ge c$; hence
$\widehat\mu_{ix}(1-\widehat\mu_{ix})\asymp \pi_{n,x}$.  For
$u=sa_i$, direct algebra from \eqref{eq:bnb-qs} gives
\[
 e^{-|u|}\widehat\mu_{ix}(1-\widehat\mu_{ix})
 \le \mu_{ix,s}(1-\mu_{ix,s})
 \le e^{|u|}\widehat\mu_{ix}(1-\widehat\mu_{ix})\max\{e^{|u|},1\}.
\]
The maximal-residual bound in~\eqref{eq:nb-empirical-events} gives
$\max_i|a_i|\le C(1+\eta_{n,y}\mu_{n,y})H$.  Hence, on \eqref{eq:bnb-s0},
\[
 \max_i|sa_i|\le Cc,
 \qquad
 \mu_{ix,s}(1-\mu_{ix,s})\asymp \pi_{n,x}
\]
uniformly in $i$ and $s$.  Substituting this comparison and the second
residual-moment bound in \eqref{eq:nb-empirical-events} into
\eqref{eq:bnb-K2-exact} gives the explicit variance bound
\[
 c\pi_{n,x}\mu_{n,y}(1+\eta_{n,y}\mu_{n,y})
 \le K_n''(s)
 \le C\pi_{n,x}\mu_{n,y}(1+\eta_{n,y}\mu_{n,y}),
\]
which is \eqref{eq:bnb-K2-bound}.  Similarly,
\eqref{eq:bnb-K3-exact} and the third residual-moment bound give
\[
 \frac1n\sum_i\widetilde\E_s|W_{in}-\widetilde\E_sW_{in}|^3
 \le C\pi_{n,x}\mu_{n,y}(1+\eta_{n,y}\mu_{n,y})^2,
\]
which is \eqref{eq:bnb-K3-bound}.  Dividing the latter bound by the lower
variance bound to the power $3/2$ gives explicitly
\[
 \sup_{|s|\le s_{0n}}\kappa_n(s)
 \le
 \frac{C\pi_{n,x}\mu_{n,y}(1+\eta_{n,y}\mu_{n,y})^2}
 {[c\pi_{n,x}\mu_{n,y}(1+\eta_{n,y}\mu_{n,y})]^{3/2}}
 \le C\sqrt{\frac{1+\eta_{n,y}\mu_{n,y}}{\pi_{n,x}\mu_{n,y}}},
\]
which is \eqref{eq:bnb-kappa} and closes all three moment bounds.
\end{proof}

\subsection{Proof of Lemma~\ref{lem:le-cam-elementary}}

\begin{proof}
This is Proposition~1 of \citet{LeCam1960}.
\end{proof}

\subsection{Proof of Lemma~\ref{lem:poisson-jump-edgeworth}}

\begin{proof}
Put
\[
 Y_L\equiv\sum_{j=1}^Jd_jN_{jL},\qquad
 s_L^2\equiv\Var(Y_L)=\sigma_L^2/h^2,\qquad
 \gamma_L\equiv\frac{\operatorname{cum}_3(Y_L)}{s_L^3}
 =\frac{\kappa_{3,L}}{\sigma_L^3}.
\]
The mean bounds imply $s_L^2\asymp L$ and $\gamma_L=O(L^{-1/2})$, uniformly
over the permitted means.  If $a_L$ is a support point, then
$m_L\equiv a_L/h+\E Y_L$ is an integer and
$z_L\equiv(m_L-\E Y_L)/s_L=a_L/\sigma_L$.

The centered characteristic function of $Y_L$ is
\[
 \psi_L(t)
 =\exp\left\{
   \sum_{j=1}^J\mu_{jL}(e^{id_jt}-1-id_jt)
  \right\}.
\]
Because $\gcd(d_1,\ldots,d_J)=1$, the function
$\sum_j\{1-\cos(d_jt)\}$ vanishes on $[-\pi,\pi]$ only at zero.
Consequently, for every sufficiently small fixed $\delta>0$,
\begin{equation}
 |\psi_L(t)|\le e^{-c_1Lt^2}\quad(|t|\le\delta),
 \qquad
 |\psi_L(t)|\le e^{-c_2L}\quad(\delta\le|t|\le\pi),
 \label{eq:poisson-jump-fourier-decay}
\end{equation}
with constants uniform over the means.

For $|t|\le\delta$, Taylor expansion of the exponent gives
\[
 \log\psi_L(t)
 =-\frac{s_L^2t^2}{2}
  -\frac{i\operatorname{cum}_3(Y_L)t^3}{6}
  +R_L(t),
 \qquad |R_L(t)|\le C L|t|^4.
\]
Define $\mathcal R_L(u)$ for $|u|\le\pi s_L$ by
\[
 \psi_L(u/s_L)
 =e^{-u^2/2}\left(1-\frac{i\gamma_L u^3}{6}\right)
  +\mathcal R_L(u),
\]
and use the preceding Taylor expansion when $|u|\le\delta s_L$.  The decay
bounds above control the remaining values of $u$.  Together they imply
\begin{equation}
 \int_{-\pi s_L}^{\pi s_L}|\mathcal R_L(u)|\,du
 +\int_{-\pi s_L}^{\pi s_L}\frac{|\mathcal R_L(u)|}{|u|}\,du
 \le \frac{C}{L}.
 \label{eq:poisson-jump-fourier-remainder}
\end{equation}
At $u=0$ the second integrand is read by continuity.  To see the powers in
this bound explicitly, on the central interval the remainder is bounded by
\[
 C L^{-1}(u^4+u^6)e^{-c u^2},
\]
and outside that interval \eqref{eq:poisson-jump-fourier-decay} is
exponentially small.  Thus both integrals in
\eqref{eq:poisson-jump-fourier-remainder} are finite and have the stated
order.

Lattice Fourier inversion and \eqref{eq:poisson-jump-fourier-remainder}
now give, uniformly for $|z_L|\le M$,
\[
\begin{aligned}
 \P(A_L=a_L)
 &=\P(Y_L=m_L)\\
 &=\frac1{s_L}\left[
   \phi(z_L)+\frac{\gamma_L}{6}(z_L^3-3z_L)\phi(z_L)+O_M(L^{-1})
  \right].
\end{aligned}
\]
Since $s_L=\sigma_L/h$ and $s_L^{-1}\gamma_L=O(L^{-1})$, this ends the local
calculation with
\[
 \P(A_L=a_L)=\frac{h}{\sigma_L}\phi(z_L)+O_M(L^{-1}),
\]
which is \eqref{eq:poisson-jump-local}.

For completeness, the midpoint tail follows from the same Fourier
expansion rather than from an i.i.d. lattice theorem.  Symmetrically summing
the lattice inversion formula above and assigning half weight at $m_L$ gives
the following identity, interpreted as a Cauchy principal value at $t=0$:
\[
 \P(Y_L>m_L)+\frac12\P(Y_L=m_L)
 =\frac12+\frac1{2\pi}\operatorname{p.v.}
  \int_{-\pi}^{\pi}
  \psi_L(t)e^{-it(m_L-\E Y_L)}
  \left\{-\frac{i}{2}\cot(t/2)\right\}\,dt.
\]
Specifically, if $f$ is singular only at zero, then
\[
 \operatorname{p.v.}\int_{-b}^{b}f(t)\,dt
 \equiv\lim_{\varepsilon\downarrow0}
 \left\{\int_{-b}^{-\varepsilon}f(t)\,dt
       +\int_{\varepsilon}^{b}f(t)\,dt\right\}.
\]
This symmetric limit is needed because the cotangent factor has a $1/t$
singularity at zero; the singular contributions from the two sides cancel.
Near zero,
\[
 -\frac{i}{2}\cot(t/2)=-\frac{i}{t}+O(t).
\]
Substitute $u=s_Lt$.  The $O(t)$ term contributes $O(s_L^{-2})=O(L^{-1})$;
the two integrals in \eqref{eq:poisson-jump-fourier-remainder} show that
the characteristic-function remainder also contributes $O(L^{-1})$.
The two retained terms in the characteristic-function expansion satisfy
\[
\begin{aligned}
 \frac12+\frac1{2\pi}\operatorname{p.v.}
 \int_{\mathbb R}e^{-u^2/2-iuz_L}\left(-\frac{i}{u}\right)\,du
 &=\Phibar(z_L),\\
 \frac1{2\pi}\int_{\mathbb R}e^{-u^2/2-iuz_L}
 \left(-\frac{i\gamma_Lu^3}{6}\right)
 \left(-\frac{i}{u}\right)\,du
 &=\frac{\gamma_L}{6}(z_L^2-1)\phi(z_L).
\end{aligned}
\]
Combining these identities with the two $O(L^{-1})$ bounds above gives
\[
 \P(A_L>a_L)+\frac12\P(A_L=a_L)
 =\Phibar(z_L)+\frac{\kappa_{3,L}}{6\sigma_L^3}(z_L^2-1)\phi(z_L)
  +O_M(L^{-1}),
\]
which is \eqref{eq:poisson-jump-midtail}.

This completes both uniform expansions.
\end{proof}

\subsection{Proof of Lemma~\ref{lem:poisson-central-tail}}

\begin{proof}
We first establish the local-mass expansion.
Stirling's formula, expanded uniformly for
$a=\lambda+z\sqrt\lambda$ with $|z|\le M$, gives
\[
 \log\P(P_\lambda=a)
 =-\frac12\log(2\pi\lambda)-\frac{z^2}{2}
 +O_M(\lambda^{-1/2}),
\]
and therefore closes the local-mass calculation with
\[
 \P(P_\lambda=a)=\lambda^{-1/2}\phi(z)+O_M(\lambda^{-1}),
\]
which is \eqref{eq:poisson-local-compact}.

For the midpoint tail, apply
Lemma~\ref{lem:poisson-jump-edgeworth} with $J=1$, $d_1=h=1$,
$L=\mu_1=\lambda$.  This is a varying-mean Poisson triangular array covered
directly by that lemma.  Here $\sigma^2=\kappa_3=\lambda$, and the lemma's
midpoint convention is exactly the definition of $Q_\lambda(a)$.  Therefore,
uniformly for $|z|\le M$,
\[
 Q_\lambda(a)
 =\Phibar(z)+\frac{z^2-1}{6\sqrt\lambda}\phi(z)
 +O_M(\lambda^{-1}),
\]
which is \eqref{eq:poisson-midtail-compact}.

It remains to derive the two central formulas.  Set $a=\lambda=k$, so
$z=0$.  The sharper central form of Stirling's
expansion gives
\[
 \P(Z_k=k)=\frac1{\sqrt{2\pi k}}\{1+O(k^{-1})\}
 =\frac1{\sqrt{2\pi k}}+O(k^{-3/2}),
\]
which is \eqref{eq:poisson-mass}.  The exact identity
\[
 \P(Z_k\ge k)=Q_k(k)+\frac12\P(Z_k=k)
\]
gives
\[
 \P(Z_k\ge k)
 =\frac12-\frac{\phi(0)}{6\sqrt k}
 +\frac{\phi(0)}{2\sqrt k}+O(k^{-1})
 =\frac12+\frac1{3\sqrt{2\pi k}}+O(k^{-1}),
\]
which is \eqref{eq:poisson-tail}.
\end{proof}

\subsection{Proof of Lemma~\ref{lem:poisson-like-lr}}

\begin{proof}
We first locate the saddlepoint.  Put
$s_\lambda\equiv(M+2)\lambda^{-1/2}$.  Uniformly for
$|s|\le s_\lambda$, Taylor's theorem together with
\eqref{eq:poisson-like-cumulant-assumptions} and
\eqref{eq:poisson-like-derivative-assumptions} gives
\[
 K_\lambda''(s)=\lambda\{1+O_M(\lambda^{-1/2})\},
 \qquad
 K_\lambda'(\pm s_\lambda)
 =\pm(M+2)\sqrt\lambda+O_M(1).
\]
Thus $K_\lambda'$ is strictly increasing on this interval and crosses
$w$ by \eqref{eq:poisson-like-cutoff-assumption}, so the saddlepoint is
unique and satisfies
$\widehat s=O_M(\lambda^{-1/2})$.  Thus this part ends with the explicit
location bound
\[
 |\widehat s|\le C_M\lambda^{-1/2}.
\]

We next expand the quantities entering the LR formula.  Use the
actual variance standardization
\[
 \widetilde z\equiv\frac{w}{\sqrt{\kappa_2}},
 \qquad y\equiv\widehat s\sqrt{\kappa_2},
 \qquad \gamma\equiv\frac{\kappa_3}{\kappa_2^{3/2}}.
\]
By \eqref{eq:poisson-like-cumulant-assumptions},
\[
 \kappa_2=\lambda\{1+O(\lambda^{-2})\},
 \qquad
 \kappa_3=\lambda\{1+O(\lambda^{-2})\},
 \qquad
 \sqrt{\kappa_2}=\sqrt\lambda\{1+O(\lambda^{-2})\}.
\]
Combining these relations with the cutoff expansion
\eqref{eq:poisson-like-cutoff-assumption} gives
\[
\begin{aligned}
 \widetilde z
 &=\frac{z\sqrt\lambda+O(\lambda^{-1/2})}
         {\sqrt\lambda\{1+O(\lambda^{-2})\}}
 =z+O_M(\lambda^{-1}),\\
 \gamma
 &=\frac{\lambda\{1+O(\lambda^{-2})\}}
         {\lambda^{3/2}\{1+O(\lambda^{-2})\}}
 =\lambda^{-1/2}+O(\lambda^{-5/2}).
\end{aligned}
\]
Taylor expansion of the saddlepoint equation, with each remainder retaining
its power of $y$, gives
\[
 \widetilde z
 =y+\frac{\gamma y^2}{2}+O_M(\lambda^{-1}y^3),
 \qquad
 y=\widetilde z-\frac{\gamma\widetilde z^2}{2}
   +O_M(\lambda^{-1}\widetilde z^3).
\]
Similarly,
\[
 K_\lambda(\widehat s)
 =\frac{y^2}{2}+\frac{\gamma y^3}{6}
   +O_M(\lambda^{-1}y^4),
 \qquad
 \frac{K_\lambda''(\widehat s)}{\kappa_2}
 =1+\gamma y+O_M(\lambda^{-1}y^2).
\]
Substitution into the definitions of $r$ and $u$ now yields the explicit
factorizations
\[
 r=\widetilde z\left\{1-\frac{\gamma\widetilde z}{6}
      +O_M(\lambda^{-1}\widetilde z^2)\right\},
 \qquad
 u=\widetilde z\{1+O_M(\lambda^{-1}\widetilde z^2)\}.
\]
For $\widetilde z\ne0$, removing the common factor and taking reciprocals
and squares gives
\[
 \frac1u-\frac1r=-\frac{\gamma}{6}+O_M(\lambda^{-1})
 =-\frac1{6\sqrt\lambda}+O_M(\lambda^{-1}),
 \qquad
 \frac{u^2-r^2}{2}
 =\frac{\gamma\widetilde z^3}{6}+O_M(\lambda^{-1})
 =\frac{z^3}{6\sqrt\lambda}+O_M(\lambda^{-1}).
\]
The factorizations also show that these two bounds extend continuously to
$\widetilde z=0$.  Finally, they imply the additive bounds
\[
 r=z-\frac{z^2}{6\sqrt\lambda}+O_M(\lambda^{-1}),
 \qquad
 u=z+O_M(\lambda^{-1}).
\]
These are the explicit bounds on the two correction terms used below.

For LR, Taylor expansion in~\eqref{eq:new-lr} gives
\[
 \Phibar(r)
 =\Phibar(z)+\frac{z^2\phi(z)}{6\sqrt\lambda}
  +O_M(\lambda^{-1}),
 \qquad
 \phi(r)\left(\frac1u-\frac1r\right)
 =-\frac{\phi(z)}{6\sqrt\lambda}+O_M(\lambda^{-1}).
\]
Adding these terms yields
\[
 p_{\LR}(w)=\Phibar(z)
 +\frac{\phi(z)(z^2-1)}{6\sqrt\lambda}+O_M(\lambda^{-1}),
\]
which is \eqref{eq:poisson-like-lr}.

\end{proof}

\section{Detailed proofs of the supporting lemmas for
Theorem~\ref{thm:spacrt-validity}}
\label{sec:spacrt-detailed-proofs}

\subsection{Supporting lemmas for Theorem~\ref{thm:spacrt-validity}:
fitted-target validity}
\label{sec:spacrt-supporting-proofs}

\subsubsection{Proof of Lemma~\ref{lem:spacrt-cb-consequences}}

\begin{proof}
Taking $k=3$ gives the first inequality in \eqref{eq:spacrt-cb-third-variance}, while conditional Lyapunov's inequality gives $v^{3/2}\leq\E(|\xi|^3\mid\mathcal G)\leq3bv$, and hence the second, with the case $v=0$ immediate.

For the multiplier claim, put $a_n\equiv\max_i|a_i|$.  Conditional
independence and centering are preserved, and, for every $k\geq3$,
\begin{align*}
  \E(|W_i|^k\mid\mathcal G)\leq\frac{k!}{2}b_n^{k-2}|a_i|^k v_i\leq\frac{k!}{2}(b_na_n)^{k-2}a_i^2v_i=\frac{k!}{2}(b_na_n)^{k-2}\V(W_i\mid\mathcal G).
\end{align*}
This is (CB) in~\eqref{eq:cb-property} with the stated scale.
\end{proof}

\subsubsection{Proof of Lemma~\ref{lem:spacrt-oracle-score-transfer}}

\begin{proof}
By~\eqref{eq:spacrt-oracle-score}, $U_{n,\dCRT}$ is the sum of the
products $e^0_{x,i}e^0_{y,i}$.  Because the triples
$(\ex{i},\ey{i},\rz{i})$ are independent across $i$ and
$\ex{}\independent\ey{}\mid\rz{}$ under the null, these products are,
conditionally on $\mathcal Z_n$, independent and centered, with total
conditional variance
\[
  \sum_{i=1}^n
  \E\{(e^0_{x,i})^2\mid\rz{i}\}
  \E\{(e^0_{y,i})^2\mid\rz{i}\}
  =
  n\sigma_{n,\dCRT}^2.
\]
Conditional independence, the two true-residual Bernstein bounds of
Assumption~\ref{assu:spacrt-regularity}(ii), and
\eqref{eq:spacrt-cb-third-variance}, applied conditionally on
$\mathcal Z_n$, give
\[
  \frac{
    \sum_{i=1}^n
    \E(|e^0_{x,i}e^0_{y,i}|^3\mid\mathcal Z_n)
  }{
    n^{3/2}\sigma_{n,\dCRT}^3
  }
  \leq
  \frac{
    9b^0_{n,x}b^0_{n,y}
    \sum_i v^0_{x,i}v^0_{y,i}}
  {n^{3/2}\sigma_{n,\dCRT}^3}
  =
  \frac{9b^0_{n,x}b^0_{n,y}}
       {\sqrt n\sigma_{n,\dCRT}}
  \convp0.
\]
The conditional Berry--Esseen inequality and
Assumption~\ref{assu:spacrt-regularity}(ii) therefore yield
\begin{equation*}
  \sup_{t\in\R}
  \left|
    \P\left(
      \left.
      \frac{U_{n,\dCRT}}{\sqrt n\sigma_{n,\dCRT}}\leq t
      \ \right|\ \mathcal Z_n
    \right)
    -\Phi(t)
  \right|
  \convp0.
\end{equation*}
On $\{\sigma_{n,\dCRT}>0\}$ the displayed supremum is bounded by a
universal constant times the preceding third-moment ratio, while the
complement has probability tending to zero.  Since the supremum is at
most one, boundedness and convergence in probability imply convergence
in $L^1$; integration of the conditional law then gives
$U_{n,\dCRT}/(\sqrt n\sigma_{n,\dCRT})\convd N(0,1)$.

It remains to transfer this limit to the fitted score.  The exact
identity
\begin{equation}
  S_n-U_{n,\dCRT}
  =
  -\sum_{i=1}^n e^0_{x,i}\Delta_{y,i}
  -\sum_{i=1}^n e^0_{y,i}\Delta_{x,i}
  +\sum_{i=1}^n\Delta_{x,i}\Delta_{y,i}
  \label{eq:spacrt-score-decomposition}
\end{equation}
holds without approximation.  The separate-fit measurability and
conditional independence under the null give, for
$(r,s,\mathcal H_s)=(x,y,\mathcal Y_n)$ or
$(y,x,\mathcal X_n)$,
\[
  \E\left(
    \left.\sum_i e^0_{r,i}\Delta_{s,i}\ \right|\ \mathcal H_s
  \right)=0,
  \qquad
  \V\left(
    \left.\sum_i e^0_{r,i}\Delta_{s,i}\ \right|\ \mathcal H_s
  \right)
  =
  n(E'_{n,s})^2.
\]
Thus, for either pair and every fixed $\varepsilon>0$, conditional
Chebyshev's inequality gives
\begin{align*}
  \P\left(
    \left|\sum_i e^0_{r,i}\Delta_{s,i}\right|
    >\varepsilon\sqrt n\sigma_{n,\dCRT}
  \right)\quad\leq
  \P(\sigma_{n,\dCRT}=0)
  +\E\left[
    \min\left\{
      1,
      \frac{(E'_{n,s})^2}
           {\varepsilon^2\sigma_{n,\dCRT}^2}
    \right\}
    \indicator\{\sigma_{n,\dCRT}>0\}
  \right]
  \longrightarrow0.
\end{align*}
The integrand is bounded and converges to zero in probability by
clause~\textup{(i)} of Assumption~\ref{assu:spacrt-regularity}, so its
expectation vanishes.  Hence both linear nuisance terms are
$o_{\P_n}(\sqrt n\sigma_{n,\dCRT})$.  Finally,
Cauchy--Schwarz gives
\[
  \left|\sum_i\Delta_{x,i}\Delta_{y,i}\right|
  \leq
  nE_{n,x}E_{n,y}
  =
  o_{\P_n}(\sqrt n\sigma_{n,\dCRT}).
\]
Substitution into~\eqref{eq:spacrt-score-decomposition} proves the
second assertion in~\eqref{eq:spacrt-oracle-score-transfer}.
\end{proof}

\subsubsection{Proof of Lemma~\ref{lem:spacrt-variance-matching}}

\begin{proof}
Put
$T_n\equiv\sum_i\widehat v_{x,i}\{(e^0_{y,i})^2-v^0_{y,i}\}$.
Under the null, conditionally on $\mathcal X_n$, its summands are
independent and centered, and the $\rz{i}$-conditional response-moment
bounds remain valid.  The conditional Bernstein bound with $k=4$
and~\eqref{eq:spacrt-cb-third-variance} give
\[
  \E\left[
    \{(e^0_{y,i})^2-v^0_{y,i}\}^2
    \mid\rz{i}
  \right]
  \leq
  \E(|e^0_{y,i}|^4\mid\rz{i})
  \leq12(b^0_{n,y})^2v^0_{y,i},
  \qquad
  \widehat v_{x,i}^2
  \leq9\widehat b_{n,x}^2\widehat v_{x,i}.
\]
Consequently,
\[
  \V(T_n\mid\mathcal X_n)
  \leq
  108\widehat b_{n,x}^2(b^0_{n,y})^2
  \sum_i\widehat v_{x,i}v^0_{y,i}.
\]
For every fixed $\varepsilon>0$, conditional Chebyshev's inequality
therefore bounds $\P(|T_n|>\varepsilon n\sigma_{n,\dCRT}^2)$ by
\[
  \P(\sigma_{n,\dCRT}=0)
  +\E\left[
    \min\left\{
      1,
      \frac{
        108\widehat b_{n,x}^2(b^0_{n,y})^2
        \sum_i\widehat v_{x,i}v^0_{y,i}}
      {\varepsilon^2n^2\sigma_{n,\dCRT}^4}
    \right\}
    \indicator\{\sigma_{n,\dCRT}>0\}
  \right].
\]
By the variance-calibration condition in
Assumption~\ref{assu:spacrt-regularity}(i), the ratio inside the minimum is
\[
  O_{\P_n}\left[
    \left\{
      \frac{\widehat b_{n,x}b^0_{n,y}}
           {\sqrt n\sigma_{n,\dCRT}}
    \right\}^2
  \right]
  =o_{\P_n}(1)
\]
by clause~\textup{(ii)} of Assumption~\ref{assu:spacrt-regularity}. Boundedness plus convergence in probability makes its expectation vanish, and $\P(\sigma_{n,\dCRT}=0)\to0$, so $T_n=o_{\P_n}(n\sigma_{n,\dCRT}^2)$. It follows that
\begin{align}
  \sum_{i=1}^n\widehat v_{x,i}(e^0_{y,i})^2
  &=
  \sum_{i=1}^n\widehat v_{x,i}v^0_{y,i}
  +o_{\P_n}(n\sigma_{n,\dCRT}^2)\notag\\
  &=
  n\sigma_{n,\dCRT}^2
  +\sum_{i=1}^n
  (\widehat v_{x,i}-v^0_{x,i})v^0_{y,i}
  +o_{\P_n}(n\sigma_{n,\dCRT}^2)\notag\\
  &=
  n\sigma_{n,\dCRT}^2\{1+o_{\P_n}(1)\}.
  \label{eq:spacrt-oracle-residual-square}
\end{align}
For the fitted response residuals,
\begin{align*}
  nV_{n,\dCRT}^2=\sum_i\widehat v_{x,i}(e^0_{y,i}-\Delta_{y,i})^2=\sum_i\widehat v_{x,i}(e^0_{y,i})^2-2\sum_i\widehat v_{x,i}e^0_{y,i}\Delta_{y,i}+\sum_i\widehat v_{x,i}\Delta_{y,i}^2.
\end{align*}
The last term equals $n(\widehat E'_{n,y})^2$ and is
$o_{\P_n}(n\sigma_{n,\dCRT}^2)$ by
Assumption~\ref{assu:spacrt-regularity}(i).  By Cauchy--Schwarz, the
absolute value of the middle sum is at most
\[
  \left\{
    \sum_i\widehat v_{x,i}(e^0_{y,i})^2
  \right\}^{1/2}
  \left\{
    n(\widehat E'_{n,y})^2
  \right\}^{1/2}
  =
  o_{\P_n}(n\sigma_{n,\dCRT}^2),
\]
where~\eqref{eq:spacrt-oracle-residual-square} controls the first factor.
Combining these bounds proves~\eqref{eq:spacrt-variance-matching}.
\end{proof}

\subsubsection{Proof of Lemma~\ref{lem:spacrt-resampling-clt}}

\begin{proof}
For the resampling summands
$W_{in}=\widehat e_{y,i}(\widetilde X_i-\widehat\mu_{x,i})$
in~\eqref{eq:dcrt-resampling-summand}, conditionally on $\mathcal F_n$ the
variables $W_{in}$ are independent and centered, with total variance
$nV_{n,\dCRT}^2$.  Put
$\bar b_n\equiv\widehat b_{n,x}\max_i|\widehat e_{y,i}|$.  The multiplier
scale~\eqref{eq:spacrt-cb-multiplier-scale} and the $k=3$
bound~\eqref{eq:spacrt-cb-third-variance} give
\[
  \sum_i\widetilde\E(|W_{in}|^3\mid\mathcal F_n)
  \leq
  3\bar b_n\sum_i\widetilde\E(W_{in}^2\mid\mathcal F_n)
  =3\bar b_n nV_{n,\dCRT}^2.
\]
Thus, on $\{V_{n,\dCRT}>0\}$, conditional Berry--Esseen gives
\[
  \sup_{t\in\R}
  \left|
    \widetilde \P\left(
      \left.
      \frac{\sum_iW_{in}}{\sqrt nV_{n,\dCRT}}\leq t
      \ \right|\ \mathcal F_n
    \right)
    -\Phi(t)
  \right|
  \leq
  3C\frac{\bar b_n}{\sqrt nV_{n,\dCRT}}
  \convp0
\]
by Assumption~\ref{assu:spacrt-regularity}(iii), proving
\eqref{eq:spacrt-conditional-resampling-clt}.

It remains only to account for the weak inequality in the upper tail.
Let $F_n^*$ be the conditional distribution function of
$\sum_iW_{in}/(\sqrt nV_{n,\dCRT})$ and let
$\Delta_n\equiv\sup_t|F_n^*(t)-\Phi(t)|$.  For every fixed $h>0$,
\[
  \sup_t\widetilde \P\left(
    \left.
    \frac{\sum_iW_{in}}{\sqrt nV_{n,\dCRT}}=t
    \ \right|\ \mathcal F_n
  \right)
  \leq
  2\Delta_n+\sup_{t\in\R}\{\Phi(t)-\Phi(t-h)\}.
\]
First letting $n\to\infty$ and then $h\downarrow0$ shows that the
largest conditional atom is $o_{\P_n}(1)$.  A non-strict upper survival
probability differs from $1-F_n^*(t)$ by that atom, proving the asserted
uniform survival-function convergence.
\end{proof}

\section{High-dimensional nuisance fits}
\label{sec:spacrt-high-dimensional}

We give a separate high-dimensional example with two binary variables
and fully in-sample logistic-lasso nuisance fits. Let
$\bm Z=(1,\bm U^{\T})^{\T}\in\R^{p_n}$. Under the
baseline null, suppose $X\independent Y\mid \bm Z$, so that the two
conditionals below determine the joint law of $(X,Y)$ given $\bm Z$, and
\begin{equation}\label{eq:spacrt-logistic-association-model}
\begin{aligned}
 X\mid \bm Z&\sim\Ber\{\mu^0_{n,x}(\bm Z)\},
 &Y\mid \bm Z&\sim\Ber\{\mu^0_{n,y}(\bm Z)\},\\
 \mu^0_{n,j}(\bm Z)&=\expit(\alpha_{n,j}+\bm U^{\T}\bm{\beta}_{n,j}),
 &j&\in\{x,y\}.
\end{aligned}
\end{equation}
Write $\pi_{n,j}=\expit(\alpha_{n,j})$ and
$s_{n,j}=1+\|\bm{\beta}_{n,j}\|_0$ for $j\in\{x,y\}$. For the given
constants $K,B,B_{\alpha}<\infty$ and $c_\Sigma>0$, independent of $n$, the
envelope conditions are
\begin{align}
 &\|\bm U\|_\infty\le K\quad\text{a.s.};
 &&\tag{L1}\label{cond:spacrt-lasso-design-bound}\\
 &\|\bm{\beta}_{n,j}\|_1\le B,\quad \alpha_{n,j}\le B_{\alpha},\quad j\in\{x,y\};
 &&\tag{L2}\label{cond:spacrt-lasso-coefficients}\\
 &\lambda_{\min}\{\E(\bm Z\bm Z^\top)\}\ge c_\Sigma.
 &&\tag{L3}\label{cond:spacrt-lasso-eigenvalue}
\end{align}
The regime conditions are $p_n\to\infty$ together with the rate conditions
\begin{equation}\label{eq:spacrt-logistic-lasso-rates}
 n\pi_{n,x}\pi_{n,y}\to\infty,
 \qquad
 \frac{s_{n,x}^2\log p_n}{n\pi_{n,x}}
 +\frac{s_{n,y}^2\log p_n}{n\pi_{n,y}}\to0,
 \qquad
 \frac{s_{n,x}s_{n,y}\log^2p_n}{n}\to0.
\end{equation}
Fit the two logistic regressions separately on the same observations, omitting
$X$ from the response fit, leaving both intercepts unpenalized---the explicit
penalized program is~\eqref{eq:spacrt-insample-logistic-fit} below---and using
\[
 \lambda_{n,j}
 =A\left\{\left(\frac{\pi_{n,j}\log p_n}{n}\right)^{1/2}
 +\frac{\log p_n}{n}\right\}
\]
for a sufficiently large fixed $A$. This prevalence-dependent penalty is part
of the theoretical construction, not an empirical tuning recommendation. If
either penalized objective has no finite minimizer, both reported p-values
are one; the zero coefficient vector is only an auxiliary extension for
proof quantities, as in Supplementary Section~\ref{sec:spacrt-logistic-estimator}.
Under alternatives, replace the
response logit by
$\logit\mu^0_{n,y}(\bm Z)+\rho_nX$ and write $\P_n$ for the
resulting joint law of the sample ($\P_n^0$ denoting the baseline
null law); the remaining regime conditions are the
vanishing-alternative conditions
\begin{equation}\label{eq:spacrt-logistic-local-scale}
 \rho_n\to0,
 \qquad
 \rho_n^2\pi_{n,x}^2
 \left(\frac{n\pi_{n,y}}{\log p_n}\right)^{1/2}\to0.
\end{equation}
Under $\P_n$ the restricted $Y$-on-$\bm Z$ logistic lasso is
misspecified: the conditional response mean acquires a perturbation of
relative size $O(\pi_{n,x}|\rho_n|)$. The second condition in
\eqref{eq:spacrt-logistic-local-scale} makes the resulting excess $\ell_1$
error negligible. It is implied by the effective-local scale
$|\rho_n|\sqrt{n\pi_{n,x}\pi_{n,y}}=O(1)$, but also permits larger signals; when
$\pi_{n,x}$ and $\pi_{n,y}$ are constant it reads
$|\rho_n|=o\{(\log p_n/n)^{1/4}\}$.

For deterministic $\delta_n\to\infty$, write $\PLL_n=\PLL_n(K,B,B_{\alpha},c_\Sigma,A)$ for the class of
data-generating processes---dimension $p_n$, covariate law, nuisance
parameters $(\alpha_{n,x},\alpha_{n,y},\bm{\beta}_{n,x},\bm{\beta}_{n,y})$ with
sparsities $(s_{n,x},s_{n,y})$, and association $\rho_n$---defined by
\[
\begin{aligned}
  \PLL_n
  =\Bigl\{\P_n:\;
  &\text{\textup{(L1)}--\textup{(L3)} hold with these
  }(K,B,B_{\alpha},c_\Sigma),\ \text{penalty constant }A;\\
  &p_n\ge\delta_n,\quad n\pi_{n,x}\pi_{n,y}\ge\delta_n,\quad
  |\rho_n|\le\frac1{\delta_n},\quad
  \rho_n^2\pi_{n,x}^2\Bigl(\frac{n\pi_{n,y}}{\log p_n}\Bigr)^{1/2}
  \le\frac1{\delta_n},\\
  &\frac{s_{n,x}^2\log p_n}{n\pi_{n,x}}
  +\frac{s_{n,y}^2\log p_n}{n\pi_{n,y}}\le\frac1{\delta_n},\quad
  \frac{s_{n,x}s_{n,y}\log^2p_n}{n}\le\frac1{\delta_n}
  \Bigr\}.
\end{aligned}
\]
Its null slice is $\PLLnull_n=\{\P_n\in\PLL_n:\rho_n=0\}$.

\begin{theorem}[Logistic-lasso \spacrt{} under
vanishing alternatives]
\label{thm:new-lasso}\label{thm:spacrt-logistic-lasso-local-alternative}
Uniformly over $\P_n\in\PLL_n$, the nuisance fits and saddlepoint exist with
probability tending to one.  Moreover,
\begin{equation}\label{eq:spacrt-logistic-local-relative-error}
 \frac{p_{\infty,n}^{\dCRT}}{p_{\mathrm{SPA},n}^{\dCRT}}-1
 =\OP{\PLL}\{(n\pi_{n,x}\pi_{n,y})^{-1/2}+|\rho_n|\}.
\end{equation}
\end{theorem}

Under the null $\rho_n=0$, both $p_{\infty,n}^{\dCRT}$ and $p_{\mathrm{SPA},n}^{\dCRT}$
are uniformly asymptotically valid over the null slice:
\begin{equation}\label{eq:lasso-null-uniformity}
 \sup_{\P_n\in\PLLnull_n}\sup_{t\in[0,1]}
 |\P_n(p_{\infty,n}^{\dCRT}\le t)-t|\longrightarrow0,
 \qquad
 \sup_{\P_n\in\PLLnull_n}\sup_{t\in[0,1]}
 |\P_n(p_{\mathrm{SPA},n}^{\dCRT}\le t)-t|\longrightarrow0.
\end{equation}
(Supplementary Section~\ref{sec:spacrt-logistic-null-validity}).
The squared-sparsity terms in~\eqref{eq:spacrt-logistic-lasso-rates} localize the
two rare-event regressions; the product term controls the interaction of their
in-sample errors. Thus the same effective-information rate survives fully
in-sample, high-dimensional nuisance estimation without sample splitting.
General logistic-lasso oracle inequalities are available under standard
high-dimensional conditions \citep{vandeGeer2008,Kwemou2016}, and rare-event
logistic theory identifies the number of events as the information scale
\citep{Wang2020RareEvents}. The contribution here is the
joint same-sample verification for two nuisance fits with diverging negative
intercepts, relative fitted-probability control, and coupling to both the
fitted \dCRT{} target and its saddlepoint approximation.

\subsection{Proof of Theorem~\ref{thm:spacrt-logistic-lasso-local-alternative}: logistic-lasso spaCRT under vanishing alternatives}
\label{sec:proof-spacrt-logistic-lasso}
\mbox{}\par

The supporting lemmas are stated at the point where they enter the argument; all of their proofs are deferred to Supplementary Section~\ref{sec:spacrt-logistic-supporting-proofs}.

\subsubsection{Model notation and envelope bounds}
\label{sec:spacrt-logistic-model-setup}

Throughout this subsection and Supplementary Section~\ref{sec:spacrt-logistic-supporting-proofs}, the \emph{sequence conditions of Theorem~\ref{thm:spacrt-logistic-lasso-local-alternative}} are the envelope conditions \textup{(L1)}--\textup{(L3)} with the lasso fits at the penalty constant $A$, together with the limit form of the regime conditions: $p_n\to\infty$, the rate conditions~\eqref{eq:spacrt-logistic-lasso-rates}, and the vanishing-alternative conditions~\eqref{eq:spacrt-logistic-local-scale}. Every array with $\P_n\in\PLL_n$ for all $n$ satisfies them, since $\delta_n\to\infty$; the upgrade to the class-uniform forms of the theorem is by the diagonalization lemmas of Supplementary Section~\ref{sec:uniform-null-diagonal}.

For $j\in\{x,y\}$, write $S_{n,j}\equiv\{0\}\cup\operatorname{supp}(\bm\beta_{n,j})$, $s_{n,j}\equiv|S_{n,j}|$, and
\[
  v^0_{j,i}
  \equiv
  \mu^0_{n,j}(\rz{i})\{1-\mu^0_{n,j}(\rz{i})\}.
\]
Here and in Supplementary Section~\ref{sec:spacrt-logistic-supporting-proofs}, the
oracle quantities of Supplementary Section~\ref{sec:spacrt-notation} are instantiated
at the logistic model, with $\mu^0_{j,i}=\mu^0_{n,j}(\rz{i})$ and
$v^0_{j,i}$ as above; in particular, the error functionals $E_{n,j}$,
$E'_{n,x}$, $E'_{n,y}$, and $\widehat E'_{n,y}$ and the scales
$\sigma_{n,\dCRT}$ and $V_{n,\dCRT}$ defined there are evaluated at
the lasso fits constructed in
Supplementary Section~\ref{sec:spacrt-logistic-estimator}.
Here $K$, $B$, $B_{\alpha}$, and $c_\Sigma$ are the fixed envelope constants
of conditions \textup{(L1)}--\textup{(L3)} in Supplementary Section~\ref{sec:spacrt-high-dimensional}: $K$
bounds the design, $B$ the coefficient $\ell_1$-norms, $B_{\alpha}$ the
intercepts $\alpha_{n,j}$, and $c_\Sigma$ the design eigenvalues.  The
design bound~\eqref{cond:spacrt-lasso-design-bound} and the
coefficient bounds~\eqref{cond:spacrt-lasso-coefficients} imply, by
H\"older's inequality
$|\bm u^{\T}\bm\beta_{n,j}|
\leq\|\bm u\|_\infty\|\bm\beta_{n,j}\|_1$,
the elementary ratio bound
$e^{-|h|}\leq\expit(u+h)/\expit(u)\leq e^{|h|}$,
and monotonicity of $\expit$,
\begin{equation}
  |\bm u^{\T}\bm\beta_{n,j}|\leq KB,
  \qquad
  e^{-KB}\pi_{n,j}
  \leq
  \mu^0_{n,j}(\bm z)
  \leq
  e^{KB}\pi_{n,j},
  \qquad
  \mu^0_{n,j}(\bm z)
  \leq
  \expit(B_{\alpha}+KB)<1,
  \label{eq:spacrt-logistic-intercept-envelope}
\end{equation}
uniformly over $\bm z=(1,\bm u^{\T})^{\T}$ in the support of $\rz{}$.
In particular,
$v^0_{j,i}\asymp \pi_{n,j}$ uniformly in $i$ and $j$.

\subsubsection{The in-sample logistic-lasso estimator}
\label{sec:spacrt-logistic-estimator}

For $j\in\{x,y\}$ and generic binary responses $A_{j,i}\in\{0,1\}$
observed alongside $\rz{i}$, the in-sample fit is the
$\ell_1$-penalized logistic regression of $A_{j,\cdot}$ on
$\rz{\cdot}$, with the intercept left unpenalized:
\begin{equation}
  \operatorname*{arg\,min}_{(a,\bm b)\in\R\times\R^{p_n-1}}
  \left[
    \frac1n\sum_{i=1}^n
    \left\{
      \log(1+e^{a+\bm U_i^{\T}\bm b})
      -A_{j,i}(a+\bm U_i^{\T}\bm b)
    \right\}
    +\lambda_{n,j}\|\bm b\|_1
  \right],
  \label{eq:spacrt-insample-logistic-fit}
\end{equation}
where
$\lambda_{n,j}\equiv A\{(\pi_{n,j}\log p_n/n)^{1/2}+\log p_n/n\}$ for a
sufficiently large fixed constant $A$.  Whenever the argmin is
nonempty, choose a measurable element
$\widehat{\bm\theta}_{n,j}$; otherwise set
$\widehat{\bm\theta}_{n,j}=\bm0$ as an auxiliary extension for proof quantities.
On an empty-argmin event the reported exact and SPA p-values are both one;
they are not computed from this zero-vector extension. On nonempty-argmin
events the estimator and procedure are unchanged. Define
$\widehat\mu_{j,i}\equiv\expit(\rz{i}^{\T}\widehat{\bm\theta}_{n,j})$ and,
with $\bm\theta^0_{n,j}\equiv(\alpha_{n,j},\bm\beta_{n,j}^{\T})^{\T}$ the
baseline coefficient vector, so that
$\mu^0_{n,j}(\rz{i})=\expit(\rz{i}^{\T}\bm\theta^0_{n,j})$, write
\[
  \widehat{\bm\delta}_{n,j}\equiv\widehat{\bm\theta}_{n,j}-\bm\theta^0_{n,j},
  \qquad
  h_{j,i}\equiv\rz{i}^{\T}\widehat{\bm\delta}_{n,j},
  \qquad
  \widehat v_{j,i}\equiv\widehat\mu_{j,i}(1-\widehat\mu_{j,i}).
\]
For the two fits in Theorem~\ref{thm:spacrt-logistic-lasso-local-alternative}, take
\[
 A_{x,i}=X_i,\qquad A_{y,i}=Y_i.
\]
Neither the program nor these derived quantities presuppose a
distribution for the responses; the null and alternative analyses
below differ only in the conditional law they assign to $A_{j,i}$.

\subsubsection{Auxiliary logistic-lasso verification results}\label{sec:spacrt-logistic-verification}

We present several important lemmas that will be useful when proving Theorem~\ref{thm:new-lasso}. Before presenting the lemmas, we first introduce a sequence of useful notations. For observation $i$, abbreviate the oracle means of Supplementary Section~\ref{sec:spacrt-notation} and their $\rho$-tilt by
\[
  p_i\equiv\mu^0_{x,i}=\mu^0_{n,x}(\rz{i}),
  \qquad
  q_i\equiv\mu^0_{y,i}=\mu^0_{n,y}(\rz{i}),
  \qquad
  q_i(\rho)\equiv\expit\{\logit(q_i)+\rho\}.
\]
For $\rho\in\R$, let $\P_{n,\rho}$ denote the sample law $\P_n$ introduced in Supplementary Section~\ref{sec:spacrt-high-dimensional} preceding~\eqref{eq:spacrt-logistic-local-scale}, with $\rho_n$ replaced by $\rho$. Thus $\P_n=\P_{n,\rho_n}$ and $\P_n^0=\P_{n,0}$. Explicitly, under $\P_{n,\rho}$ the triples $\{(\ex{i},\ey{i},\rz{i}):1\leq i\leq n\}$ are independent and identically distributed across $i$, the pair $(\ex{i},\rz{i})$ is distributed as under the baseline model~\eqref{eq:spacrt-logistic-association-model}, and $\ey{i}\mid(\ex{i},\rz{i})\sim\Ber\{q_i(\rho\ex{i})\}$.  Thus the response logit is $\logit(q_i)+\rho\ex{i}$.  At $\rho=0$ this reduces to the baseline model, so $\P_{n,0}$ is the null law, under which $\ex{}\independent\ey{}\mid\rz{}$.  Since $\ex{i}\in\{0,1\}$, the conditional response mean is exactly
\begin{equation}
  \E_{n,\rho}(\ey{i}\mid\ex{i},\rz{i})
  =
  q_i+\ex{i}\Delta_i(\rho),
  \qquad
  \Delta_i(\rho)\equiv q_i(\rho)-q_i.
  \label{eq:spacrt-logistic-alt-conditional-mean}
\end{equation}

Now we introduce the first result.

\begin{lemma}[In-sample prevalence-adaptive logistic-lasso bounds]
\label{lem:spacrt-logistic-lasso}
\label{lem:spacrt-insample-logistic-oracle}
Retain the design and coefficient
conditions~\eqref{cond:spacrt-lasso-design-bound}--\eqref{cond:spacrt-lasso-eigenvalue},
$p_n\to\infty$, and the rate
conditions~\eqref{eq:spacrt-logistic-lasso-rates}, all stated before
Theorem~\ref{thm:spacrt-logistic-lasso-local-alternative}, and fit by
the estimator of Supplementary Section~\ref{sec:spacrt-logistic-estimator}.
\begin{enumerate}[label=\textup{(\alph*)}]
\item\label{part:spacrt-oracle} Fix $j\in\{x,y\}$ and suppose that,
conditionally on $\mathcal Z_n$, the responses $A_{j,i}\in\{0,1\}$ are
independent with means
\begin{equation*}
  m_{j,i}
  \equiv
  \E(A_{j,i}\mid\rz{i})
  =
  \mu^0_{n,j}(\rz{i})+w_{j,i},
  \qquad
  \omega_{n,j}
  \equiv
  \left(\frac1n\sum_{i=1}^nw_{j,i}^2\right)^{1/2},
\end{equation*}
so that for constants $0<c_m\leq C_m$ and almost
surely uniformly in $i$,
\begin{equation}
  c_m\pi_{n,j}\leq m_{j,i}\leq C_m\pi_{n,j},
  \qquad
  m_{j,i}\leq\expit(B_{\alpha}+KB+1)<1,
  \label{eq:spacrt-logistic-perturbed-envelope}
\end{equation}
and that the localization scale
\begin{equation*}
  \mathfrak b_{n,j}
  \equiv
  \frac{s_{n,j}\lambda_{n,j}}{\pi_{n,j}}
  +\frac{\sqrt{s_{n,j}}\,\omega_{n,j}}{\pi_{n,j}}
  +\frac{\omega_{n,j}^2}{\pi_{n,j}\lambda_{n,j}}
\end{equation*}
converges to zero in $\P_n$-probability.  In this lemma, for a positive,
possibly random rate $a_n$, the notation $X_n=O_{\P_n}(a_n)$ means
$X_n/a_n=O_{\P_n}(1)$.  Then a finite minimizer
in~\eqref{eq:spacrt-insample-logistic-fit} exists with probability
tending to one, and
\begin{align}\label{eq:spacrt-logistic-oracle-bounds}
  \|\widehat{\bm\delta}_{n,j}\|_1
  &=O_{\P_n}(\mathfrak b_{n,j})=o_{\P_n}(1),\notag\\
  \frac1n\sum_{i=1}^nh_{j,i}^2
  &=O_{\P_n}\left(\frac{s_{n,j}\lambda_{n,j}^2}{\pi_{n,j}^2}+\frac{\omega_{n,j}^2}{\pi_{n,j}^2}
  \right),
  \qquad
  \max_{1\leq i\leq n}|h_{j,i}|=o_{\P_n}(1).
\end{align}
Consequently,
\begin{equation}
  \max_{1\leq i\leq n}
  \left|
    \frac{\widehat\mu_{j,i}}{\mu^0_{n,j}(\rz{i})}-1
  \right|
  +
  \max_{1\leq i\leq n}
  \left|
    \frac{\widehat v_{j,i}}{v^0_{j,i}}-1
  \right|
  =o_{\P_n}(1),
  \label{eq:spacrt-logistic-relative-consistency}
\end{equation}
and
\begin{equation}
  \frac1n\sum_{i=1}^n
  \{\widehat\mu_{j,i}-\mu^0_{n,j}(\rz{i})\}^2
  =
  O_{\P_n}\left(
    \frac{\pi_{n,j}s_{n,j}\log p_n}{n}
    +\omega_{n,j}^2
  \right).
  \label{eq:spacrt-logistic-mean-square-rate}
\end{equation}
\item\label{part:spacrt-null-verification} Under $\P_{n,0}$, retain the
baseline model~\eqref{eq:spacrt-logistic-association-model} and consider the estimators defined in Supplementary Section~\ref{sec:spacrt-logistic-estimator}. Then, for $j\in\{x,y\}$,
\[
  E_{n,j}^2
  =
  O_{\P_{n,0}}\left(\frac{\pi_{n,j}s_{n,j}\log p_n}{n}\right),
  \qquad
  \max_{1\leq i\leq n}
  \left|
    \frac{\widehat\mu_{j,i}}{\mu^0_{n,j}(\rz{i})}-1
  \right|
  \convp0.
\]
Moreover, Assumption~\ref{assu:spacrt-regularity} holds with
$b^0_{n,x}=b^0_{n,y}=\widehat b_{n,x}=1$, with
$\sigma_{n,\dCRT}^2\asymp \pi_{n,x}\pi_{n,y}$ and
$V_{n,\dCRT}^2/\sigma_{n,\dCRT}^2\convp1$.
\end{enumerate}
\end{lemma}

\begin{remark}[Zero-bias and biased instantiations]
\label{rem:spacrt-zero-bias}
The correctly specified fit is the zero-bias specialization
$w_{j,i}\equiv0$ of part~\ref{part:spacrt-oracle}, in which
$\omega_{n,j}=0$ and
$\mathfrak b_{n,j}=s_{n,j}\lambda_{n,j}/\pi_{n,j}$.  Both fits under
$\P_{n,0}$---whence part~\ref{part:spacrt-null-verification}---and the
$X$-fit under every $\P_{n,\rho}$ are of that kind; only the $Y$-fit
under an alternative carries $\omega_{n,y}>0$.
\end{remark}

\begin{remark}[Envelope-built constants and tail levels]
\label{rem:spacrt-envelope-constants}
Throughout the alternative-law analysis, constants may depend on the
fixed bounds $K$, $B$, $B_{\alpha}$, $c_\Sigma$, and the penalty constant $A$,
and $|\rho_n|\leq1$ may be assumed
because $\rho_n\to0$
under~\eqref{eq:spacrt-logistic-local-scale}.  For the uniformity
clause of the theorem, every stochastic bound in this module and in
Supplementary Section~\ref{sec:spacrt-logistic-supporting-proofs} has
envelope-built constants: its constant and tail level depend only on
$(K,B,B_{\alpha},c_\Sigma,A)$, and every ``for all sufficiently large $n$''
onset is a deterministic function of the sequence's trajectories
in \eqref{eq:spacrt-logistic-lasso-rates}
and~\eqref{eq:spacrt-logistic-local-scale}.
\end{remark}

The next two lemmas replace the null verification lemmas under the alternative laws.

\begin{lemma}[In-sample logistic-lasso bounds under vanishing alternatives]
\label{lem:spacrt-logistic-alt-fit}
Retain the sequence conditions of
Theorem~\ref{thm:spacrt-logistic-lasso-local-alternative}, including
the vanishing-alternative
conditions~\eqref{eq:spacrt-logistic-local-scale}, and work under
$\P_n$.
\begin{enumerate}[label=\textup{(\roman*)}]
\item The joint law of $\{(\ex{i},\rz{i}):1\leq i\leq n\}$ is the same
under $\P_n$ as under $\P_{n,0}$, and
part~\ref{part:spacrt-oracle} of
Lemma~\ref{lem:spacrt-insample-logistic-oracle} applies to the
$X$-fit with $w_{x,i}\equiv0$.  Consequently every $j=x$ conclusion of
that lemma---existence of the fit, the
bounds~\eqref{eq:spacrt-logistic-oracle-bounds}, the relative
consistency~\eqref{eq:spacrt-logistic-relative-consistency}, and the
mean-square rate~\eqref{eq:spacrt-logistic-mean-square-rate}---holds
verbatim under $\P_n$, with $\omega_{n,x}=0$.
\item part~\ref{part:spacrt-oracle} of
Lemma~\ref{lem:spacrt-insample-logistic-oracle} applies to the
$Y$-fit with $w_{y,i}=p_i\Delta_i(\rho_n)$ and
$\omega_{n,y}\leq C_2\pi_{n,x}\pi_{n,y}|\rho_n|$.  In particular a finite
minimizer of~\eqref{eq:spacrt-insample-logistic-fit} with $j=y$ exists
with probability tending to one, and
\begin{align*}
  \|\widehat{\bm\delta}_{n,y}\|_1
  &=
  O_{\P_n}\left(
    \frac{s_{n,y}\lambda_{n,y}}{\pi_{n,y}}
    +s_{n,y}^{1/2}\pi_{n,x}|\rho_n|
    +\frac{\pi_{n,x}^2\pi_{n,y}\rho_n^2}{\lambda_{n,y}}
  \right)
  =o_{\P_n}(1),
  \notag\\
  \frac1n\sum_{i=1}^nh_{y,i}^2
  &=
  O_{\P_n}\left(
    \frac{s_{n,y}\lambda_{n,y}^2}{\pi_{n,y}^2}
    +\pi_{n,x}^2\rho_n^2
  \right),
  \qquad
  \max_{1\leq i\leq n}|h_{y,i}|=o_{\P_n}(1).
\end{align*}
\item Consequently,
\[
  \max_{1\leq i\leq n}
  \left|\frac{\widehat\mu_{y,i}}{q_i}-1\right|
  +
  \max_{1\leq i\leq n}
  \left|\frac{\widehat v_{y,i}}{v^0_{y,i}}-1\right|
  =o_{\P_n}(1)
\]
and
\begin{equation*}
  E_{n,y}^2
  =
  \frac1n\sum_{i=1}^n(\widehat\mu_{y,i}-q_i)^2
  =
  O_{\P_n}\left(
    \frac{\pi_{n,y}s_{n,y}\log p_n}{n}
    +\pi_{n,x}^2\pi_{n,y}^2\rho_n^2
  \right).
\end{equation*}
\end{enumerate}
All the stochastic orders above are computed under $\P_n$.
\end{lemma}

\begin{lemma}[Score and variance coordinates under vanishing
alternatives]
\label{lem:spacrt-logistic-alt-coordinates}
Under the sequence conditions of
Theorem~\ref{thm:spacrt-logistic-lasso-local-alternative}, working
under $\P_n$,
\begin{equation}
  nA_{2n}
  =\sum_{i=1}^n\widehat v_{x,i}\widehat e_{y,i}^2
  \asymp_{\P_n}n\pi_{n,x}\pi_{n,y},
  \qquad
  |S_n|
  =O_{\P_n}\left\{
    (n\pi_{n,x}\pi_{n,y})^{1/2}
    +n\pi_{n,x}\pi_{n,y}|\rho_n|
  \right\}.
  \label{eq:spacrt-logistic-alt-coordinates}
\end{equation}
\end{lemma}

\subsubsection{Existence and quantitative relative accuracy}
\label{sec:proof-main-spacrt-logistic-lasso}

The proof of the theorem is carried out directly under $\P_n$, using the alternative-law Lemmas~\ref{lem:spacrt-logistic-alt-fit} and~\ref{lem:spacrt-logistic-alt-coordinates}.

\begin{proof}[of Theorem~\ref{thm:spacrt-logistic-lasso-local-alternative}]

All stochastic orders are computed under $\P_n$, along an arbitrary sequence satisfying the sequence conditions of the theorem. Lemma~\ref{lem:spacrt-logistic-alt-fit}(i)--(ii) shows that both nuisance fits exist with probability tending to one. The quantitative conclusions are an application of Corollary~\ref{thm:spacrt-approximation}, the observable specialization of Theorem~\ref{thm:conditional-lr}: conditionally on the data, the resampling summands are $W_{in}=\widehat e_{y,i}(\widetilde X_i-\widehat\mu_{x,i})$ with cutoff $t_n=S_n$, and $nA_{2n}=\sum_{i=1}^n\widehat v_{x,i}\widehat e_{y,i}^2$ is the total resampling variance (denoted by $V_n$ in Theorem~\ref{thm:conditional-lr}), as shown in Supplementary Section~\ref{sec:proof-spacrt-approximation}. Lemma~\ref{lem:spacrt-logistic-alt-coordinates} gives $nA_{2n}=\Theta_{\P_n}(n\pi_{n,x}\pi_{n,y})\to\infty$, so $\P_n(nA_{2n}>0)\to1$ and
\[
  \frac1{\sqrt{nA_{2n}}}
  =O_{\P_n}\{(n\pi_{n,x}\pi_{n,y})^{-1/2}\}.
\]
Since $\ey{i}$ and $\widehat\mu_{y,i}$ lie in $[0,1]$ and each
$\widetilde X_i$ is Bernoulli, the support-diameter hypothesis of
Corollary~\ref{thm:spacrt-approximation} holds with $D_0=1$, and
$|\widehat e_{y,i}|\le1$ implies $nA_{3n}\le nA_{2n}$ and
$e_n^\star\le1$, and therefore
\[
  \frac{B_n}{\sqrt n}
  \le (nA_{2n})^{-1/2}
  =O_{\P_n}\{(n\pi_{n,x}\pi_{n,y})^{-1/2}\}.
\]
The cutoff bound in~\eqref{eq:spacrt-logistic-alt-coordinates} gives
\[
  Z_n=\frac{|S_n|}{\sqrt{nA_{2n}}}
  =O_{\P_n}
  \left\{
    \frac{(n\pi_{n,x}\pi_{n,y})^{1/2}+n\pi_{n,x}\pi_{n,y}|\rho_n|}
         {(n\pi_{n,x}\pi_{n,y})^{1/2}}
  \right\}
  =O_{\P_n}\{1+|\rho_n|\sqrt{n\pi_{n,x}\pi_{n,y}}\},
\]
which is the second coordinate bound.  Combining the last two
displays,
\[
  \frac{B_n(1+Z_n)}{\sqrt n}
  \le\frac{1+Z_n}{\sqrt{nA_{2n}}}
  =O_{\P_n}\{(n\pi_{n,x}\pi_{n,y})^{-1/2}+|\rho_n|\}
  =o_{\P_n}(1),
\]
and, since $D_0=1$ and $e_n^\star\le1$,
\[
  D_n=\frac{e_n^\star|S_n|}{nA_{2n}}
  \le\frac{Z_n}{\sqrt{nA_{2n}}}
  =O_{\P_n}\{(n\pi_{n,x}\pi_{n,y})^{-1/2}+|\rho_n|\}
  =o_{\P_n}(1),
\]
because $n\pi_{n,x}\pi_{n,y}\to\infty$ and $\rho_n\to0$.
Together with $\P_n(nA_{2n}>0)\to1$, the preceding two
$o_{\P_n}(1)$ bounds verify with probability tending to one
the three event conditions $nA_{2n}>0$,
$B_n(1+Z_n)/\sqrt n\le c_\star$, and $D_n\le c_\star$ in
Corollary~\ref{thm:spacrt-approximation}.  Then we have
\[
  \left|\frac{p_{\infty,n}^{\dCRT}}{p_{\mathrm{SPA},n}^{\dCRT}}-1\right|
  =O_{\P_n}\{(n\pi_{n,x}\pi_{n,y})^{-1/2}+|\rho_n|\},
\]
which is~\eqref{eq:spacrt-logistic-local-relative-error}, together with saddlepoint existence and $p_{\mathrm{SPA},n}^{\dCRT}>0$ with probability tending to one. The ratio uses the shared exceptional-set convention stated at the start of the proofs.  This establishes the sequence-wise form of the quantitative conclusions of Theorem~\ref{thm:spacrt-logistic-lasso-local-alternative}.  Every constant and tail level above is envelope-built (Remark~\ref{rem:spacrt-envelope-constants}), so the $\OP{\PLL}$ form follows by Corollary~\ref{cor:op-sup-seq-use}.
Applying the same corollary to the indicator of the complement of the
joint existence event, with deterministic rate $n^{-1}$, gives the
asserted class-uniform existence probability.
\end{proof}

\subsubsection{Null validity}
\label{sec:spacrt-logistic-null-validity}

Let $\P_n\in\PLLnull_n$ for each $n$.  We show that $p_{\infty,n}^{\dCRT}\convd\Unif(0,1), p_{\mathrm{SPA},n}^{\dCRT}\convd\Unif(0,1)$. Also, both limits are uniform over the null models in the sense of~\eqref{eq:lasso-null-uniformity}.

Part~\ref{part:spacrt-null-verification} of
Lemma~\ref{lem:spacrt-logistic-lasso} verifies
Assumption~\ref{assu:spacrt-regularity}.  Under the null, the
feasible-score CLT~\eqref{eq:spacrt-feasible-score-clt} gives
\eqref{eq:spacrt-intrinsic-local-alternative}. Apply
Theorem~\ref{thm:spacrt-validity}. This proves the two sequence-wise
limits, and the argument covers every array satisfying the sequence
conditions of
Theorem~\ref{thm:spacrt-logistic-lasso-local-alternative} with
$\rho_n=0$.  The uniform null clause now follows from
Corollary~\ref{cor:uniform-null-diagonal-use}.

\section{Detailed proofs of the supporting lemmas for
Theorem~\ref{thm:spacrt-logistic-lasso-local-alternative}}
\label{sec:spacrt-logistic-supporting-proofs}

Throughout this section, the oracle quantities, error functionals,
variance scales, and sigma-fields are those defined in
Supplementary Section~\ref{sec:spacrt-notation}; the logistic-model notation is from
Supplementary Sections~\ref{sec:spacrt-logistic-model-setup}--
\ref{sec:spacrt-logistic-verification}.

\subsection{Proof of part~\ref{part:spacrt-oracle} of
Lemma~\ref{lem:spacrt-insample-logistic-oracle}}

\begin{proof}
Fix $j$ and use the following abbreviations within this proof: $A_i\equiv A_{j,i}$,
$m_i\equiv m_{j,i}$, $w_i\equiv w_{j,i}$,
$\omega_n\equiv\omega_{n,j}$,
$s_n\equiv s_{n,j}$, $\mathcal S_n\equiv S_{n,j}$,
$\bm\theta^0\equiv\bm\theta^0_{n,j}$, $\lambda_n\equiv\lambda_{n,j}$,
$\widehat{\bm\delta}\equiv\widehat{\bm\delta}_{n,j}$,
$h_i\equiv h_{j,i}$, $\mu_i^0\equiv\mu^0_{n,j}(\rz{i})$, and
$\mathfrak b_n\equiv\mathfrak b_{n,j}$.  In this proof only, write
\[
  b(t)\equiv\log(1+e^t),
  \qquad
  \ell_n(\bm\theta)
  \equiv
  \frac1n\sum_{i=1}^n
  \{b(\rz{i}^{\T}\bm\theta)-A_i\rz{i}^{\T}\bm\theta\},
  \qquad
  Q_n(\bm\delta)
  \equiv
  \left\{
    \frac1n\sum_{i=1}^n(\rz{i}^{\T}\bm\delta)^2
  \right\}^{1/2}.
\]
Let $K_0\equiv1\vee K$, where $K$ is the design bound in~\eqref{cond:spacrt-lasso-design-bound}; since the intercept coordinate of $\rz{i}$ equals one and $\|\bm U_i\|_\infty\leq K$ almost surely, $\|\rz{i}\|_\infty\leq K_0$ almost surely.

\paragraph{Proof of existence.}  The envelope~\eqref{eq:spacrt-logistic-perturbed-envelope} and $n\pi_{n,j}\to\infty$ imply that the sample contains at least one zero and at least one one with probability tending to one, because
\[
  \P\left(\sum_{i=1}^nA_i=0\right)
  \leq(1-c_m\pi_{n,j})^n
  \leq\exp(-c_mn\pi_{n,j}),
  \qquad
  \P\left(\sum_{i=1}^nA_i=n\right)
  \leq\{\expit(B_{\alpha}+KB+1)\}^n,
\]
and $n\pi_{n,j}\geq n\pi_{n,x}\pi_{n,y}\to\infty$
by~\eqref{eq:spacrt-logistic-lasso-rates}.  On the complementary
event, if $\|\bm b\|_1\to\infty$ then the penalty diverges because
$\lambda_n>0$, while if $\bm b$ remains bounded and $|a|\to\infty$
then the presence of both outcomes makes the likelihood diverge.  Thus
the continuous convex objective is coercive and has a finite
minimizer.  The prescribed zero fallback is used only on an event
whose probability tends to zero and therefore does not alter any of
the probability bounds below.

\paragraph{Proof of rate: score decomposition and score event.}  The score at the baseline parameter splits into a conditionally centered part and a bias:
\[
  \bm g_n
  \equiv
  \nabla\ell_n(\bm\theta^0)
  =
  \frac1n\sum_{i=1}^n\rz{i}(\mu_i^0-A_i)
  =
  \bm g_n^{\mathrm{cen}}-\bm b_n.
\]
Here
\[
  \bm g_n^{\mathrm{cen}}
  \equiv\frac1n\sum_{i=1}^n\rz{i}(m_i-A_i)
  \text{ and }
  \bm b_n
  \equiv\frac1n\sum_{i=1}^n\rz{i}w_i.
\]
Conditionally on $\mathcal Z_n$, each coordinate summand of
$\bm g_n^{\mathrm{cen}}$ is centered, bounded by $K_0$, and has
conditional variance at most $K_0^2m_i\leq C_mK_0^2\pi_{n,j}$
by~\eqref{eq:spacrt-logistic-perturbed-envelope}.  Bernstein's
inequality applied conditionally, a union bound over all $p_n$
coordinates, and integration therefore give constants $c_0,C_0>0$
depending only on $K$ and $C_m$ such that, for every $t>0$,
\begin{align*}
  \P\left[
    \|\bm g_n^{\mathrm{cen}}\|_\infty
    >
    C_0\left\{
      \sqrt{\frac{\pi_{n,j}(t+\log p_n)}{n}}
      +\frac{t+\log p_n}{n}
    \right\}
  \right]
  \leq2e^{-c_0t}.
\end{align*}
This bound includes coordinate zero, even though the intercept is not
penalized. Taking $t=\log p_n$ and choosing the fixed penalty constant
$A$ in the definition of $\lambda_{n,j}$
following~\eqref{eq:spacrt-insample-logistic-fit} sufficiently large
bounds the failure probability by $2p_n^{-c_0}$, which vanishes because
$p_n\to\infty$. This yields the event
\begin{equation}
  \mathcal E_{n,\mathrm{score}}
  \equiv
  \{\|\bm g_n^{\mathrm{cen}}\|_\infty\leq\lambda_n/4\},
  \qquad
  \P(\mathcal E_{n,\mathrm{score}})\longrightarrow1.
  \label{eq:spacrt-logistic-score-event}
\end{equation}
The bias is not paired with the $\ell_1$ norm but with the empirical
prediction seminorm: for every $\bm\delta$, Cauchy--Schwarz gives,
almost surely,
\begin{equation}
  |\bm b_n^{\T}\bm\delta|
  \leq
  \left(\frac1n\sum_{i=1}^nw_i^2\right)^{1/2}Q_n(\bm\delta)
  =
  \omega_nQ_n(\bm\delta).
  \label{eq:spacrt-logistic-bias-pairing}
\end{equation}

\paragraph{Proof of rate: design event.}  Put
$\bm\Sigma_n\equiv\E(\rz{}\rz{}^{\T})$,
$\widehat{\bm\Sigma}_n\equiv n^{-1}\sum_{i=1}^n\rz{i}\rz{i}^{\T}$, and
$\kappa_\Sigma\equiv c_\Sigma^{1/2}$, where $c_\Sigma$ is the eigenvalue
lower bound in~\eqref{cond:spacrt-lasso-eigenvalue}.  Define also
$\zeta_n\equiv\sqrt{\log(p_n)/n}+\log(p_n)/n$.  Bernstein's inequality
and a union bound over the entries give
\begin{equation*}
  \widehat\zeta_n
  \equiv
  \|\widehat{\bm\Sigma}_n-\bm\Sigma_n\|_{\max}
  =
  O_{\P_n}\left(
    \sqrt{\frac{\log p_n}{n}}+\frac{\log p_n}{n}
  \right)
  =O_{\P_n}(\zeta_n),
\end{equation*}
This concerns $\{\rz{i}\}$ only, whose law does not involve the
perturbation.  Since $s_n^2\log p_n/n\to0$
by~\eqref{eq:spacrt-logistic-lasso-rates}, we have $s_n\zeta_n\to0$
and hence $s_n\widehat\zeta_n=o_{\P_n}(1)$; let $\mathcal E_{n,\mathrm{design}}\equiv\{(s_n\widehat\zeta_n)^{1/2}\leq\kappa_\Sigma/10\}$ and $\P(\mathcal E_{n,\mathrm{design}})\longrightarrow1$. On $\mathcal E_{n,\mathrm{design}}$, for \emph{every}
$\bm\delta\in\R^{p_n}$---no cone restriction is available once
$w_i\not\equiv0$, so none is used---the eigenvalue
bound~\eqref{cond:spacrt-lasso-eigenvalue} gives
$c_\Sigma\|\bm\delta\|_2^2
\leq Q_n(\bm\delta)^2+\widehat\zeta_n\|\bm\delta\|_1^2$, whence
\begin{equation}
  \|\bm\delta_{\mathcal S_n}\|_1
  \leq
  \sqrt{s_n}\|\bm\delta\|_2
  \leq
  \frac{\sqrt{s_n}}{\kappa_\Sigma}Q_n(\bm\delta)
  +\frac{(s_n\widehat\zeta_n)^{1/2}}{\kappa_\Sigma}\|\bm\delta\|_1
  \leq
  \frac{\sqrt{s_n}}{\kappa_\Sigma}Q_n(\bm\delta)
  +\frac1{10}\|\bm\delta\|_1,
  \label{eq:spacrt-logistic-active-control}
\end{equation}
and therefore, adding $\|\bm\delta_{\mathcal S_n^c}\|_1$ to both sides
and rearranging,
\begin{equation}
  \|\bm\delta\|_1
  \leq
  \frac{2\sqrt{s_n}}{\kappa_\Sigma}Q_n(\bm\delta)
  +2\|\bm\delta_{\mathcal S_n^c}\|_1.
  \label{eq:spacrt-logistic-l1-control}
\end{equation}
This replaces, and is weaker than, the compatibility condition
available in the correctly specified case.

\paragraph{Proof of rate: curvature.}  We record three deterministic
consequences of the rate condition~\eqref{eq:spacrt-logistic-lasso-rates}.
Since $s_n\geq1$ and $\pi_{n,j}\leq1$,
\begin{equation}
  \frac{s_n\lambda_n}{\pi_{n,j}}=o(1),
  \qquad
  \frac{\log p_n}{n\pi_{n,j}}=o(1),
  \qquad
  \lambda_n^2
  =O\left(\frac{\pi_{n,j}\log p_n}{n}\right).
  \label{eq:spacrt-logistic-rate-consequences}
\end{equation}
The first two follow from $s_n^2\log p_n/(n\pi_{n,j})\to0$, and the
third makes the squared linear term in $\lambda_n$ negligible relative
to its variance term.  For a coefficient increment $\bm\delta$, define
the empirical Bregman divergence
\[
  \mathfrak D_n(\bm\delta)
  \equiv
  \ell_n(\bm\theta^0+\bm\delta)
  -\ell_n(\bm\theta^0)
  -\bm g_n^{\T}\bm\delta
  =
  \frac1n\sum_{i=1}^n
  \left\{
    b(\rz{i}^{\T}\bm\theta^0+h_i)
    -b(\rz{i}^{\T}\bm\theta^0)
    -b'(\rz{i}^{\T}\bm\theta^0)h_i
  \right\},
\]
with $h_i=\rz{i}^{\T}\bm\delta$.  It is a deterministic function of
$\{\rz{i}\}$ and $\bm\delta$ alone; in particular the responses, and
with them the perturbation, have cancelled, so the curvature bound
below is the same under every law satisfying the hypotheses.  The
logistic curvature satisfies, for every $u,t\in\R$,
\begin{equation}
  \frac{b''(u+t)}{b''(u)}\geq e^{-|t|}.
  \label{eq:spacrt-logistic-curvature-ratio}
\end{equation}
Because $b''(\rz{i}^{\T}\bm\theta^0)\geq c_1\pi_{n,j}$
by~\eqref{eq:spacrt-logistic-intercept-envelope}---an envelope on the
unperturbed mean $\mu_i^0$, which is what parameterizes the
criterion---Taylor's integral formula
and~\eqref{eq:spacrt-logistic-curvature-ratio} yield
\begin{equation}
  \mathfrak D_n(\bm\delta)
  \geq
  \frac{c_1\pi_{n,j}}{2}
  \exp(-K_0\|\bm\delta\|_1)
  \,Q_n(\bm\delta)^2.
  \label{eq:spacrt-logistic-local-curvature}
\end{equation}

\paragraph{Proof of rate: basic inequality.} Define the penalized
objective
$F_n(\bm\theta)\equiv\ell_n(\bm\theta)
+\lambda_n\|\bm\theta_{-0}\|_1$, where coordinate zero is the
unpenalized intercept.  Fix any $\bm\delta$ in the sublevel set
$F_n(\bm\theta^0+\bm\delta)\leq F_n(\bm\theta^0)$.  On the event that
the finite minimizer exists, the estimation error
$\widehat{\bm\delta}$ belongs to this set by optimality.  Writing
$\bm b^0$ for the slope part of $\bm\theta^0$, the objective comparison
and the definition of $\mathfrak D_n$ give
\[
  \mathfrak D_n(\bm\delta)
  +\lambda_n\left\{
    \|\bm b^0+\bm\delta_{-0}\|_1-\|\bm b^0\|_1
  \right\}
  \leq
  -\bm g_n^{\T}\bm\delta.
\]
Let $A_n\equiv\mathcal S_n\setminus\{0\}
=\operatorname{supp}(\bm b^0)$.  Since $\bm b^0$ vanishes on
$\mathcal S_n^c$, the reverse triangle inequality yields
\begin{align*}
  \|\bm b^0+\bm\delta_{-0}\|_1-\|\bm b^0\|_1
  &=
  \|\bm b^0_{A_n}+\bm\delta_{A_n}\|_1
  -\|\bm b^0_{A_n}\|_1
  +\|\bm\delta_{\mathcal S_n^c}\|_1\\
  &\geq
  \|\bm\delta_{\mathcal S_n^c}\|_1
  -\|\bm\delta_{A_n}\|_1.
\end{align*}
Moreover, on $\mathcal E_{n,\mathrm{score}}$, the splitting
$\bm g_n=\bm g_n^{\mathrm{cen}}-\bm b_n$, H\"older's inequality,
and~\eqref{eq:spacrt-logistic-score-event}--
\eqref{eq:spacrt-logistic-bias-pairing} give
\begin{align*}
  -\bm g_n^{\T}\bm\delta
  &\leq
  \|\bm g_n^{\mathrm{cen}}\|_\infty\|\bm\delta\|_1
  +|\bm b_n^{\T}\bm\delta|\\
  &\leq
  \frac{\lambda_n}{4}
  \left\{
    |\delta_0|+\|\bm\delta_{A_n}\|_1
    +\|\bm\delta_{\mathcal S_n^c}\|_1
  \right\}
  +\omega_nQ_n(\bm\delta).
\end{align*}
Combining these bounds and collecting the active and inactive terms
gives
\begin{equation}
\begin{aligned}
  \mathfrak D_n(\bm\delta)
  +\frac{3\lambda_n}{4}\|\bm\delta_{\mathcal S_n^c}\|_1
  &\leq
  \frac{\lambda_n}{4}|\delta_0|
  +\frac{5\lambda_n}{4}\|\bm\delta_{A_n}\|_1
  +\omega_nQ_n(\bm\delta)\\
  &\leq
  \frac{5\lambda_n}{4}\|\bm\delta_{\mathcal S_n}\|_1
  +\omega_nQ_n(\bm\delta).
\end{aligned}
\label{eq:spacrt-logistic-basic-inequality}
\end{equation}
The intercept is included in the score norm and in the active part on
the right; it is not included in the penalty.

\paragraph{Proof of rate: localization and rates.}
The curvature lower bound~\eqref{eq:spacrt-logistic-local-curvature}
contains the factor $\exp(-K_0\|\bm\delta\|_1)$ and is therefore
uniformly useful only on a small $\ell_1$-ball.  Applying it directly
to $\widehat{\bm\delta}$ would be circular: the curvature needed to
prove that $\|\widehat{\bm\delta}\|_1$ is small would itself require
that smallness.  We remove this circularity using the standard convex
localization, or boundary-rescaling, argument of
\citet[Lemma~9.21, pp.~282--283]{Wainwright2019}.  Namely, we contract
$\widehat{\bm\delta}$ to a prescribed radius only when necessary,
prove the rate for the contracted increment, and then show that the
contraction cannot have been active.

Put $R_n\equiv2C_6\mathfrak b_n$, with $C_6$ the constant produced
below.  Since $\mathfrak b_n\convp0$, the localization event
\[
  \mathcal E_{n,\mathrm{loc}}\equiv\{K_0R_n\leq1\}
  =\{2K_0C_6\mathfrak b_n\leq1\}
\]
has probability tending to one.  Work throughout this paragraph on
its intersection with $\mathcal E_{n,\mathrm{score}}$,
$\mathcal E_{n,\mathrm{design}}$, and the event that a finite minimizer
exists; this intersection also has probability tending to one.
Set $\tau_n\equiv1\wedge\frac{R_n}{\|\widehat{\bm\delta}\|_1}$ and
$\widetilde{\bm\delta}\equiv\tau_n\widehat{\bm\delta}$, with $\tau_n=1$
when $\widehat{\bm\delta}=\bm0$. Optimality and convexity give
\[
  F_n(\bm\theta^0+\widetilde{\bm\delta})
  \leq
  (1-\tau_n)F_n(\bm\theta^0)
  +\tau_nF_n(\bm\theta^0+\widehat{\bm\delta})
  \leq
  F_n(\bm\theta^0).
\]
Thus~\eqref{eq:spacrt-logistic-basic-inequality} applies to
$\widetilde{\bm\delta}$ without assuming that
$\widehat{\bm\delta}$ is already local, whereas
$\|\widetilde{\bm\delta}\|_1\leq R_n$ makes the curvature bound
effective.  Indeed, on $\mathcal E_{n,\mathrm{loc}}$, $K_0R_n\leq1$, so
the curvature bound~\eqref{eq:spacrt-logistic-local-curvature} gives
$\mathfrak D_n(\widetilde{\bm\delta})\geq c_2\pi_{n,j}Q^2$, where
$c_2\equiv c_1/(2e)$ and $Q\equiv Q_n(\widetilde{\bm\delta})$.
Bounding $\|\widetilde{\bm\delta}_{\mathcal S_n}\|_1$
by~\eqref{eq:spacrt-logistic-active-control}, the
inequality~\eqref{eq:spacrt-logistic-basic-inequality} becomes
\[
  c_2\pi_{n,j}Q^2
  +\frac{3\lambda_n}{4}
   \|\widetilde{\bm\delta}_{\mathcal S_n^c}\|_1
  \leq
  \frac{5\lambda_n}{4}
  \left\{
    \frac{\sqrt{s_n}}{\kappa_\Sigma}Q
    +\frac1{10}\|\widetilde{\bm\delta}\|_1
  \right\}
  +\omega_nQ
  \leq
  \frac{3\lambda_n}{2\kappa_\Sigma}\sqrt{s_n}\,Q
  +\frac{\lambda_n}{4}
   \|\widetilde{\bm\delta}_{\mathcal S_n^c}\|_1
  +\omega_nQ,
\]
where the second step used~\eqref{eq:spacrt-logistic-l1-control}.
Absorbing the middle term into the left side and applying Young's
inequality $aQ\leq(c_2\pi_{n,j}/4)Q^2+a^2/(c_2\pi_{n,j})$ to both
remaining $Q$-terms on the right yields
\begin{equation*}
  \frac{c_2\pi_{n,j}}{2}Q^2
  +\frac{\lambda_n}{2}
   \|\widetilde{\bm\delta}_{\mathcal S_n^c}\|_1
  \leq
  C_4\left(
    \frac{s_n\lambda_n^2}{\pi_{n,j}}
    +\frac{\omega_n^2}{\pi_{n,j}}
  \right),
\end{equation*}
whence
\begin{equation}
  Q
  \leq
  C_5\left(
    \frac{\sqrt{s_n}\lambda_n}{\pi_{n,j}}
    +\frac{\omega_n}{\pi_{n,j}}
  \right),
  \qquad
  \|\widetilde{\bm\delta}_{\mathcal S_n^c}\|_1
  \leq
  C_5\left(
    \frac{s_n\lambda_n}{\pi_{n,j}}
    +\frac{\omega_n^2}{\pi_{n,j}\lambda_n}
  \right),
  \label{eq:spacrt-logistic-prediction-bound}
\end{equation}
and then, by~\eqref{eq:spacrt-logistic-l1-control},
$\|\widetilde{\bm\delta}\|_1\leq C_6\mathfrak b_n=R_n/2$ for a
deterministic $C_6$ depending only on $c_2$, $\kappa_\Sigma$, and
$C_5$.  If $\|\widehat{\bm\delta}\|_1>R_n$ then
$\|\widetilde{\bm\delta}\|_1=R_n$, a contradiction; hence
$\|\widehat{\bm\delta}\|_1\leq R_n=2C_6\mathfrak b_n$ on an event of
probability tending to one, which is the first bound
in~\eqref{eq:spacrt-logistic-oracle-bounds}; its $o_{\P_n}(1)$ conclusion
follows from $\mathfrak b_n\convp0$.  In particular
$\widetilde{\bm\delta}=\widehat{\bm\delta}$ with probability tending
to one, so squaring the first bound
in~\eqref{eq:spacrt-logistic-prediction-bound} gives
\[
  \frac1n\sum_{i=1}^nh_i^2
  =
  Q_n(\widehat{\bm\delta})^2
  =
  O_{\P_n}\left(
    \frac{s_n\lambda_n^2}{\pi_{n,j}^2}
    +\frac{\omega_n^2}{\pi_{n,j}^2}
  \right),
\]
the second bound in~\eqref{eq:spacrt-logistic-oracle-bounds}; the
third follows from
$\max_i|h_i|\leq K_0\|\widehat{\bm\delta}\|_1=o_{\P_n}(1)$.

\paragraph{Proof of rate: consequences.}  If $\mu(t)\equiv\expit(t)$ then $e^{-|h|}\leq\mu(u+h)/\mu(u)\leq e^{|h|}$, and the same calculation applied to $1-\mu$ gives $e^{-|h|}\leq\{1-\mu(u+h)\}/\{1-\mu(u)\}\leq e^{|h|}$; multiplying the two yields
\[
  e^{-2|h_i|}
  \leq
  \frac{\widehat v_{j,i}}{v^0_{j,i}}
  \leq
  e^{2|h_i|}.
\]
Together with $\max_i|h_i|=o_{\P_n}(1)$, these inequalities
prove~\eqref{eq:spacrt-logistic-relative-consistency}.  On the event
$\max_i|h_i|\leq1$, the mean-value theorem and
$\expit'(\rz{i}^{\T}\bm\theta^0+t)
\leq e^{|t|}\expit(\rz{i}^{\T}\bm\theta^0)\leq C\pi_{n,j}$ for
$|t|\leq1$ give $|\widehat\mu_{j,i}-\mu_i^0|\leq C\pi_{n,j}|h_i|$, so
that, by the second bound
in~\eqref{eq:spacrt-logistic-oracle-bounds}
and~\eqref{eq:spacrt-logistic-rate-consequences},
\[
  \frac1n\sum_{i=1}^n(\widehat\mu_{j,i}-\mu_i^0)^2
  \leq
  C\pi_{n,j}^2\frac1n\sum_{i=1}^nh_i^2
  =
  O_{\P_n}(s_n\lambda_n^2+\omega_n^2)
  =
  O_{\P_n}\left(
    \frac{\pi_{n,j}s_n\log p_n}{n}
    +\omega_n^2
  \right).
\]
This is~\eqref{eq:spacrt-logistic-mean-square-rate} and completes the
proof.
\end{proof}

\subsection{Proof of part~\ref{part:spacrt-null-verification} of
Lemma~\ref{lem:spacrt-logistic-lasso}}

\begin{proof}
\paragraph{Null-fit rates.}
All probabilities and stochastic limits in this proof are under $\P_{n,0}$.
By Remark~\ref{rem:spacrt-zero-bias},
part~\ref{part:spacrt-oracle} of
Lemma~\ref{lem:spacrt-insample-logistic-oracle} applies with $j=x,y$
under $\P_{n,0}$; the required localization scales vanish by
\eqref{eq:spacrt-logistic-lasso-rates}.
Equation~\eqref{eq:spacrt-logistic-mean-square-rate}
gives the two empirical mean-square rates stated in the lemma,
and~\eqref{eq:spacrt-logistic-relative-consistency} gives its uniform
relative-consistency conclusion.

\paragraph{Verification of Assumption~\ref{assu:spacrt-regularity}.}
It remains to verify Assumption~\ref{assu:spacrt-regularity} and the
two intrinsic-variance assertions in part~\ref{part:spacrt-null-verification}.
We establish the more precise relations
\begin{equation}
  \sigma_{n,\dCRT}^2\asymp \pi_{n,x}\pi_{n,y},
  \qquad
  \frac{V_{n,\dCRT}^2}{\sigma_{n,\dCRT}^2}\convp1,
  \label{eq:spacrt-logistic-intrinsic-variances}
\end{equation}
and
\begin{align}\label{eq:spacrt-logistic-weighted-rates}
  (E'_{n,x})^2=O_{\P_{n,0}}\left(\frac{\pi_{n,x}\pi_{n,y}s_{n,x}\log p_n}{n}\right),\qquad
  (E'_{n,y})^2+(\widehat E'_{n,y})^2=O_{\P_{n,0}}\left(\frac{\pi_{n,x}\pi_{n,y}s_{n,y}\log p_n}{n}\right).
\end{align}
\paragraph{Clause~\textup{(i)}: nuisance rates and variance calibration.}
The intercept envelope~\eqref{eq:spacrt-logistic-intercept-envelope}
gives
\begin{equation*}
  v^0_{x,i}\asymp \pi_{n,x},
  \qquad
  v^0_{y,i}\asymp \pi_{n,y},
  \qquad
  \sigma_{n,\dCRT}^2
  =\frac1n\sum_{i=1}^n v^0_{x,i}v^0_{y,i}
  \asymp \pi_{n,x}\pi_{n,y}.
\end{equation*}
In particular, $\P_{n,0}(\sigma_{n,\dCRT}>0)\to1$. We first verify Assumption~\ref{assu:spacrt-regularity}(i).  By
Part~\ref{part:spacrt-oracle} of
Lemma~\ref{lem:spacrt-insample-logistic-oracle}, applied under
$\P_{n,0}$ in the zero-bias case $w_{j,i}\equiv0$,
\begin{align*}
  \frac{\sqrt nE_{n,x}E_{n,y}}{\sigma_{n,\dCRT}}
  &=
  O_{\P_{n,0}}\left(
    \sqrt{\frac{s_{n,x}s_{n,y}\log^2p_n}{n}}
  \right)
  =o_{\P_{n,0}}(1),\\
  (E'_{n,x})^2
  &\leq C\pi_{n,y}E_{n,x}^2
  =
  O_{\P_{n,0}}\left(
    \frac{\pi_{n,x}\pi_{n,y}s_{n,x}\log p_n}{n}
  \right),\\
  (E'_{n,y})^2
  &\leq C\pi_{n,x}E_{n,y}^2
  =
  O_{\P_{n,0}}\left(
    \frac{\pi_{n,x}\pi_{n,y}s_{n,y}\log p_n}{n}
  \right).
\end{align*}
Uniform relative variance consistency in
\eqref{eq:spacrt-logistic-relative-consistency} also gives
$\max_i\widehat v_{x,i}=O_{\P_{n,0}}(\pi_{n,x})$ and hence
\[
  (\widehat E'_{n,y})^2
  \leq
  \max_i\widehat v_{x,i}E_{n,y}^2
  =
  O_{\P_{n,0}}\left(
    \frac{\pi_{n,x}\pi_{n,y}s_{n,y}\log p_n}{n}
  \right).
\]
After division by $\sigma_{n,\dCRT}$, each weighted rate
is respectively
$O_{\P_{n,0}}\{(s_{n,j}\log p_n/n)^{1/2}\}=o_{\P_{n,0}}(1)$.
The last convergence follows from
\eqref{eq:spacrt-logistic-lasso-rates}, because $s_{n,j}\geq1$ and
$\pi_{n,j}\leq1$.  This proves~\eqref{eq:spacrt-logistic-weighted-rates}
and the nuisance-rate part of clause~\textup{(i)}.

For the remaining variance-calibration part of clause~\textup{(i)}, use
\eqref{eq:spacrt-logistic-relative-consistency} to obtain
\begin{align*}
  \left|
    \frac1n\sum_{i=1}^n
    (\widehat v_{x,i}-v^0_{x,i})v^0_{y,i}
  \right|
  &\leq
  \max_i
  \left|
    \frac{\widehat v_{x,i}}{v^0_{x,i}}-1
  \right|
  \frac1n\sum_{i=1}^n v^0_{x,i}v^0_{y,i}\\
  &=o_{\P_{n,0}}(\sigma_{n,\dCRT}^2).
\end{align*}
\paragraph{Clause~\textup{(ii)}: Bernstein moment control.}
Every centered Bernoulli variable $B-p$ satisfies, for every integer $k\geq3, \E|B-p|^k\leq\E(B-p)^2\leq\frac{k!}{2}\E(B-p)^2$. Consequently, the true predictor and response laws and the fitted
predictor-resampling law satisfy the three conditional Bernstein bounds
with $b^0_{n,x}=b^0_{n,y}=\widehat b_{n,x}=1$.  Moreover,
\[
  \frac{
    b^0_{n,y}(b^0_{n,x}+\widehat b_{n,x})}
  {\sqrt n\sigma_{n,\dCRT}}
  =
  O_{\P_{n,0}}\!\left(\frac1{\sqrt{n\pi_{n,x}\pi_{n,y}}}\right)
  \longrightarrow0
  \quad\text{by~\eqref{eq:spacrt-logistic-lasso-rates}}.
\]
Thus clause~\textup{(ii)} holds.

\paragraph{Variance matching.}
The $X$- and $Y$-fits have the separate-fit measurability required by
Lemma~\ref{lem:spacrt-variance-matching}.  Having verified
clauses~\textup{(i)}--\textup{(ii)}, that lemma gives $\frac{V_{n,\dCRT}^2}{\sigma_{n,\dCRT}^2}\convp1$, which is the second assertion in
\eqref{eq:spacrt-logistic-intrinsic-variances} and does not presuppose
clause~\textup{(iii)}.

\paragraph{Clause~\textup{(iii)}: nondegeneracy and maximal residual.}
Finally, \eqref{eq:spacrt-logistic-lasso-rates} gives
$n\pi_{n,x}\pi_{n,y}\to\infty$; together with
\eqref{eq:spacrt-logistic-intrinsic-variances}, this implies
$\sqrt nV_{n,\dCRT}\to\infty$ in probability.  Since both $Y_i$ and
$\widehat\mu_{y,i}$ belong to $[0,1]$,
$\max_i|\widehat e_{y,i}|\leq1$.  Hence
\[
  \P_{n,0}(V_{n,\dCRT}>0)\longrightarrow1,
  \qquad
  \frac{
    \widehat b_{n,x}\max_i|\widehat e_{y,i}|}
  {\sqrt nV_{n,\dCRT}}
  \convp0,
\]
which is clause~\textup{(iii)}.

\paragraph{Clause~\textup{(iv)}: successful construction.}
Part~\ref{part:spacrt-oracle} of
Lemma~\ref{lem:spacrt-insample-logistic-oracle}, applied to both fits
under the null, ensures finite minimizers with probability tending to one.
On their joint existence event, the fitted Bernoulli probabilities define
the conditional resampling law. This verifies clause~\textup{(iv)} and
completes the proof of part~\ref{part:spacrt-null-verification}.
\end{proof}

\subsection{Statement and proof of
Lemma~\ref{lem:spacrt-logistic-alt-perturbation}}

\begin{lemma}[Mean and posterior perturbations under vanishing
alternatives]
\label{lem:spacrt-logistic-alt-perturbation}
Retain the model of
Theorem~\ref{thm:spacrt-logistic-lasso-local-alternative} with
conditions~\eqref{cond:spacrt-lasso-design-bound}--\eqref{cond:spacrt-lasso-eigenvalue},
and let $|\rho|\leq1$.  Then, almost surely and uniformly in $i$:
\begin{enumerate}[label=\textup{(\alph*)}]
\item the logistic increment has the exact form
\begin{equation*}
  \Delta_i(\rho)
  =
  \frac{q_i(1-q_i)(e^\rho-1)}
       {1+q_i(e^\rho-1)},
\end{equation*}
and satisfies
$|\Delta_i(\rho)|\leq C_1\pi_{n,y}|\rho|$,
$e^{-1}q_i\leq q_i(\rho)\leq eq_i$, and
$q_i(\rho)\leq\expit(B_{\alpha}+KB+1)<1$;
\item the $Z$-conditional response mean
$m_i\equiv\E_{n,\rho}(\ey{i}\mid\rz{i})=q_i+w_i$, with
$w_i\equiv p_i\Delta_i(\rho)$, satisfies
\[
  |w_i|\leq C_2\pi_{n,x}\pi_{n,y}|\rho|,
  \qquad
  e^{-1-KB}\pi_{n,y}\leq m_i\leq e^{1+KB}\pi_{n,y},
  \qquad
  m_i\leq\expit(B_{\alpha}+KB+1);
\]
\item conditionally on $(\ey{i},\rz{i})$, the predictor $\ex{i}$ is
Bernoulli with mean
$\widetilde p_i\equiv\P_{n,\rho}(\ex{i}=1\mid\ey{i},\rz{i})$;
this posterior mean satisfies
\[
  \logit\widetilde p_i-\logit p_i
  =
  \rho\ey{i}-\log\{1+q_i(e^\rho-1)\},
\]
and
\[
  |\widetilde p_i-p_i|
  \leq
  C_3\pi_{n,x}|\rho|(\ey{i}+\pi_{n,y}),
  \qquad
  \widetilde p_i\leq C_3\pi_{n,x}.
\]
\end{enumerate}
\end{lemma}

\begin{proof}
For all three parts, put $D_i(\rho)\equiv1+q_i(e^\rho-1)=(1-q_i)+q_ie^\rho$. For $|\rho|\leq1$, $e^{-1}\leq D_i(\rho)\leq e$ and $|e^\rho-1|\leq e|\rho|$.

\paragraph{Part~\textup{(a)}: logistic response-mean increment.}
Direct calculation gives
\[
  q_i(\rho)=\frac{q_ie^\rho}{D_i(\rho)},
  \qquad
  \Delta_i(\rho)
  =\frac{q_i(1-q_i)(e^\rho-1)}{D_i(\rho)}.
\]
Thus $|\Delta_i(\rho)|\leq Cq_i|\rho|\leq
C_1\pi_{n,y}|\rho|$ by
\eqref{eq:spacrt-logistic-intercept-envelope}.  The same representation
and the fact that $D_i(\rho)$ is a convex combination of $1$ and
$e^\rho$ give $e^{-1}q_i\leq q_i(\rho)\leq eq_i$, while monotonicity of
$\expit$ gives $q_i(\rho)\leq\expit(B_{\alpha}+KB+1)$.  This proves part~(a).

\paragraph{Part~\textup{(b)}: the $Z$-conditional response mean.}
Taking conditional expectations in
\eqref{eq:spacrt-logistic-alt-conditional-mean} gives
$m_i=q_i+p_i\Delta_i(\rho)=q_i+w_i$.  Since
$p_i\leq e^{KB}\pi_{n,x}$ by
\eqref{eq:spacrt-logistic-intercept-envelope}, part~(a) gives
\[
  |w_i|
  =p_i|\Delta_i(\rho)|
  \leq C_1e^{KB}\pi_{n,x}\pi_{n,y}|\rho|
  \leq C_2\pi_{n,x}\pi_{n,y}|\rho|.
\]
Also, $m_i=(1-p_i)q_i+p_iq_i(\rho)$, so part~(a) and the intercept
envelope give the remaining bounds in part~(b).

\paragraph{Part~\textup{(c)}: the posterior predictor mean.}
Bayes' rule gives
\[
  \frac{\widetilde p_i}{1-\widetilde p_i}
  =
  \frac{p_i}{1-p_i}
  \left\{\frac{q_i(\rho)}{q_i}\right\}^{\ey{i}}
  \left\{\frac{1-q_i(\rho)}{1-q_i}\right\}^{1-\ey{i}},
\]
and $q_i(\rho)/q_i=e^\rho/D_i(\rho)$ and
$\{1-q_i(\rho)\}/(1-q_i)=1/D_i(\rho)$.  This proves the stated exact logit identity.  Moreover, $D_i(0)=1$ and $\frac{d}{dt}\log D_i(t)=\frac{q_ie^t}{D_i(t)}=q_i(t)$. For every $t$ between $0$ and $\rho$, part~(a) gives $q_i(t)\leq eq_i$.  Therefore,
\[
  |\log D_i(\rho)|
  =\left|\int_0^\rho q_i(t)\,dt\right|
  \leq eq_i|\rho|
  \leq e^{1+KB}\pi_{n,y}|\rho|.
\]
Here
$\widetilde\Delta_i\equiv\logit\widetilde p_i-\logit p_i$ is the
posterior log-odds increment, distinct from the response-mean increment
$\Delta_i(\rho)$ in part~(a).  The exact logit identity and the last
display give
\[
  |\widetilde\Delta_i|
  =|\rho\ey{i}-\log D_i(\rho)|
  \leq|\rho|\{\ey{i}+e^{1+KB}\pi_{n,y}\}
  \leq1+e^{1+KB}.
\]
By the mean-value theorem, there is some $t_i$ between $0$ and
$\widetilde\Delta_i$.  Since
$\expit'(u)=\expit(u)\{1-\expit(u)\}\leq\expit(u)$ and the logistic
ratio bound gives
$\expit(\logit p_i+t)\leq e^{|t|}\expit(\logit p_i)=e^{|t|}p_i$,
we have
\begin{align*}
  |\widetilde p_i-p_i|
  &=\expit'(\logit p_i+t_i)|\widetilde\Delta_i|
  \leq e^{|t_i|}p_i|\widetilde\Delta_i|\leq C p_i|\rho|(\ey{i}+\pi_{n,y})
  \leq C_3\pi_{n,x}|\rho|(\ey{i}+\pi_{n,y}),\\
  \widetilde p_i
  &=\expit(\logit p_i+\widetilde\Delta_i)
  \leq e^{|\widetilde\Delta_i|}p_i
  \leq C_3\pi_{n,x},
\end{align*}
where the last inequalities use $p_i\leq e^{KB}\pi_{n,x}$ and the same
logistic ratio bound as above.
This proves part~(c).
\end{proof}

\subsection{Proof of Lemma~\ref{lem:spacrt-logistic-alt-fit}}
\label{sec:spacrt-logistic-alt-fit-proof}

\begin{proof}
All probabilities and stochastic orders are under $\P_n$,
with $n$ large enough that $|\rho_n|\leq1$
by~\eqref{eq:spacrt-logistic-local-scale}.

\paragraph{Part~\textup{(i)}: invariance of the $X$-fit.}
The joint law of $\{(\ex{i},\rz{i})\}_{i=1}^n$ is unchanged from
$\P_{n,0}$, and
$\E_{n,\rho_n}(\ex{i}\mid\rz{i})=\mu^0_{n,x}(\rz{i})$.  Thus the
$X$-fit has $w_{x,i}=0$, $\omega_{n,x}=0$, and
$\mathfrak b_{n,x}=s_{n,x}\lambda_{n,x}/\pi_{n,x}=o(1)$
by~\eqref{eq:spacrt-logistic-rate-consequences}.
Part~\ref{part:spacrt-oracle} of
Lemma~\ref{lem:spacrt-insample-logistic-oracle} proves part~(i).

\paragraph{Part~\textup{(ii)}: localization of the $Y$-fit.}
By Lemma~\ref{lem:spacrt-logistic-alt-perturbation}(b), conditionally
on $\mathcal Z_n$ the responses are independent Bernoulli variables
with
\[
  \E_{n,\rho_n}(\ey{i}\mid\rz{i})=q_i+w_i,
  \qquad
  w_i=p_i\Delta_i(\rho_n),
  \qquad
  \omega_{n,y}\leq C_2\pi_{n,x}\pi_{n,y}|\rho_n|,
\]
and their means satisfy
\eqref{eq:spacrt-logistic-perturbed-envelope} with constants depending
only on $(K,B)$.  It remains only to check $\mathfrak b_{n,y}\to0$.
Its first term tends to zero by
\eqref{eq:spacrt-logistic-rate-consequences}, while
\[
  \frac{\omega_{n,y}^2}{\pi_{n,y}\lambda_{n,y}}
  \leq
  C \pi_{n,x}^2\rho_n^2
  \left(\frac{n\pi_{n,y}}{\log p_n}\right)^{1/2}
  =o(1)
\]
by~\eqref{eq:spacrt-logistic-local-scale}.  The middle term is their
geometric mean:
\[
  \frac{\sqrt{s_{n,y}}\,\omega_{n,y}}{\pi_{n,y}}
  =
  \left\{
    \frac{s_{n,y}\lambda_{n,y}}{\pi_{n,y}}
    \cdot
    \frac{\omega_{n,y}^2}{\pi_{n,y}\lambda_{n,y}}
  \right\}^{1/2}
  =o(1).
\]
Thus $\mathfrak b_{n,y}\to0$, and
Part~\ref{part:spacrt-oracle} of
Lemma~\ref{lem:spacrt-insample-logistic-oracle} applies to the $Y$-fit.
Substitution of the displayed bound on $\omega_{n,y}$ into
\eqref{eq:spacrt-logistic-oracle-bounds} gives part~(ii).

\paragraph{Part~\textup{(iii)}: fitted mean and variance rates.}
The same substitution into
\eqref{eq:spacrt-logistic-relative-consistency} and
\eqref{eq:spacrt-logistic-mean-square-rate} gives part~(iii).
\end{proof}

\subsection{Proof of Lemma~\ref{lem:spacrt-logistic-alt-coordinates}}
\label{sec:spacrt-logistic-alt-coordinates-proof}

\begin{proof}
All statements are under $\P_n$, with $n$ large enough that
$|\rho_n|\leq1$ by~\eqref{eq:spacrt-logistic-local-scale}.  Put
$\xi_i\equiv\ey{i}-q_i-\ex{i}\Delta_i(\rho_n)$, the response residual
centered conditionally on $(\ex{i},\rz{i})$.  Lemma~\ref{lem:spacrt-logistic-alt-perturbation}(a)
and the intercept envelope~\eqref{eq:spacrt-logistic-intercept-envelope}
give, almost surely and uniformly in $1\leq i\leq n$,
\begin{equation}
  \V(\xi_i\mid\mathcal X_n)
  =\{q_i+\ex{i}\Delta_i(\rho_n)\}
   \{1-q_i-\ex{i}\Delta_i(\rho_n)\}
  \asymp \pi_{n,y},
  \qquad
  |\Delta_i(\rho_n)|\leq C\pi_{n,y}|\rho_n|.
  \label{eq:spacrt-logistic-alt-variance-envelope}
\end{equation}
By Lemma~\ref{lem:spacrt-logistic-alt-fit}(i),(iii) and the same intercept envelope, with probability tending to one $\widehat v_{x,i}\asymp \pi_{n,x}$ uniformly in $i$, while
\begin{align}
  \sum_i\Delta_{x,i}^2
  &=O_{\P_n}(\pi_{n,x}s_{n,x}\log p_n),
  \notag\\
  \sum_i\Delta_{y,i}^2
  &=O_{\P_n}\left(
    \pi_{n,y}s_{n,y}\log p_n
    +n\pi_{n,x}^2\pi_{n,y}^2\rho_n^2
  \right).
  \label{eq:spacrt-logistic-alt-fit-error-sums}
\end{align}
Markov's inequality and the model envelopes also give
\begin{align}
  \sum_i\ex{i}
  &=O_{\P_n}(n\pi_{n,x}),
  \qquad
  \sum_i\ey{i}=O_{\P_n}(n\pi_{n,y}),
  \notag\\
  \sum_i(e^0_{x,i})^2
  &=O_{\P_n}(n\pi_{n,x}).
  \label{eq:spacrt-logistic-alt-basic-sums}
\end{align}

\paragraph{Proof of the $nA_{2n}$ bound.}
By the definitions
$\widehat e_{y,i}=\ey{i}-\widehat\mu_{y,i}$ and
$\Delta_{y,i}=\widehat\mu_{y,i}-q_i$,
\begin{align*}
  \widehat e_{y,i}-\xi_i=(\ey{i}-\widehat\mu_{y,i})-\{\ey{i}-q_i-\ex{i}\Delta_i(\rho_n)\}=\ex{i}\Delta_i(\rho_n)-(\widehat\mu_{y,i}-q_i)=\ex{i}\Delta_i(\rho_n)-\Delta_{y,i}.
\end{align*}
Thus \eqref{eq:spacrt-logistic-alt-variance-envelope},
\eqref{eq:spacrt-logistic-alt-fit-error-sums}, and
\eqref{eq:spacrt-logistic-alt-basic-sums} imply
\begin{align}
  \sum_i(\widehat e_{y,i}-\xi_i)^2
  &\leq
  C\pi_{n,y}^2\rho_n^2\sum_i\ex{i}
  +2\sum_i\Delta_{y,i}^2
  \notag\\
  &=O_{\P_n}\left(
    n\pi_{n,x}\pi_{n,y}^2\rho_n^2
    +\pi_{n,y}s_{n,y}\log p_n
    +n\pi_{n,x}^2\pi_{n,y}^2\rho_n^2
  \right)
  =o_{\P_n}(n\pi_{n,y}),
  \label{eq:spacrt-logistic-alt-nuisance-square}
\end{align}
where the final equality follows from
\eqref{eq:spacrt-logistic-local-scale},
\eqref{eq:spacrt-logistic-lasso-rates}, and
$\pi_{n,x},\pi_{n,y}\leq1$.  Conditionally on $\mathcal X_n$, the
$\xi_i$ are independent and centered, $|\xi_i|\leq1$, and
\eqref{eq:spacrt-logistic-alt-variance-envelope} gives
\[
  \E\left(\sum_i\xi_i^2\mid\mathcal X_n\right)
  \asymp n\pi_{n,y},
  \qquad
  \V\left(\sum_i\xi_i^2\mid\mathcal X_n\right)
  \leq\sum_i\E(\xi_i^4\mid\mathcal X_n)
  \leq\sum_i\V(\xi_i\mid\mathcal X_n)
  =O(n\pi_{n,y}).
\]
Since $n\pi_{n,y}\geq n\pi_{n,x}\pi_{n,y}\to\infty$
by~\eqref{eq:spacrt-logistic-lasso-rates}, conditional Chebyshev gives
\begin{equation*}
  \sum_i\xi_i^2\asymp_{\P_n}n\pi_{n,y}.
\end{equation*}
The Euclidean triangle inequality and
\eqref{eq:spacrt-logistic-alt-nuisance-square} therefore imply
$\sum_i\widehat e_{y,i}^2\asymp_{\P_n}n\pi_{n,y}$.  The uniform fitted-variance
conclusion of Lemma~\ref{lem:spacrt-logistic-alt-fit}(i) now yields
\[
  nA_{2n}=\sum_i\widehat v_{x,i}\widehat e_{y,i}^2
  \asymp_{\P_n}n\pi_{n,x}\pi_{n,y}.
\]

\paragraph{Proof of the $S_n$ bound.}
Put $\nu_n\equiv(n\pi_{n,x}\pi_{n,y})^{1/2}$ and $\mu_n\equiv n\pi_{n,x}\pi_{n,y}|\rho_n|$. Using $\widehat e_{x,i}=e^0_{x,i}-\Delta_{x,i}$ and
$\widehat e_{y,i}=\xi_i+\ex{i}\Delta_i(\rho_n)-\Delta_{y,i}$ gives
the three-block identity
\[
  S_n
  =\sum_i\widehat e_{x,i}\xi_i
   +\sum_i\widehat e_{x,i}\ex{i}\Delta_i(\rho_n)
   -\sum_i\widehat e_{x,i}\Delta_{y,i}.
\]
Conditionally on $\mathcal X_n$, the first sum is centered and
\begin{align*}
  \V\left(\sum_i\widehat e_{x,i}\xi_i\ \middle|\ \mathcal X_n\right)\leq C\pi_{n,y}\sum_i\widehat e_{x,i}^2\leq
  2C\pi_{n,y}\left\{
    \sum_i(e^0_{x,i})^2+\sum_i\Delta_{x,i}^2
  \right\}
  =O_{\P_n}(\nu_n^2).
\end{align*}
Here the last equality uses
\eqref{eq:spacrt-logistic-alt-fit-error-sums} and
\eqref{eq:spacrt-logistic-alt-basic-sums}.
Conditional Chebyshev therefore gives $\sum_i\widehat e_{x,i}\xi_i=O_{\P_n}(\nu_n)$. Since $|\widehat e_{x,i}|\leq1$, the second sum in the three-block
identity satisfies
\[
  \left|\sum_i\widehat e_{x,i}\ex{i}\Delta_i(\rho_n)\right|
  \leq C\pi_{n,y}|\rho_n|\sum_i\ex{i}
  =O_{\P_n}(\mu_n).
\]
The last equality follows from
\eqref{eq:spacrt-logistic-alt-basic-sums}.
For the remaining sum $\sum_i\widehat e_{x,i}\Delta_{y,i}$, we consider the following identity
\begin{align*}
  -\sum_i\widehat e_{x,i}\Delta_{y,i}
  ={}&-\sum_i(\ex{i}-\widetilde p_i)\Delta_{y,i}
      -\sum_i(\widetilde p_i-p_i)\Delta_{y,i}
      +\sum_i\Delta_{x,i}\Delta_{y,i}
\end{align*}
This identity isolates the dependence created by the alternative. First,
Cauchy--Schwarz and the fit-error bounds
in~\eqref{eq:spacrt-logistic-alt-fit-error-sums} give
\begin{align*}
  \left|\sum_i\Delta_{x,i}\Delta_{y,i}\right|
  &\leq
  \left(\sum_i\Delta_{x,i}^2\right)^{1/2}
  \left(\sum_i\Delta_{y,i}^2\right)^{1/2}\\
  &=O_{\P_n}\left[
    \{\pi_{n,x}s_{n,x}\log p_n\}^{1/2}
    \left\{
      \{\pi_{n,y}s_{n,y}\log p_n\}^{1/2}
      +\{n\pi_{n,x}^2\pi_{n,y}^2\rho_n^2\}^{1/2}
    \right\}
  \right]\\
  &=O_{\P_n}\left\{
    \{\pi_{n,x}\pi_{n,y}s_{n,x}s_{n,y}\log^2p_n\}^{1/2}
    +\{n\pi_{n,x}^3\pi_{n,y}^2s_{n,x}\log p_n\,\rho_n^2\}^{1/2}
  \right\}\\
  &=O_{\P_n}\left\{
    \nu_n\left(\frac{s_{n,x}s_{n,y}\log^2p_n}{n}\right)^{1/2}
    +\mu_n\left(
      \frac{\pi_{n,x}s_{n,x}\log p_n}{n}
    \right)^{1/2}
  \right\}.
\end{align*}
The second line uses $\sqrt{a+b}\leq\sqrt a+\sqrt b$, and the last
line uses $\nu_n^2=n\pi_{n,x}\pi_{n,y}$ and
$\mu_n=n\pi_{n,x}\pi_{n,y}|\rho_n|$.  Both deterministic multipliers in
the last line vanish.  Indeed, the first does so directly by
\eqref{eq:spacrt-logistic-lasso-rates}, while, since $s_{n,x}\geq1$
and $0<\pi_{n,x}\leq1$,
\[
  \frac{\pi_{n,x}s_{n,x}\log p_n}{n}
  =
  \frac{s_{n,x}^2\log p_n}{n\pi_{n,x}}
  \frac{\pi_{n,x}^2}{s_{n,x}}
  \leq
  \frac{s_{n,x}^2\log p_n}{n\pi_{n,x}}
  \longrightarrow0.
\]
Consequently, $\left|\sum_i\Delta_{x,i}\Delta_{y,i}\right|=o_{\P_n}(\nu_n+\mu_n)$. Conditionally on $\mathcal Y_n$, the predictors are independent $\Ber(\widetilde p_i)$ variables by the product structure of their conditional law, and $\Delta_{y,i}$ is measurable. Put $T_n\equiv\sum_i(\ex{i}-\widetilde p_i)\Delta_{y,i}$. Then $\E(T_n\mid\mathcal Y_n)=0$, and conditional independence together with Lemma~\ref{lem:spacrt-logistic-alt-perturbation}(c) gives
\begin{align*}
  \V(T_n\mid\mathcal Y_n)=\sum_i\widetilde p_i(1-\widetilde p_i)\Delta_{y,i}^2\leq C\pi_{n,x}\sum_i\Delta_{y,i}^2.
\end{align*}
Thus, for every $M>0$, conditional Chebyshev's inequality yields
\[
  \P\left\{
    |T_n|>M\left(\pi_{n,x}\sum_i\Delta_{y,i}^2\right)^{1/2}
    \ \middle|\ \mathcal Y_n
  \right\}
  \leq\frac{C}{M^2}.
\]
After averaging over $\mathcal Y_n$ and substituting
\eqref{eq:spacrt-logistic-alt-fit-error-sums}, this implies
\begin{align*}
  T_n
  &=O_{\P_n}\left[
    \left\{
      \pi_{n,x}\pi_{n,y}s_{n,y}\log p_n
      +n\pi_{n,x}^3\pi_{n,y}^2\rho_n^2
    \right\}^{1/2}
  \right]\\
  &=O_{\P_n}\left\{
    \{\pi_{n,x}\pi_{n,y}s_{n,y}\log p_n\}^{1/2}
    +\{n\pi_{n,x}^3\pi_{n,y}^2\rho_n^2\}^{1/2}
  \right\}\\
  &=O_{\P_n}\left\{
    \nu_n\left(\frac{s_{n,y}\log p_n}{n}\right)^{1/2}
    +\mu_n\left(\frac{\pi_{n,x}}{n}\right)^{1/2}
  \right\}.
\end{align*}
The two multipliers vanish; the first uses
\eqref{eq:spacrt-logistic-lasso-rates}, while the second is elementary:
\[
  \frac{s_{n,y}\log p_n}{n}
  =
  \frac{s_{n,y}^2\log p_n}{n\pi_{n,y}}
  \frac{\pi_{n,y}}{s_{n,y}}
  \leq
  \frac{s_{n,y}^2\log p_n}{n\pi_{n,y}}
  \longrightarrow0,
  \qquad
  \frac{\pi_{n,x}}{n}\leq\frac1n\longrightarrow0.
\]
Substituting the two convergences just displayed into the preceding
$O_{\P_n}$ bound gives $T_n=o_{\P_n}(\nu_n+\mu_n)$.
Finally, the posterior mean-shift bound in
Lemma~\ref{lem:spacrt-logistic-alt-perturbation}(c),
Cauchy--Schwarz, and the bounds in
\eqref{eq:spacrt-logistic-alt-fit-error-sums} and
\eqref{eq:spacrt-logistic-alt-basic-sums} give
\begin{align*}
  \left|\sum_i(\widetilde p_i-p_i)\Delta_{y,i}\right|
  &\leq
  C\pi_{n,x}|\rho_n|
  \left\{
    \left(\sum_i\ey{i}\right)^{1/2}
    +\pi_{n,y}\sqrt n
  \right\}
  \left(\sum_i\Delta_{y,i}^2\right)^{1/2}\\
  &=O_{\P_n}\left\{
    \mu_n\left(\frac{s_{n,y}\log p_n}{n}\right)^{1/2}
    +\mu_n\pi_{n,x}\pi_{n,y}^{1/2}|\rho_n|
  \right\}
  =o_{\P_n}(\nu_n+\mu_n),
\end{align*}
where $\pi_{n,y}\sqrt n\leq(n\pi_{n,y})^{1/2}$ was used in the first line,
and the final $o_{\P_n}$ step uses
\eqref{eq:spacrt-logistic-lasso-rates}
and~\eqref{eq:spacrt-logistic-local-scale}.  Combining the three blocks proves
$|S_n|=O_{\P_n}(\nu_n+\mu_n)$, which is
\eqref{eq:spacrt-logistic-alt-coordinates}.
\end{proof}

\clearpage
\endgroup

\end{document}